# Ancient human genomes suggest three ancestral populations for present-day Europeans

Iosif Lazaridis[1,2], Nick Patterson[2], Alissa Mittnik[3], Gabriel Renaud[4], Swapan Mallick[1,2], Karola Kirsanow[5], Peter H. Sudmant[6], Joshua G. Schraiber[7], Sergi Castellano[4], Mark Lipson[8], Bonnie Berger[2,8], Christos Economou[9], Ruth Bollongino[5], Qiaomei Fu[1,4,10], Kirsten I. Bos[3], Susanne Nordenfelt[1,2], Heng Li[1,2], Cesare de Filippo[4], Kay Prüfer[4], Susanna Sawyer[4], Cosimo Posth[3], Wolfgang Haak[11], Fredrik Hallgren[12], Elin Fornander[12], Nadin Rohland[1,2], Dominique Delsate[13,14], Michael Francken[15], Jean-Michel Guinet[13], Joachim Wahl[16], George Ayodo[17], Hamza A. Babiker[18,19], Graciela Bailliet[20], Elena Balanovska[21], Oleg Balanovsky[21,22], Ramiro Barrantes[23], Gabriel Bedoya[24], Haim Ben-Ami[25], Judit Bene[26], Fouad Berrada[27], Claudio M. Bravi[20], Francesca Brisighelli[28], George Busby[29,30], Francesco Cali[31], Mikhail Churnosov[32], David E. C. Cole[33], Daniel Corach[34], Larissa Damba[35], George van Driem[36], Stanislav Dryomov[37], Jean-Michel Dugoujon[38], Sardana A. Fedorova[39], Irene Gallego Romero[40], Marina Gubina[35], Michael Hammer[41], Brenna Henn[42], Tor Hervig[43], Ugur Hodoglugil[44], Aashish R. Jha[40], Sena Karachanak-Yankova[45], Rita Khusainova[46,47], Elza Khusnutdinova[46,47], Rick Kittles[48], Toomas Kivisild[49], William Klitz[50], Vaidutis Kučinskas[51], Alena Kushniarevich[52], Leila Laredj[53], Sergey Litvinov[46,47,52], Theologos Loukidis[54], Robert W. Mahley[55], Béla Melegh[26], Ene Metspalu[56], Julio Molina[57], Joanna Mountain[58], Klemetti Näkkäläjärvi[59], Desislava Nesheva[45,] Thomas Nyambo[60], Ludmila Osipova[35], Jüri Parik[56], Fedor Platonov[61], Olga Posukh[35], Valentino Romano[62], Francisco Rothhammer[63,64,65], Igor Rudan[66], Ruslan Ruizbakiev[67], Hovhannes Sahakyan[52,68], Antti Sajantila[69,70], Antonio Salas[71], Elena B. Starikovskaya[37], Ayele Tarekegn[72], Draga Toncheva[45], Shahlo Turdikulova[73], Ingrida Uktveryte[51], Olga Utevska[74], René Vasquez[75], Mercedes Villena[75], Mikhail Voevoda[35,76] , Cheryl Winkler[77], Levon Yepiskoposyan[68], Pierre Zalloua[78,79], Tatijana Zemunik[80], Alan Cooper[11], Cristian Capelli[29], Mark G. Thomas[81], Andres Ruiz-Linares[81], Sarah A. Tishkoff[82], Lalji Singh[83,84], Kumarasamy Thangaraj[83], Richard Villems[52,56,85], David Comas[86], Rem Sukernik[37], Mait Metspalu[52], Matthias Meyer[4], Evan E. Eichler[6,87] , Joachim Burger[5], Montgomery Slatkin[7], Svante Pääbo[4], Janet Kelso[4], David Reich[1,2,88,†] and Johannes Krause[3,89,†]

† Co-senior authors

[1] Department of Genetics, Harvard Medical School, Boston, MA, 02115, USA
[2] Broad Institute of Harvard and MIT, Cambridge, MA, 02142, USA
[3] Institute for Archaeological Sciences, University of Tübingen, Tübingen, 72074, Germany

[4] Max Planck Institute for Evolutionary Anthropology, Leipzig, 04103, Germany
[5] Johannes Gutenberg University Mainz, Institute of Anthropology, Mainz, D-55128, Germany
[6] Department of Genome Sciences, University of Washington, Seattle, WA, 98195, USA
[7] Department of Integrative Biology, University of California, Berkeley, CA, 94720-3140, USA
[8] Department of Mathematics and Computer Science and Artificial Intelligence Laboratory, Massachusetts Institute of Technology, Cambridge, MA 02139, USA
[9] Archaeological Research Laboratory, Stockholm University, 114 18, Sweden
[10] Key Laboratory of Vertebrate Evolution and Human Origins of Chinese Academy of Sciences, IVPP, CAS, Beijing, 100049, China
[11] Australian Centre for Ancient DNA, School of Earth and Environmental Sciences, University of Adelaide, Adelaide, South Australia, SA 5005, Australia
[12] The Cultural Heritage Foundation, Västerås, 722 12, Sweden
[13] National Museum of Natural History, L-2160, Luxembourg
[14] National Center of Archaeological Research, National Museum of History and Art, L-2345, Luxembourg
[15] Department of Paleoanthropology, Senckenberg Center for Human Evolution and Paleoenvironment, University of Tübingen, Tübingen, D-72070, Germany
[16] State Office for Cultural Heritage Management Baden-Württemberg, Osteology, Konstanz, D-78467, Germany
[17] Center for Global Health and Child Development, Kisumu, 40100, Kenya
[18] Institutes of Evolution, Immunology and Infection Research, School of Biological Sciences, University of Edinburgh, Edinburgh, EH9 3JT, UK
[19] Biochemistry Department, Faculty of Medicine, Sultan Qaboos University, Alkhod, Muscat, 123, Oman
[20] Laboratorio de Genética Molecular Poblacional, Instituto Multidisciplinario de Biología Celular (IMBICE), CCT-CONICET & CICPBA, La Plata, B1906APO, Argentina
[21] Research Centre for Medical Genetics, Moscow, 115478, Russia
[22] Vavilov Institute for General Genetics, Moscow, 119991, Russia
[23] Escuela de Biología, Universidad de Costa Rica, San José, 2060, Costa Rica
[24] Institute of Biology, Research group GENMOL, Universidad de Antioquia, Medellín, Colombia
[25] Rambam Health Care Campus, Haifa, 31096, Israel
[26] Department of Medical Genetics and Szentagothai Research Center, University of Pécs, Pécs, H-7624 Hungary
[27] Al Akhawayn University in Ifrane (AUI), School of Science and Engineering, Ifrane, 53000, Morocco
[28] Forensic Genetics Laboratory, Institute of Legal Medicine, Università Cattolica del Sacro Cuore, Rome, 00168, Italy
[29] Department of Zoology, University of Oxford, Oxford, OX1 3PS, UK
[30] Wellcome Trust Centre for Human Genetics, University of Oxford, Oxford, OX3 7BN, UK
[31] Laboratorio di Genetica Molecolare, IRCCS Associazione Oasi Maria SS, Troina, 94018, Italy
[32] Belgorod State University, Belgorod, 308015, Russia
[33] Department of Laboratory Medicine and Pathobiology, University of Toronto, Toronto, Ontario, M5G 1L5, Canada
[34] Servicio de Huellas Digitales Genéticas, School of Pharmacy and Biochemistry, Universidad de Buenos Aires, 1113 CABA, Argentina
[35] Institute of Cytology and Genetics, Siberian Branch of Russian Academy of Sciences, Novosibirsk, 630090, Russia
[36] Institute of Linguistics, University of Bern, Bern, CH-3012, Switzerland
[37] Laboratory of Human Molecular Genetics, Institute of Molecular and Cellular Biology, Russian Academy of Science, Siberian Branch, Novosibirsk, 630090, Russia
[38] Anthropologie Moléculaire et Imagerie de Synthèse, CNRS UMR 5288, Université Paul Sabatier Toulouse III, Toulouse, 31000, France
[39] Yakut Research Center of Complex Medical Problems and North-Eastern Federal University, Yakutsk, 677010, Russia
[40] Department of Human Genetics, University of Chicago, Chicago, IL, 60637, USA
[41] ARL Division of Biotechnology, University of Arizona, Tucson, AZ, 85721, USA
[42] Department of Ecology and Evolution, Stony Brook University, Stony Brook, NY, 11794, USA
[43] Department of Clinical Science, University of Bergen, Bergen, 5021, Norway
[44] NextBio, part of Illumina, Santa Clara, CA, USA 95050
[45] Dept. of Medical Genetics, National Human Genome Center, Medical University Sofia, Sofia, 1431, Bulgaria

[46] Institute of Biochemistry and Genetics, Ufa Research Centre, Russian Academy of Sciences, Ufa, 450054, Russia
[47] Department of Genetics and Fundamental Medicine, Bashkir State University, Ufa, 450074, Russia
[48] College of Medicine, University of Illinois at Chicago, Chicago, IL, 60607, USA
[49] Division of Biological Anthropology, University of Cambridge, Cambridge, United Kingdom CB2 1QH
[50] School of Public Health, University of California, Berkeley, CA, 94720, USA
[51] Department of Human and Medical Genetics, Vilnius University, Vilnius, LT-08661, Lithuania
[52] Estonian Biocentre, Evolutionary Biology group, Tartu, 51010, Estonia
[53] Translational Medicine and Neurogenetics, Institut de Génétique et de Biologie Moléculaire et Cellulaire, Illkirch, 67404, France
[54] Currently employed by AMGEN; 33 Kazantzaki Str, Ilioupolis 16342, Athens, Greece
[55] Gladstone Institutes, San Francisco, CA, 94158, USA
[56] Department of Evolutionary Biology, University of Tartu, Tartu, 51010, Estonia
[57] Centro de Investigaciones Biomédicas de Guatemala, Ciudad de Guatemala, Guatemala
[58] Research Department, 23andMe, Inc. Mountain View, CA, 94043, USA
[59] Cultural Anthropology Program, University of Oulu, Oulu, 90014, Finland
[60] Department of Biochemistry, Muhimbili University of Health and Allied Sciences, Dar es Salaam, Tanzania
[61] Research Institute of Health, North-Eastern Federal University, Yakutsk, 677000, Russia
[62] Dipartimento di Fisica e Chimica, Università di Palermo, Palermo, 90128, Italy
[63] Instituto de Alta Investigación, Universidad de Tarapacá, Arica, Chile
[64] Programa de Genética Humana ICBM Facultad de Medicina Universidad de Chile, Santiago, Chile
[65] Centro de Investigaciones del Hombre en el Desierto, Arica, Chile
[66] Centre for Population Health Sciences, The University of Edinburgh Medical School, Edinburgh, Scotland, EH8 9AG, UK
[67] Deceased: formerly of the Institute of Immunology, Academy of Science, Tashkent, 70000, Uzbekistan
[68] Laboratory of Ethnogenomics, Institute of Molecular Biology, National Academy of Sciences of Armenia, Yerevan, 0014, Armenia
[69] Department of Forensic Medicine, Hjelt Institute, University of Helsinki, Helsinki, 00014, Finland
[70] Institute of Applied Genetics, Department of Molecular and Medical Genetics, University of North Texas Health Science Center, Fort Worth, Texas 76107, USA
[71] Unidade de Xenética, Departamento de Anatomía Patolóxica e Ciencias Forenses, and Instituto de Ciencias Forenses, Grupo de Medicina Xenómica (GMX), Facultade de Medicina, Universidade de Santiago de Compostela, Galcia, 15872, Spain
[72] Research Fellow, Henry Stewart Group, Russell House, London WC1A 2HN, UK
[73] Institute of Bioorganic Chemistry Academy of Sciences Republic of Uzbekistan, Tashkent, 100125, Uzbekistan
[74] Department of Genetics and Cytology, V.N. Karazin Kharkiv National University, Kharkiv, 61077, Ukraine
[75] Instituto Boliviano de Biología de la Altura, Universidad Autonoma Tomás Frías, Potosí, Bolivia
[76] Inst. of Internal Medicine, Siberian Branch of Russian Acad. of Medical Sciences, Novosibirsk, 630089, Russia
[77] Molecular Genetics Epidemiology Section, Frederick Natoinal Lab for Cancer Research, NCI, NIH, Frederick, MD 21702, USA
[78] Lebanese American University, School of Medicine, Beirut, 13-5053, Lebanon
[79] Harvard School of Public Health, Boston, 02115, USA
[80] Department of Medical Biology, University of Split, School of Medicine, Split, 21000, Croatia
[81] Department of Genetics, Evolution and Environment, University College London, WC1E 6BT, UK
[82] Department of Biology and Genetics. University of Pennsylvania, Philadelphia, Pennsylvania, 19104, USA
[83] CSIR-Centre for Cellular and Molecular Biology, Hyderabad, 500 007, India
[84] Present address: Banaras Hindu University, Varanasi, 221 005, India
[85] Estonian Academy of Sciences, Tallinn, 10130, Estonia
[86] Institut de Biologia Evolutiva (CSIC-UPF), Departament de Ciències Experimentals i de la Salut, Universitat Pompeu Fabra, Barcelona, 08003, Spain
[87] Howard Hughes Medical Institute, University of Washington, Seattle, WA, USA 98195
[88] Howard Hughes Medical Institute, Harvard Medical School, Boston, MA, 02115, USA
[89] Senckenberg Centre for Human Evolution and Palaeoenvironment, University of Tübingen, 72070, Germany

**We sequenced genomes from a ~7,000 year old early farmer from Stuttgart in Germany, an ~8,000 year old hunter-gatherer from Luxembourg, and seven ~8,000 year old hunter-gatherers from southern Sweden. We analyzed these data together with other ancient genomes and 2,345 contemporary humans to show that the great majority of present-day Europeans derive from at least three highly differentiated populations: West European Hunter-Gatherers (WHG), who contributed ancestry to all Europeans but not to Near Easterners; Ancient North Eurasians (ANE), who were most closely related to Upper Paleolithic Siberians and contributed to both Europeans and Near Easterners; and Early European Farmers (EEF), who were mainly of Near Eastern origin but also harbored WHG-related ancestry. We model these populations' deep relationships and show that EEF had ~44% ancestry from a "Basal Eurasian" lineage that split prior to the diversification of all other non-African lineages.**

Ancient DNA studies have demonstrated that migration played a major role in the introduction of agriculture to Europe, as early farmers were genetically distinct from hunter-gatherers[1,2] and closer to present-day Near Easterners[2,3]. Modelling the ancestry of present-day Europeans as a simple mixture of two ancestral populations[2], however, does not take into account their genetic affinity to an Ancient North Eurasian (ANE) population[4,5] who also contributed genetically to Native Americans[6]. To better understand the deep ancestry of present-day Europeans, we sequenced nine ancient genomes that span the transition from hunting and gathering to agriculture in Europe (Fig. 1A; Extended Data Fig. 1): "Stuttgart" (19-fold coverage), a ~7,000 year old skeleton found in Germany in the context of artifacts from the first widespread Neolithic farming culture of central Europe, the *Linearbandkeramik*; "Loschbour" (22-fold coverage), an ~8,000 year old skeleton from the Loschbour rock shelter in Heffingen, Luxembourg, discovered in the context of Mesolithic hunter-gatherer artifacts (SI1; SI2); and seven samples (0.01-2.4-fold coverage) from an ~8,000 year old Mesolithic hunter-gatherer burial in Motala, Sweden.

A central challenge is to show that DNA sequences retrieved from ancient samples are authentic and not due to present-day human contamination. The rate of C→T and G→A mismatches to the human genome at the ends of the molecules in libraries from each of the ancient samples exceeds 20%, a signature that suggests the DNA is largely ancient[7,8] (SI3). We inferred mitochondrial

DNA (mtDNA) consensus sequences, and based on the number of sites that differed, estimated contamination rates of 0.3% for Loschbour, 0.4% for Stuttgart, and 0.01%-5% for the Motala individuals (SI3). We inferred similar levels of contamination for the nuclear DNA of Loschbour (0.4%) and Stuttgart (0.3%) using a maximum-likelihood-based test (SI3). The effective contamination rate for the high coverage samples is likely to be far lower, as consensus diploid genotype calling (SI2) tends to reduce the effects of a small fraction of contaminating reads.

Stuttgart belongs to mtDNA haplogroup T2, typical of Neolithic Europeans[9], while Loschbour and all Motala individuals belong to haplogroups U5 and U2, typical of pre-agricultural Europeans[1,7] (SI4). Based on the ratio of reads aligning to chromosomes X and Y, Stuttgart is female, while Loschbour and five of seven Motala individuals are male[10] (SI5). Loschbour and the four Motala males whose haplogroups we could determine all belong to Y-chromosome haplogroup I, suggesting that this was a predominant haplogroup in pre-agricultural northern Europeans analogous to mtDNA haplogroup U[11] (SI5).

We carried out most of our sequencing on libraries prepared in the presence of uracil DNA glycosylase (UDG), which reduces C→T and G→A errors due to ancient DNA damage (SI3). We first confirm that the ancient samples had statistically indistinguishable levels of Neandertal ancestry to each other (~2%) and to present-day Eurasians (SI6), and so we do not consider this further in our analyses of population relationships. We report analyses that leverage the type of information that can only be obtained from deep coverage genomes, mostly focusing on Loschbour and Stuttgart, and for some analyses also including Motala12 (2.4×) and La Braña from Mesolithic Iberia (3.4×)[12]. Heterozygosity, the number of differences per nucleotide between an individual's two chromosomes, is 0.00074 for Stuttgart, at the high end of present-day Europeans, and 0.00048 for Loschbour, lower than in any present-day humans (SI2). Through comparison of Loschbour's two chromosomes we find that this low diversity is not due to recent inbreeding but instead due to a population bottleneck in this individual's more distant ancestors (Extended Data Fig. 2). Regarding alleles that affect phenotype, we find that the *AMY1* gene coding for salivary amylase had 5, 6, 13, and 16 copies in La Braña[12], Motala12, Loschbour and Stuttgart respectively; these numbers are within the range of present-day Europeans (SI7), suggesting that high copy counts of *AMY1* are not entirely due to selection since the switch to agriculture[13]. The genotypes at SNPs associated with lactase persistence indicate that Stuttgart, Loschbour, and Motala12 were unable to digest milk as adults. Both Loschbour and Stuttgart likely had dark hair (>99% probability); Loschbour, like La Braña and Motala12, likely had blue or intermediate-colored eyes (>75% probability), while Stuttgart most likely had brown eyes (>99% probability) (SI8). Neither Loschbour nor La Braña carries the skin-lightening allele in *SLC24A5* that is homozygous in Stuttgart and nearly fixed in Europeans today, indicating that they probably had darker skin[12]. However, Motala12 carries at least one copy of the derived allele, indicating that this locus was already polymorphic in Europeans prior to the advent of agriculture.

To place the ancient European genomes in the context of present-day human genetic variation, we assembled a dataset of 2,345 present-day humans from 203 populations genotyped at 594,924 autosomal single nucleotide polymorphisms (SNPs)[5] (SI9) (Extended Data Table 1). We used ADMIXTURE[14] to identify 59 "West Eurasian" populations (777 individuals) that cluster with Europe and the Near East (SI9 and Extended Data Fig. 3). Principal component analysis (PCA)[15] (SI10) (Fig. 1B) reveals a discontinuity between the Near East and Europe, with each showing north-south clines bridged only by a few populations of mainly Mediterranean origin. Our PCA differs from previous studies that showed a correlation with the map of Europe[16,17], which we determined is due to our study having relatively fewer central and northwestern Europeans, and more Near Easterners and eastern Europeans (SI10). We projected[18] the newly sequenced and previously published[2,6,12,19] ancient genomes onto the first two PCs inferred from present-day samples (Fig. 1B). MA1 and AG2, both Upper Paleolithic hunter-gatherers from Lake Baikal[6] in Siberia, project at the northern end of the PCA, suggesting an "Ancient North Eurasian" meta-population (ANE). European hunter-gatherers from Spain, Luxembourg, and Sweden fall outside the genetic variation of West Eurasians in the direction of European differentiation from the Near East, with a "West European Hunter-Gatherer" (WHG) cluster including Loschbour and La Braña[12], and a "Scandinavian Hunter-Gatherer" (SHG) cluster including the Motala individuals and ~5,000 year old hunter-gatherers from the Swedish Pitted Ware Culture[2]. An "Early European Farmer" (EEF) cluster includes Stuttgart, the ~5,300 year old Tyrolean Iceman[19] and a ~5,000 year old southern Swedish farmer[2], and is near present-day Sardinians[2,19].

PCA gradients of genetic variation may arise under very different histories[20]. To test if they reflect population mixture events or are entirely due to genetic drift within West Eurasia, we computed an $f_4$-statistic[18] that tests whether the ancient MA1 from Siberia shares more alleles with a *Test* West Eurasian population or with Stuttgart. We find that $f_4(Test, Stuttgart; MA1, Chimp)$ is positive for many West Eurasians, which must be due to variable degrees of admixture with ancient populations related to MA1 (Extended Data Fig. 4). We also find that $f_4(Test, Stuttgart; Loschbour, Chimp)$ is nearly always positive in Europeans and always negative in Near Easterners, indicating that Europeans have more ancestry from ancient populations related to Loschbour than do Near Easterners (Extended Data Fig. 4). To investigate systematically the history of population mixture in West Eurasia, we computed all possible statistics of the form $f_3(X; Ref_1, Ref_2)$ (SI11). An $f_3$-statistic is expected to be positive if no admixture has taken place, but if *X* is admixed between populations related to $Ref_1$ and $Ref_2$, it can be negative[5]. We tested all possible pairs of $Ref_1$, $Ref_2$ chosen from the list of 192 present-day populations with at least four individuals, and five ancient genomes (SI11). The lowest $f_3$-statistics for Europeans are usually negative (93% are >4 standard errors below zero using a standard error from a block jackknife[5,21]). The most negative statistic (Table 1) always involves at least one ancient individual as a reference, and for Europeans it is nearly always significantly lower than the most negative statistic obtained using only present-day populations as references (SI11). MA1 is a better surrogate (Extended Data Fig. 5) for Ancient North Eurasian ancestry than the Native American Karitiana who were first used to represent this component of ancestry in Europe[4,5]. Motala12 never appears as one of the references, suggesting that SHG may not be a source for Europeans. Instead, present-day European populations usually have their lowest $f_3$ with either the (EEF, ANE) or (WHG, Near East) pair (SI11, Extended Data Table 1). For Near Easterners, the lowest $f_3$-statistic always takes as references Stuttgart and a population from Africa, the Americas, South Asia, or MA1 (Table 1), reflecting the fact that both Stuttgart and present-day Near Easterners harbor ancestry from ancient Near Easterners. Extended Data Fig. 6 plots statistics of the form $f_4(West\ Eurasian\ X, Chimp; Ancient_1, Ancient_2)$ onto a map, showing strong gradients in the relatedness to Stuttgart (EEF), Loschbour (WHG) and MA1 (ANE).

We determined formally that a minimum of three source ancestral populations are needed to explain the data for many European populations taken together by studying the correlation patterns of all possible statistics of the form $f_4(Test_{base}, Test_i; O_{base}, O_j)$ (SI12). Here $Test_{base}$ is a reference European population and $Test_i$ the set of all other European *Test* populations; $O_{base}$ is a reference outgroup population, and $O_i$ the set of other outgroups (ancient DNA samples, Onge, Karitiana, and Mbuti). The rank of the *(i, j)* matrix reflects the minimum number of source populations that contributed to the *Test* populations[22,23]. For a pool of 23 *Test* populations comprising most present-day Europeans, this analysis rejects descent from just two sources ($P<10^{-12}$ by a Hotelling T-test[23]). However, three source populations are consistent with the data after excluding the Spanish who have evidence for African admixture[24-26] (P=0.019, not significant after multiple-hypothesis correction). Our finding of at least three source populations is also qualitatively consistent with the results from ADMIXTURE (SI9), PCA (Fig. 1B, SI10) and *f*-statistics (Extended Data Table 1, Extended Data Fig. 6, SI11, SI12). We caution that the finding of three sources could be consistent with a larger number of mixture events, as the method cannot distinguish between one or more mixture events if they are from the same set of sources. Our analysis also does not assume that the inferred source populations were themselves unadmixed; indeed, the positive $f_4$*(Stuttgart, X; Loschbour, Chimp)* statistics obtained when *X* is a Near Eastern population (Extended Data Table 1) implies that EEF had some WHG-related ancestry, which we show in SI13 was at least 0% and less than 45%.

Motivated by the evidence of at least three source populations for present-day Europeans, we set out to develop a model consistent with our data. To constrain our search space for modeling, we first studied $f_4$-statistics comparing the ancient individuals from Europe and Siberia and diverse eastern non-African groups (Oceanians, East Asians, Siberians, Native Americans, and Onge from the Andaman Islands[27]) (SI14). We find that: (1) Loschbour (WHG) and Stuttgart (EEF) share more alleles with each other than either does with MA1 (ANE), as might be expected by geography, but MA1 shares more alleles with Loschbour than with Stuttgart, indicating a link between Eurasian hunter-gatherers to the exclusion of European farmers; (2) Eastern non-Africans share more alleles with Eurasian hunter-gatherers (MA1, Loschbour, La Braña, and Motala12) than with Stuttgart; (3) Every eastern non-African population except for Native Americans and Siberians is equally closely related to diverse Eurasian hunter-gatherers, but Native Americans and Siberians share more alleles with MA1 than with European hunter-gatherers; and (4) Eurasian hunter-gatherers and Stuttgart both share more alleles with Native

Americans than with other eastern non-Africans. We use the ADMIXTUREGRAPH[18] software to search for a model of population relationships (a tree structure augmented by admixture events) that is consistent with these observations. We explored models with 0, 1, or 2 admixture events in the ancestry of the three ancient source populations and eastern non-Africans, and identified a single model with two admixture events that fit the data. The successful model (Fig. 2A) includes the previously reported gene flow into Native Americans from an MA1-like population[6], as well as the novel inference that Stuttgart is partially (44 ± 10%) derived from a "Basal Eurasian" lineage that split prior to the separation of eastern non-Africans from the common ancestor of WHG and ANE. If this model is accurate, the ANE/WHG split must have occurred >24,000 years ago since this is the age[6] of MA1 and this individual is on the ANE lineage. The WHG must then have split from eastern non-Africans >40,000 years ago, as this is the age of the Chinese Tianyuan sample which clusters with eastern non-Africans to the exclusion of Europeans[28]. The Basal Eurasian split would then have to be even older. A Basal Eurasian lineage in the Near East is plausible given the presence of anatomically modern humans in the Levant[29] ~100 thousand years ago and African-related tools likely made by modern humans in Arabia[30,31]. Alternatively, evidence for gene flow between the Near East and Africa[32], and African morphology in pre-farming Natufians[33] from Israel, may also be consistent with the population representing a later movement of humans out of Africa and into the Near East.

We tested the robustness of the ADMIXTUREGRAPH model in various ways. First, we verified that Stuttgart and the Iceman (EEF), and Loschbour and LaBraña (WHG) can be formally fit as clades (SI14). We also used the unsupervised MixMapper[4] (SI15) and TreeMix[34] software (SI16) to fit graph models; both found all the same admixture events. The statistics supporting our key inferences about history also provide consistent results when restricted to transversions polymorphisms not affected by ancient DNA damage, and when repeated with whole-genome sequencing data that is not affected by SNP ascertainment bias[35] (Extended Data Table 2).

We next fit present-day European populations into our working model. We found that few European populations could be fit as 2-way mixtures, but nearly all were compatible with being 3-way mixtures of ANE/EEF/WHG (SI14). Mixture proportions (Fig. 2B; Extended Data Table 3) inferred via our model are consistent with those from an independent method that relates

European populations to diverse outgroups using $f_4$-statistics while making much weaker modeling assumptions (only assuming that MA1 is an unmixed descendent of ANE, Loschbour of WHG, and Stuttgart of EEF; SI17). These analyses allow us to infer that EEF ancestry in Europe today ranges from ~30% in the Baltic region to ~90% in the Mediterranean, a gradient that is also consistent with patterns of identity-by-descent (IBD) sharing[36] (SI18) and chromosome painting[37] (SI19) in which Loschbour shares more segments with northern Europeans and Stuttgart with southern Europeans. Our estimates suggest that Southern Europeans inherited their European hunter-gatherer ancestry mostly via EEF ancestors (Extended Data Fig. 6), while Northern Europeans acquired up to 50% additional WHG ancestry. Europeans have a larger proportion of WHG than ANE ancestry (WHG/(WHG+ANE) = 0.6-0.8) with the ANE ancestry never being larger than ~20%. (By contrast, in the Near East there is no detectible WHG ancestry, but substantial ANE ancestry, up to ~29% in the North Caucasus) (SI14). While ANE ancestry was not as pervasive in Europe during the agricultural transition as it is today (we do not detect it in either Loschbour or Stuttgart), it was already present, since MA1 shares more alleles with Motala12 (SHG) than with Loschbour, and Motala12 fits as a mixture of 81% WHG and 19% ANE (SI14).

Two sets of European populations are poor fits. Sicilians, Maltese, and Ashkenazi Jews have EEF estimates beyond the 0-100% interval (SI17) and cannot be jointly fit with other Europeans (SI14). These populations may have more Near Eastern ancestry than can be explained via EEF admixture (SI14), consistent with their falling in the gap between European and Near Eastern populations in Fig. 1B. Finns, Mordovians and Russians from northeastern Europe also do not fit (SI14; Extended Data Table 3). To better understand this, we plotted $f_4(X, Bedouin2; Han, Mbuti)$ against $f_4(X, Bedouin2; MA1, Mbuti)$. These statistics measure the degree of a European population's allele sharing with Han Chinese or MA1 (Extended Data Fig. 7). Europeans fall on a line of slope >1 in the plot of these two statistics. However, northeastern Europeans including Chuvash and Saami (which we add in to the analysis) fall away from this line in the direction of East Asians. This is consistent with East Asian (most likely Siberian) gene flow into northeastern Europeans, some of which may be more recent[38] than the original ANE admixture (SI14).

Three questions seem particularly important to address in follow-up work. Where did the EEF obtain their WHG ancestry? Southeastern Europe is a candidate as it lies along the path from Anatolia into central Europe[39]. When and where the ancestors of present-day Europeans first acquire their ANE ancestry? Based on discontinuity in mtDNA haplogroup frequencies, this may have occurred ~5,500-4,000 years ago[40] in Central Europe. When and where did Basal Eurasians mix into the ancestors of the EEF? An important aim for future work should be to collect DNA from additional ancient samples to illuminate these transformations.

**Methods Summary**

We extracted DNA from nine sets of ancient human remains and converted the extracts into Illumina sequencing libraries in dedicated clean rooms. We assessed whether sequences for these libraries were consistent with genuine ancient DNA by searching for characteristic deaminations at the ends of molecules[7,8]. We also tested for contamination by searching for evidence of mixture of DNA from multiple individuals. For large-scale shotgun sequencing we used libraries that we made in the presence of the enzymes Uracil-DNA-glycosylase and endonuclease VIII, which reduce the rate of ancient DNA-induced errors. After removal of duplicated molecules, we called consensus genotypes for the high coverage samples using the Genome Analysis Toolkit[41]. We merged the data with published ancient genomes, as well as with 2,345 present-day humans from 203 populations genotyped at 594,924 autosomal single nucleotide polymorphisms. We visualized population structure using Principal Component Analysis[15] and ADMIXTURE[14]. To make inferences about population history, we used methods that can analyze allele frequency correlation statistics to detect population mixture[5]; that can estimate mixture proportions in the absence of accurate ancestral populations; that can infer the minimum number of source populations for a collection of tests population[23]; and that can assess formally the fit of genetic data to models of population history[5].

**Supplementary Information** is linked to the online version of the paper. The fully public version of the Human Origins dataset can be found at http://genetics.med.harvard.edu/reichlab/Reich_Lab/Datasets.html. The full version of the dataset (including additional samples) is available to researchers who send a signed letter to DR indicating that they will abide by specified usage conditions.

**Acknowledgments** We are grateful to Cynthia Beall, Neil Bradman, Amha Gebremedhin, Damian Labuda, Maria Nelis and Anna Di Rienzo for sharing DNA samples; to Detlef Weigel, Christa Lanz, Verena Schünemann, Peter Bauer and Olaf Riess for support and access to DNA sequencing facilities; to Philip Johnson for advice on contamination estimation; and to Pontus Skoglund for sharing the graphics software that we used to generate Extended Data Fig. 6. We thank Kenneth Nordtvedt for alerting us about the existence of newly discovered Y-chromosome SNPs. The collections and methods for the Population Reference Sample (POPRES) discussed in SI18 are described in ref.[42], and the dataset used for our analyses was obtained from dbGaP at http://www.ncbi.nlm.nih.gov/projects/gap/cgi-bin/study.cgi?study_id=phs000145.v4.p2 through dbGaP accession number phs000145.v1.p2. We thank all the volunteers who donated DNA; the staff of the Unità Operativa Complessa di Medicina Trasfusionale, Azienda Ospedaliera Umberto I, Siracusa, Italy for assistance in sample collection; and The National Laboratory for the Genetics of Israeli Populations for facilitating access to DNA. We thank colleagues at the Applied Genomics at the Children's Hospital of Philadelphia, especially Hakon Hakonarson, Cecilia Kim, Kelly Thomas, and Cuiping Hou, for genotyping samples on the Human Origins array. JK is grateful for support from DFG grant # KR 4015/1-1, the Carl-Zeiss Foundation and the Baden Württemberg Foundation. SP acknowledges support from the Presidential Innovation Fund of the Max Planck Society. JGS acknowledges use of the Extreme Science and Engineering Discovery Environment (XSEDE), which is supported by NSF grant number OCI-1053575. EB and OB were supported by RFBR grants 13-06-00670, 13-04-01711, 13-04-90420 and by the Molecular and Cell Biology Program of the Presidium, Russian Academy of Sciences. BM was supported by grants OTKA 73430 and 103983. ASaj was supported by a Finnish Professorpool (Paulo Foundation) Grant. The Lithuanian sampling was supported by the LITGEN project (VP1-3.1-ŠMM-07-K-01-013), funded by the European Social Fund under the Global Grant Measure. AS was supported by Spanish grants SAF2008-02971 and EM 2012/045. OU was supported by Ukrainian SFFS grant F53.4/071. SAT was supported by NIH Pioneer Award 8DP1ES022577-04 and NSF HOMINID award BCS-0827436. KT was supported by an Indian CSIR Network Project (GENESIS: BSC0121). LS was supported by an Indian CSIR Bhatnagar Fellowship. RV, MM, JP and EM were supported by the European Union Regional Development Fund through the Centre of Excellence in Genomics to the Estonian Biocentre and University of Tartu and by a Estonian Basic Research grant SF0270177As08. MM was additionally supported by Estonian Science Foundation grant #8973. JGS and MS were supported by NIH grant GM40282. PHS and EEE were supported by NIH grants HG004120 and HG002385. DR and NP were supported by NSF HOMINID award BCS-1032255 and NIH grant GM100233. DR and EEE are Howard Hughes Medical Institute investigators.

**Author contributions**

BB, EEE, JBu, MS, SP, JKe, DR and JKr supervised the study. IL, NP, AM, GR, SM, KK, PHS, JGS, SC, ML, QF, HL, CdF, KP, WH, MMey and DR analyzed genetic data. FH, EF, DD, MF, J-MG, JW, AC and JKr obtained human remains. AM, CE, RBo, KB, SS, CP, NR and JKr processed ancient DNA. IL, NP, SN, NR, GA, HAB, GBa, EB, OB, RBa, GBe, HB-A, JBe, FBe, CMB, FBr, GBJB, FC, MC, DECC, DCor, LD, GvD, SD, J-MD, SAF, IGR, MG, MH, BH, TH, UH, ARJ, SK-Y, RKh, EK, RKi, TK, WK, VK, AK, LL, SL, TL, RWM, BM, EM, JMol, JMou,

KN, DN, TN, LO, JP, FP, OLP, VR, FR, IR, RR, HS, ASaj, ASal, EBS, ATar, DT, ST, IU, OU, RVa, MVi, MVo, CW, LY, PZ, TZ, CC, MGT, AR-L, SAT, LS, KT, RVi, DCom, RS, MMet, SP and DR assembled the genotyping dataset. IL, NP, DR and JKr wrote the manuscript with help from all co-authors.

## Author information

The aligned sequences are available through the Sequence Read Archive (SRA) under accession numbers that will be made available upon publication. The authors declare competing financial interests: UH is an employee of Illumina, TL is an employee of AMGEN, and JM is an employee of 23andMe. Correspondence and requests for materials should be addressed to David Reich (reich@genetics.med.harvard.edu) or Johannes Krause (johannes.krause@uni-tuebingen.de).

**Table 1: Lowest $f_3$-statistics for each West Eurasian population**

| $Ref_1$ | $Ref_2$ | **Target for which these two references give the lowest $f_3(X; Ref_1, Ref_2)$** |
|---|---|---|
| **WHG** | **EEF** | Sardinian[***] |
| **WHG** | **Near East** | Basque, Belarusian, Czech, English, Estonian, Finnish, French_South, Icelandic, Lithuanian, Mordovian, Norwegian, Orcadian, Scottish, Spanish, Spanish_North, Ukrainian |
| **EEF** | **ANE** | Abkhasian[***], Albanian, Ashkenazi_Jew[****], Bergamo, Bulgarian, Chechen[****], Croatian, Cypriot[****], Druze[**], French, Greek, Hungarian, Lezgin, MA1, Maltese, Sicilian, Turkish_Jew, Tuscan |
| **EEF** | **Native American** | Adygei, Balkar, Iranian, Kumyk, North_Ossetian, Turkish |
| **EEF** | **African** | BedouinA, BedouinB†, Jordanian, Lebanese, Libyan_Jew, Moroccan_Jew, Palestinian, Saudi[****], Syrian, Tunisian_Jew[***], Yemenite_Jew[***] |
| **EEF** | **South Asian** | Armenian, Georgian[****], Georgian_Jew[*], Iranian_Jew[***], Iraqi_Jew[***] |

Note: WHG = Loschbour or LaBraña; EEF=Stuttgart; ANE=MA1; Native American=Piapoco; African=Esan, Gambian, or Kgalagadi; South Asian=GujaratiC or Vishwabrahmin. Statistics are negative with Z<-4 unless otherwise noted: † (positive) or [*], [**], [***], [****], to indicate Z less than 0, -1, -2, and -3 respectively. The complete list of statistics can be found in Extended Data Table 1.

# Figure Legends

**Figure 1: Map of West Eurasian populations and Principal Component Analysis.** (a) Geographical locations of ancient and present-day samples, with color coding matching the PCA. We show all sampling locations for each population, which results in multiple points for some populations (e.g., Spain). (b) PCA on all present-day West Eurasians, with the ancient and selected eastern non-Africans projected. European hunter-gatherers fall beyond present-day Europeans in the direction of European differentiation from the Near East. Stuttgart clusters with other Neolithic Europeans and present-day Sardinians. MA1 falls outside the variation of present-day West Eurasians in the direction of southern-northern differentiation along dimension 2 and between the European and Near Eastern clines along dimension 1.

**Figure 2: Modeling of West Eurasian population history.** (a) A three-way mixture model that is a statistical fit to the data for many European populations, ancient DNA samples, and non-European populations. Present-day samples are colored in blue, ancient samples in red, and reconstructed ancestral populations in green. Solid lines represent descent without mixture, and dashed lines represent admixture events. For the two mixture events relating the highly divergent ancestral populations, we print estimates for the mixture proportions as well as one standard error. (b) We plot the proportions of ancestry from each of three inferred ancestral populations (EEF, ANE and WHG) as inferred from the model-based analysis.

## Methods

### Archeological context, sampling and DNA extraction

The Loschbour sample stems from a male skeleton excavated in 1935 at the Loschbour rock shelter in Heffingen, Luxembourg. The skeleton was AMS radiocarbon dated to 7,205 ± 50 years before present (OxA-7738; 6,220-5,990 cal BC)[43]. At the Palaeogenetics Laboratory in Mainz, material for DNA extraction was sampled from a molar (M48) after irradiation with UV-light, surface removal, and pulverization in a mixer mill. DNA extraction took place in the palaeogenetics facilities in the Institute for Archaeological Sciences at the University of Tübingen. Three extracts were made in total, one from 80 mg of powder using an established silica based protocol[44] and two additional extracts from 90 mg of powder each with a protocol optimized for the recovery of short DNA molecules[45].

The Stuttgart sample was taken from a female skeleton excavated in 1982 at the site Viesenhäuser Hof, Stuttgart-Mühlhausen, Germany. It was attributed to the Linearbandkeramik (5,500-4,800 BC) through associated pottery artifacts and the chronology was corroborated by radiocarbon dating of the stratigraphy[46]. Both sampling and DNA extraction took place in the Institute for Archaeological Sciences at the University of Tübingen. The M47 molar was removed and material from the inner part was sampled with a sterile dentistry drill. An extract was made using 40 mg of bone powder[45].

The Motala individuals were recovered from the site of Kanaljorden in the town of Motala, Östergötland, Sweden, excavated between 2009 and 2013. The human remains at this site are represented by several adult skulls and one infant skeleton. All individuals are part of a ritual deposition at the bottom of a small lake. Direct radiocarbon dates on the remains range between 7,013 ± 76 and 6,701 ± 64 BP (6,361-5,516 cal BC), corresponding to the late Middle Mesolithic of Scandinavia. Samples were taken from the teeth of the nine best preserved skulls, as well as a femur and tibia. Bone powder was removed from the inner parts of the teeth or bones with a sterile dentistry drill. DNA from 100 mg of bone powder was extracted[47] in the ancient DNA laboratory of the Archaeological Research Laboratory, Stockholm.

**Library preparation**

Illumina sequencing libraries were prepared using either double- or single-stranded library preparation protocols[48,49] (SI1). For high-coverage shotgun sequencing libraries, a DNA repair step with Uracil-DNA-glycosylase (UDG) and endonuclease VIII (endo VIII) treatment was included in order to remove uracil residues[50]. Size fractionation on a PAGE gel was also performed in order to remove longer DNA molecules that are more likely to be contaminants[49]. Positive and blank controls were carried along during every step of library preparation.

**Shotgun sequencing and read processing**

All non-UDG-treated libraries were sequenced either on an Illumina Genome Analyzer IIx with 2×76 + 7 cycles for the Loschbour and Motala libraries, or on an Illumina MiSeq with 2×150 + 8 + 8 cycles for the Stuttgart library. We followed the manufacturer's protocol for multiplex sequencing. Raw overlapping forward and reverse reads were merged and filtered for quality[51] and mapped to the human reference genome (hg19/GRCh37/1000Genomes) using the Burrows-Wheeler Aligner (BWA)[52] (SI2). For deeper sequencing, UDG-treated libraries of Loschbour were sequenced on 3 Illumina HiSeq 2000 lanes with 50-bp single-end reads, 8 Illumina HiSeq 2000 lanes of 100-bp paired-end reads and 8 Illumina HiSeq 2500 lanes of 101-bp paired-end reads. The UDG-treated library for Stuttgart was sequenced on 8 HiSeq 2000 lanes and 101-bp paired-end reads. The UDG-treated libraries for Motala were sequenced on 8 HiSeq 2000 lanes of 100-bp paired-end reads, with 4 lanes each for two pools (one of 3 individuals and one of 4 individuals). We also sequenced an additional 8 HiSeq 2000 lanes for Motala12, the Motala sample with the highest percentage of endogenous human DNA.

**Enrichment of mitochondrial DNA and sequencing**

Non-UDG-treated libraries of Loschbour and all Motala samples were enriched for human mitochondrial DNA using a bead-based capture approach with present-day human DNA as bait[53] to test for DNA preservation and mtDNA contamination. UDG-treatment was omitted in order to allow characterization of damage patterns typical for ancient DNA[8]. The captured libraries were sequenced on an llumina Genome Analyzer IIx platform with 2 × 76 + 7 cycles and the resulting reads were merged and quality filtered[51]. The sequences were mapped to the Reconstructed Sapiens Reference Sequence, RSRS[54], using a custom iterative mapping assembler, MIA[55] (SI4).

### Contamination estimates

We assessed if the sequences had the characteristics of authentic ancient DNA using four approaches. First we searched for evidence of contamination by determining whether the sequences mapping to the mitochondrial genome were consistent with deriving from more than one individual[55,56]. Second, for the high-coverage Loschbour and Stuttgart genomes, we used a maximum-likelihood-based estimate of autosomal contamination that uses variation at sites that are fixed in the 1000 Genomes data to estimate error, heterozygosity and contamination[57] simultaneously. Third, we estimated contamination based on the rate of polymorphic sites on the X chromosome of the male Loschbour individual[58] (SI3) Fourth, we analyzed non-UDG treated reads mapping to the RSRS to search for aDNA-typical damage patterns resulting in C→T changes at the 5'-end of the molecule[8] (SI3).

### Phylogenetic analysis of the mitochondrial genomes

All nine complete mitochondrial genomes that fulfilled the criteria of authenticity were assigned to haplogroups using Haplofind[59]. A Maximum Parsimony tree including present day humans and previously published ancient mtDNA sequences was generated with MEGA[60]. The effect of branch shortening due to a lower number of substitutions in ancient lineages was studied by calculating the nucleotide edit distance to the root for all haplogroup R sequences (SI4).

### Sex Determination and Y-chromosome Analysis

We assessed the sex of all sequenced individuals by using the ratio of (chrY) to (chrY+chrX) aligned reads[10]. We downloaded a list of Y-chromosome SNPs curated by the International Society of Genetic Genealogy (ISOGG, http://www.isogg.org) v. 9.22 (accessed Feb. 18, 2014) and determined the state of the ancient individuals at positions where a single allele was observed and MAPQ≥30. We excluded C/G or A/T SNPs due to uncertainty about the polarity of the mutation in the database. The ancient individuals were assigned haplogroups based on their derived state (SI5). We also used BEAST v1.7.51[61] to assess the phylogenetic position of Loschbour using 623 males from around the world with 2,799 variant sites across 500kb of non-recombining Y-chromosome sequence[62] (SI5).

**Estimation of Neandertal admixture**

We estimate Neandertal admixture in ancient individuals with the $f_4$-ratio or $S$-statistic[5,63,64] $\hat{\alpha} = f_4(Altai,\ Denisova; Test,\ Yoruba) / f_4(Altai,\ Denisova; Vindija,\ Yoruba)$ which uses whole genome data from Altai, a high coverage (52×) Neanderthal genome sequence[35], Denisova, a high coverage sequence[49] from another archaic human population (31×), and Vindija, a low coverage (1.3×) Neanderthal genome from a mixture of three Neanderthal individuals from Vindija Cave in Croatia[63].

**Inference of demographic history and inbreeding**

We used the Pairwise Sequentially Markovian Coalescent (PSMC)[65] to infer the size of the ancestral population of Stuttgart and Loschbour. This analysis requires high quality diploid genotype calls and cannot be performed in the low-coverage Motala samples. To determine whether the low effective population size inferred for Loschbour is due to recent inbreeding, we plotted the time-to-most-recent common ancestor (TMRCA) along each of chr1-22 to detect runs of low TMRCA.

**Analysis of segmental duplications and copy number variants**

We built read-depth based copy number maps for the Loschbour, Stuttgart and Motala12 genomes in addition to the Denisova and Altai Neanderthal genome and 25 deeply sequenced modern genomes[35] (SI7). We built these maps by aligning reads, subdivided into their non-overlapping 36-bp constituents, against the reference genome using the mrsFAST aligner[66], and renormalizing read-depth for local GC content. We estimated copy numbers in windows of 500 unmasked base pairs slid at 100 bp intervals across the genome. We called copy number variants using a scale space filter algorithm. We genotyped variants of interest and compared the genotypes to those from individuals sequenced as part of the 1000 Genomes Project[67].

**Phenotypic inference**

We inferred likely phenotypes (SI8) by analyzing DNA polymorphism data in the VCF format[68] using VCFtools (http://vcftoools.sourceforge.net/). For the Loschbour and Stuttgart individuals, we included data from sites not flagged as LowQuality, with genotype quality (GQ) of ≥30, and SNP quality (QUAL) of ≥50. For Motala12, which is of lower coverage, we included sites having at least 2× coverage and passed visual inspection of the local alignment using samtools tview (http://samtools.sourceforge.net)[69]

## Human Origins dataset curation

The Human Origins array consists of 14 panels of SNPs for which the ascertainment is well known[5,70]. All population genetics analysis were carried out on a set of 594,924 autosomal SNPs, after restricting to sites that had >90% completeness across 7 different batches of sequencing, and that had >97.5% concordance with at least one of two subsets of samples for which whole genome sequencing data was also available. The total dataset consists of 2,722 individuals, which we filtered to 2,345 individuals (203 populations) after removing outlier individuals or relatives based on visual inspection of PCA plots[15,71] or model-based clustering analysis[14]. Whole genome amplified (WGA) individuals were not used in analysis, except for a Saami individual who we forced in because of the special interest of this population for Northeastern European population history (Extended Data Fig. 7).

## ADMIXTURE analysis

We merged all Human Origins genotype data with whole genome sequencing data from Loschbour, Stuttgart, MA1, Motala12, Motala_merge, and LaBrana. We then thinned the resulting dataset to remove SNPs in linkage-disequilibrium with PLINK 1.07[72], using a window size of 200 SNPs advanced by 25 SNPs and an $r^2$ threshold of 0.4. We ran ADMIXTURE 1.23[14,73] for 100 replicates with different starting random seeds, default 5-fold cross-validation, and varying the number of ancestral populations K between 2 and 20. We assessed clustering quality using CLUMPP[74]. We used the ADMIXTURE results to identify a set of 59 "West Eurasian" (European/Near Eastern) populations based on values of a "West Eurasian" ancestral population at K=3 (SI9). We also identified 15 populations for use as "non-West Eurasian outgroups" based on their having at least 10 individuals and no evidence of European or Near Eastern admixture at K=11, the lowest K for which Near Eastern/European-maximized ancestral populations appeared consistently across all 100 replicates.

## Principal Components Analysis

We used *smartpca*[15] (version: 10210) from EIGENSOFT[71,75] 5.0.1 to carry out Principal Components Analysis (PCA) (SI10). We performed PCA on a subset on individuals and then projected others using the *lsqproject: YES* option that gives an unbiased inference of the position of samples even in the presence of missing data (especially important for ancient DNA).

### $f_3$-statistics

We use the $f_3$-statistic[5] $f_3(Test;\ Ref_1, Ref_2) = \frac{1}{N}\sum_{i=1}^{N}(t_i - r_{1,i})\,(t_i - r_{2,i})$, where $t_i$, $r_{1,i}$ and $r_{2,i}$ are the allele frequencies for the $i^{th}$ SNP in populations *Test*, $Ref_1$, $Ref_2$, respectively, to determine if there is evidence that the *Test* population is derived from admixture of populations related to $Ref_1$ and $Ref_2$ (SI11). A significantly negative statistic provides unambiguous evidence of mixture in the *Test* population[5]. We allow $Ref_1$ and $Ref_2$ to be any Human Origins population with 4 or more individuals, or Loschbour, Stuttgart, MA1, Motala12, LaBrana. We assess significance of the $f_3$-statistics using a block jackknife[21] and a block size of 5cM. We report significance as the number of standard errors by which the statistic differs from zero (Z-score). We also perform an analysis in which we constrain the reference populations to be (i) EEF (Stuttgart) and WHG (Loschbour or LaBrana), (ii) EEF and a Near Eastern population, (iii) EEF and ANE (MA1), or (iv) any two present-day populations, and compute a $Z_{diff}$ score between the lowest $f_3$-statistic observed in the dataset, and the $f_3$-statistic observed for the specified pair.

### $f_4$-statistics

We analyze $f_4$-statistics[5] of the form $f_4(A, B;\ C, D) = \frac{1}{N}\sum_{i=1}^{N}(a_i - b_i)\,(c_i - d_i)$ to assess if populations A, B are consistent with forming a clade in an unrooted tree with respect to C, D. If they form a clade, the allele frequency differences between the two pairs should be uncorrelated and the statistic has an expected value of 0. We set the outgroup *D* to be a sub-Saharan African population or Chimpanzee. We systematically tried all possible combinations of the ancient samples or 15 “non-West Eurasian outgroups” identified by ADMIXTURE analysis as A, B, C to determine their genetic affinities (SI14). Setting A as a present-day test population and B as either Stuttgart or BedouinB, we documented relatedness to C=(Loschbour or MA1) or C=(MA1 and Karitiana) or C=(MA1 or Han) (Extended Data Figs. 4, 5, 7). Setting C as a test population and (A, B) a pair from (Loschbour, Stuttgart, MA1) we documented differential relatedness to ancient populations (Extended Data Fig. 6). We computed *D*-statistics[63] using transversion polymorphisms in whole genome sequence data[35] to confirm robustness to ascertainment and ancient DNA damage (Extended Data Table 2).

**Minimum number of source populations for Europeans**

We used *qpWave*[22,23] to study the minimum number of source populations for a designated set of Europeans (SI12). We use $f_4$-statistics of the form $X(l, r) = f_4(l_0, l; r_0, r)$ where $l_0, r_0$ are arbitrarily chosen "base" populations, and $l$, $r$ are other populations from two sets $L$ and $R$ respectively. If $X(l, r)$ has rank $r$ and there were $n$ waves of immigration into $R$ with no back-migration from $R$ to $L$, then $r+1 \leq n$. We set $L$ to include *Stuttgart, Loschbour, MA1, Onge, Karitiana, Mbuti* and $R$ to include 23 modern European populations who fit the model of SI14 and had admixture proportions within the interval [0,1] for the method with minimal modeling assumptions (SI17).

**Admixture proportions for Stuttgart in the absence of a Near Eastern ancient genome**

We used Loschbour and BedouinB as surrogates for "Unknown hunter-gatherer" and Near Eastern (NE) farmer populations that contributed to Stuttgart (SI13). Ancient Near Eastern ancestry in Stuttgart is estimated by the $f_4$-ratio[5,18] $f_4(Outgroup, X; Loschbour, Stuttgart) / f_4(Outgroup, X; Loschbour, NE)$. A complication is that BedouinB is a mixture of NE and African ancestry. We therefore subtracted[23] the effects of African ancestry using estimates of the BedouinB African admixture proportion from ADMIXTURE (SI9) or ALDER[76].

**Admixture graph modeling**

We used ADMIXTUREGRAPH[5] (version 3110) to model population relationships between Loschbour, Stuttgart, Onge, and Karitiana using Mbuti as an African outgroup. We assessed model fit using a block jackknife of differences between estimated and fitted *f*-statistics for the set of included populations (we expressed the fit as a Z score). We determined that a model failed if |Z|>3 for at least one *f*-statistic. A basic tree model failed and we manually amended the model to test all possible models with a single admixture event, which also failed. Further manual amendment to include 2 admixture events resulted in 8 successful models, only one of which could be amended to also fit MA1 as an additional constraint. We successfully fit both the Iceman and LaBrana into this model as simple clades and Motala12 as a 2-way mixture. We also fit present-day West Eurasians as clades, 2-way mixtures, or 3-way mixtures in this basic model, achieving a successful fit for a larger number of European populations (n=26) as 3-way mixtures. We estimated the individual admixture proportions from the fitted model parameters. To test if fitted parameters for different populations are consistent with each other, we jointly fit all pairs of populations *A* and *B* by modifying ADMIXTUREGRAPH to add a large constant (10,000) to the variance term $f_3(A_0, A, B)$. By doing this, we can safely ignore recent gene flow within Europe that affects statistics that include both *A* and *B*.

**Ancestry estimates from $f_4$-ratios**

We estimate EEF ancestry using the $f_4$-ratio[5,18] $f_4$*(Mbuti, Onge; Loschbour, European)* / $f_4$*(Mbuti, Onge; Loschbour, Stuttgart),* which produces consistent results with ADMIXTUREGRAPH (SI14). We use $f_4$*(Stuttgart, Loschbour; Onge MA1)* / $f_4$*(Mbuti, MA1; Onge, Loschbour)* to estimate Basal Eurasian admixture into Stuttgart. We use $f_4$*(Stuttgart, Loschbour; Onge Karitiana)* / $f_4$*(Stuttgart, Loschbour; Onge MA1)* to estimate ANE mixture in Karitiana (Fig. 2B). We use $f_4$*(Test, Stuttgart; Karitiana, Onge)* / $f_4$*(MA1, Stuttgart; Karitiana, Onge)* to lower bound ANE mixture into North Caucasian populations.

***MixMapper* analysis**

We carried out *MixMapper* 2.0[4] analysis, a semi-supervised admixture graph fitting technique. First, we infer a scaffold tree of populations without strong evidence of mixture relative to each other (Mbuti, Onge, Loschbour and MA1). We do not include European populations in the scaffold as all had significantly negative $f_3$-statistics indicating admixture. We then ran *MixMapper* to infer the relatedness of the other ancient and present-day samples, fitting them onto the scaffold as 2- or 3-way mixtures. The uncertainty in all parameter estimates is measured by block bootstrap resampling of the SNP set (100 replicates with 50 blocks).

***TreeMix* analysis**

We applied *TreeMix*[34] to Loschbour, Stuttgart, Motala12, and MA1[6], LaBrana[12] and the Iceman[19], along with the present-day samples of Karitiana, Onge and Mbuti. We restricted the analysis to 265,521 Human Origins array sites after excluding any SNPs where there were no-calls in any of the studied individuals. The tree was rooted with Mbuti and standard errors were estimated using blocks of 500 SNPs. We repeated the analysis on whole-genome sequence data, rooting with Chimp and replacing Onge with Dai since we did not have Onge whole genome sequence data[35]. We varied the number of migration events (*m*) between 0 and 5.

### Inferring admixture proportions with minimal modeling assumptions

We devised a method to infer ancestry proportions from three ancestral populations (EEF, WHG, and ANE) without strong phylogenetic assumptions (SI17). We rely on 15 "non-West Eurasian" outgroups and study $f_4(European, Stuttgart; O_1, O_2)$ which equals $\alpha\beta\, f_4(Loschbour, Stuttgart; O_1, O_2) + \alpha(1\text{-}\beta)\, f_4(MA1, Stuttgart; O_1, O_2)$ if *European* has 1-*a* ancestry from EEF and $\beta$, 1-$\beta$ ancestry from WHG and ANE respectively. This defines a system of $\binom{15}{2} = 105$ equations with unknowns $\alpha\beta$, $\alpha(1\text{-}\beta)$, which we solve with least squares implemented in the function *lsfit* in *R* to obtain estimates of $\alpha$ and $\beta$. We repeated this computation 22 times dropping one chromosome at a time[26] to obtain block jackknife[21] estimates of the ancestry proportions and standard errors, with block size equal to the number of SNPs per chromosome. We assessed consistency of the inferred admixture proportions with those derived from the ADMIXTUREGRAPH model based on the number of standard errors between the two (Extended Data Table 1).

### Haplotype-based analyses

We used RefinedIBD from BEAGLE 4[77] with the settings *ibdtrim*=20 and *ibdwindow*=25 to study IBD sharing between Loschbour and Stuttgart and populations from the POPRES dataset[42]. We kept all IBD tracts spanning at least 0.5 centimorgans (cM) and with a LOD score >3 (SI18) .We also used ChromoPainter[37] to study haplotype sharing between Loschbour and Stuttgart and present-day West Eurasian populations (SI19). We identified 495,357 SNPs that were complete in all individuals and phased the data using Beagle 4[77] with parameters *phase-its*=50 and *impute-its*=10. We did not keep sites with missing data to avoid imputing modern alleles into the ancient individuals. We combined ChromoPainter output for chromosomes 1-22 using ChromoCombine[37]. We carried out a PCA of the co-ancestry matrix using fineSTRUCTURE[37].

# Figure 1

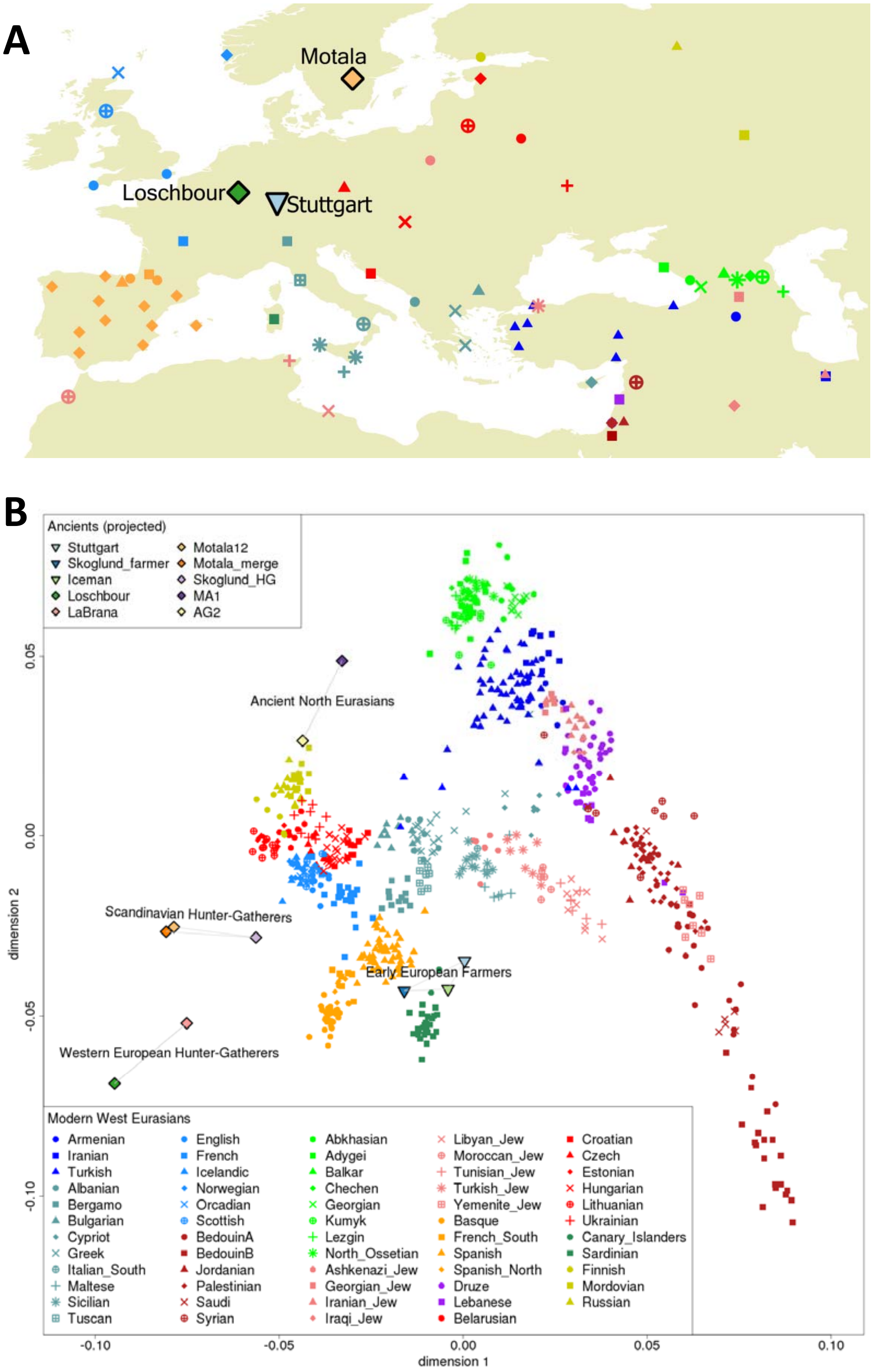

# Figure 2

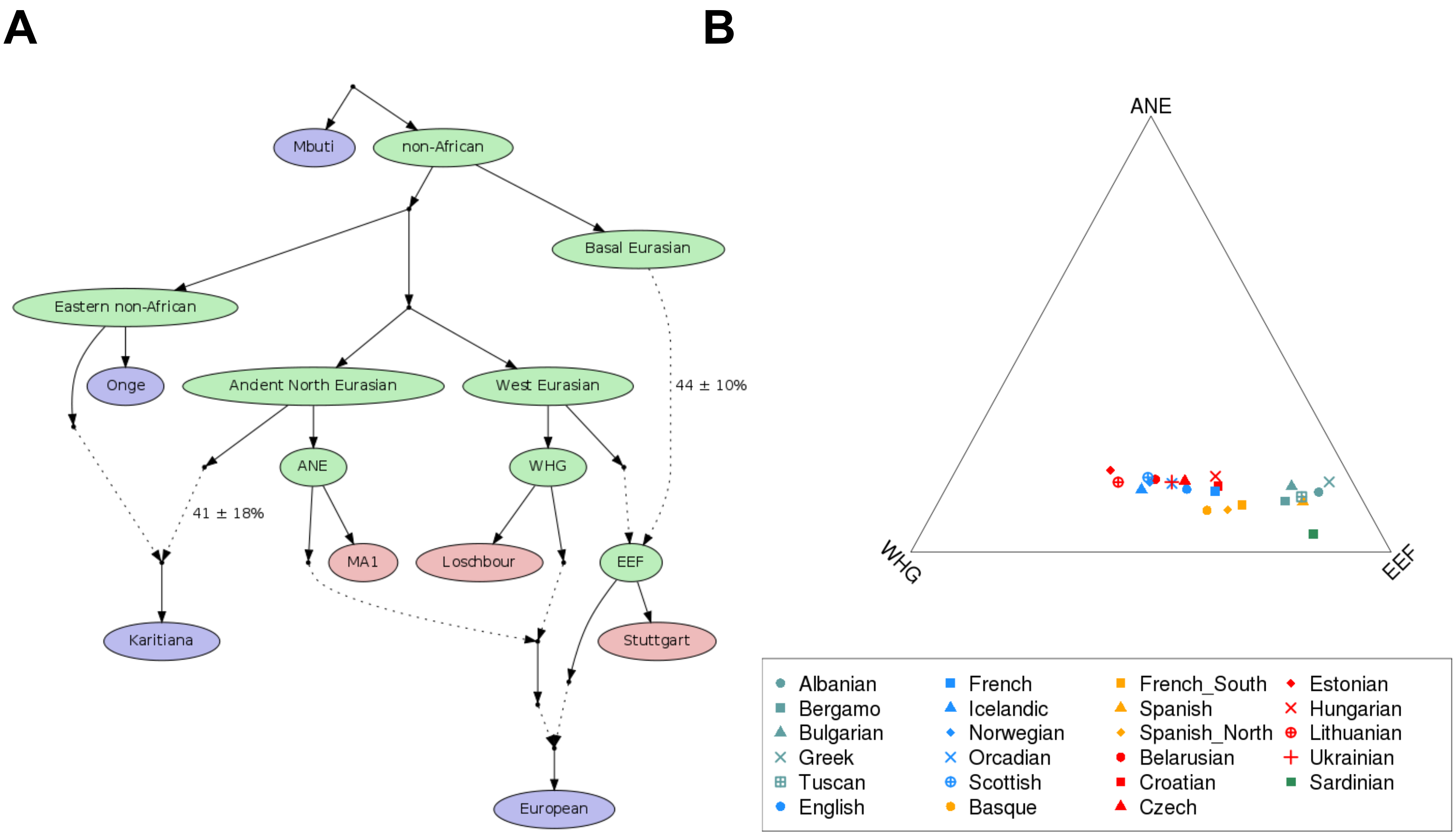

# Extended Data

**Extended Data Figure 1: Photographs of analyzed ancient samples.** (A) Loschbour skull; (B) Stuttgart skull, missing the lower right M2 we sampled; (C) excavation at Kanaljorden in Motala, Sweden; (D) Motala 1 in situ.

**A**

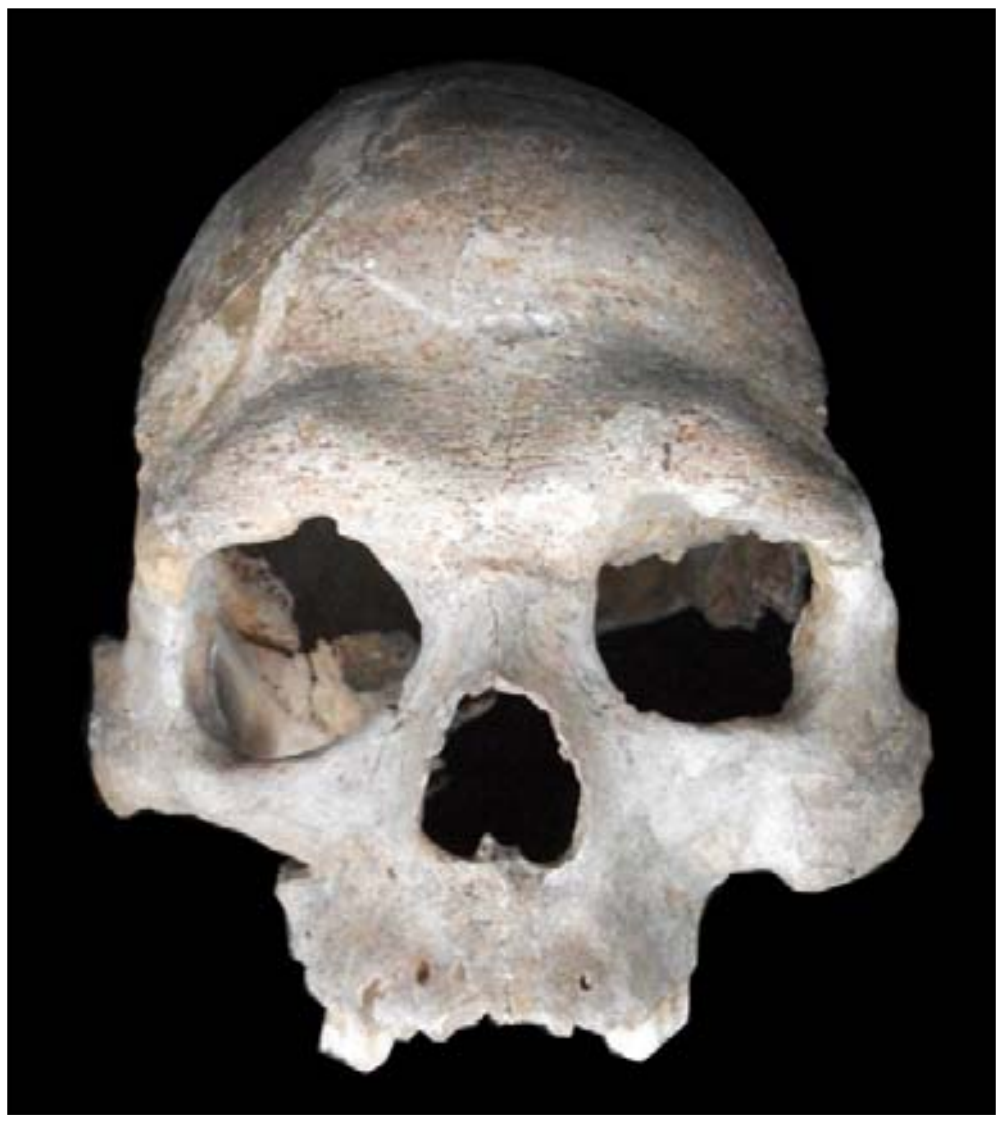

**B**

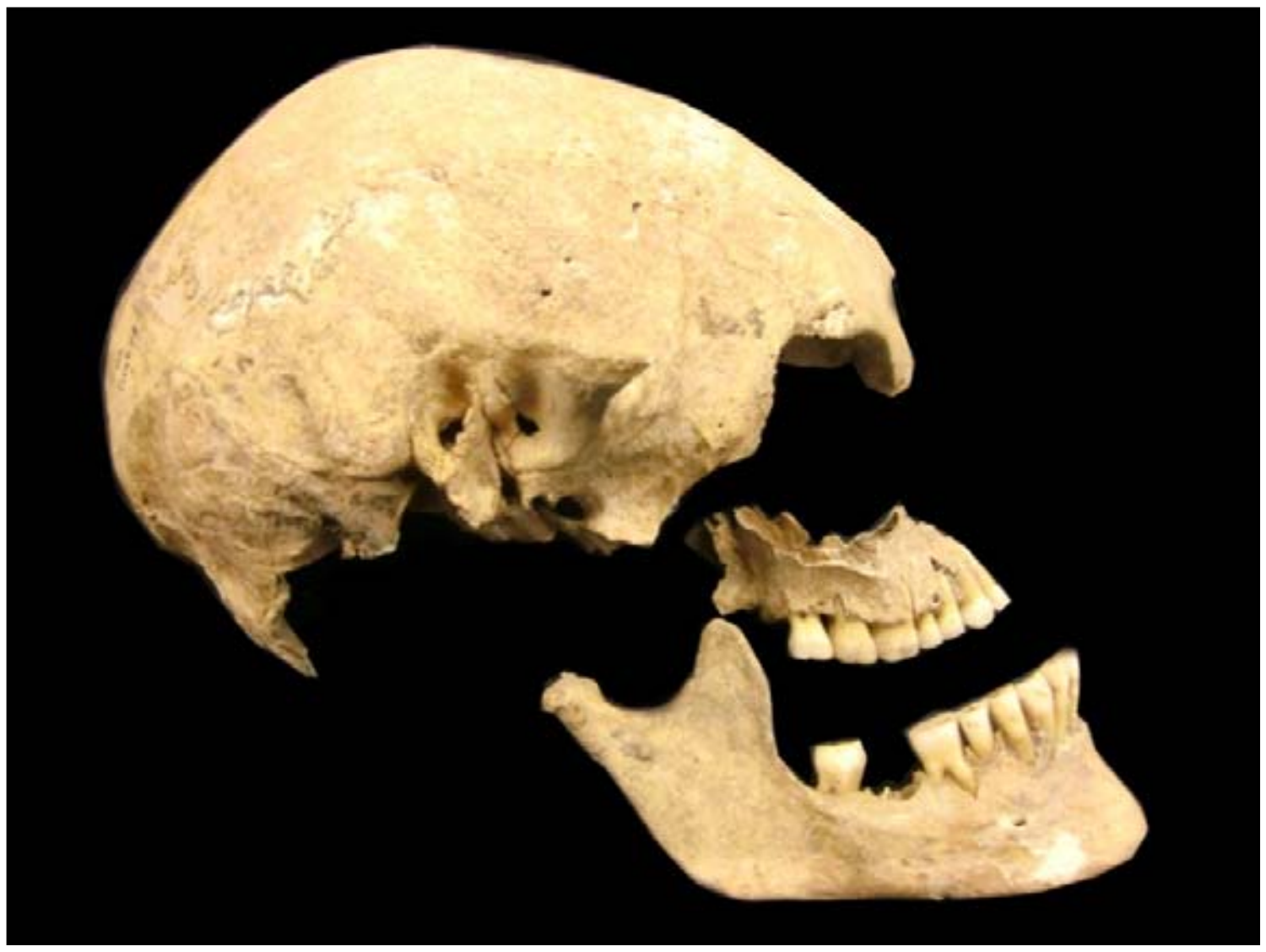

**C**

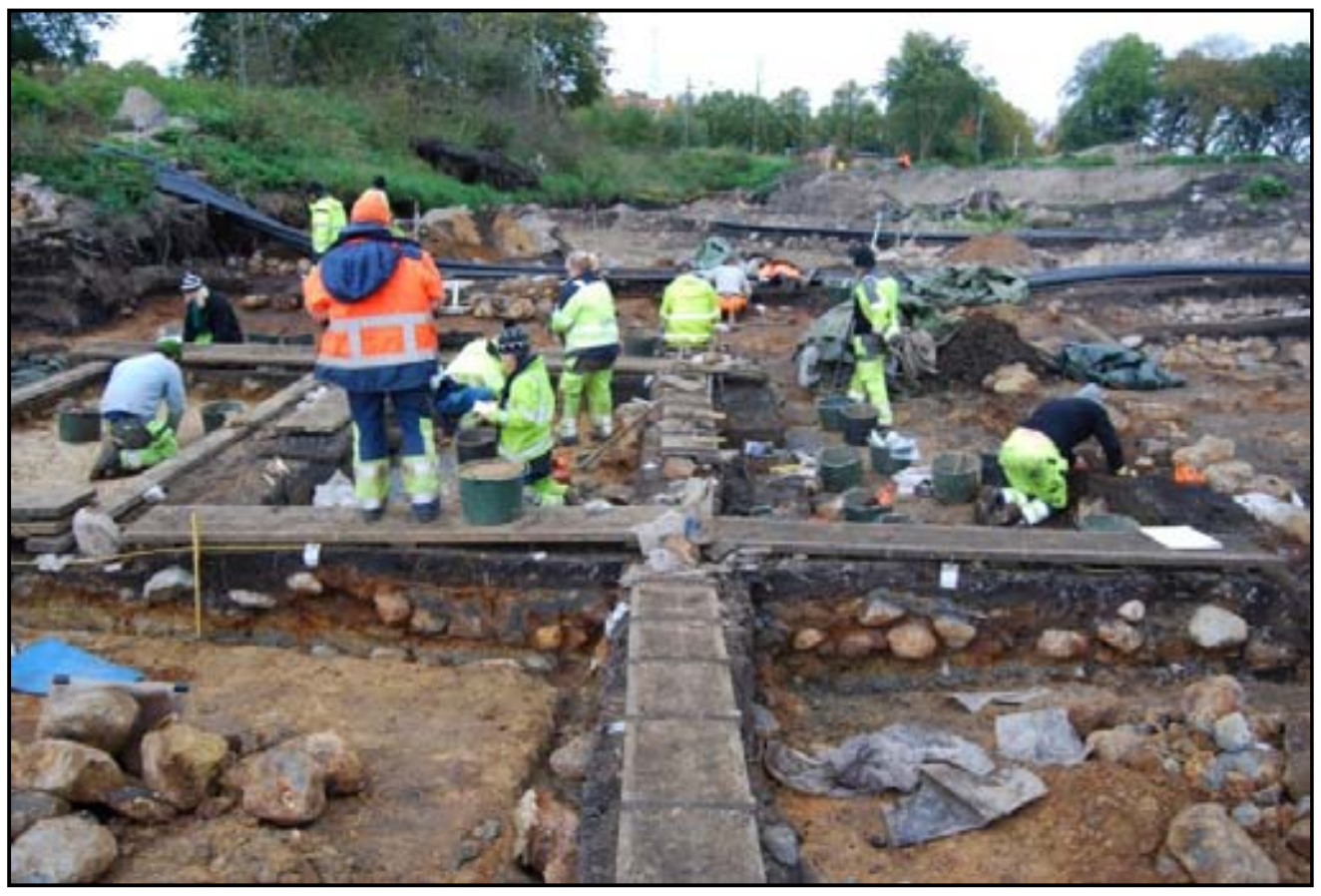

**D**

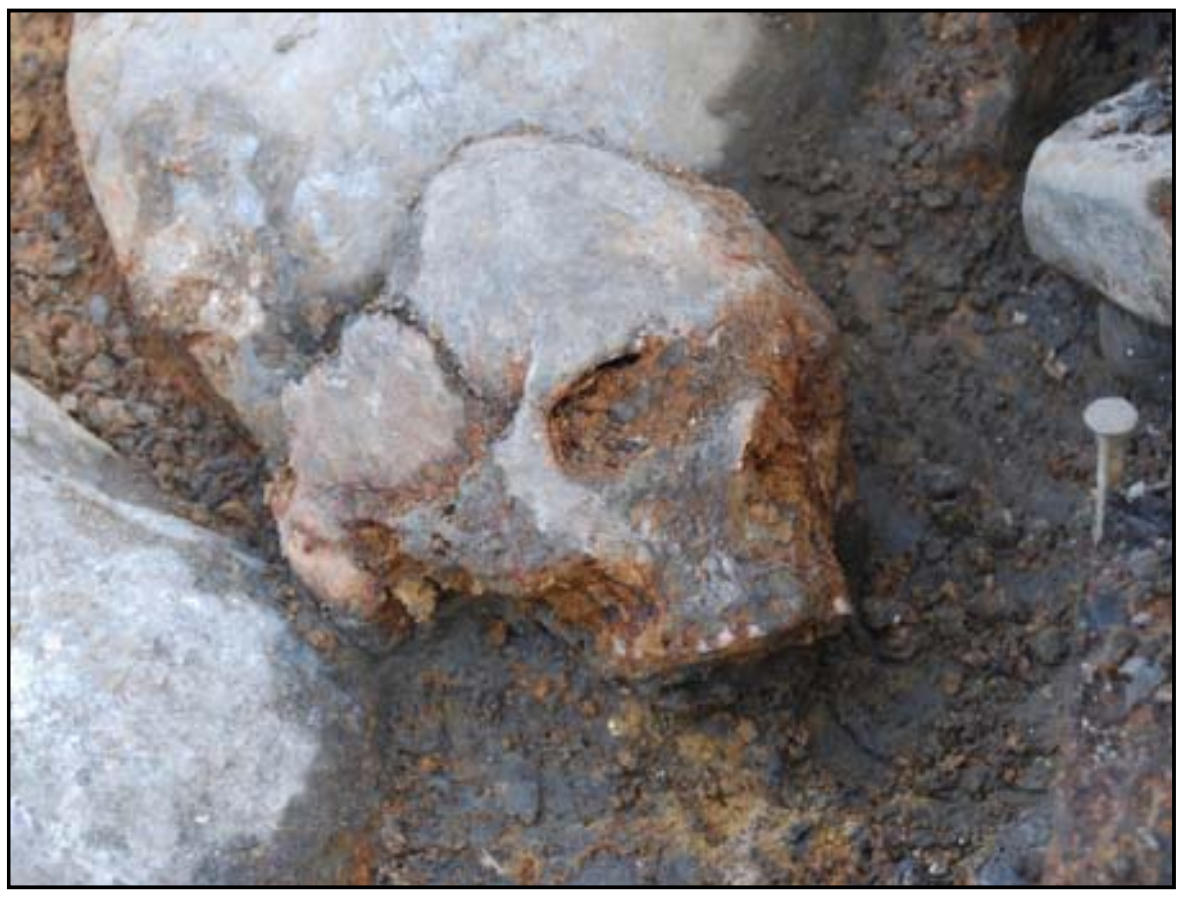

**Extended Data Figure 2: Pairwise Sequential Markovian Coalescent (PSMC) analysis.** (A) Inference of population size as a function of time, showing a very small recent population size over the most recent period in the ancestry of Loschbour (at least the last 5-10 thousand years). (B) Inferred time since the most recent common ancestor from the PSMC for chromosomes 20, 21, 22 (top to bottom); Stuttgart is plotted on top and Loschbour at bottom.

**A**

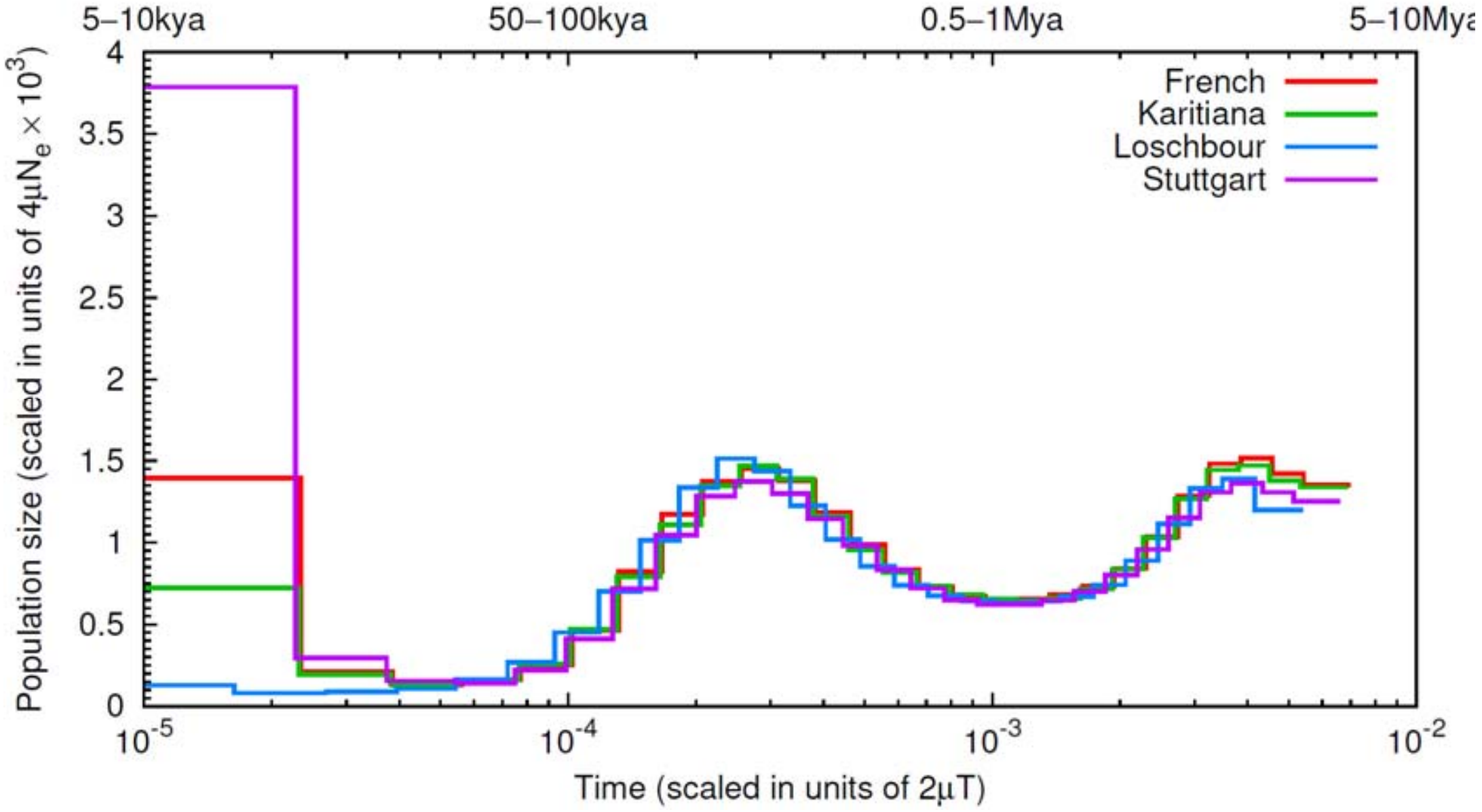

**B**

**B**

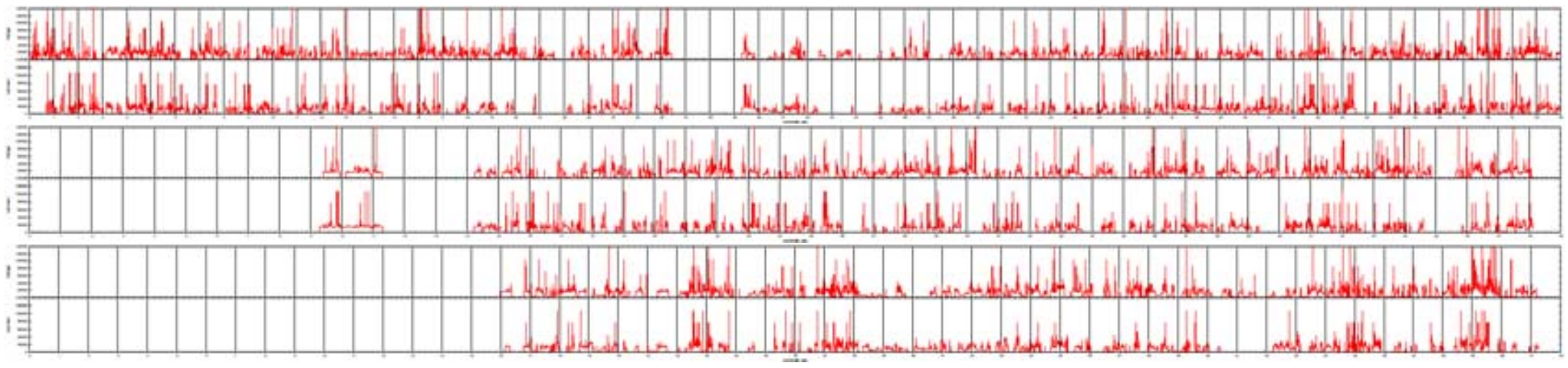

**Extended Data Figure 3: ADMIXTURE analysis (K=2 to K=20).** Ancient samples (Loschbour, Stuttgart, Motala_merge, Motala12, MA1, and LaBrana) are at left.

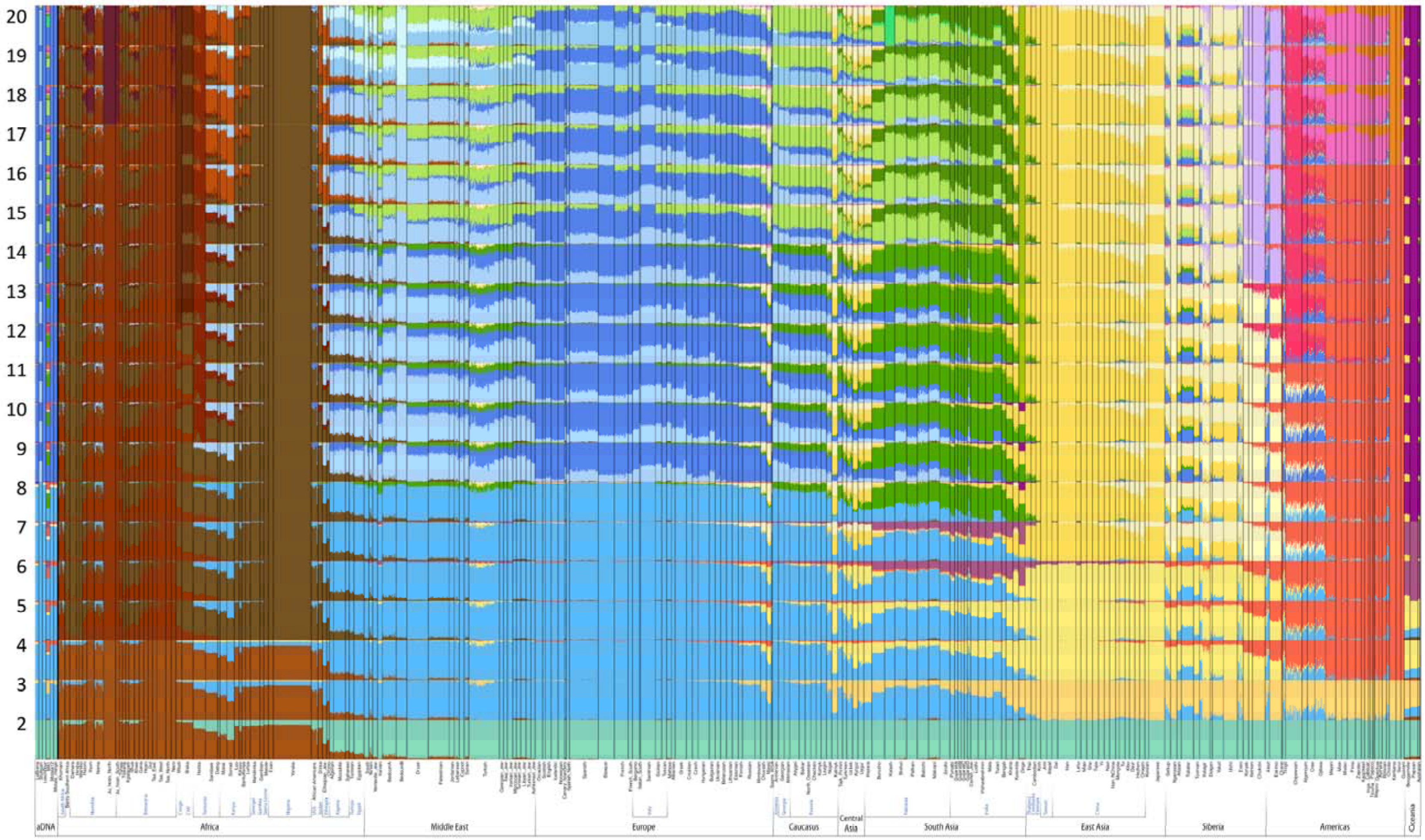

**Extended Data Figure 4: ANE ancestry is present in both Europe and the Near East but WHG ancestry is restricted to Europe, which cannot be due to a single admixture event.** (*x*-axis) We computed the statistic *f*$_4$*(Test, Stuttgart; MA1, Chimp)*, which measures where MA1 shares more alleles with a test population than with Stuttgart. It is positive for most European and Near Eastern populations, consistent with ANE (MA1-related) gene flow into both regions. (*y*-axis) We computed the statistic *f*$_4$*(Test, Stuttgart; Loschbour, Chimp),* which measures whether Loschbour shares more alleles with a test sample than with Stuttgart. Only European populations show positive values of this statistic, providing evidence of WHG (Loschbour-related) admixture only in Europeans.

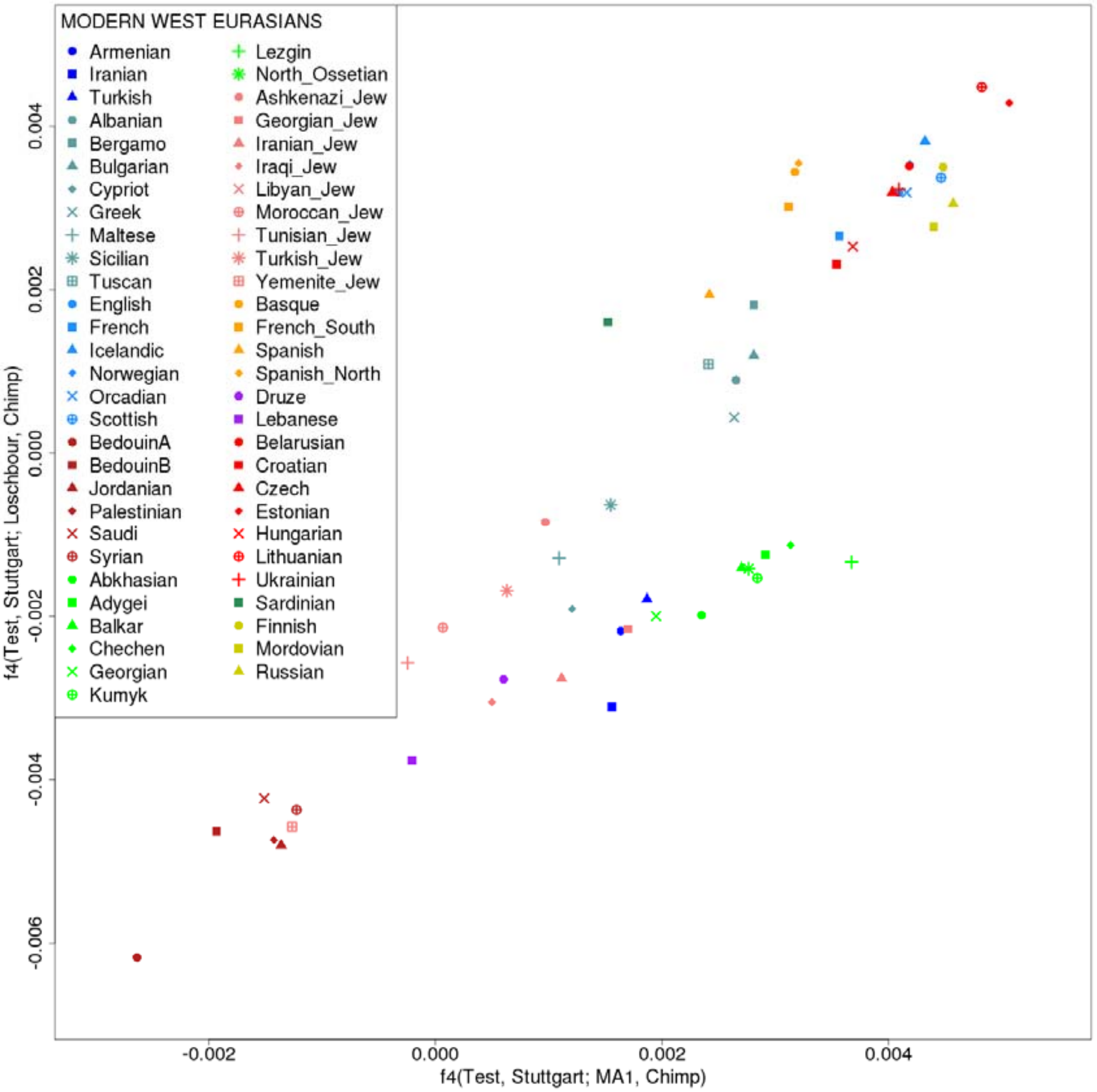

**Extended Data Figure 5: MA1 is the best surrogate for ANE for which we have data.** Europeans share more alleles with MA1 than with Karitiana, as we see from the fact that in a plot of $f_4$*(Test, BedouinB; MA1, Chimp)* and $f_4$*(Test, BedouinB; Karitiana, Chimp)*, the European cline deviates in the direction of MA1, rather than Karitiana (the slope is >1 and European populations are above the line indicating equality of these two statistics).

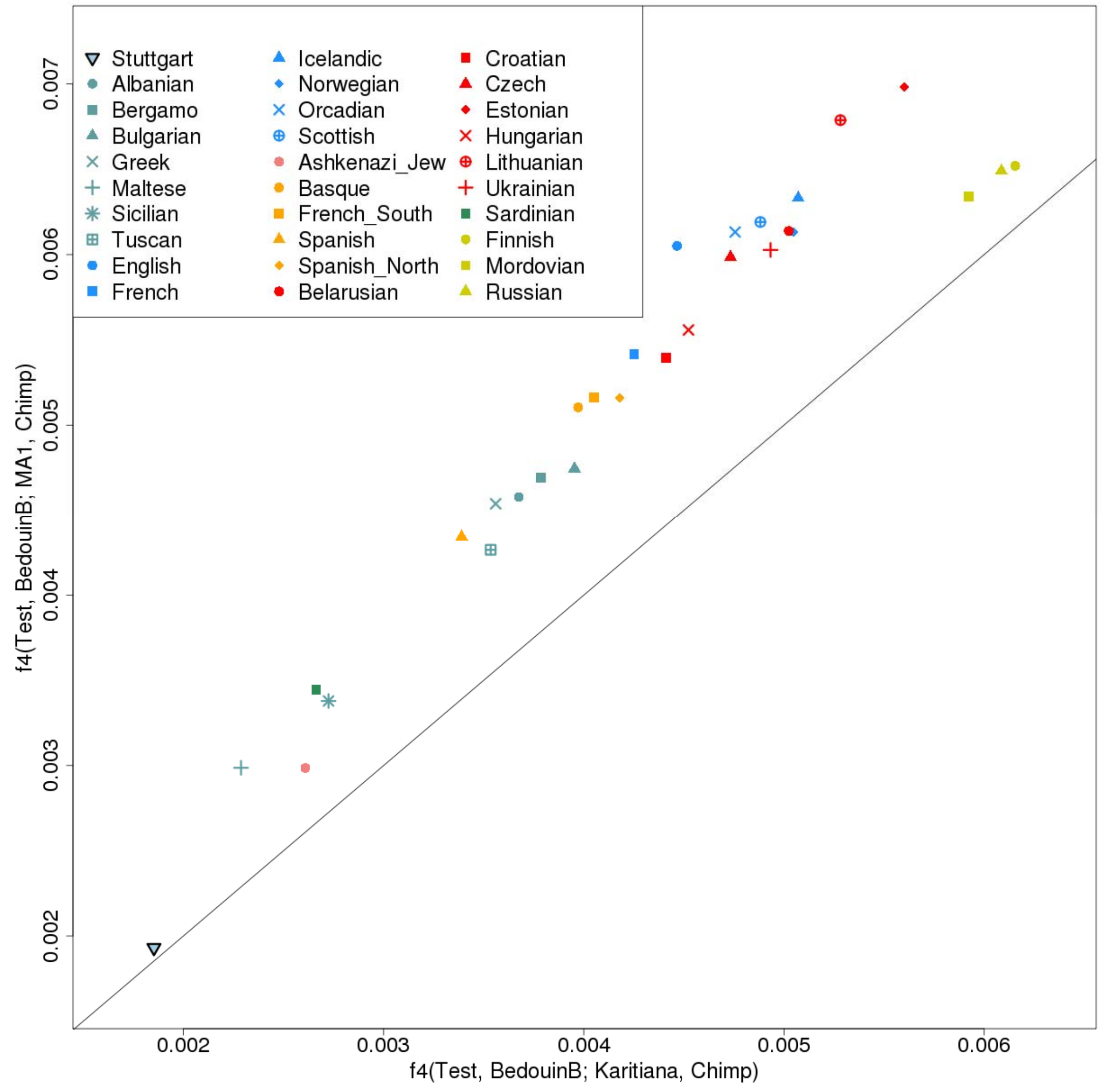

**Extended Data Figure 6: The differential relatedness of West Eurasians to Stuttgart (EEF), Loschbour (WHG), and MA1 (ANE) cannot be explained by two-way mixture.** We plot on a West Eurasian map the statistic *f*$_4$*(Test, Chimp; A*$_1$*, A*$_2$*)*, where $A_1$ and $A_2$ are a pair of the three ancient samples representing the three ancestral populations of Europe. (A) In both Europe and the Near East/Caucasus, populations from the south have more relatedness to Stuttgart than those from the north where ANE influence is also important. (B) Northern European populations share more alleles with Loschbour than with Stuttgart, as they have additional WHG ancestry beyond what was already present in EEF. (C) We observe a striking contrast between Europe west of the Caucasus and the Near East in degree of relatedness to WHG. In Europe, there is a much higher degree of allele sharing with Loschbour than with MA1, which we ascribe to the 60-80% WHG/(WHG+ANE) ratio in most Europeans that we report in SI14. In contrast, the Near East has no appreciable WHG ancestry but some ANE ancestry, especially in the northern Caucasus. (Jewish populations are marked with a square in this figure to assist in interpretation as their ancestry is often anomalous for their geographic regions.)

**A**

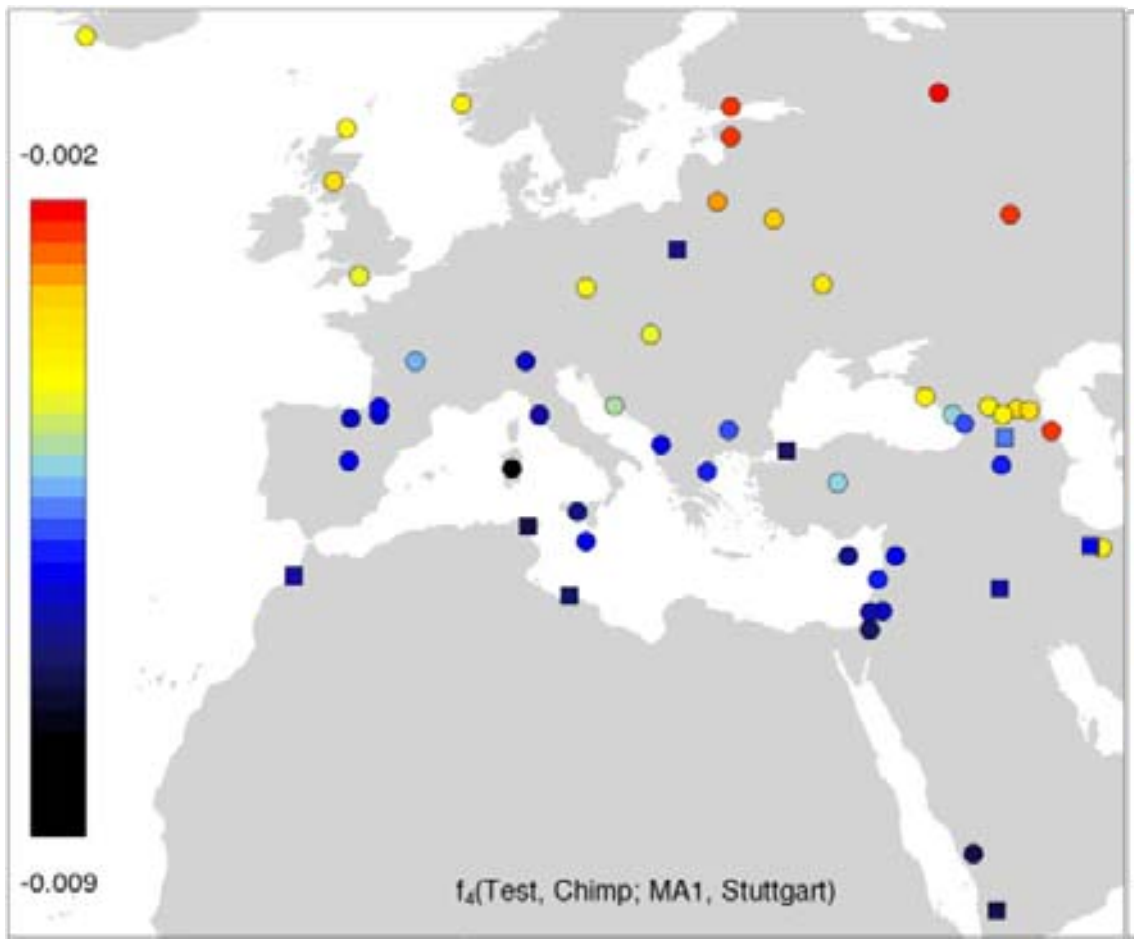

**B**

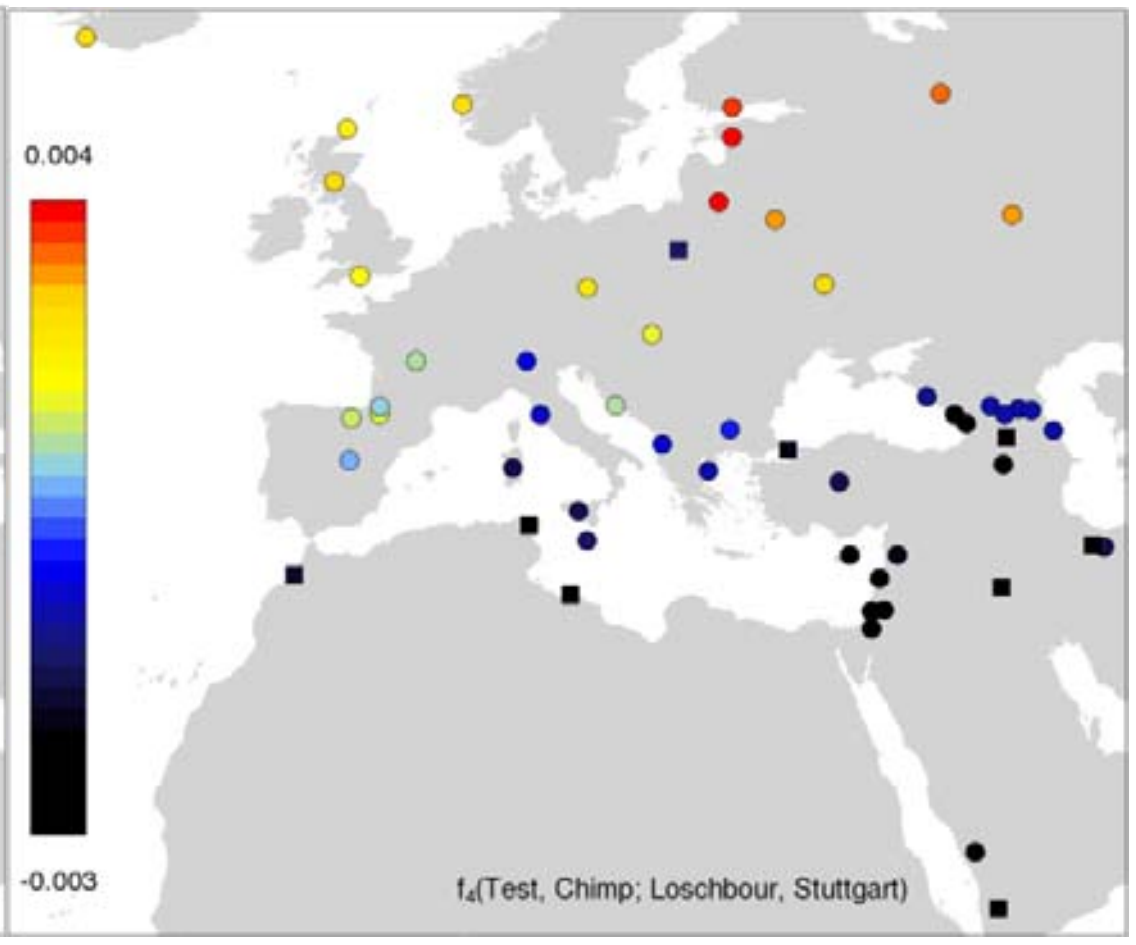

**C**

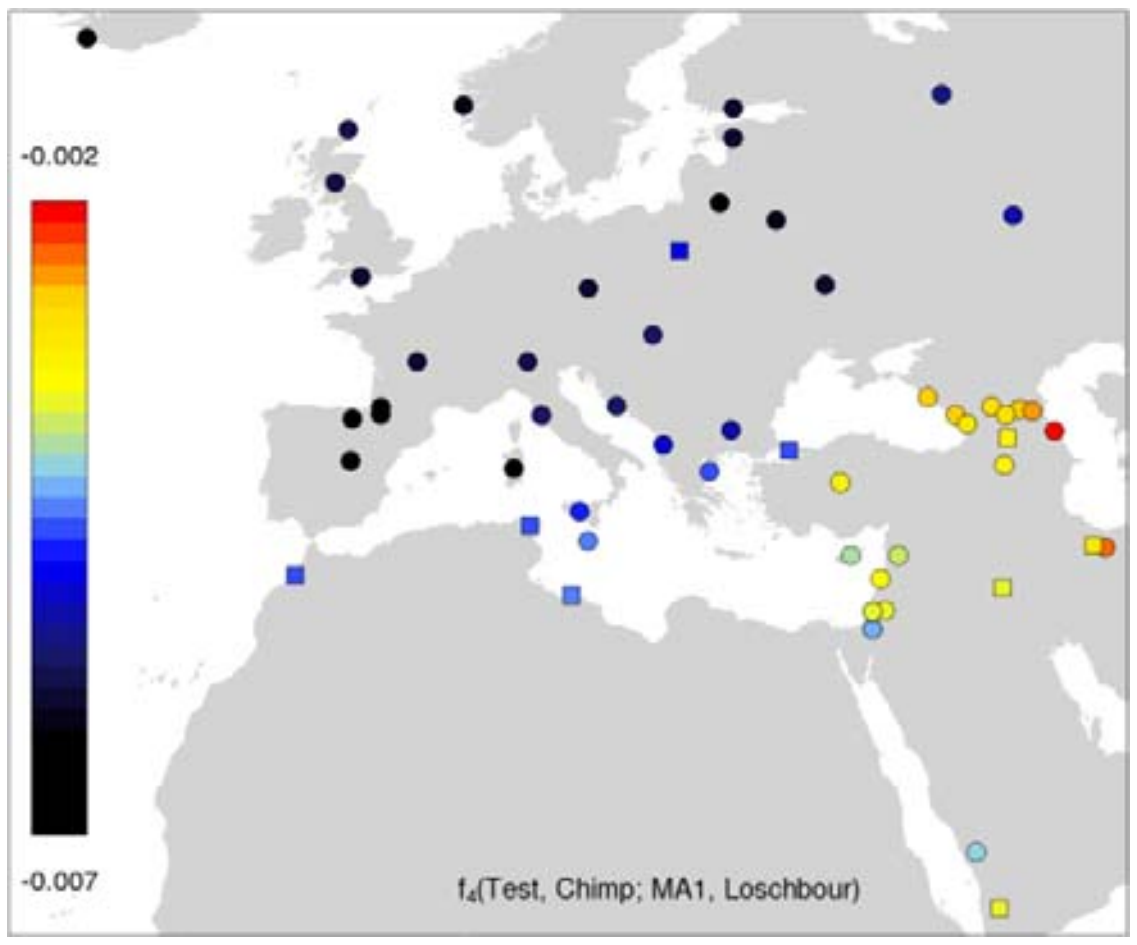

**Extended Data Figure 7: Evidence for Siberian gene flow into far northeastern Europe.** Some northeastern European populations (Chuvash, Finnish, Russian, Mordovian, Saami) share more alleles with Han Chinese than with other Europeans who are arrayed in a cline from Stuttgart to Lithuanians/Estonians in a plot of *$f_4$(Test, BedouinB; Han, Mbuti)* against *$f_4$(Test, BedouinB; MA1, Mbuti)*.

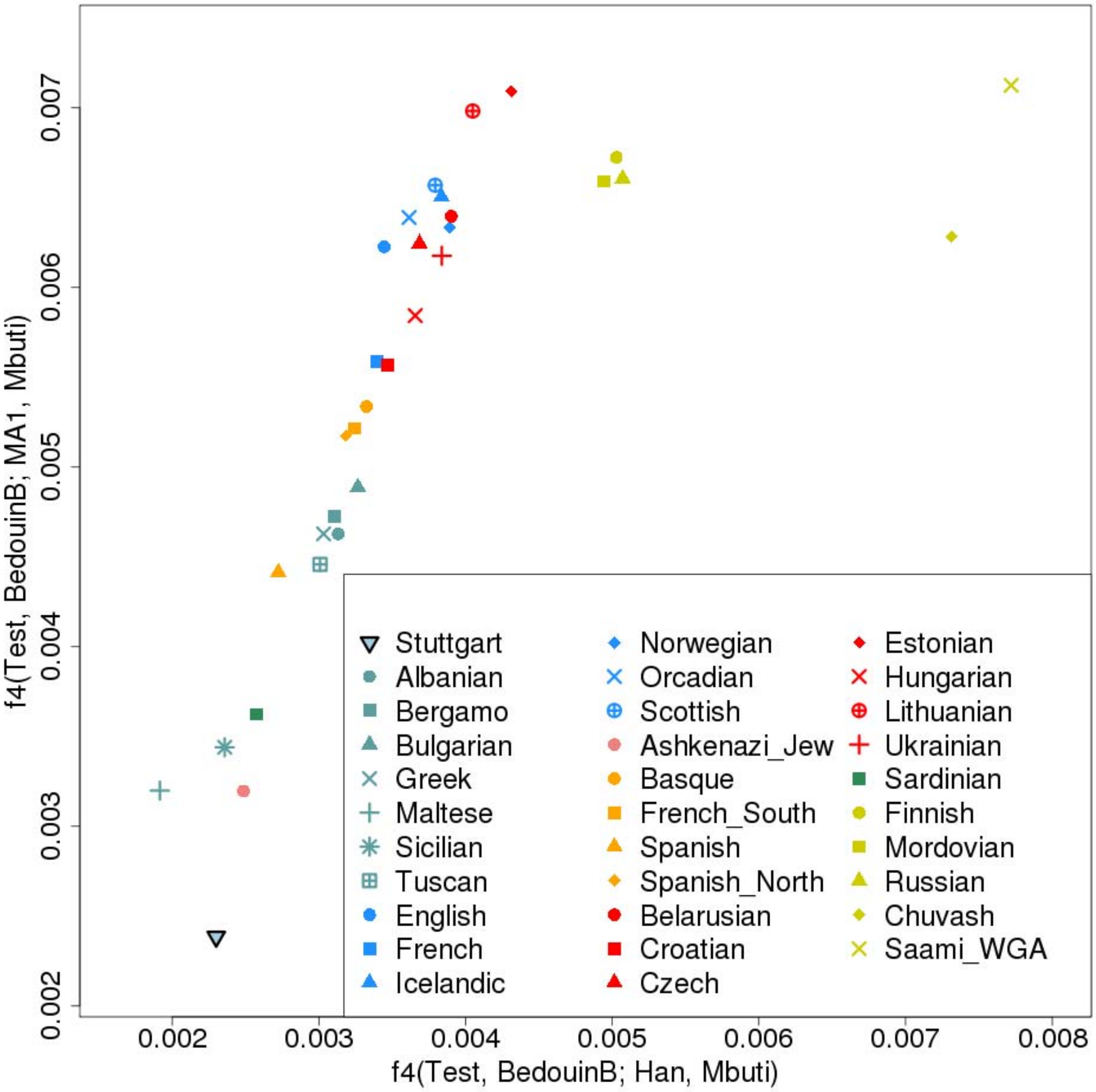

**Extended Data Table 1:** West Eurasians genotyped on the Human Origins array and key *f*-statistics.

| | Sampling Location | | | Lowest $f_3(X; Ref_1, Ref_2)$ | | | | Lowest $f_3(X; EEF, WHG)$ (Z<0 and Zdiff<3 reported) | | | | | Lowest $f_3(X; Near\ East, WHG)$ (Z<0 and Zdiff<3 reported) | | | | | Lowest $f_3(X; EEF, ANE)$ (Z<0 and Zdiff<3 reported) | | | | | $f_4(Stuttgart, X; Loschbour, Chimp)$ | | $f_4(Stuttgart, X; MA1, Chimp)$ | |
|---|---|---|---|---|---|---|---|---|---|---|---|---|---|---|---|---|---|---|---|---|---|---|---|---|---|---|
| X | N | Lat. | Long. | $Ref_1$ | $Ref_2$ | statistic | Z | $Ref_1$ | $Ref_2$ | statistic | Z | Zdiff | $Ref_1$ | $Ref_2$ | statistic | Z | Zdiff | $Ref_1$ | $Ref_2$ | statistic | Z | Zdiff | statistic | Z | statistic | Z |
| Abkhasian | 9 | 43 | 41.02 | Stu | MA1 | -0.0053 | -2.9 | | | | | | Georgian | LaB | -0.0004 | -0.5 | 2.6 | Stu | MA1 | -0.0053 | -2.9 | 0.0 | 0.0020 | 4.2 | -0.0023 | -4.7 |
| Adygei | 17 | 44 | 39 | Piapoco | Stu | -0.0073 | -5.9 | | | | | | | | | | | Stu | MA1 | -0.0067 | -4.1 | 0.3 | 0.0013 | 2.6 | -0.0029 | -6.0 |
| Albanian | 6 | 41.33 | 19.83 | Stu | MA1 | -0.0121 | -7.0 | | | | | | Iraqi_Jew | Los | -0.0090 | -9.1 | 1.7 | Stu | MA1 | -0.0121 | -7.0 | 0.0 | -0.0009 | -1.8 | -0.0027 | -5.4 |
| Armenian | 10 | 40.19 | 44.55 | GujaratiC | Stu | -0.0070 | -8.2 | | | | | | | | | | | Stu | MA1 | -0.0068 | -4.1 | 0.1 | 0.0022 | 4.5 | -0.0016 | -3.3 |
| Ashkenazi_Jew | 7 | 52.23 | 21.02 | Stu | MA1 | -0.0057 | -3.4 | | | | | | Iraqi_Jew | Los | -0.0042 | -4.7 | 1.0 | Stu | MA1 | -0.0057 | -3.4 | 0.0 | 0.0008 | 1.7 | -0.0010 | -2.0 |
| Balkar | 10 | 43.48 | 43.62 | Piapoco | Stu | -0.0113 | -8.9 | | | | | | | | | | | Stu | MA1 | -0.0092 | -5.5 | 1.1 | 0.0014 | 2.9 | -0.0027 | -5.6 |
| Basque | 29 | 43.04 | -0.65 | Iraqi_Jew | Los | -0.0083 | -10.3 | Stu | Los | -0.0061 | -3.8 | 1.3 | Iraqi_Jew | Los | -0.0083 | -10.3 | 0.0 | Stu | MA1 | -0.0041 | -2.4 | 2.2 | -0.0034 | -7.2 | -0.0032 | -6.7 |
| BedouinA | 25 | 31 | 35 | Esan | Stu | -0.0162 | -18.2 | | | | | | | | | | | | | | | | 0.0062 | 13.0 | 0.0026 | 5.4 |
| BedouinB | 19 | 31 | 35 | Esan | Stu | 0.0089 | 7.8 | | | | | | | | | | | | | | | | 0.0046 | 9.3 | 0.0019 | 3.9 |
| Belarusian | 10 | 53.92 | 28.01 | Georgian | Los | -0.0133 | -17.6 | | | | | | Georgian | Los | -0.0133 | -17.6 | 0.0 | Stu | MA1 | -0.0102 | -6.1 | 1.9 | -0.0035 | -6.9 | -0.0042 | -8.6 |
| Bergamo | 12 | 46 | 10 | Stu | MA1 | -0.0106 | -6.2 | Stu | Los | -0.0068 | -4.2 | 1.7 | Iraqi_Jew | Los | -0.0100 | -11.9 | 0.3 | Stu | MA1 | -0.0106 | -6.2 | 0.0 | -0.0018 | -3.9 | -0.0028 | -5.8 |
| Bulgarian | 10 | 42.16 | 24.74 | Stu | MA1 | -0.0130 | -8.2 | Stu | LaB | -0.0074 | -4.5 | 2.8 | Iraqi_Jew | Los | -0.0106 | -12.4 | 1.5 | Stu | MA1 | -0.0130 | -8.2 | 0.0 | -0.0012 | -2.5 | -0.0028 | -5.9 |
| Chechen | 9 | 43.33 | 45.65 | Stu | MA1 | -0.0056 | -3.2 | | | | | | Georgian | Los | -0.0002 | -0.3 | 2.8 | Stu | MA1 | -0.0056 | -3.2 | 0.0 | 0.0011 | 2.3 | -0.0031 | -6.2 |
| Croatian | 10 | 43.51 | 16.45 | Stu | MA1 | -0.0114 | -6.7 | Stu | Los | -0.0065 | -3.8 | 2.1 | Iraqi_Jew | Los | -0.0112 | -13.0 | 0.2 | Stu | MA1 | -0.0114 | -6.7 | 0.0 | -0.0023 | -4.7 | -0.0035 | -7.4 |
| Cypriot | 8 | 35.13 | 33.43 | Stu | MA1 | -0.0057 | -3.2 | | | | | | Yemenite_Jew | Los | -0.0013 | -1.5 | 2.5 | Stu | MA1 | -0.0057 | -3.2 | 0.0 | 0.0019 | 3.9 | -0.0012 | -2.5 |
| Czech | 10 | 50.1 | 14.4 | Georgian | Los | -0.0137 | -17.9 | Stu | Los | -0.0088 | -5.3 | 3.0 | Georgian | Los | -0.0137 | -17.9 | 0.0 | Stu | MA1 | -0.0121 | -7.2 | 0.9 | -0.0032 | -6.6 | -0.0040 | -8.2 |
| Druze | 39 | 32 | 35 | Stu | MA1 | -0.0024 | -1.5 | | | | | | | | | | | Stu | MA1 | -0.0024 | -1.5 | 0.0 | 0.0028 | 5.9 | -0.0006 | -1.3 |
| English | 10 | 50.75 | -2.09 | Iraqi_Jew | Los | -0.0129 | -14.8 | Stu | Los | -0.0090 | -5.5 | 2.2 | Iraqi_Jew | Los | -0.0129 | -14.8 | 0.0 | Stu | MA1 | -0.0125 | -7.4 | 0.1 | -0.0032 | -6.5 | -0.0041 | -8.5 |
| Estonian | 10 | 58.54 | 24.89 | Abkhasian | Los | -0.0124 | -15.1 | | | | | | Abkhasian | Los | -0.0124 | -15.1 | 0.0 | Stu | MA1 | -0.0094 | -5.6 | 1.9 | -0.0043 | -8.5 | -0.0051 | - |
| Finnish | 7 | 60.2 | 24.9 | Abkhasian | Los | -0.0102 | -11.3 | | | | | | Abkhasian | Los | -0.0102 | -11.3 | 0.0 | Stu | MA1 | -0.0078 | -4.4 | 1.4 | -0.0035 | -6.9 | -0.0045 | -9.1 |
| French | 25 | 46 | 2 | Stu | MA1 | -0.0131 | -8.4 | Stu | Los | -0.0098 | -6.3 | 1.5 | Iraqi_Jew | Los | -0.0129 | -16.8 | 0.2 | Stu | MA1 | -0.0131 | -8.4 | 0.0 | -0.0027 | -5.6 | -0.0036 | -7.7 |
| French_South | 7 | 43.44 | -0.62 | Iraqi_Jew | Los | -0.0095 | -9.5 | Stu | LaB | -0.0089 | -5.0 | 0.3 | Iraqi_Jew | Los | -0.0095 | -9.5 | 0.0 | Stu | MA1 | -0.0086 | -4.8 | 0.4 | -0.0030 | -6.2 | -0.0031 | -6.2 |
| Georgian | 10 | 42.5 | 41.85 | GujaratiC | Stu | -0.0036 | -4.0 | | | | | | | | | | | Stu | MA1 | -0.0036 | -2.1 | -0.2 | 0.0020 | 4.2 | -0.0019 | -3.9 |
| Georgian_Jew | 7 | 41.72 | 44.78 | GujaratiC | Stu | -0.0009 | -0.9 | | | | | | | | | | | Stu | MA1 | -0.0002 | -0.1 | 0.3 | 0.0022 | 4.3 | -0.0017 | -3.4 |
| Greek | 20 | 39.84 | 23.17 | Stu | MA1 | -0.0118 | -7.4 | | | | | | Iraqi_Jew | Los | -0.0080 | -11.1 | 2.3 | Stu | MA1 | -0.0118 | -7.4 | 0.0 | -0.0004 | -0.9 | -0.0026 | -5.6 |
| Hungarian | 20 | 47.49 | 19.08 | Stu | MA1 | -0.0133 | -8.4 | Stu | Los | -0.0087 | -5.6 | 2.2 | Iraqi_Jew | Los | -0.0127 | -15.9 | 0.4 | Stu | MA1 | -0.0133 | -8.4 | 0.0 | -0.0025 | -5.3 | -0.0037 | -7.8 |
| Icelandic | 12 | 64.13 | -21.93 | Abkhasian | Los | -0.0121 | -15.6 | Stu | Los | -0.0078 | -4.8 | 2.7 | Abkhasian | Los | -0.0121 | -15.6 | 0.0 | Stu | MA1 | -0.0097 | -5.9 | 1.5 | -0.0038 | -7.7 | -0.0043 | -8.9 |
| Iranian | 8 | 35.59 | 51.46 | Piapoco | Stu | -0.0094 | -7.2 | | | | | | | | | | | Stu | MA1 | -0.0087 | -5.2 | 0.4 | 0.0031 | 6.3 | -0.0016 | -3.2 |
| Iranian_Jew | 9 | 35.7 | 51.42 | GujaratiC | Stu | -0.0018 | -2.0 | | | | | | | | | | | Stu | MA1 | -0.0012 | -0.6 | 0.2 | 0.0028 | 5.7 | -0.0011 | -2.2 |
| Iraqi_Jew | 6 | 33.33 | 44.42 | Vishwabrahmin | Stu | -0.0026 | -2.6 | | | | | | | | | | | Stu | MA1 | -0.0009 | -0.5 | 0.9 | 0.0030 | 6.1 | -0.0005 | -1.0 |
| Jordanian | 9 | 32.05 | 35.91 | Esan | Stu | -0.0145 | -14.3 | | | | | | | | | | | | | | | | 0.0048 | 9.6 | 0.0014 | 2.8 |
| Kumyk | 8 | 43.25 | 46.58 | Piapoco | Stu | -0.0111 | -8.2 | | | | | | | | | | | Stu | MA1 | -0.0109 | -6.5 | 0.1 | 0.0015 | 3.1 | -0.0028 | -5.7 |
| Lebanese | 8 | 33.82 | 35.57 | Esan | Stu | -0.0105 | -9.4 | | | | | | | | | | | Stu | MA1 | -0.0068 | -3.9 | 1.9 | 0.0038 | 7.7 | 0.0002 | 0.4 |
| Lezgin | 9 | 42.12 | 48.18 | Stu | MA1 | -0.0100 | -6.0 | | | | | | | | | | | Stu | MA1 | -0.0100 | -6.0 | 0.0 | 0.0013 | 2.7 | -0.0037 | -7.5 |
| Libyan_Jew | 9 | 32.92 | 13.18 | Esan | Stu | -0.0051 | -4.4 | | | | | | | | | | | Stu | MA1 | 0.0000 | 0.0 | 2.7 | 0.0030 | 6.2 | 0.0004 | 0.9 |
| Lithuanian | 10 | 54.9 | 23.92 | Abkhasian | Los | -0.0119 | -14.9 | | | | | | Abkhasian | Los | -0.0119 | -14.9 | 0.0 | Stu | MA1 | -0.0069 | -3.9 | 2.8 | -0.0045 | -9.0 | -0.0048 | -9.9 |
| Maltese | 8 | 35.94 | 14.38 | Stu | MA1 | -0.0086 | -4.9 | | | | | | Yemenite_Jew | Los | -0.0051 | -6.0 | 2.0 | Stu | MA1 | -0.0086 | -4.9 | 0.0 | 0.0013 | 2.7 | -0.0011 | -2.3 |
| Mordovian | 10 | 54.18 | 45.18 | Abkhasian | Los | -0.0115 | -14.4 | | | | | | Abkhasian | Los | -0.0115 | -14.4 | 0.0 | Stu | MA1 | -0.0113 | -6.6 | 0.3 | -0.0028 | -5.5 | -0.0044 | -9.0 |
| Moroccan_Jew | 6 | 34.02 | -6.84 | Esan | Stu | -0.0062 | -5.2 | | | | | | Yemenite_Jew | Los | -0.0021 | -2.2 | 2.9 | Stu | MA1 | -0.0032 | -1.7 | 1.4 | 0.0021 | 4.3 | -0.0001 | -0.1 |
| North_Ossetian | 10 | 43.02 | 44.65 | Piapoco | Stu | -0.0093 | -7.2 | | | | | | | | | | | Stu | MA1 | -0.0076 | -4.4 | 1.0 | 0.0014 | 2.9 | -0.0028 | -5.6 |
| Norwegian | 11 | 60.36 | 5.36 | Georgian | Los | -0.0120 | -14.8 | | | | | | Georgian | Los | -0.0120 | -14.8 | 0.0 | Stu | MA1 | -0.0093 | -5.4 | 1.4 | -0.0035 | -7.3 | -0.0042 | -8.7 |
| Orcadian | 13 | 59 | -3 | Armenian | Los | -0.0102 | -13.4 | Stu | Los | -0.0059 | -3.6 | 2.5 | Armenian | Los | -0.0102 | -13.4 | 0.0 | Stu | MA1 | -0.0098 | -5.9 | 0.5 | -0.0032 | -6.7 | -0.0042 | -8.6 |
| Palestinian | 38 | 32 | 35 | Esan | Stu | -0.0120 | -13.2 | | | | | | | | | | | | | | | | 0.0047 | 10.2 | 0.0014 | 3.1 |
| Russian | 22 | 61 | 40 | Chukchi | Los | -0.0119 | -11.3 | | | | | | Abkhasian | Los | -0.0119 | -17.1 | 0.0 | Stu | MA1 | -0.0106 | -6.6 | 0.8 | -0.0030 | -6.2 | -0.0046 | -9.4 |
| Sardinian | 27 | 40 | 9 | Stu | LaB | -0.0044 | -2.6 | Stu | LaB | -0.0044 | -2.6 | 0.0 | Iraqi_Jew | Los | -0.0033 | -4.2 | 0.0 | Stu | MA1 | -0.0035 | -2.1 | 0.3 | -0.0016 | -3.4 | -0.0015 | -3.3 |
| Saudi | 8 | 18.49 | 42.52 | Kgalagadi | Stu | -0.0042 | -3.6 | | | | | | | | | | | | | | | | 0.0042 | 8.6 | 0.0015 | 3.1 |
| Scottish | 4 | 56.04 | -3.94 | Iraqi_Jew | Los | -0.0103 | -8.3 | | | | | | Iraqi_Jew | Los | -0.0103 | -8.3 | 0.0 | Stu | MA1 | -0.0090 | -4.7 | 0.7 | -0.0034 | -6.4 | -0.0045 | -8.7 |
| Sicilian | 11 | 37.59 | 13.77 | Stu | MA1 | -0.0108 | -6.5 | | | | | | Yemenite_Jew | Los | -0.0066 | -8.1 | 2.4 | Stu | MA1 | -0.0108 | -6.5 | 0.0 | 0.0006 | 1.3 | -0.0015 | -3.2 |
| Spanish | 53 | 40.43 | -2.83 | Iraqi_Jew | Los | -0.0126 | -17.8 | Stu | Los | -0.0104 | -6.8 | 1.4 | Iraqi_Jew | Los | -0.0126 | -17.8 | 0.0 | Stu | MA1 | -0.0120 | -7.6 | 0.3 | -0.0019 | -4.2 | -0.0024 | -5.2 |
| Spanish_North | 5 | 42.8 | -2.7 | Iraqi_Jew | Los | -0.0112 | -9.9 | Stu | Los | -0.0102 | -5.4 | 0.5 | Iraqi_Jew | Los | -0.0112 | -9.9 | 0.0 | Stu | MA1 | -0.0082 | -4.4 | 1.3 | -0.0035 | -6.9 | -0.0032 | -6.4 |
| Syrian | 8 | 35.13 | 36.87 | Esan | Stu | -0.0101 | -8.7 | | | | | | | | | | | | | | | | 0.0044 | 8.6 | 0.0012 | 2.4 |
| Tunisian_Jew | 7 | 36.8 | 10.18 | Gambian | Stu | -0.0026 | -2.0 | | | | | | | | | | | | | | | | 0.0026 | 5.2 | 0.0002 | 0.5 |
| Turkish | 56 | 39.22 | 32.66 | Piapoco | Stu | -0.0129 | -11.3 | | | | | | | | | | | Stu | MA1 | -0.0106 | -6.9 | 1.3 | 0.0018 | 3.8 | -0.0019 | -4.0 |
| Turkish_Jew | 8 | 41.02 | 28.95 | Stu | MA1 | -0.0075 | -4.3 | | | | | | Yemenite_Jew | Los | -0.0049 | -5.8 | 1.4 | Stu | MA1 | -0.0075 | -4.3 | 0.0 | 0.0017 | 3.6 | -0.0006 | -1.3 |
| Tuscan | 8 | 43 | 11 | Stu | MA1 | -0.0109 | -6.4 | Stu | Los | -0.0055 | -3.2 | 2.3 | Iraqi_Jew | Los | -0.0092 | -10.1 | 0.9 | Stu | MA1 | -0.0109 | -6.4 | 0.0 | -0.0011 | -2.2 | -0.0024 | -5.0 |
| Ukrainian | 9 | 50.29 | 31.56 | Georgian | Los | -0.0134 | -16.7 | | | | | | Georgian | Los | -0.0134 | -16.7 | 0.0 | Stu | MA1 | -0.0114 | -6.6 | 1.3 | -0.0032 | -6.4 | -0.0041 | -8.5 |
| Yemenite_Jew | 8 | 15.35 | 44.2 | Esan | Stu | -0.0027 | -2.4 | | | | | | | | | | | | | | | | 0.0046 | 9.1 | 0.0013 | 2.6 |

Note: *Zdiff* is the number of standard errors of the difference between the lowest $f_3$-statistic over all reference pairs and the lowest $f_3$-statistic for a subset of reference pairs.

## Extended Data Table 2: Confirmation of key findings on transversions and on whole genome sequence data.

| Interpretation | D(A, B; C, D) on Human Origins genotype data | | | | | | | | D(A, B; C, D) on whole genome sequence data transversions | | | | | |
|---|---|---|---|---|---|---|---|---|---|---|---|---|---|---|
| | | | | | 594,924 SNPs | | 110,817 transversions | | | | | | | |
| | A | B | C | D | statistic | Z | statistic | Z | A | B | C | D | statistic | Z |
| Stuttgart has Near Eastern ancestry | Stuttgart | Armenian | Loschbour | Chimp | 0.0219 | 4.5 | 0.0189 | 2.9 | | | | | | |
| Europeans have more WHG-related ancestry than Stuttgart | Stuttgart | French | Loschbour | Chimp | -0.0266 | -5.7 | -0.031 | -5.0 | Stuttgart | French2 | Loschbour | Chimp | -0.03 | -4.7 |
| | Lithuanian | Stuttgart | Loschbour | Chimp | 0.0446 | 9.1 | 0.0477 | 7.2 | | | | | | |
| West Eurasians have more ANE-related ancestry than Stuttgart | French | Stuttgart | MA1 | Chimp | 0.0367 | 7.7 | 0.0386 | 5.5 | French2 | Stuttgart | MA1 | Chimp | 0.037 | 6.4 |
| | Lezgin | Stuttgart | MA1 | Chimp | 0.0372 | 7.6 | 0.0409 | 5.6 | | | | | | |
| MA1 is a better surrogate of ANE ancestry than Karitiana | French | Chimp | MA1 | Karitiana | 0.0207 | 4.5 | 0.0214 | 2.8 | French2 | Chimp | MA1 | Karitiana2 | 0.026 | 3.8 |
| Eastern non-Africans closer to WHG/ANE/SHG than to EEF | Loschbour | Stuttgart | Onge | Chimp | 0.0196 | 3.5 | 0.0202 | 2.5 | | | | | | |
| | Loschbour | Stuttgart | Papuan | Chimp | 0.0142 | 2.6 | 0.0127 | 1.5 | Loschbour | Stuttgart | Papuan2 | Chimp | 0.017 | 2.7 |
| | Loschbour | Stuttgart | Dai | Chimp | 0.0164 | 3.2 | 0.021 | 2.8 | Loschbour | Stuttgart | Dai2 | Chimp | 0.018 | 2.9 |
| | MA1 | Stuttgart | Papuan | Chimp | 0.0139 | 2.2 | 0.0103 | 1.0 | MA1 | Stuttgart | Papuan2 | Chimp | 0.018 | 2.8 |
| | MA1 | Stuttgart | Dai | Chimp | 0.0174 | 3.0 | 0.016 | 1.7 | MA1 | Stuttgart | Dai2 | Chimp | 0.028 | 4.3 |
| | Motala12 | Stuttgart | Papuan | Chimp | 0.0182 | 3.2 | 0.011 | 1.1 | Motala12 | Stuttgart | Papuan2 | Chimp | 0.023 | 3.7 |
| | Motala12 | Stuttgart | Dai | Chimp | 0.0156 | 2.8 | 0.0149 | 1.6 | Motala12 | Stuttgart | Dai2 | Chimp | 0.02 | 3.2 |
| | LaBrana | Stuttgart | Papuan | Chimp | 0.0123 | 2.3 | 0.0101 | 1.1 | LaBrana | Stuttgart | Papuan2 | Chimp | 0.02 | 3.2 |
| | LaBrana | Stuttgart | Dai | Chimp | 0.0149 | 2.9 | 0.0228 | 2.5 | LaBrana | Stuttgart | Dai2 | Chimp | 0.024 | 3.7 |
| Native Americans closer to ANE than to WHG | Karitiana | Chimp | MA1 | Loschbour | 0.0467 | 7.1 | 0.0467 | 4.4 | Karitiana2 | Chimp | MA1 | Loschbour | 0.052 | 7.1 |
| West Eurasians closer to Native Americans than to other Eastern non-Africans | Stuttgart | Chimp | Karitiana | Papuan | 0.0559 | 10.9 | 0.0474 | 6.6 | Stuttgart | Chimp | Karitiana2 | Papuan2 | 0.052 | 7.6 |
| | Stuttgart | Chimp | Karitiana | Onge | 0.0237 | 5.1 | 0.0179 | 2.6 | | | | | | |
| Ancient Eurasian hunter-gatherers equally related to Eastern non-Africans other than Native Americans | Loschbour | MA1 | Dai | Chimp | -0.0015 | -0.2 | 0.0016 | 0.2 | Loschbour | MA1 | Dai2 | Chimp | -0.013 | -1.9 |
| | Loschbour | MA1 | Papuan | Chimp | 0.0002 | 0.0 | 0.0012 | 0.1 | Loschbour | MA1 | Papuan2 | Chimp | -0.003 | -0.4 |
| | Loschbour | Motala12 | Dai | Chimp | 0.0024 | 0.4 | 0.009 | 0.9 | Loschbour | Motala12 | Dai2 | Chimp | -0.002 | -0.3 |
| | Loschbour | Motala12 | Papuan | Chimp | -0.0028 | -0.4 | 0.0046 | 0.5 | Loschbour | Motala12 | Papuan2 | Chimp | -0.004 | -0.6 |
| | MA1 | Motala12 | Dai | Chimp | 0.0026 | 0.4 | 0.0047 | 0.4 | MA1 | Motala12 | Dai2 | Chimp | 0.01 | 1.5 |
| | MA1 | Motala12 | Papuan | Chimp | -0.0047 | -0.7 | -0.001 | -0.1 | MA1 | Motala12 | Papuan2 | Chimp | -0.004 | -0.5 |
| LaBrana and Loschbour are a clade | LaBrana | Loschbour | Dai | Chimp | -0.0028 | -0.5 | 0.0024 | 0.3 | LaBrana | Loschbour | Dai2 | Chimp | 0.007 | 1.1 |
| | LaBrana | Loschbour | Papuan | Chimp | -0.0031 | -0.5 | -0.0012 | -0.1 | LaBrana | Loschbour | Papuan2 | Chimp | 0.002 | 0.3 |
| | LaBrana | Loschbour | MA1 | Chimp | -0.006 | -0.8 | 0.0101 | 0.7 | LaBrana | Loschbour | MA1 | Chimp | 0.005 | 0.7 |
| SHG closer to ANE than to WHG | Motala12 | Loschbour | MA1 | Chimp | 0.0425 | 5.3 | 0.0353 | 2.6 | Motala12 | Loschbour | MA1 | Chimp | 0.042 | 5.9 |
| | Motala12 | LaBrana | MA1 | Chimp | 0.0465 | 5.8 | 0.0347 | 2.4 | Motala12 | LaBrana | MA1 | Chimp | 0.038 | 5.4 |
| LaBrana and Loschbour equally related to Stuttgart | LaBrana | Loschbour | Stuttgart | Chimp | -0.0176 | -2.6 | -0.0106 | -1.0 | LaBrana | Loschbour | Stuttgart | Chimp | -0.012 | -1.8 |

**Extended Data Table 3:** Admixture proportions for European populations. The estimates from the model with minimal assumptions are from SI17. The estimates from the full modeling are from SI14 either by single population analysis or co-fitting population pairs and averaging over fits (these averages are the results plotted in Fig. 2B). Populations that do not fit the models are not reported.

| | Full modeling of population relationships (individual fits) | | | Full modeling of population relationships (averaged fits) | | | | | | Modeling of population relationships with minimal assumptions | | | Model-based (averaged) - Model with minimal assumptions (Z-score) | | |
|---|---|---|---|---|---|---|---|---|---|---|---|---|---|---|---|
| | EEF | WHG | ANE | EEF | | WHG | | ANE | | EEF | WHG | ANE | EEF | WHG | ANE |
| | | | | Mean | Range | Mean | Range | Mean | Range | | | | | | |
| Albanian | 0.781 | 0.092 | 0.127 | 0.781 | 0.772-0.819 | 0.082 | 0.032-0.098 | 0.137 | 0.129-0.158 | 0.595 ± 0.112 | 0.353 ± 0.150 | 0.052 ± 0.049 | 1.658 | -1.807 | 1.741 |
| Ashkenazi_Jew | 0.931 | 0 | 0.069 | | | | | | | 0.938 ± 0.146 | -0.021 ± 0.185 | 0.083 ± 0.049 | | | |
| Basque | 0.593 | 0.293 | 0.114 | 0.569 | 0.527-0.616 | 0.335 | 0.255-0.392 | 0.096 | 0.076-0.129 | 0.569 ± 0.091 | 0.315 ± 0.124 | 0.115 ± 0.041 | -0.001 | 0.165 | -0.472 |
| Belarusian | 0.418 | 0.431 | 0.151 | 0.426 | 0.397-0.464 | 0.408 | 0.338-0.443 | 0.167 | 0.150-0.199 | 0.272 ± 0.094 | 0.554 ± 0.131 | 0.174 ± 0.047 | 1.637 | -1.118 | -0.158 |
| Bergamo | 0.715 | 0.177 | 0.108 | 0.721 | 0.704-0.793 | 0.163 | 0.061-0.189 | 0.117 | 0.104-0.147 | 0.644 ± 0.125 | 0.248 ± 0.170 | 0.108 ± 0.053 | 0.615 | -0.503 | 0.162 |
| Bulgarian | 0.712 | 0.147 | 0.141 | 0.718 | 0.707-0.778 | 0.132 | 0.047-0.151 | 0.151 | 0.138-0.175 | 0.556 ± 0.110 | 0.328 ± 0.143 | 0.116 ± 0.043 | 1.469 | -1.372 | 0.804 |
| Croatian | 0.561 | 0.293 | 0.145 | 0.564 | 0.548-0.586 | 0.285 | 0.242-0.310 | 0.151 | 0.137-0.172 | 0.453 ± 0.122 | 0.407 ± 0.159 | 0.140 ± 0.046 | 0.911 | -0.768 | 0.238 |
| Czech | 0.495 | 0.338 | 0.167 | 0.489 | 0.460-0.531 | 0.348 | 0.273-0.382 | 0.163 | 0.145-0.196 | 0.402 ± 0.117 | 0.400 ± 0.162 | 0.198 ± 0.050 | 0.744 | -0.322 | -0.698 |
| English | 0.495 | 0.364 | 0.141 | 0.503 | 0.476-0.536 | 0.353 | 0.296-0.382 | 0.144 | 0.130-0.169 | 0.475 ± 0.091 | 0.357 ± 0.125 | 0.168 ± 0.043 | 0.304 | -0.028 | -0.561 |
| Estonian | 0.322 | 0.495 | 0.183 | 0.323 | 0.293-0.345 | 0.49 | 0.451-0.520 | 0.187 | 0.172-0.205 | 0.072 ± 0.121 | 0.778 ± 0.176 | 0.150 ± 0.064 | 2.070 | -1.636 | 0.584 |
| French | 0.554 | 0.311 | 0.135 | 0.563 | 0.537-0.601 | 0.297 | 0.230-0.328 | 0.14 | 0.126-0.169 | 0.498 ± 0.097 | 0.359 ± 0.127 | 0.142 ± 0.039 | 0.672 | -0.487 | -0.060 |
| French_South | 0.675 | 0.195 | 0.13 | 0.636 | 0.589-0.738 | 0.256 | 0.111-0.323 | 0.108 | 0.088-0.151 | 0.636 ± 0.116 | 0.225 ± 0.165 | 0.140 ± 0.057 | -0.003 | 0.189 | -0.558 |
| Greek | 0.792 | 0.058 | 0.151 | 0.791 | 0.780-0.816 | 0.048 | 0.019-0.060 | 0.161 | 0.150-0.171 | 0.658 ± 0.098 | 0.255 ± 0.127 | 0.086 ± 0.039 | 1.357 | -1.627 | 1.915 |
| Hungarian | 0.558 | 0.264 | 0.179 | 0.548 | 0.520-0.590 | 0.279 | 0.199-0.313 | 0.174 | 0.156-0.210 | 0.391 ± 0.109 | 0.454 ± 0.153 | 0.155 ± 0.050 | 1.437 | -1.145 | 0.371 |
| Icelandic | 0.394 | 0.456 | 0.15 | 0.409 | 0.386-0.424 | 0.448 | 0.409-0.473 | 0.143 | 0.126-0.170 | 0.342 ± 0.102 | 0.476 ± 0.137 | 0.182 ± 0.045 | 0.654 | -0.204 | -0.861 |
| Lithuanian | 0.364 | 0.464 | 0.172 | 0.352 | 0.327-0.384 | 0.488 | 0.433-0.527 | 0.16 | 0.135-0.184 | 0.248 ± 0.117 | 0.548 ± 0.163 | 0.205 ± 0.052 | 0.886 | -0.367 | -0.864 |
| Maltese | 0.932 | 0 | 0.068 | | | | | | | 1.298 ± 0.185 | -0.509 ± 0.248 | 0.211 ± 0.079 | | | |
| Norwegian | 0.411 | 0.428 | 0.161 | 0.417 | 0.388-0.438 | 0.423 | 0.383-0.450 | 0.16 | 0.140-0.181 | 0.273 ± 0.115 | 0.557 ± 0.161 | 0.170 ± 0.055 | 1.252 | -0.831 | -0.185 |
| Orcadian | 0.457 | 0.385 | 0.158 | 0.465 | 0.439-0.493 | 0.378 | 0.329-0.403 | 0.157 | 0.140-0.179 | 0.395 ± 0.088 | 0.437 ± 0.122 | 0.168 ± 0.041 | 0.798 | -0.487 | -0.264 |
| Sardinian | 0.817 | 0.175 | 0.008 | 0.818 | 0.791-0.874 | 0.141 | 0.058-0.182 | 0.041 | 0.026-0.068 | 0.883 ± 0.128 | 0.075 ± 0.166 | 0.042 ± 0.048 | -0.510 | 0.400 | -0.024 |
| Scottish | 0.39 | 0.428 | 0.182 | 0.408 | 0.387-0.424 | 0.421 | 0.384-0.448 | 0.171 | 0.149-0.201 | 0.286 ± 0.112 | 0.532 ± 0.156 | 0.182 ± 0.053 | 1.091 | -0.712 | -0.210 |
| Sicilian | 0.903 | 0 | 0.097 | | | | | | | 1.012 ± 0.149 | -0.131 ± 0.199 | 0.119 ± 0.060 | | | |
| Spanish | 0.809 | 0.068 | 0.123 | 0.759 | 0.736-0.804 | 0.126 | 0.066-0.170 | 0.115 | 0.091-0.151 | 0.856 ± 0.126 | -0.015 ± 0.165 | 0.160 ± 0.049 | -0.769 | 0.855 | -0.922 |
| Spanish_North | 0.713 | 0.125 | 0.163 | 0.612 | 0.561-0.660 | 0.292 | 0.214-0.365 | 0.096 | 0.072-0.126 | 0.581 ± 0.120 | 0.298 ± 0.158 | 0.121 ± 0.046 | 0.254 | -0.038 | -0.533 |
| Tuscan | 0.746 | 0.136 | 0.118 | 0.751 | 0.737-0.806 | 0.123 | 0.047-0.145 | 0.126 | 0.114-0.150 | 0.734 ± 0.118 | 0.153 ± 0.160 | 0.113 ± 0.054 | 0.141 | -0.188 | 0.249 |
| Ukrainian | 0.462 | 0.387 | 0.151 | 0.463 | 0.445-0.491 | 0.376 | 0.322-0.399 | 0.16 | 0.148-0.187 | 0.259 ± 0.123 | 0.596 ± 0.173 | 0.145 ± 0.057 | 1.661 | -1.269 | 0.269 |
| Finnish | | | | | | | | | | -0.299 ± 0.204 | 1.194 ± 0.296 | 0.105 ± 0.105 | | | |
| Mordovian | | | | | | | | | | -0.255 ± 0.173 | 1.151 ± 0.246 | 0.104 ± 0.090 | | | |
| Russian | | | | | | | | | | -0.303 ± 0.211 | 1.230 ± 0.301 | 0.072 ± 0.106 | | | |

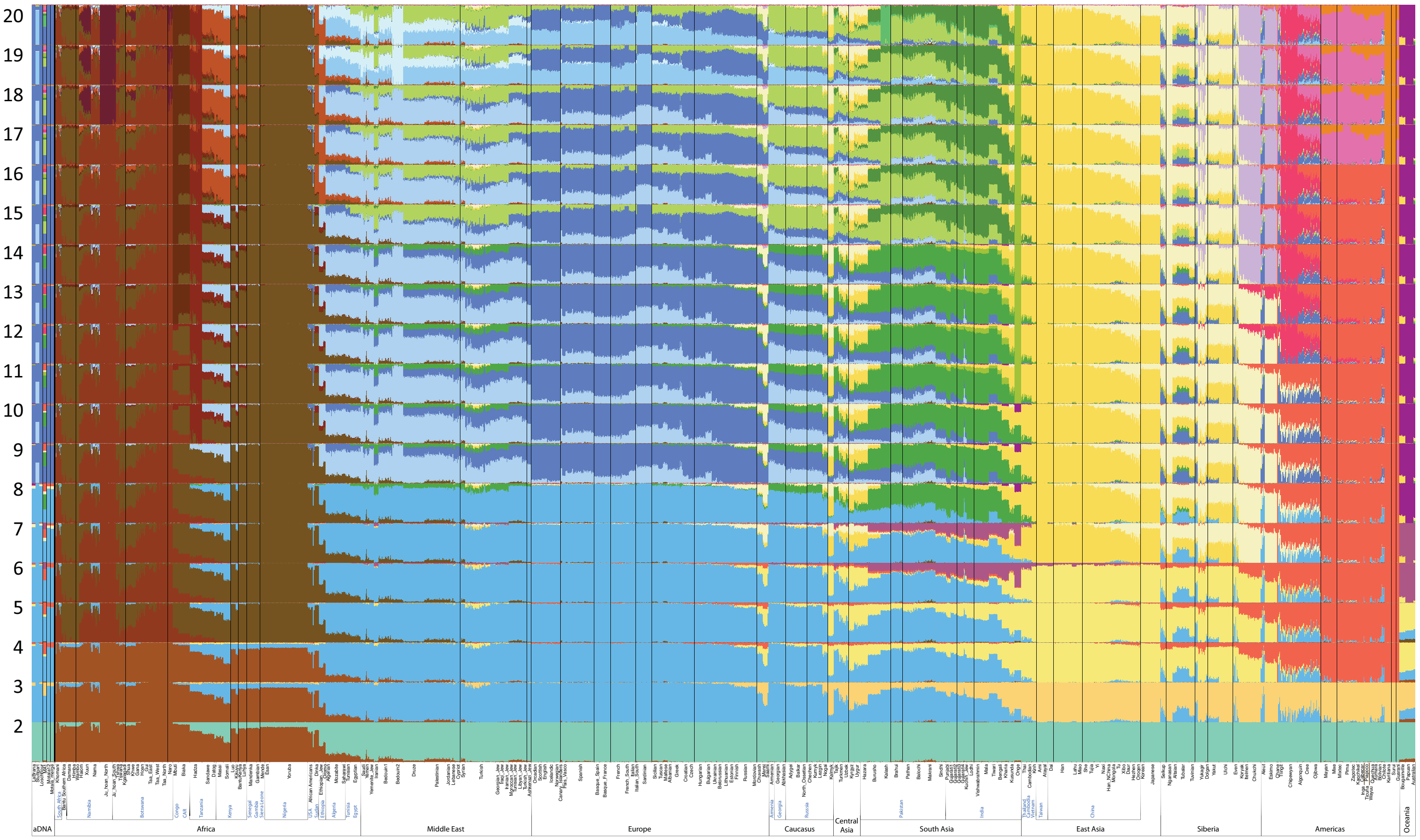

# Supplementary Information

## Ancient human genomes suggest three ancestral populations for present-day Europeans

# Supplementary Information 1

## Sampling, Library Preparation and Sequencing

Alissa Mittnik*, Susanna Sawyer, Ruth Bollongino, Christos Economou, Dominique Delsate, Michael Francken, Joachim Wahl, Johannes Krause

* To whom correspondence should be addressed (amittnik@gmail.com)

### Samples and extraction

*Loschbour*

The Late Mesolithic Loschbour sample stems from a male skeleton recovered from the Loschbour rock shelter in Heffingen, Luxembourg.

The skeleton was excavated in 1935 by Nicolas Thill. The *in situ* find is not documented, but was described retrospectively by Heuertz (1950 [1], 1969 [2]). According to his reports it seemed to be a primary burial, as the skeleton was lying on its back in a flexed position and with arms crossed over the chest[3]. The inhumation was accompanied by two ribs of *Bos primigenius*, dated in 1975 by conventional radiocarbon to 7115 ±45 BP (GrN-7177; 6,010-5,850 cal BC)[4] and a small flint scraper. The skeleton was AMS radiocarbon dated in 1998 to 7,205 ± 50 before present (BP) (OxA-7738; 6,220-5,990 cal BC)[5]. Based on morphological, radiological and histological data, the estimated age of death is 34 to 47 years[6]. Pathological finds include slight dorsal and lumbar vertebral osteoarthritic lesions, minimal unsystematized enthesopathies and an osteo-dental discharge fistula[6]. The skull seems at least partly decorated with ocher[6]. A second and older (final middle Mesolithic) burial, with a cremated individual dated in 1999 to 7,960±40 BP (Beta 132067, AMS radiocarbon method), was discovered in a nearby pit among ashes[5]. The disturbed archaeological layers in which the two burials were found contained rich lithic assemblages, including microlithic artefacts of early, middle and late Mesolithic periods (e.g. points with retouched and unretouched bases, points with bilateral retouch, an obliquely truncated point, a point with a slanted base and surface retouch, mistletoe points with surface retouch, a scalene triangle, narrow backed bladelets and a truncated bladelet with a narrow back), massive antler tools, faunal remains from aurochs, red deer, wild boar, and roe deer [4, 7, 8]and two perforated allochtonous fossilized shells of *Bayana lactea* [9]. New excavations in 1981 and 2003 revealed additional information on the stratigraphy[10,11], taphonomic processes and palaeoenvironment [12].

The DNA extraction was performed on a molar (M12) sampled from the skull, pictured in Extended Data Figure 1A, in as sterile as possible conditions in 2009. After sampling, the tooth was UV-irradiated, and the surface was removed and again irradiated with UV-light in the Palaeogenetics Laboratory in Mainz. Subsequently, the sample was pulverized in a mixer mill (Retsch).

The initial extraction was performed using 80 mg of tooth powder by a silica protocol after Rohland and Hofreiter (2007) [13], resulting in 100 µl of extract (extract LB 1, Table S1.1).

Two more extracts with a volume of 100 µl each were prepared from an additional 90 mg of tooth powder using a protocol optimized to recover short fragments [14] (extracts LB2 and LB3, Table S1.1).

*Stuttgart*

The Stuttgart sample stems from a female skeleton (LBK380, Extended Data Fig. 1B) that was excavated in 1982 at the site Viesenhäuser Hof, Stuttgart-Mühlhausen, Germany. The site reflects a long period of habitation starting from the earliest Neolithic to the Iron Age. The early Neolithic at this site is represented by a large number of well-preserved burials belonging to the Linear Pottery Culture (*Linearbandkeramik*, LBK), dated to 5,500-4,800 BC, as inferred from artifacts such as pottery associated with the graves of the female skeletons as well as surrounding graves [15]. The Neolithic part of the graveyard separates into two large areas including burials from the early (area-2) and middle and late phases (area-1) of the LBK. The relative chronology of the burials from area-1 has been corroborated by calibrated radiocarbon dates of 5,100-4,800 BC [16].

Based on morphology, Stuttgart (LBK380) is a female who died at an estimated age of 20 to 30 years. The skeleton derives from a grave (I-78, area-1) excavated among 83 others from area I of the cemetery and is well preserved but partially fragmented. The skeleton was buried in the characteristic way of the LBK, lying in a seated position on the right side. The burial was oriented from East-North-East to West-North-West with the skull facing north. Most of the body parts were represented [17]. Strontium isotope analysis suggests that the female was of local origin [18]. Noticeable pathological changes were present including multiple osseous lesions, compression fractures, and an angular kyphosis affecting several vertebrae. These may be due to a diagnosis of primary hyperparathyroidism in this individual [19].

For DNA analysis the M47 molar was removed. A total of 40 mg of tooth powder were taken from the inner part of the Stuttgart molar by a sterile dentistry drill in the clean room facilities of the University of Tübingen and extracted[14] resulting in 100 µl of DNA extract (extract LBK1, Table S1.1).

***Table S1.1:*** *Summary of extractions*

| Extract | Individual | Tissue | Amount (mg) | Extractionprotocol |
|---|---|---|---|---|
| LB1 | Loschbour | Molar | 80 | Rohland& Hofreiter (2007) |
| LB2 | Loschbour | Molar | 90 | Dabney *et al.* (2013) |
| LB3 | Loschbour | Molar | 90 | Dabney *et al.* (2013) |
| LBK1 | Stuttgart (LBK380) | Molar | 40 | Dabney *et al.* (2013) |
| MOT1 | Motala 1 | Molar | 100 | Yang *et al.* (1998) |
| MOT2 | Motala 2 | Molar | 100 | Yang *et al.* (1998) |
| MOT3 | Motala 2 | Skull | 100 | Yang *et al.* (1998) |
| MOT4 | Motala 3 | Molar | 100 | Yang *et al.* (1998) |
| MOT5 | Motala 4 | Molar | 100 | Yang *et al.* (1998) |
| MOT6 | Motala 5 | Molar | 100 | Yang *et al.* (1998) |
| MOT7 | Motala 6 | Molar | 100 | Yang *et al.* (1998) |
| MOT8 | Motala 6 | Skull | 100 | Yang *et al.* (1998) |
| MOT9 | Motala 7 | Skull | 100 | Yang *et al.* (1998) |
| MOT10 | Motala 8 | Molar | 100 | Yang *et al.* (1998) |
| MOT11 | Motala 8 | Skull | 100 | Yang *et al.* (1998) |
| MOT12 | Motala 9 | Molar | 100 | Yang *et al.* (1998) |
| MOT13 | Motala 12 | Molar | 100 | Yang *et al.* (1998) |
| MOT14 | Motala 12 | Maxilla | 100 | Yang *et al.* (1998) |
| MOT15 | Motala 4170 | Tibia | 100 | Yang *et al.* (1998) |
| MOT16 | Motala MkA | Femur | 100 | Yang *et al.* (1998) |

*Motala*

The Motala samples come from the site of Kanaljorden in the town of Motala, Östergötland, Sweden. The site was excavated between 2009 and 2013. The samples that we analyzed in the present study were retrieved in 2009 and 2010 (Extended Data Figs. 1C, 1D).

The human remains are part of ritual depositions that were made on a 14 × 14 meter stone-packing, constructed on the bottom of a small lake. The stone-packing was completely submerged and covered by at least 0.5m of water at the time of use. The ritual depositions include human bones: mostly skulls and fragments of skulls but also some stray bones from other parts of the body. The minimal number of individuals is inferred to be ten adults and one infant. The infant is the only individual that has bone representation from the entire body. Two of the skulls were mounted on wooden stakes still imbedded in the crania at the time of discovery. Apparently the skulls were put on display prior to the deposition into the lake. In addition to human bones, the ritual depositions also includes artifacts of antler, bone, wood and stone, animal carcasses/bones, as well as nuts, mushrooms and berries.

Direct dates on 11 human bones range between 7,013 ± 76 and 6,701 ± 64 BP (6,361-5,516 cal BC), with a twelfth outlier at 7,212 ± 109 BP. Dates on animal bones (N=11) and resin, bark and worked wood (N=6) range between 6,954 ± 50 and 6,634 ± 45 BP (5,898 - 5,531 cal BC). These dates correspond to a late phase of the Middle Mesolithic of Scandinavia.

***Table S1.2:*** *Summary of libraries sequenced as part of this study*

| **Library** | **From extract** | **Extract vol. in lib. (ul)** | **Library prep. protocol** | **UDG treatment** | **Insert size fractionation** |
|---|---|---|---|---|---|
| ALB1 | LB1 | 20 | Meyer & Kircher (2010) | no | none |
| ALB2-10 | LB1 | 28.5 (total) | Meyer *et al.* (2012) | yes | 55-300bp |
| ALB11-12 | LB2 | 25 | Briggs *et al.* (2010) | yes | 80-180bp |
| ALB13-14 | LB3 | 25 | Briggs *et al.* (2010) | yes | 80-180bp |
| ALBK1 | LBK1 | 5 | Meyer & Kircher (2010) | no | none |
| ALBK2 | LBK1 | 50 | Briggs *et al.* (2010) | yes | 70-180bp |
| AMOT1 | MOT1 | 10 | Meyer & Kircher (2010) | no | none |
| AMOT2 | MOT2 | 10 | Meyer & Kircher (2010) | no | none |
| AMOT3 | MOT3 | 10 | Meyer & Kircher (2010) | no | none |
| AMOT4 | MOT4 | 10 | Meyer & Kircher (2010) | no | none |
| AMOT5 | MOT5 | 10 | Meyer & Kircher (2010) | no | none |
| AMOT6 | MOT6 | 10 | Meyer & Kircher (2010) | no | none |
| AMOT7 | MOT7 | 10 | Meyer & Kircher (2010) | no | none |
| AMOT8 | MOT8 | 10 | Meyer & Kircher (2010) | no | none |
| AMOT9 | MOT9 | 10 | Meyer & Kircher (2010) | no | none |
| AMOT10 | MOT10 | 10 | Meyer & Kircher (2010) | no | none |
| AMOT11 | MOT11 | 10 | Meyer & Kircher (2010) | no | none |
| AMOT12 | MOT12 | 10 | Meyer & Kircher (2010) | no | none |
| AMOT13 | MOT13 | 10 | Meyer & Kircher (2010) | no | none |
| AMOT14 | MOT14 | 10 | Meyer & Kircher (2010) | no | none |
| AMOT15 | MOT15 | 10 | Meyer & Kircher (2010) | no | none |
| AMOT16 | MOT16 | 10 | Meyer & Kircher (2010) | no | none |
| AMOT17 | MOT1 | 15 | Briggs *et al.* (2010) | yes | none |
| AMOT18 | MOT2 | 15 | Briggs *et al.* (2010) | yes | none |
| AMOT19 | MOT4 | 15 | Briggs *et al.* (2010) | yes | none |
| AMOT20 | MOT5 | 15 | Briggs *et al.* (2010) | yes | none |
| AMOT21 | MOT7 | 15 | Briggs *et al.* (2010) | yes | none |
| AMOT22 | MOT12 | 15 | Briggs *et al.* (2010) | yes | none |
| AMOT23 | MOT13 | 15 | Briggs *et al.* (2010) | yes | none |

Teeth from nine of the better-preserved skulls were selected for DNA analysis, as well as a femur and a tibia (Motala MkA and 4170, from the first two human bones found in 2009). Extraction of the samples from Motala took place in the clean-room facilities of the Ancient DNA laboratory at the Archaeological Research Laboratory, Stockholm University. Bone powder was removed from the inner parts of the bones or teeth with a sterile dentistry drill and extracted according to a protocol by Yang *et al.* (1998) [20] resulting in 16 extracts (MOT 1 to 16, Table S1.1).

**Library Preparation**

For screening and mtDNA capture, libraries for all samples were prepared using either double- or single-stranded library preparation protocols (Table S1.2) [21, 22].

For large scale shotgun sequencing, additional libraries were produced including a DNA repair step with Uracil-DNA-glycosylase (UDG) and endonuclease VIII (endo VIII) treatment [23]. Libraries ALB 2-14 and ALBK1 were furthermore size fractionated on a PAGE gel [22] (Table S1.2).

**Shotgun Sequencing**

All non-UDG-treated libraries were random shotgun sequenced. For libraries ALB1 and AMOT 1, 2, 3, 4, 6, 9, and 12, sequencing was carried out on an Illumina Genome Analyzer IIx with 2×76 + 7 cycles. For library ALBK1 we carried out the sequencing on an Illumina MiSeq with 2×150 + 8 + 8 cycles. In all cases, we followed the manufacturer's protocol for multiplex sequencing.

Raw reads were analyzed as described in Kircher (2012) [24] and were mapped with BWA 0.6. [25] to the human reference genome (hg19/GRCh37/1000Genomes) in order to calculate the fraction of endogenous human DNA. After duplicate removal, 0.82% to 66.4% of reads were estimated to map to the human reference genome with a mapping quality of at least > 30 (Table S1.3).

Based on the results, the extracts LB 1-3, LBK1, MOT 1, 2, 4, 5, 7, 12 and 13—representing individuals Loschbour, Stuttgart, and Motala 1, 2, 3, 4, 6, 9, and 12, respectively—were chosen for UDG-treatment and possible further deep sequencing.

***Table S1.3:*** *Summary of whole-genome deep sequencing runs*

| **Library** | **Pooled** | **No. lanes** | **Read length** | **Facility** |
|---|---|---|---|---|
| ALB2 | no | 3 | 50 bp | HMS, Boston |
| ALB2 | no | 3 | 100 bp | Illumina, San Diego |
| ALB3-10 | LB Pool 1 | 5 | 100 bp | Illumina, San Diego |
| ALB11-14 | LB Pool 2 | 8 | 101 bp | MPI, Leipzig |
| ALBK2 | no | 8 | 101 bp | MPI, Tübingen |
| AMOT17, 18, 23 | Motala Pool 1 | 4 | 100 bp | Illumina, San Diego |
| AMOT19-22 | Motala Pool 2 | 4 | 100 bp | Illumina, San Diego |
| AMOT23 | no | 8 | 100 bp | Illumina, San Diego |

The UDG treated library ALB2 was sequenced on 3 Illumina HiSeq 2000 lanes with 50-bp single-end reads in the Harvard Medical School Biopolymers Facility, followed by 3 Illumina HiSeq 2000 lanes of 100-bp paired-end reads at Illumina, San Diego. We also carried out sequencing at Illumina, San Diego of 5 HiSeq 2000 lanes of 100-bp paired-end reads of pooled libraries ALB3-10.

***Figure S1.1: Visualization of sample preparation process. (Top)*** *Loschbour,* ***(Middle)*** *Stuttgart and* ***(Bottom)*** *Motala. The responsible research group for each step is marked in green.*

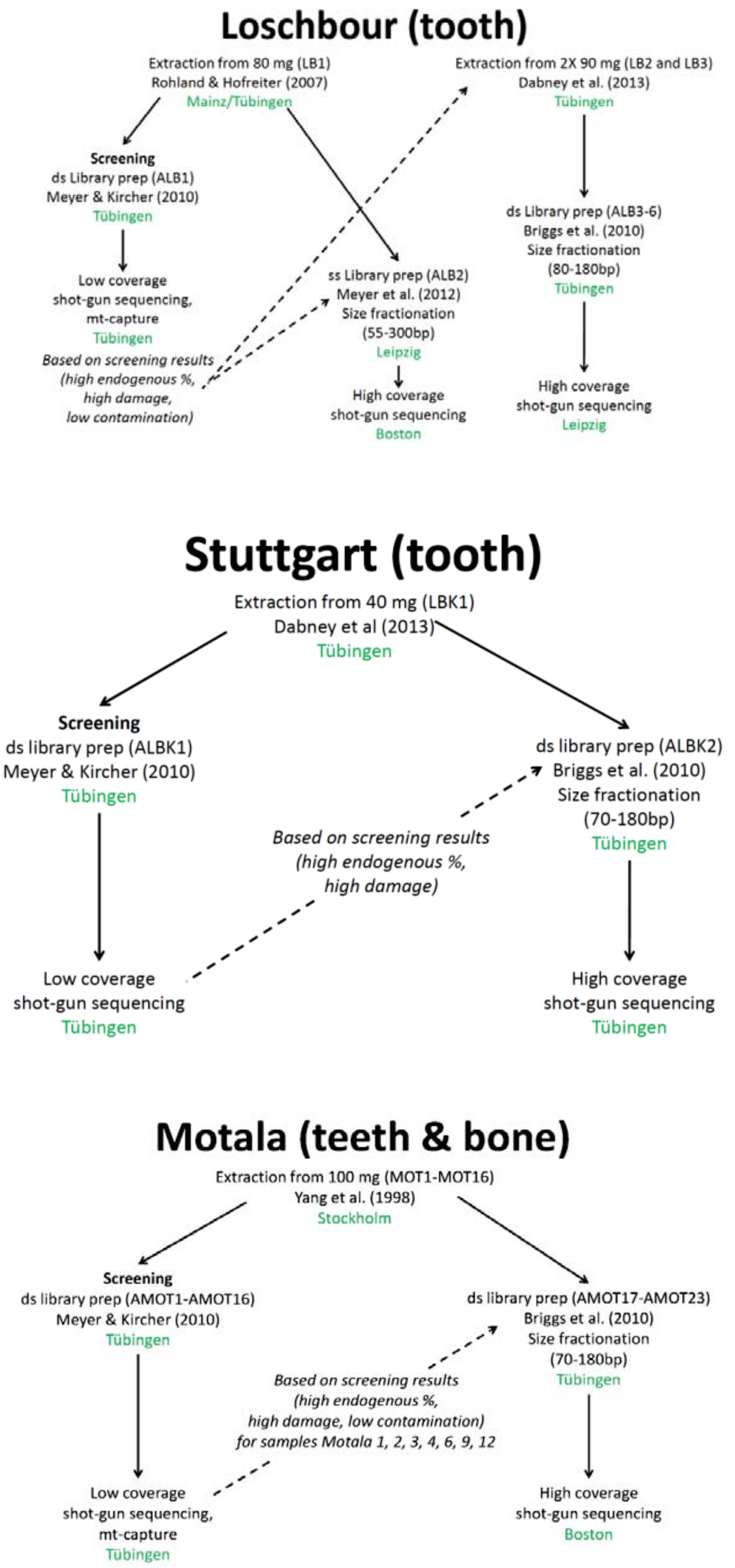

We prepared UDG and endo VIII damage repaired libraries (USER-enzyme)[23] for ALB 3-6. We then size fractionated these libraries to an insert size of about 80-180bp. We performed 8 lanes of shotgun sequencing of these libraries on an Illumina HiSeq 2000 at the Max Planck Institute for Evolutionary Anthropology in Leipzig. Sequences were produced using 101-bp paired-end reads using CR2 forward (5' – TCTTTCCCTACACGACGCTCTTCCGATCTGTCT) and CR2 reverse (5' – GTGACTGGAGTTCAGACGTGTGCTCTTCCGATCTGTCT) custom primers. In addition, seven cycles were sequenced for a P7 index using the P7 Illumina Mulitplex primer. The P5 index was not sequenced. The instructions from the manufacturers were followed for multiplex sequencing on the HiSeq platform with a TruSeq PE Cluster Kit v3 - cBot – HS cluster generation kit and a TruSeq SBS Kit v3 sequencing chemistry. An indexed control library of ϕX 174 was spiked into each library prior to sequencing, contributing to 0.5% of the sequences from each lane.

We also prepared UDG-treated libraries for the Stuttgart sample, and size fractionated them to an insert size of about 70-180bp. ALBK2 was sequenced on 8 HiSeq 2000 lanes and 101-bp paired-end reads plus seven cycles for a P7 index using the P7 Illumina Mulitplex sequencing primer at the Max Planck Institute for Developmental Biology in Tübingen. Instructions from the manufacturers were followed using a TruSeq PE Cluster Kit v3 - cBot – HS cluster generation kit and a TruSeq SBS Kit v3 sequencing chemistry.

The UDG-treated libraries for Motala (AMOT17-23) were sequenced on 8 HiSeq 2000 lanes of 100-bp paired-end reads, with 4 lanes each for two pools (one of 3 individuals and one of 4 individuals), through contract sequencing at Illumina, San Diego of 100-bp paired-end reads. We also sequenced an additional 8 HiSeq 2000 lanes of AMOT23 at Illumina, San Diego through contract sequencing. This was the library with the highest percentage of endogenous human DNA (from Motala12).

A visual overview of sample processing, including library preparation, capture methods and sequencing results is shown in Figure S1.1.

# Supplementary Information 2

## Processing of sequencing data and estimation of heterozygosity

Gabriel Renaud*, Cesare de Filippo, Swapan Mallick, Janet Kelso and Kay Pruefer

* To whom correspondence should be addressed (gabriel_renaud@eva.mpg.de)

### Overview

This note describes the processing of the sequence data for the Loschbour, Stuttgart and Motala samples. It also describes the estimation of heterozygosity for the high coverage Stuttgart and Loschbour individuals. For Stuttgart, heterozygosity was estimated to be higher than in any of 15 present-day non-African samples and lower than any of 10 present-day African samples. For Loschbour, heterozygosity was estimated to be lower than in any of 25 present-day samples.

### Sequencing data

All ancient DNA (aDNA) libraries were sequenced on the Illumina HiSeq platform. Base-calling was carried out using the default Illumina basecaller, Bustard, except where noted. The following data were generated (summarized in Table S2.1):

Loschbour:

1. Four double-stranded libraries (ALB11-14) were sequenced for 101-cycles, paired-end, on a HiSeq 2500 platform (8 lanes). Base-calling was performed using freeIbis[1].
2. Nine single-stranded libraries (ALB2-10) were sequenced for 100-cycles, paired-end, on a HiSeq 2000 platform. This consisted of 3 lanes for ALB2 and 5 lanes for a pool of ALB3-10.

Stuttgart:

A double-stranded library (ALBK2) was sequenced for 101-cycles, paired-end, on a HiSeq 2000 platform (8 lanes).

Motala:

Seven double-stranded libraries (AMOT17-23) were sequenced for 100-cycles, paired-end, on a HiSeq 2000 platform (8 lanes). Motala12 (AMOT23), the sample with the highest percentage of endogenous DNA, was then sequenced on a further 8 lanes.

### Processing of sequencing data prior to genotyping

Ancient DNA molecules are often short enough that the paired-end reads carry the flanking sequencing adaptors at the ends. The reads were therefore pre-processed to trim adaptors and to merge overlapping paired-end reads using the merger program in aLib[2] (*-mergeoverlap* option). The merged sequences and unmerged read pairs were then mapped. Sequences with more than five bases with quality less than 10 were flagged as "QC failed" and were removed.

The sequences from Loschbour and Stuttgart were mapped to the *hg19* genome assembly (1000 Genomes version) using BWA[3] version 0.5.10, parameters "-n 0.01 -o 2", with the seed disabled. Sequences that were merged and pairs that were flagged as properly paired were retained for analysis. The mappings were sorted and duplicates were removed using bam-rmdup2 version 0.4.9. Indel realignment was performed using GATK[4] version 1.3-25. To restore the MD field in the BAM files, "samtools fillmd" was used (samtools[5] 0.1.18). Sequences produced from libraries prepared using the single-stranded protocol still carry uracils at the first or last two bases of the molecules. These are read as thymine during sequencing, and cannot be identified or corrected using metrics such as base quality. Since they can influence variant calling we reduced the base quality of any 'T' in the first bases or last two bases of sequence reads to a PHRED score of 2 for all single-stranded libraries. Similarly, sequences produced from libraries prepared using the double-stranded protocol may carry uracils at the first base causing C→T changes and G→A changes on the last base. Qualities of thymines in the first and adenines in the last base were reduced to a PHRED score of 2[6].

The seven Motala samples had to be treated slightly differently. Initial light shotgun sequencing of seven Motala libraries was performed to determine candidate libraries for deeper sequencing. The samples were sequenced as a pool, so we de-multiplexed the data by searching among the sequences for ones that had no more than one mismatch compared with each of the expected P7 and P5 indices for the seven samples. Reads were stripped of adapters, merged using SeqPrep[7], and aligned with BWA[3] version 0.5.10, with parameters "-n 0.01 –o 2" (seed disabled). Duplicates were removed using samtools[5] 0.1.18. PCA indicated that the Motala samples were relatively homogenous in ancestry and we therefore merged the data for all of the samples except for Motala3 and Motala12 (using samtools *merge*5) to increase coverage for population genetic analysis (labeled 'Motala_merge' in Fig. 1B).

Comparisons of the endogenous rates for all Motala samples indicated that the library from Motala12 had the highest percentage of endogenous DNA, and thus a further eight lanes of sequencing were generated for this individual.

Table S2.1 reports summary statistics for all the libraries we sequenced. Figure S2.1 reports base-specific substitution patterns per library.

***Table S2.1. Sequencing results by library for Loschbour, Stuttgart and Motala***

| Sample | Library ID | Library type | Mapped sequences | Mean insert size (bp) | Std. Dev. in insert size (bp) | Genome coverage |
|---|---|---|---|---|---|---|
| Loschbour1 | ALB11 | Double strand + UDG | 93,342,792 | 87 | 23 | 2.8 |
| Loschbour2 | ALB12 | Double strand + UDG | 111,474,060 | 80 | 22 | 3.1 |
| Loschbour3 | ALB13 | Double strand + UDG | 146,593,852 | 78 | 23 | 4.0 |
| Loschbour4 | ALB14 | Double strand + UDG | 161,736,672 | 80 | 23 | 4.5 |
| Loschbour | ALB2-10 | Single strand + UDG | 345,350,969 | 61 | 20 | 7.2 |
| Stuttgart | ALBK2 | Double strand + UDG | 788,244,122 | 70 | 17 | 19.1 |
| Motala1 | AMOT17 | Double strand + UDG | 8,050,873 | 68 | 29 | 0.18 |
| Motala2 | AMOT18 | Double strand + UDG | 6,670,241 | 70 | 31 | 0.15 |
| Motala3 | AMOT19 | Double strand + UDG | 23,622,338 | 73 | 32 | 0.55 |
| Motala4 | AMOT20 | Double strand + UDG | 3,369,460 | 64 | 29 | 0.070 |
| Motala6 | AMOT21 | Double strand + UDG | 1,032,460 | 71 | 31 | 0.024 |
| Motala9 | AMOT22 | Double strand + UDG | 484,149 | 64 | 27 | 0.010 |
| Motala12 | AMOT23 | Double strand + UDG | 94,818,771 | 73 | 32 | 2.4 |

Note: "Mapped" refers to the number of merged and properly paired sequences after duplicate removal.

**Diploid genotyping**

For the Loschbour and Stuttgart high coverage individuals, diploid genotype calls were obtained using the Genome Analysis Toolkit (GATK) [8] version 1.3-25, using the parameters: "--output_mode EMIT_ALL_SITES --genotype_likelihoods_model BOTH --baq OFF". Because GATK does not call heterozygous sites in cases in which neither allele matches the reference genome, a second round of genotyping was carried out, providing as input a modified reference sequence that carried the bases called in the first round of genotyping. The genotype calls from both rounds were then combined to obtain a final variant call format (VCF) file.

***Figure S2.1: Substitution patterns for Loschbour, Stuttgart and Motala 12 (measured on chromosome 21).*** *Single- and double-stranded Loschbour libraries are reported separately.*

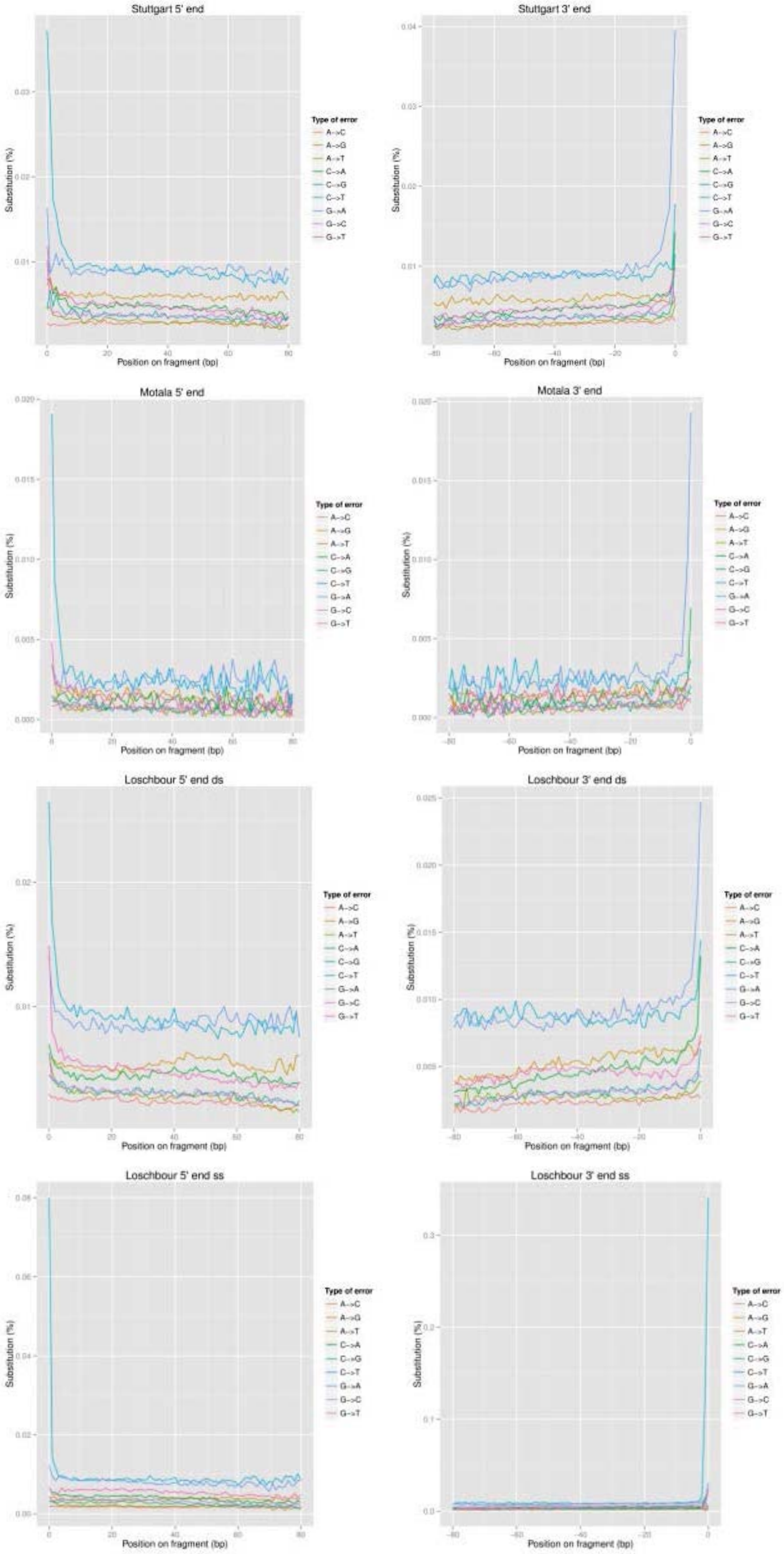

**Estimation of heterozygosity**

Two tools were used to estimate heterozygosity:

1. *mlRho*[9], which estimates heterozygosity as the maximum likelihood of the population mutation parameter (θ) from high-coverage data of one individual, assuming an infinite sites model of mutation. The program also estimates the sequencing error rate per site (ε).
2. *GATK*. The GATK genotype calls are viewed as correct, and the number of called heterozygous sites divided by the total number of screened nucleotides is interpreted as the heterozygosity.

Heterozygosity was estimated in the high-coverage genome sequences from 29 individuals: 25 diverse present-day humans, Altai Neandertal, Denisova, Stuttgart and Loschbour. This is the same dataset previously described on the high coverage Neandertal genome, here supplemented by Stuttgart and Loschbour[10]. Analysis was restricted to ~629 million sites on the autosomes that passed the following filters in all 29 individuals (the filters are described in more detail in ref. 10):

1. Fall in the most stringent mappability track (Map35_100%): positions where all overlapping 35mers align only to one location in the genome allowing for up to one mismatch.
2. A mapping quality (MQ) of 30.
3. In the 2.5% - 97.5% interval of the coverage distribution specific to each sample. For the ancient samples, coverage is computed by binning sites according to their local GC content (i.e. the number of GC bases in a 51 bp window centered at the site).
4. Do not overlap insertion / deletion polymorphisms (indels).
5. Not a simple repeat as specified by the UCSC Tandem Repeat Finder track[11] for *hg19*.

***Table S2.2: Heterozygosity and error estimates per 10,000 screened sites***

| **Sample*** | ***mlRho*** | **GATK** | ***mlRho*/GATK ratio** |
|---|---|---|---|
| Altai Neandertal | 1.68 | 1.75 | 0.96 |
| Denisova | 1.82 | 2.14 | 0.85 |
| Loschbour | 4.75 | 6.62 | 0.72 |
| Karitiana_B | 4.99 | 5.52 | 0.90 |
| Papuan_B | 5.02 | 5.98 | 0.84 |
| Mixe_B | 5.85 | 6.12 | 0.96 |
| Karitiana_A | 5.87 | 5.76 | 1.02 |
| Australian1_B | 6.03 | 6.59 | 0.91 |
| Papuan_A | 6.03 | 6.38 | 0.94 |
| Australian2_B | 6.42 | 6.66 | 0.96 |
| Dai_A | 6.46 | 7.44 | 0.87 |
| Han_B | 6.62 | 7.23 | 0.92 |
| Dai_B | 6.67 | 7.19 | 0.93 |
| Sardinian_B | 6.69 | 7.34 | 0.91 |
| French_B | 6.92 | 7.58 | 0.91 |
| Han_A | 7.04 | 7.45 | 0.95 |
| Sardinian_A | 7.07 | 7.79 | 0.91 |
| French_A | 7.38 | 7.81 | 0.94 |
| Stuttgart | 7.42 | 10.59 | 0.70 |
| Dinka_B | 8.26 | 9.68 | 0.85 |
| Mandenka_B | 9.14 | 10.01 | 0.91 |
| Dinka_A | 9.23 | 9.99 | 0.92 |
| Mbuti_B | 9.35 | 10.09 | 0.93 |
| Mbuti_A | 9.38 | 10.23 | 0.92 |
| San_B | 9.44 | 10.21 | 0.92 |
| Mandenka_A | 9.50 | 10.31 | 0.92 |
| Yoruba_B | 9.50 | 10.06 | 0.94 |
| San_A | 9.64 | 10.69 | 0.90 |
| Yoruba_A | 9.78 | 10.18 | 0.96 |

* The suffix for the 25 present-day samples indicates whether the individual is from the A or B panel.

GATK calls were extracted from the VCF files[8]. GATK heterozygosity was defined as the number of heterozygous sites divided by the number of bases screened.

*mlRho* was run directly on the BWA alignments, restricting to sites that passed the filters above and additionally restricting to DNA sequencing data with a minimum base quality of 30.

Inspection of Table S2.2 indicates that the *mlRho* estimates are smaller than the GATK estimates for nearly all samples. However, the reduction below one is most marked for Stuttgart and Loschbour:

0.92 Mean of 25 present-day humans
0.96 Altai Neandertal
0.84 Denisova
0.72 Loschbour
0.70 Stuttgart

The discrepancy between GATK and *mlRho* estimates is plausibly due to a higher error rate in the Stuttgart and Loschbour diploid genotype calls due to these two genomes' lower sequencing coverage. Specifically, the Stuttgart and Loschbour sequencing coverage is about 20× compared with >30× for most other samples. The GATK estimates do not correct for the genotyping error that occurs in the context of low coverage, and hence may produce artifactual overestimates of heterozygosity.

It is important to note that although the diploid genotype calls for both Stuttgart and Loschbour have a higher error rate than for the other genomes, these error rates are not likely to be sufficient to bias the analyses of population history reported in this study. The reason for this is that the Loschbour and Stuttgart diploid genotypes are used in this study largely for the purpose of determining allelic state at sites that are already known to be polymorphic in present-day humans: SNPs that are part of the Affymetrix Human Origins array (SI9). At these sites, the probability of polymorphism is much higher than the likely error rate of 1/1000 to 1/10000 in the Stuttgart and Loschbour data, and thus error does not contribute much to the observed variability of the inferred allelic state at these sites.

Using the *mlRho* estimates of heterozygosity that are likely to be more accurate than those from GATK because *mlRho* co-estimates and corrects for error, Loschbour is inferred to have an average of 4.75 heterozygous sites per 10,000 base pairs. This is lower than in any of 25 diverse present-day human samples to which Loschbour was compared although it is still about three times higher than the heterozygosity reported for the Denisovan and Altai Neandertal (Table S2.2). In contrast, Stuttgart has 7.42 heterozygous sites per 10,000 base pairs. This is higher than the heterozygosity measured any of 15 diverse non-Africans, although only slightly higher than the most diverse present-day non-African in the panel (French_A at 7.38 heterozygous sites per 10,000 base pairs) (Table S2.2).

**Effect of residual deamination**

To investigate whether residual deamination affects genotyping quality, we considered sites labeled as heterozygous by GATK with coverage of 16. We picked one allele at random and plotted the allelic distribution (Figure S2.2). For true heterozygous sites, we expect a binomial distribution centered in the middle. However, for sites likely to be heterozygous, two additional peaks will be seen at low allele frequencies due to sequencing errors, contamination and deamination. While considering transitions and transversions separately for the Loschbour does not seem to affect the allelic distribution, it has a significant impact on the Stuttgart sample, for which the rate of transitions is inflated at low frequency alleles.

To investigate as to whether this is due to sequencing error, contamination or residual deamination, we plotted the position with respect to the 5' or 3' end of the minor allele for heterozygous sites representing a transition (Figure S2.3). For Loschbour, there is no spike around the 5'/3' ends. However, for Stuttgart, there is a noticeable spike. For both samples, a dip in the amount of alleles representing transitions can be seen at position 1 as a result of our decreasing of quality scores. Thus, there is probably a greater contribution of residual deamination for the Stuttgart sample.

***Figure S2.2: Allele count for heterozygous sites with coverage 16 for Loschbour and Stuttgart.*** *Heterozygous sites are separated into transitions or transversions.*

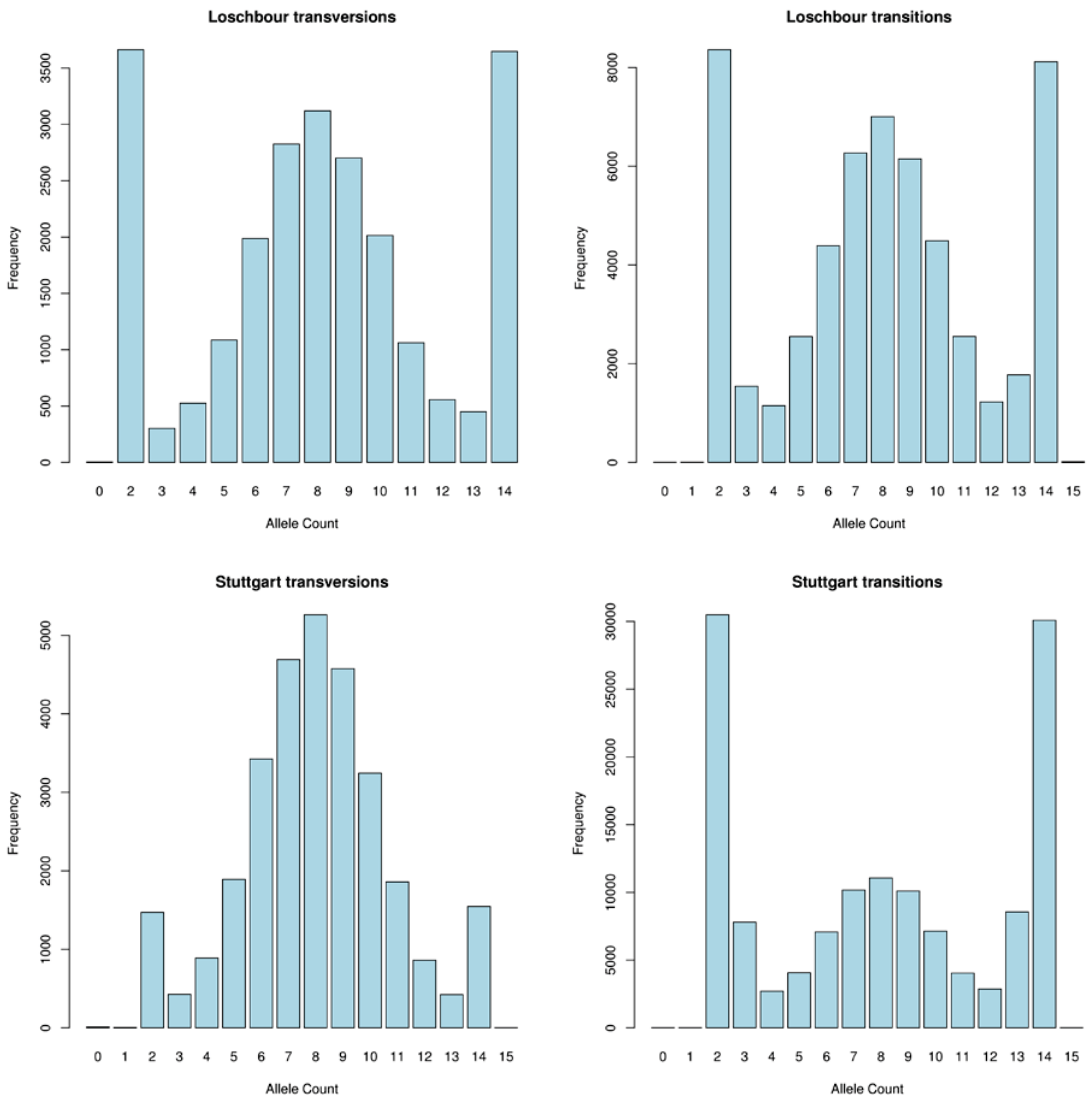

***Figure S2.3: Positions on the original sequence of the minor allele for transitions at heterozygous sites for Loschbour and Stuttgart.*** *The positions are separated according to the position with respect to the 5' or 3' end of the original sequence.*

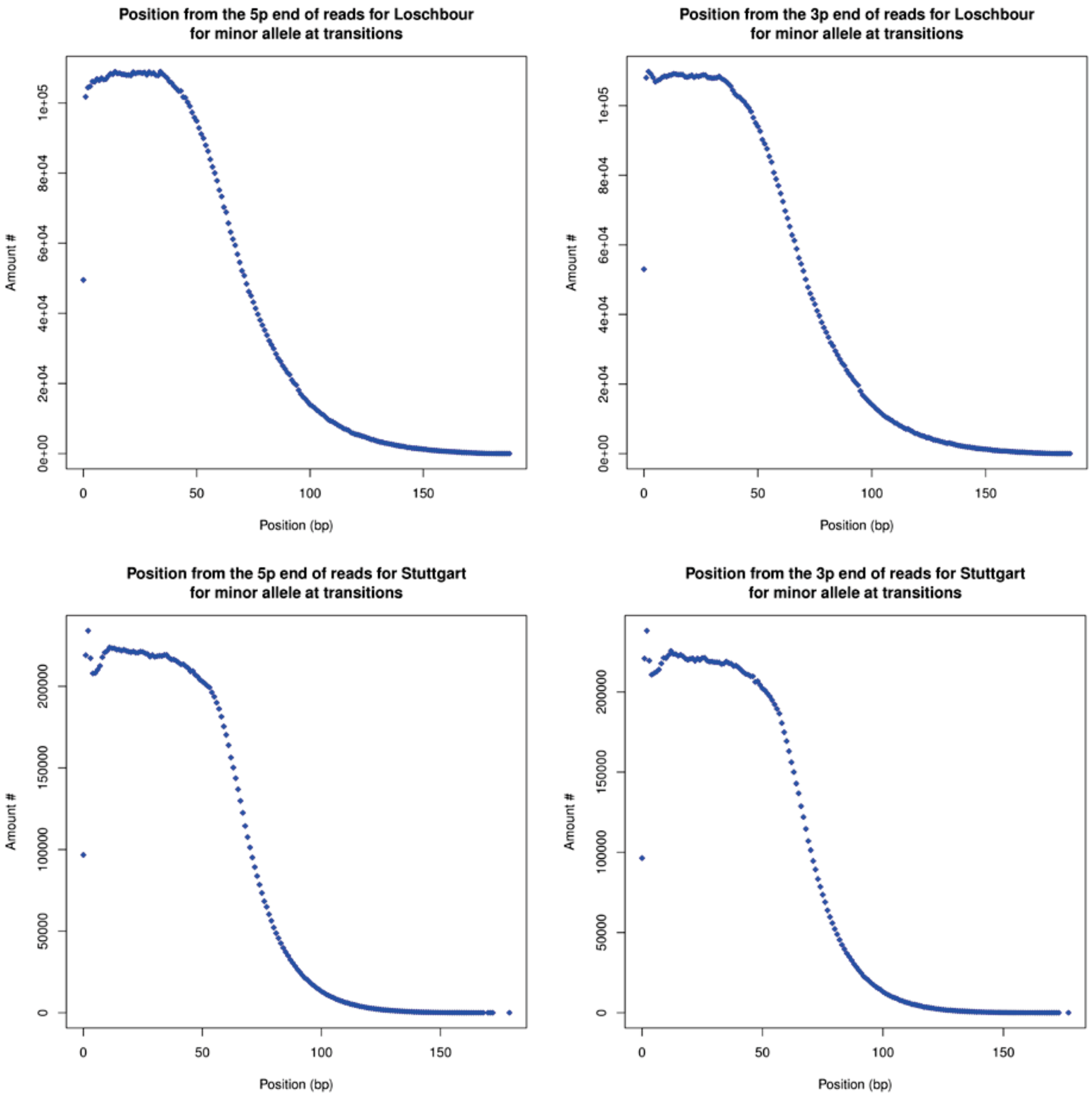

[6] decrQualDeaminated and decrQualDeaminatedDoubleStranded from https://github.com/grenaud/libbam

[7] https://github.com/jstjohn/SeqPrep

[8] McKenna A, Hanna M, Banks E, Sivachenko A, Cibulskis K, Kernytsky A, Garimella K, Altshuler D, Gabriel S, Daly M, DePristo MA (2010) The Genome Analysis Toolkit: a MapReduce framework for analyzing next-generation DNA sequencing data. Genome Res. 20, 1297-303.

[9] Haubold, B., Pfaffelhuber, P. & Lynch, M. mlRho - a program for estimating the population mutation and recombination rates from shotgun-sequenced diploid genomes. *Molecular ecology* 19 Suppl 1, 277-284, doi:10.1111/j.1365-294X.2009.04482.x (2010).

[10] Prüfer, K. et al. (2013) The complete genome sequence of a Neanderthal from the Altai Mountains. *Nature* Advance online publication December 18 2013.

[11] http://genome.ucsc.edu

# Supplementary Information 3
## Ancient DNA authenticity

AlissaMittnik*, Gabriel Renaud, Qiaomei Fu, Janet Kelso and Johannes Krause

* To whom correspondence should be addressed (amittnik@gmail.com)

**Overview**
This note describes the analyses that were performed to test the authenticity of the ancient DNA obtained from each of the ancient modern human samples. Contamination estimates were carried out for the mitochondrial DNA as well as for nuclear DNA sequences.

To identify suitable ancient human samples for deep sequencing, libraries for targeted mtDNA capture from Loschbour and all Motala individuals were prepared without the use of uracil DNA glycosylase (UDG) in order to preserve DNA damage patterns that are an indication of authentic ancient DNA [1].

MtDNA capture, sequencing and processing was performed as described in SI 4. DNA extracts that showed high proportions of apparently authentic mtDNA were used for preparation of UDG-treated libraries for deep sequencing as described in SI 1. The mtDNA contamination rates for the deeply sequenced shotgun data from UDG-treated libraries were estimated by direct comparison to the mtDNA consensus from the targeted mtDNA enrichment.

No mtDNA capture was performed for Stuttgart. For this sample, deep sequencing data was used to analyze DNA damage patterns from a non-UDG-treated library (ALBK1). The mtDNA contamination estimate was obtained from high coverage shotgun data from a UDG-treated library (ALBK2).

**Assessment of ancient DNA authenticity**
Authenticity of aDNA from ancient human DNA extracts was assessed as part of the screening procedure described in SI 1. To assess authenticity the following criteria were applied.

1. **Consistency of reads mapping to the mitochondrial genome consensus sequence** [2] showing that the majority of reads (>95%) derive from a single biological source.
2. **Presence of aDNA-typical C-to-T damage patterns at the 5'-ends of DNA fragments**, caused by post-mortem miscoding lesions [2].
3. In the case of the male sample, Loschbour, an **absence of polymorphic sites on chromosome X** [3].
4. **A maximum-likelihood-based estimate of autosomal contamination** for Loschbour and Stuttgart that uses variation at sites that are fixed in the 1000 genomes humans to estimate error, heterozygosity and contamination [4].
5. Plausibility of mitochondrial sequences in the broader context of the human mitochondrial phylogeny and contemporary population diversity, e.g. branch shortening, due to missing substitutions in ancient mtDNA [5](**see SI 4**.)

**1. Consistency of reads mapping to the mitochondrial genome consensus sequence**
Non-UDG-treated libraries from 17 ancient humans were used for estimating mtDNA contamination levels as described previously [5]. In total 8 of 17 samples show 5% or less inconsistent fragments (Table S3.1), suggesting that the DNA largely originated from a single biological source. Using a Bayesian approach that compares the read sequences to a set of 311 modern human mtDNAs and checks for consistency among the reads [5], similar results are obtained for the 8 samples (Table S3.1).

Stuttgart showed a very high percentage of endogenous DNA (Table S3.1) no enrichment of mtDNA was therefore carried out.

Nine samples were more deeply sequenced: Loschbour, Stuttgart and Motala 1, 2, 3, 4, 6, 9 and 12.

***Figure S3.1: Proportional support for the consensus base at each position of the mtDNA.*** *For UDG-treated libraries the majority of all positions show a consensus support higher than 0.98*

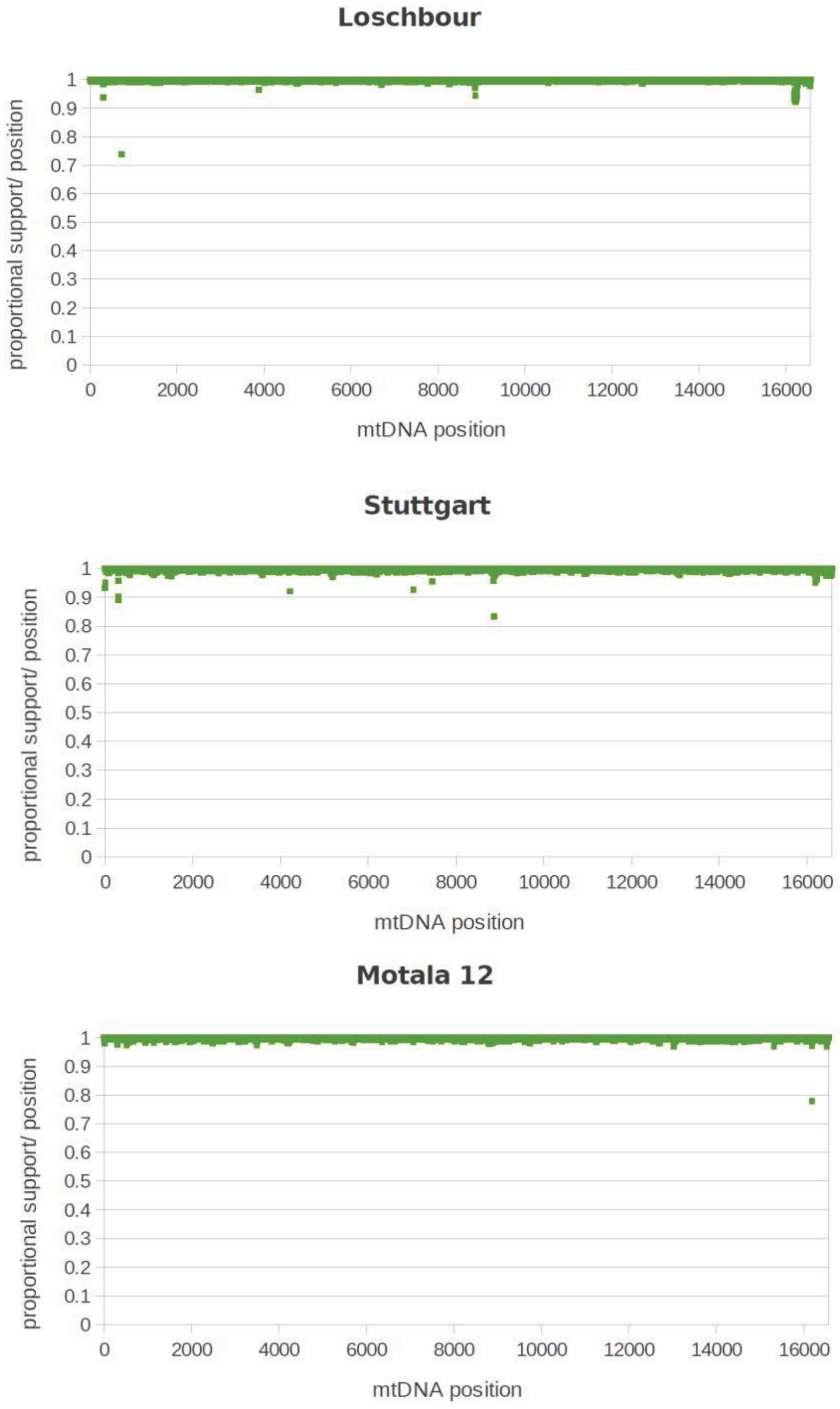

***Table S3.1.Summary of screening results from non-UDG-treated libraries***. *Samples with high levels of authentic DNA that were chosen for deep sequencing are marked in grey. Samples with relatively high amounts of endogenous DNA are marked in bold.*

| | | Capture for human mtDNA | | | | | | | Shotgun screening data | |
|---|---|---|---|---|---|---|---|---|---|---|
| library | sample | Total reads | Unique mapping reads | Average coverageof mtDNA | nt covered at 5-fold coverage(% of mtDNA) | Damage at 5' (%) | Green et al. 2008 Contamination estimate (%)[6] | Fu et al. 2013 Contamination estimate (%)[5] | Total reads | endogenous DNA (%) |
| ALB1 | **Loschbour** | 3916672 | 74435 | 320.8 | 16,569 (100%) | 27.6 | 0 - 0.5 | 1.3 - 1.9 | 5901087 | **22.5** |
| ALBK1 | **Stuttgart** | n/a | n/a | n/a | n/a | n/a | n/a | n/a | 824725 | **66.4** |
| AMOT1 | Motala 1 | 3557100 | 84947 | 321.1 | 16568 (100%) | 31.5 | 0 - 0.6 | 0.7 - 2.2 | 531393 | 3.25 |
| AMOT2 | **Motala 2** | 1590655 | 42962 | 172 | 16572 (100%) | 27.5 | 0 - 1.4 | 1.7 - 4.0 | 1502464 | **9.62** |
| AMOT3 | Motala 2 | 3114034 | 774 | 2.98 | 3610 (21.8%) | 18.1 | 30 - 90.3 | 5.5 - 64.5 | - | - |
| AMOT4 | Motala 3 | 5268145 | 107797 | 412.8 | 16567 (100%) | 29.9 | 0 - 0.3 | 5.0 - 7.8 | 1576248 | 1.07 |
| AMOT5 | Motala 4 | 4692878 | 72187 | 248.5 | 16569 (100%) | 34.7 | 0 - 0.6 | 0 - 1.5 | 860818 | 2.3 |
| AMOT6 | Motala 5 | 3628834 | 1592 | 6.1 | 10608 (64.3%) | 17.3 | 0 - 27.8 | 1.8 - 19.1 | - | - |
| AMOT7 | Motala 6 | 2253825 | 57614 | 252.7 | 16570 (100%) | 28.5 | 0 - 1.1 | 0 - 0.8 | 799225 | 0.82 |
| AMOT8 | Motala 6 | 1837405 | 23306 | 116.7 | 16570 (100%) | 14.1 | 1 - 4.7 | 0.6 - 2.7 | - | - |
| AMOT9 | Motala 7 | 4265963 | 364 | 1.3 | 231 (1.4%) | 12.9 | 4.5 - 32.1 | 0.8 - 70.9 | - | - |
| AMOT10 | Motala 8 | 948122 | 206 | 0.8 | 203 (1.23%) | 7.5 | 4.4 - 16.1 | 0.4 - 41.4 | - | - |
| AMOT11 | Motala 8 | 1265744 | 1403 | 6.4 | 129 (0.78%) | n/a | n/a | 0.4 - 45.8 | - | - |
| AMOT12 | Motala 9 | 1754892 | 30147 | 115.9 | 16569 (100%) | 35 | 0.6 - 2.9 | 1.2 - 4.3 | 555139 | 1.92 |
| AMOT13 | **Motala 12** | 1622517 | 99154 | 355.5 | 16570 (100%) | 20.4 | 0.0 - 1.0 | 0.8 - 2.3 | 416757 | **9.3** |
| AMOT14 | Motala 12 | 1207779 | 2552 | 11.6 | 16088 (97.1%) | 27.7 | 0.7 - 20.2 | 0.4 - 6.4 | - | - |
| AMOT15 | Motala 4170 | 2323981 | 142 | 0.5 | 117 (0.72%) | 0 | 1.4 - 16.5 | 0.8 - 70.6 | - | - |
| AMOT16 | Motala MKA | 3905092 | 717 | 2.3 | 1984 (12.1%) | 8.8 | 2.4 - 7.2 | 0.3 - 30.9 | - | - |

The mtDNA consensus from the targeted mtDNA enrichment for Loschbour, Motala 1, 2, 3, 4, 6, 9 and 12 was used to estimate mtDNA contamination levels from the deep sequencing of UDG-treated libraries. Reads that mapped to the human mtDNA genome with a mapping quality of at least 30 were extracted from the deep-sequencing data for all above-mentioned samples. Based on the rate at which reads mismatched the consensus, we estimated contamination rates of 0.3% for Loschbour (0.24% - 0.39%, 95% HPD) and 0.02%-3.35% for the Motala individuals (Table S3.2). The contamination for Motala 3 and 9 could not be accurately estimated due to low mtDNA coverage. For Stuttgart, the mtDNA consensus sequence was directly built from high coverage shotgun data (Figure S3.1) and used to estimate the number of reads that mismatch the consensus. The contamination estimate for Stuttgart for the deep-sequencing data was found to be 0.43% (0.29% - 0.62%, 95% HPD) showing that less than 1% of the human mitochondrial DNA sequences for Loschbour, Stuttgart and Motala 1, 2, 4, 6 and 12 are likely to come from a contaminating source with a different mitochondrial DNA.

***Table S3.2. Summary of contamination estimates for UDG-treated libraries.***

| library | sample | mtDNA contamination estimate | average mtDNA coverage | ratio mtDNA/ nuclear DNA | autosomal estimates | X Chr estimates |
|---|---|---|---|---|---|---|
| ALB2-14 | Loschbour | 0.3% (0.24-0.39, 95% HPD) | 1519.6 | 76.9 | 0.44% (CI: 0.35-0.53%) | 0.45% |
| ALBK2 | Stuttgart | 0.43% (0.29-0.62, 95% HPD) | 371.5 | 20.9 | 0.30% (CI: 0.22-0.39%) | - |
| AMOT17 | Motala1 | 0.6% (0.16-1.65, 95% HPD) | 32.8 | 241.2 | - | - |
| AMOT18 | Motala2 | 0.02% (0.00-0.29, 95% HPD) | 85 | 525.2 | - | - |
| AMOT19 | Motala3 | - | <1 | 78.9 | - | - |
| AMOT20 | Motala4 | 0.91% (0.22-4.03, 95% HPD) | 9.12 | 111.9 | - | - |
| AMOT21 | Motala6 | 0.18% (0.03-3.81, 95% HPD) | 4.79 | 171 | - | - |
| AMOT22 | Motala9 | - | 2.35 | 212.4 | - | - |
| AMOT23 | Motala12 | 0.34% (0.18-0.71 95% HPD) | 144.7 | 64.4 | - | - |

**2. Presence of aDNA-typical C-to-T damage patterns at the 5'-ends of DNA fragments**

The percentage of C-to-T changes at the 5'-ends of endogenous mtDNA fragments from libraries that were non-UDG-treated was estimated using mapDamage2.0 [7] (the exception is Stuttgart (ALBK1), where nuclear DNA from shotgun data was used to estimate damage patterns). Figure S3.1 shows the difference in DNA damage patterns of DNA mapping to the human genome (hg19) for libraries from Loschbour and Stuttgart with and without UDG-treatment. Based on previous evidence that samples older than 100 years typically have at least 20% deamination at the 5' ends [8], only samples that show more than 20% damage were considered as good candidates for harboring authentic ancient DNA (Table S3.1). All 8 samples that show internally consistent mtDNA show more than 20% damage at the 5'-ends and therefore meet our criteria of aDNA authenticity for further processing.

**mtDNA to nuclear DNA ratio**

The ratio of mtDNA to nuclear DNA was calculated by dividing the average coverage of the mtDNA by the average coverage across all autosomes, effectively giving the number of copies of the mitochondrial genome per cell. The copy number ranges from 21 to 525 between samples, and is substantially lower than previous aDNA studies on bone [6]. This could be due to differential mitochondrial density in various tissues [9]. All samples were taken from molars and suggest that dental tissue may have a comparatively low mitochondrial copy number compared to cortical bone.

***Figure S3.2: Frequency of nucleotide misincorporations at the 5'end of DNA fragments mapping to the human genome.*** *UDG-treated libraries **(right)** show low frequencies of C-to-T changes at the 5' end as UDG removes uracils that result in C-to-T substitutions.*

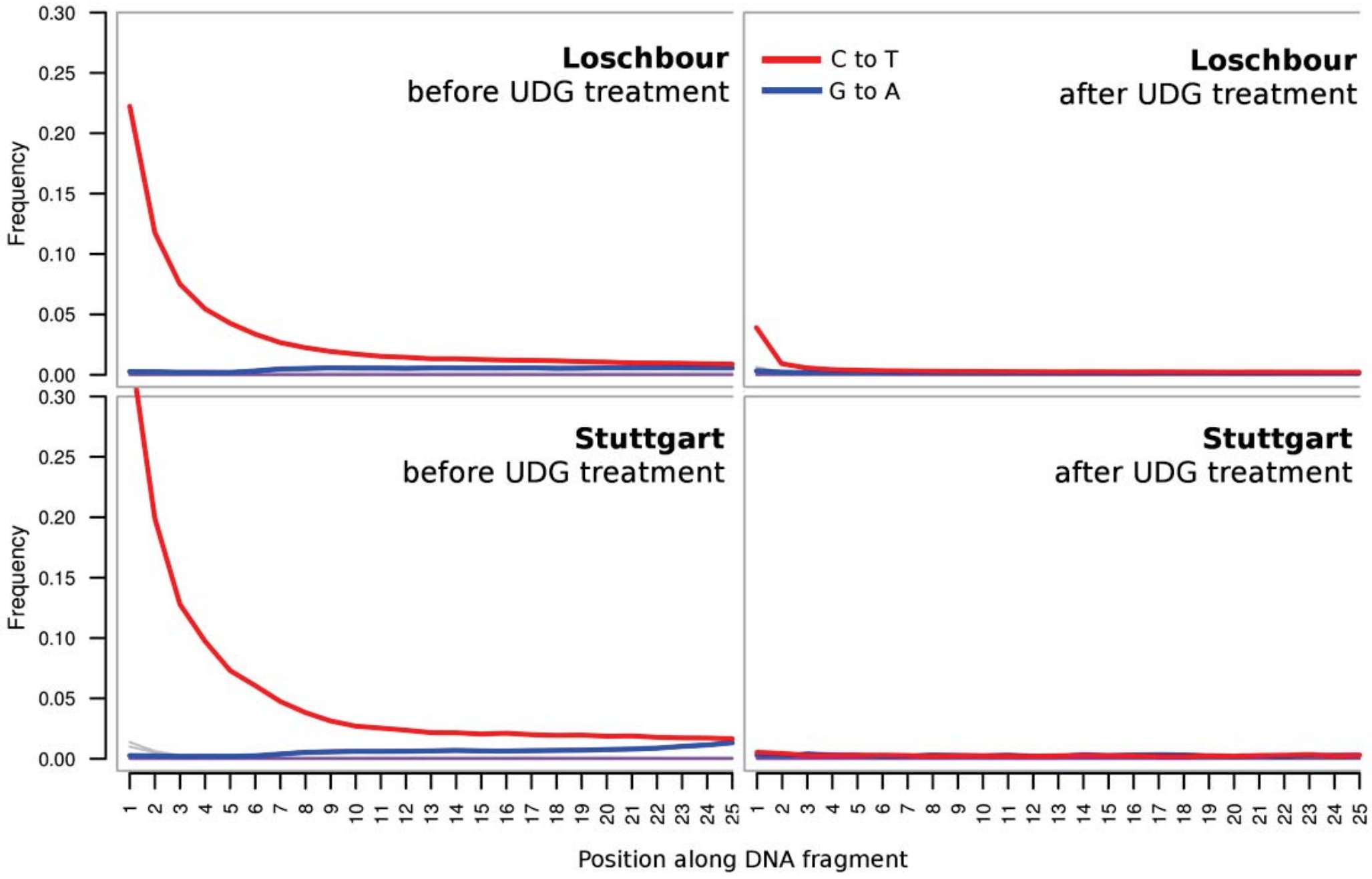

**NuclearDNA contamination estimates**
We used two approaches to estimate the proportion of nuclear contamination in Loschbour and Stuttgart

3. In the case of the male sample, Loschbour, the **absence of polymorphic sites on chromosome X** was used to estimate contamination (similar to the approach taken in ref. [3]).
4. For Loschbour and Stuttgart, we used a **maximum-likelihood-based estimate of autosomal contamination** that uses variation at sites fixed in the 1000 Genomes project humans to co-estimate error, heterozygosity and contamination [4].

**3. Absence of polymorphic sites on chromosome X**
As Loschbour is very likely a male (SI5), heterozygous sites along the X chromosome are not expected. Sites where a second allele is observed are then due to:

1. Contamination
2. Sequencing errors
3. Mismapping

In an approach similar to that used for the Australian Aboriginal Genome [3], we computed the frequency of each base at positions that are polymorphic on chromosome X in the 1000 Genomes [10] dataset.

To reduce the effect of mismapping, only genomic regions with high mappability (SI2) were analyzed. Reads were required to have a mapping quality of at least 30, and only bases with a quality of at least 30 were considered for this analysis. Sites were required to fall within the 95$^{th}$ percentile of the coverage distribution for chromosome X, resulting in a minimum coverage of 4× and a maximum of 21×.

Assuming that contamination and error are both low, the true Loschbour allele will be the majority allele at each site. The observation of minor alleles on chromosome X may arise from either contamination or error. To determine the contamination we recorded the allele frequencies at each site for the Eurasian 1000 genomes populations: British (GBR), Tuscan (TSI), Chinese (CHB, CHS), Japanese (JPT), Iberian (IBS), Finnish (FIN), and Central European (CEU). To determine the sequencing error rate, the nucleotides adjacent to each tested site were considered likely to be monomorphic. The observation of multiple alleles at these sites was assumed to approximate the background sequencing error rate.

For each site that is polymorphic among the 1000 genomes individuals the numbers of major and minor alleles were computed. Triallelic or tetraallelic sites were discarded. The number of major and minor alleles was computed for adjacent sites. The tally of minor and major alleles is presented in Table S3.3. The background probability of error is determined by the base quality cutoff.

***Table S3.3. Divergence at assumed polymorphic and monomorphic sites for Loschbour***

| **Type** | **Sample** | **Computed value** | **Observed probability of error** |
|---|---|---|---|
| Polymorphic | $h_i$<br>$e_i$ | 7,138,322<br>20,697 | 0.002891 |
| Adjacent | $h'_i$<br>$e'_i$ | 7,123,114<br>9,934 | 0.001393 |

For any polymorphic sites we use the observed probability of error ($\varepsilon$) at adjacent sites ($\varepsilon$ = 0.001393) as the background error rate. For a given contamination rate, the probability of an allele occurring at frequency $f_i$ at position $i$ is given by$cf_i.Therefore$, the probability of observing one minor allele is given by:

$$cf_i + (1 - c)\varepsilon$$

The total probability of seeing the aforementioned read distribution at position $i$ where $h_i$ is the major count and minor allele count is given by:

$$[cf_i + (1 - c)\varepsilon]^{e_i}[1 - (cf_i + (1 - c)\varepsilon)]^{h_i}$$

We compute the likelihood of the data given the parameters. The total likelihood for all sites is then given by:

$$p(data|c) = \prod_i [cf_i + (1 - c)\varepsilon]^{e_i}[1 - (cf_i + (1 - c)\varepsilon)]^{h_i}$$

By analyzing the logarithm of the likelihood surface, we infer a maximum of 0.45% contamination in Loschbour.

**4. A maximum-likelihood-based estimate of autosomal contamination**

For all samples, contamination rates on the autosomes were estimated using a method based on that of ref. [4] that is based on the observation that some sites are more susceptible to error than others. The method is a maximum likelihood-based co-estimation of sequence error, contamination and two population parameters, and assumes that present-day human contaminants will contribute derived alleles to the archaic human sequences. The analysisis conditioned on sites where the derived allele is fixed in the 1000 Genomes individuals [11] as compared to great ape outgroups. Low frequency allele counts at these homozygous positions are used to infer contamination and sequence error.

Reads were required to have a minimal length of 35 and a mapping quality of at least 30. We condition on sites where the derived allele is fixed in the 1000 Genomes individuals as compared to great ape outgroups. Low frequency allele counts at these homozygous positions are used to infer contamination and sequence error.

Reads were required to have a minimal length of 35 and a mapping quality of at least 30. The method estimates low contamination in both samples; the estimated contamination for Loschbour and Stuttgart are 0.44% (CI: 0.35-0.53%) and 0.30% (CI: 0.22-0.39%), respectively (Table S3.2).

# Supplementary Information 4

## Mitochondrial genome analysis

Alissa Mittnik* and Johannes Krause

* To whom correspondence should be addressed (amittnik@gmail.com)

This note describes the enrichment and phylogenetic analysis of mtDNA from 17 ancient human libraries derived from the Loschbour, Stuttgart and Motala individuals.

### Enrichment of complete mtDNAs and high throughput sequencing

To test for DNA preservation and contemporary modern human contamination, mitochondrial DNA from 17 ancient human samples was analyzed using a long-range PCR-product based hybridization capture protocol [1]. Libraries (see also SI1, SI3.1) for targeted DNA capture were not treated using the UDG protocol in order to observe DNA damage patterns as additional indication for authenticity [2] (see also SI3). The mtDNA capture was carried out as described previously [3]. The resulting captured mtDNA libraries were pooled and sequenced on the Illumina Genome Analyzer IIx platform with 2 × 76 + 7 cycles. Sequencing data was treated following ref. [4]. In short; raw reads were filtered according to the individual indices, adapter and index sequences were removed, and paired-end reads overlapping by at least 11 nucleotides were collapsed to one fragment where the base with the higher quality score was called in the overlapping sequence. The sequences enriched for human mtDNA were mapped to the Reconstructed Sapiens Reference Sequence (RSRS) [5] using a custom iterative mapping assembler [6]. Between 142 and 107,797 mtDNA fragments were found to map to the reference genome, resulting in average mtDNA coverage of 0.5 to 421 fold (Table S3.1).

### Phylogenetic analysis of the mitochondrial genomes

The consensus sequences of all samples that fulfilled the authenticity criteria were assigned to haplogroups using HaploFind [7] (Table S4.1). All Mesolithic genomes belong to haplogroups U2 or U5, which are common among pre-Neolithic Europeans as has been shown earlier [2, 3, 8-14] (Figure S4.1). Motala 2 and 12 share the same haplotype, suggesting a close relationship through the maternal lineage. The Neolithic sample Stuttgart belongs to haplogroup T2, which is common among early European farmers [8, 11, 14-21] as well as present-day Europeans [22].

***Table S4.1.Haplogroup assignments.***

| Sample | haplogroup | Additional substitutions |
|---|---|---|
| Loschbour | U5b1a | T16189C!, A6701G |
| Motala 1 | U5a1 | G5460A |
| Motala 2 | U2e1 | C16527T |
| Motala 3 | U5a1 | G5460A, A9389G |
| Motala 4 | U5a2d | A13158G |
| Motala 6 | U5a2d | C152T!, G6480A |
| Motala 9 | U5a2 | G228A, G1888A, A2246G, C3756T, G6917A, A9531G |
| Motala 12 | U2e1 | C16527T |
| Stuttgart | T2c1d1 | T152C!, C6340T, T16296C! |

***Figure S4.1: Haplogroup frequencies of ancient and modern Europeans. (Left)*** *shows haplogroup frequencies in Europe before the onset of the Neolithic,* ***(Middle)*** *during the Neolithic and* ***(Right)*** *today.*

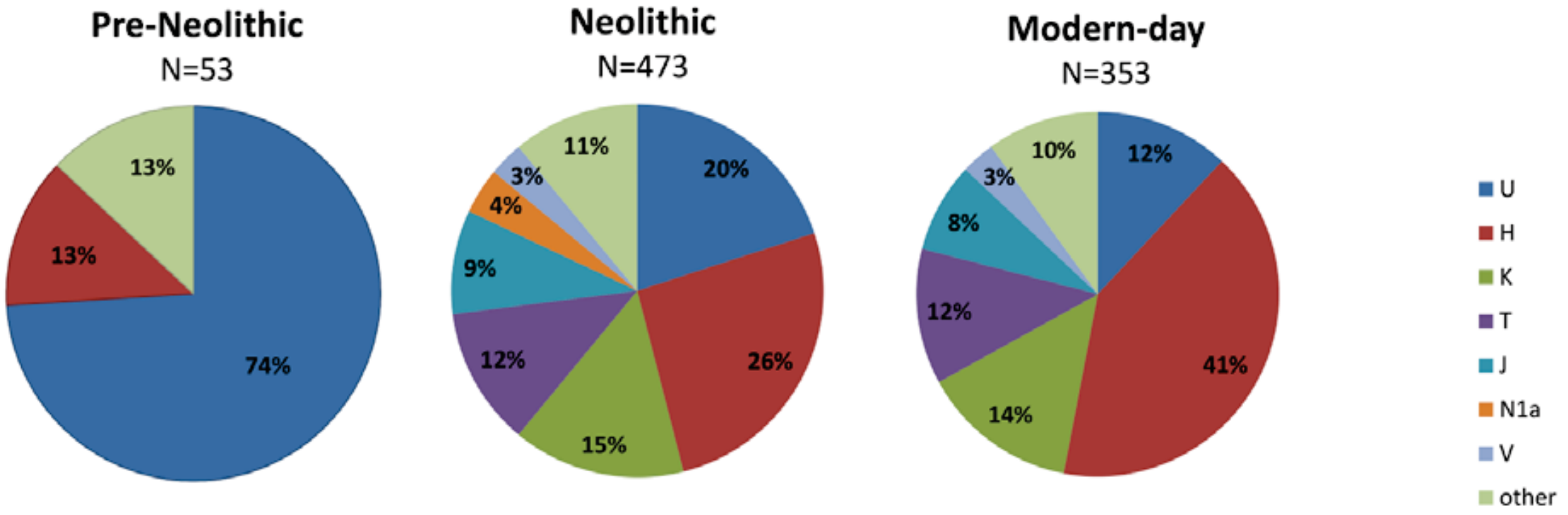

The mtDNA consensus sequences were aligned using the software MUSCLE [23]. MEGA 5.2 [24] was used to generate a Maximum Parsimony tree, which included the mtDNA sequences obtained here along with previously published complete early modern human mtDNAs [2, 3, 12, 14, 25-28] and 54 present-day human mtDNAs from a worldwide dataset [29]. Figure S4.2 shows that the Mesolithic genomes studied here cluster together with previously published pre-Neolithic mtDNAs.

MEGA 5.2 was also used to calculate nucleotide distances to the root of haplogroup R for the ancient sequences belonging to this clade as well as to 154 modern-day sequences falling into haplogroup R[6]. The mean nucleotide distance of the prehistoric samples to the most recent common ancestor of haplogroup R is significantly shorter than that of all modern mtDNAs (Mann-Whitney U test, two-tailed, $p < 0.001$, Figure S4.3), demonstrating the effect of branch shortening (ancient mtDNA has accumulated fewer substitutions over time) [3]. The early Neolithic individual Stuttgart falls at the upper end of the prehistoric distribution. Plotting the age of the samples against the pairwise nucleotide distance and calculating the slope of the regression (Figure S4.3) gives an estimate of the mitochondrial rate of $1.94 \pm 0.36 \times 10^{-8}$ substitutions per bp per year for the mtDNA genome, comparable to previous estimates [3, 30].

***Figure S4.2: Maximum Parsimony tree of 54 modern and 27 ancient mtDNA genomes.*** *The Mesolithic genomes studied here in red, the Stuttgart sample is in blue, and previously published European pre-Neolithic and Neolithic genomes are marked with red and blue asterisks, respectively. Bootstrap values above 0.9 are given at major nodes.*

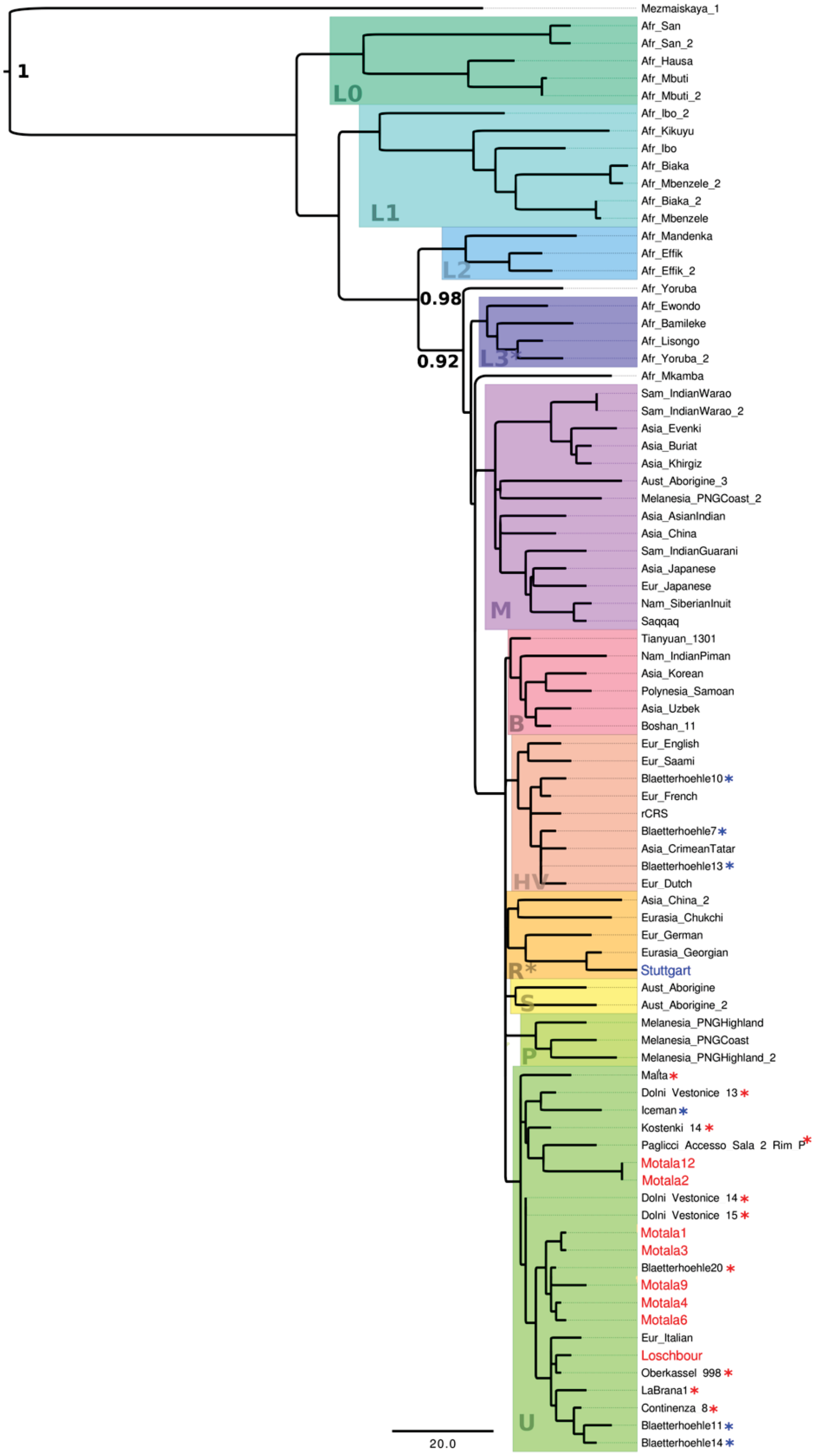

***Figure S4.3: Pairwise distance comparisons to the root of haplogroup R. (Top)*** *Pairwise nucleotide distance to the root of hg R for the complete mtDNA of 154 present-day and 20 prehistoric humans that fall inside the R clade.* ***(Bottom)*** *Plot of nucleotide distance against age of the sequence, slope of the linear regression gives a substitution rate of the whole mitochondrial genome (1.94±0.36 × $10^{-8}$ substitutions per bp per year).*

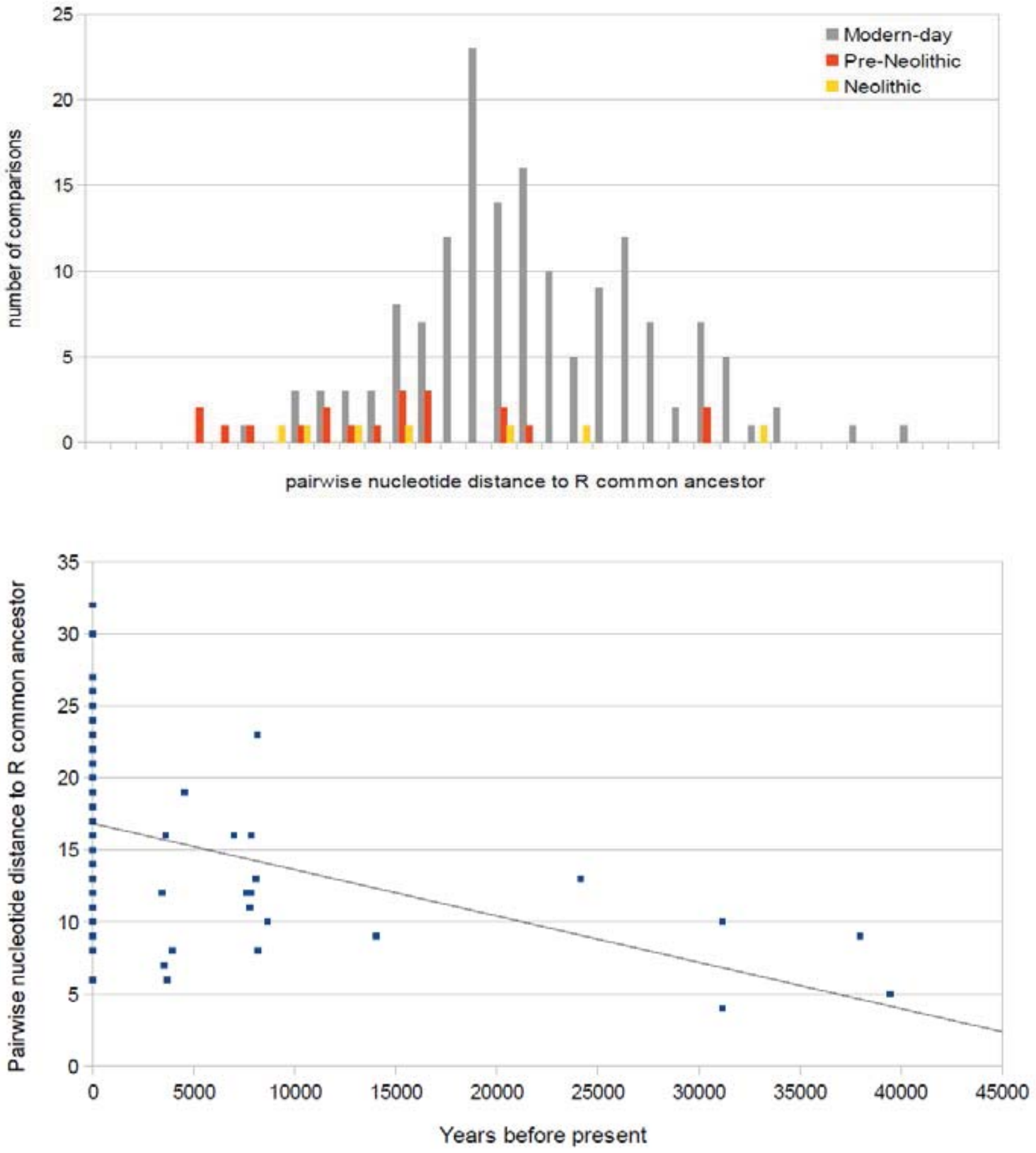

# Supplementary Information 5

## Sex determination and Y chromosome analysis

Iosif Lazaridis† and Gabriel Renaud†*

* To whom correspondence should be addressed (gabriel_renaud@eva.mpg.de)
† Contributed equally to this section

We infer which of the ancient human individuals in this study are likely to be male, and determine their Y chromosome haplogroups using publicly available Y chromosome SNPs. For Loschbour, which has the highest coverage of all the samples, we also determine its phylogenetic placement in a larger Y-SNP dataset of present-day humans.

### Sex determination

Based on morphological elements of the skeleton such as the pelvis and the skull, the sex of an individual can be inferred with high accuracy. However, ancient skeletons are often fragmentary or morphologically altered. It is therefore of interest to be able to use genetic information to determine the sex of an individual. As males have a single X and Y chromosome, the coverage of X chromosome nucleotides is expected to be about half of the autosomal coverage, and a significant number of reads are expected to map to the Y chromosome[1]. Conversely, females will have X chromosome coverage comparable to the autosomal coverage, and few reads mapping to the Y chromosome (largely due to regions of similarity between the Y and the X chromosomes). The ratio of reads mapping to the X and Y can thus be used to infer sex.

We extracted reads of high map quality (MAPQ $\geq$ 30) using *samtools* 0.1.18, and identified Loschbour and five of the Motala individuals (#2, 3, 6, 9, 12) as males by studying the ratio of chrY to (chrX+chrY) reads, using a tool recently developed for this purpose (Table S5.1).[1]

***Table S5.1: Sex determination using the number of reads (N) aligning to the X- and Y-chromosomes after MAPQ filtering.*** *The ratio (R_y) of NchrY/(NchrY+NchrX) with its standard error (SE) and 95% CI is presented.*

| Sample | NchrY+NchrX | NchrY | R_y | SE | 95% CI | Assignment |
|---|---|---|---|---|---|---|
| Loschbour | 22,068,747 | 1,873,062 | 0.0849 | 0.0001 | 0.0848−0.085 | XY |
| Stuttgart | 34,997,784 | 87,882 | 0.0025 | 0.00001 | 0.0025−0.0025 | XX |
| Motala1 | 349,043 | 1,095 | 0.0031 | 0.0001 | 0.003-0.0033 | XX |
| Motala2 | 162,747 | 13,560 | 0.0833 | 0.0007 | 0.082−0.0847 | XY |
| Motala3 | 588,788 | 49,687 | 0.0844 | 0.0004 | 0.0837−0.0851 | XY |
| Motala4 | 143,005 | 483 | 0.0034 | 0.0002 | 0.0031-0.0037 | XX |
| Motala6 | 25,549 | 2,176 | 0.0852 | 0.0017 | 0.0817−0.0886 | XY |
| Motala9 | 11,571 | 932 | 0.0805 | 0.0025 | 0.0756−0.0855 | XY |
| Motala12 | 2,384,534 | 200,346 | 0.084 | 0.0002 | 0.0837−0.0844 | XY |

### Y chromosome haplogroup determination

We used Y-chromosome SNPs included in the Y chromosome phylogeny of the International Society of Genetic Genealogy (ISOGG, http://www.isogg.org, version 9.22) to determine the haplogroups of the ancient samples. We removed SNPs with incomplete information (e.g., lacking physical position) and SNPs marked by ISOGG as "Investigation". For each SNP we examined the reference allele in the same physical position (*hg19/GRCh37*) to correct strand assignment errors. Since C/G and A/T

SNPs cannot be fixed in this manner we did not use these for further analysis. We also excluded apparently heterozygous sites since these are not expected on chromosome Y and might reflect contamination, mapping error, or deamination. We intersected the set of called sites for each individual with the physical positions of ISOGG SNPs, and used this to determine the Y-chromosome haplogroup for each individual.

Loschbour belongs to Y chromosome haplogroup I2a1b, as defined by seven mutations. Table S5.2 lists a number of upstream mutations that securely place this individual within the I haplogroup. In addition, the table lists a number of sites that are derived in present-day individuals with this haplogroup, and for which Loschbour is ancestral (Table S5.2).

***Table S5.2: Alleles assigning Loschbour to I2a1b*(xCTS5375, CTS8486. I2a1b1, I2a1b2, I2a1b3).***

| Haplogroup | SNP | anc | der | Ypos37 | Read Depth | State |
|---|---|---|---|---|---|---|
| I2a1b | CTS176 | A | G | 2,785,672 | 8 | + |
| I2a1b | CTS1293 | G | A | 7,317,227 | 4 | + |
| I2a1b | L178 | G | A | 15,574,052 | 12 | + |
| I2a1b | CTS5985 | A | G | 16,594,452 | 18 | + |
| I2a1b | CTS7218 | A | C | 17,359,886 | 8 | + |
| I2a1b | CTS8239 | A | G | 17,893,806 | 14 | + |
| I2a1b | M423 | G | A | 19,096,091 | 13 | + |
| I2a1 | P37.2 | T | C | 14,491,684 | 7 | + |
| I2a | L460 | A | C | 78,79,415 | 7 | + |
| I2 | M438 | A | G | 16,638,804 | 14 | + |
| I2 | L68 | C | T | 18,700,150 | 12 | + |
| I | P38 | A | C | 14,484,379 | 2 | + |
| I | M170 | A | C | 14,847,792 | 14 | + |
| I | M258 | T | C | 15,023,364 | 5 | + |
| I | PF3742 | G | A | 16,354,708 | 9 | + |
| I | L41 | G | A | 19,048,602 | 3 | + |
| I2a1b | CTS5375 | A | G | 16,233,135 | 16 | − |
| I2a1b | CTS8486 | C | T | 18,049,134 | 11 | − |
| I2a1b1 | M359.2 | T | C | 14,491,671 | 9 | − |
| I2a1b2 | L161.1 | C | T | 22,513,718 | 7 | − |
| I2a1b3 | L621 | G | A | 18,760,081 | 15 | − |
| I2a1b3a | L147.2 | T | C | 6,753,258 | 5 | − |

Note: State (+) here and in following tables indicates presence of the derived allele, and state (−) the ancestral allele. Of the SNPs considered to be phylogenetically equivalent for defining haplogroup I2a1b, Loschbour carried the ancestral state for two of them. Present-day individuals appear to be derived (Kenneth Nordtvedt, personal communication) for all these SNPs. It thus appears that Loschbour belonged to a branch of this haplogroup that lacked some of the mutations found in his closest relatives today.

Motala2 belongs to Y-haplogroup I on the basis of three mutations. It has the ancestral state for CTS1293 while Loschbour, Motala3, and Motala12 are derived for that SNP (Table S5.3).

***Table S5.3: Diagnostic Motala2 alleles place it in haplogroup I*(xI1, I2a2, CTS1293).***

| Haplogroup | SNP | anc | der | Ypos37 | Read Depth | State |
|---|---|---|---|---|---|---|
| I | P38 | A | C | 14,484,379 | 1 | + |
| I | PF3742 | G | A | 16,354,708 | 1 | + |
| I | L41 | G | A | 19,048,602 | 1 | + |
| I1 | S108 | T | G | 6,681,479 | 1 | − |
| I1 | L845 | T | G | 7,652,844 | 1 | − |
| I1 | M253 | C | T | 15,022,707 | 1 | − |
| I1a2a1a | S440 | G | A | 17,863,355 | 1 | − |
| I1a2b | Z2540 | C | T | 4,160,142 | 2 | − |
| I2a1b | CTS1293 | G | A | 7,317,227 | 1 | − |
| I2a1b3 | L621 | G | A | 18,760,081 | 1 | − |
| I2a2 | L37 | T | C | 17,516,123 | 1 | − |
| I2a2a1c1b | L703 | G | A | 14,288,983 | 1 | − |
| I2a2a1c2a2a1a1 | S434 | G | A | 17,147,721 | 1 | − |

Motala3 belongs to Y-haplogroup I2a1b on the basis of three mutations. It matches Loschbour for all sites for which we could make a comparison (Table S5.4).

***Table S5.4: Diagnostic Motala3 alleles place it in haplogroup I2a1b*(xI2a1b1, I2a1b3).***

| Haplogroup | SNP | anc | der | Ypos37 | Read Depth | State |
|---|---|---|---|---|---|---|
| I2a1b | CTS176 | A | G | 2,785,672 | 1 | + |
| I2a1b | CTS1293 | G | A | 7,317,227 | 1 | + |
| I2a1b | CTS7218 | A | C | 17,359,886 | 1 | + |
| I2a1 | P37.2 | T | C | 14,491,684 | 1 | + |
| I2 | L68 | C | T | 18,700,150 | 1 | + |
| I | M258 | T | C | 15,023,364 | 2 | + |
| I | PF3742 | G | A | 16,354,708 | 1 | + |
| I2a1b1 | M359.2 | T | C | 14,491,671 | 1 | − |
| I2a1b3 | L621 | G | A | 18,760,081 | 1 | − |

Motala6 has the allelic state L55+ (19413335 G>A), placing it in Y-haplogroup Q1a2a, but L232− (17516095 G>A), which contradicts the hypothesis that it belongs to haplogroup Q1. These two observations are phylogenetically inconsistent, and we thus cannot assign a haplogroup to it.

Motala9 (Table S5.5) belongs to Y-haplogroup I on the basis of P38+. However, it is not on the I1 branch on the basis of P40−. P40 is a C→T mutation and might reflect ancient DNA damage. I1 occurs at high frequencies in present-day Swedes[2], but has not been detected in prehistoric Europe, consistent with our observation here that Motala9 is probably not I1.

***Table S5.5: Diagnostic Motala9 alleles place it in haplogroup I*(xI1).***

| Haplogroup | SNP | Ancestral | Derived | GRCh37 | Read Depth | State |
|---|---|---|---|---|---|---|
| I | P38 | A | C | 14,484,379 | 1 | + |
| I1 | P40 | C | T | 14,484,394 | 1 | − |

Motala12 (Table S5.6) belongs to Y-haplogroup I2a1b on the basis of five mutations and is thus assigned to I2a1b*(xI2a1b1, I2a1b3). A number of upstream mutations securely place it in haplogroup I. Motala12 is also ancestral for the same two I2a1b-defining SNPs as Loschbour. It appears that the L178 clade was present in at least two locations of pre-Neolithic Europe, as Motala3, Motala12 and Loschbour all belong to it.

***Table S5.6: Alleles placing Motala12 in haplogroup I2a1b*(xCTS5375, CTS8486, I2a1b1, I2a1b3).***

| Haplogroup | SNP | anc | der | Ypos37 | Read Depth | State |
|---|---|---|---|---|---|---|
| I2a1b | CTS176 | A | G | 2,785,672 | 3 | + |
| I2a1b | CTS1293 | G | A | 7,317,227 | 2 | + |
| I2a1b | L178 | G | A | 15,574,052 | 2 | + |
| I2a1b | CTS5985 | A | G | 16,594,452 | 1 | + |
| I2a1b | CTS7218 | A | C | 17,359,886 | 1 | + |
| I2a1 | P37.2 | T | C | 14,491,684 | 1 | + |
| I2a | L460 | A | C | 7,879,415 | 2 | + |
| I2 | L68 | C | T | 18,700,150 | 1 | + |
| I | M170 | A | C | 14,847,792 | 1 | + |
| I | M258 | T | C | 15,023,364 | 2 | + |
| I | PF3742 | G | A | 16,354,708 | 1 | + |
| I2a1b | CTS5375 | A | G | 16,233,135 | 1 | – |
| I2a1b | CTS8486 | C | T | 18,049,134 | 3 | – |
| I2a1b1 | M359.2 | T | C | 14,491,671 | 1 | – |
| I2a1b3 | L621 | G | A | 18,760,081 | 2 | – |

**Phylogenetic analysis of the Loschbour Y chromosome**

Given that Loschbour carried a Y chromosome belonging to haplogroup I, we sought to investigate how this individual's Y-chromosome compares to the diversity of present-day humans. We used a dataset from Lippold *et al.*[3] which contains the genotype at 2,799 positions for a worldwide panel of 623 Y chromosomes. Using BEAST v1.7.51[4] with a coalescence prior of 60,000 years for all non-Africans and a tip age for Loschbour of 8,000 years, we reconstructed a Bayesian inference tree. This analysis makes no assumptions about the phylogeny, or about the Y chromosome mutation rate. Instead, the phylogeny is reconstructed based on Y chromosome polymorphism data from the analyzed samples themselves and the mutation rate is inferred based on "branch shortening"; the rate of missing mutations on the Loschbour-specific lineage due to the fact that it has evolved less.

The Y chromosome of Loschbour clusters with present-day haplogroup I individuals (Figure S5.1), confirming the placement based on diagnostic alleles for this haplogroup (Table S5.2). The coalescence of the I and J2 haplogroups is inferred to have occurred 31 kya whereas the coalescence of this group to the R haplogroup is inferred to be 40 kya. These numbers are broadly in the range of date estimates for the expansion of populations in Europe[5]. On a finer scale, a modern Russian individual (HGDP00887) was found to share a high degree of similarity to the Loschbour individual. Out of 2,790 informative positions in both individuals (9 were not covered by reads in Loschbour), only 5 sites were different, including 3 transversions and 2 transitions. We used another Russian individual (HGDP00894) to show that all five of these five mutations fell on the HGDP00887 branch rather than on the Loschbour branch. These derived mutations may either have occurred on the HGDP00887 branch after divergence from Loschbour, or they might represent errors in HGDP00887.

***Figure S5.1**: Phylogenetic position of Loschbour Y chromosome within present-day haplogroup I.*
*The highlighted branch (yellow) displays the Loschbour individual and its closest relative for the Y chromosome in the dataset, a present-day Russian. The inferred coalescence of the sub-tree here with the R haplogroup (not shown) is 40,661 years, consistent with some previous estimates[6].*

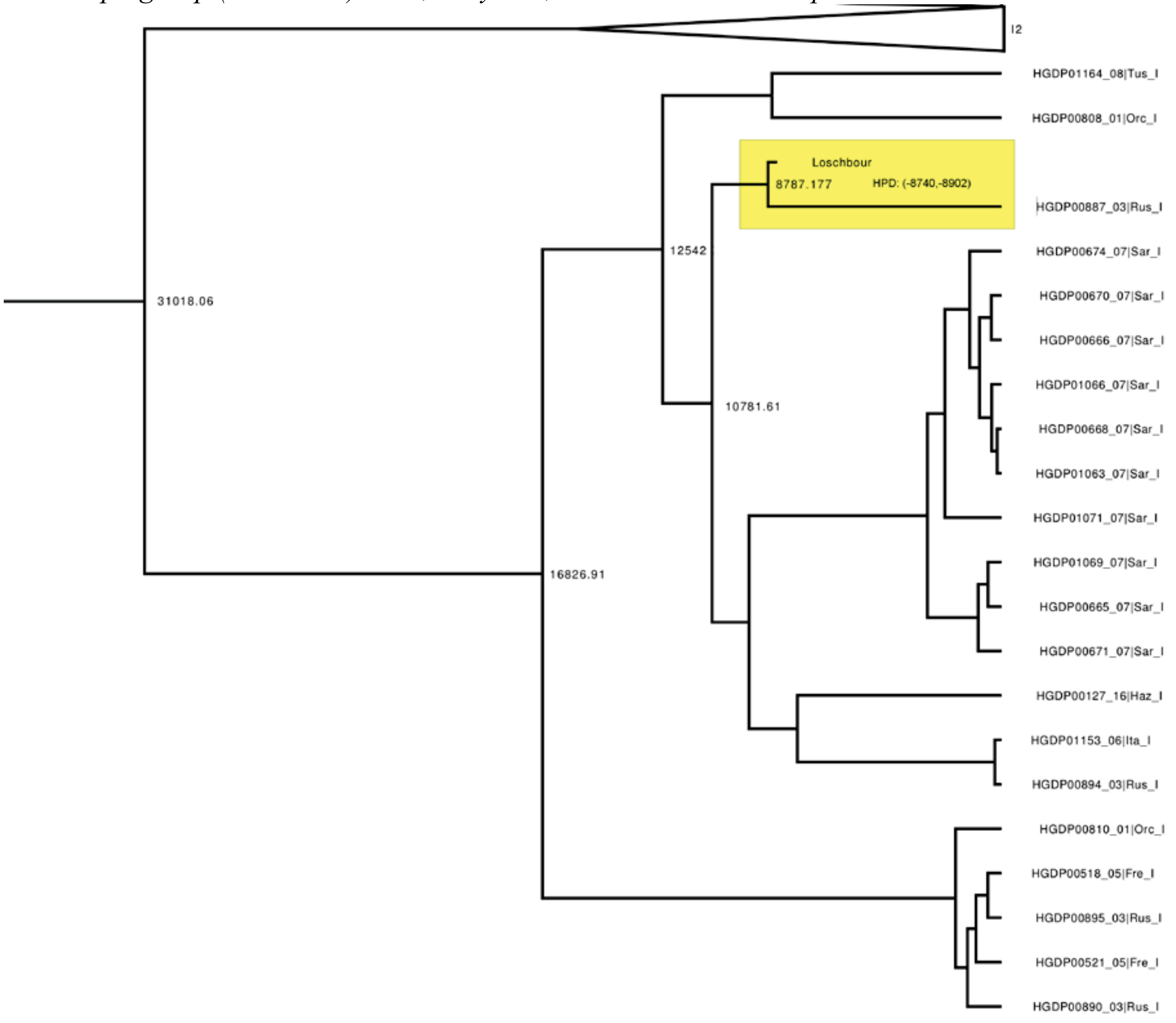

**Frequency of Y chromosome haplotypes**

All 5 of the male individuals in this study belonged to the I haplogroup. Among present-day Germans, this is found at a frequency of ~24% (Figure S5.2). At present, the limited number of ancient samples for which Y chromosome data is available makes it difficult to assess how statistically surprising it is that the Y haplogroup group occurs in all five of the ancient Mesolithic males but in only a quarter of present-day German males.

***Figure S5.2: Pie chart of Y chromosome haplogroups of the individuals in this study and present-day Germans.*** *For the ancient individuals, only haplogroup I was found. However, in present-day Europeans from Germany, I is a minority haplogroup.*

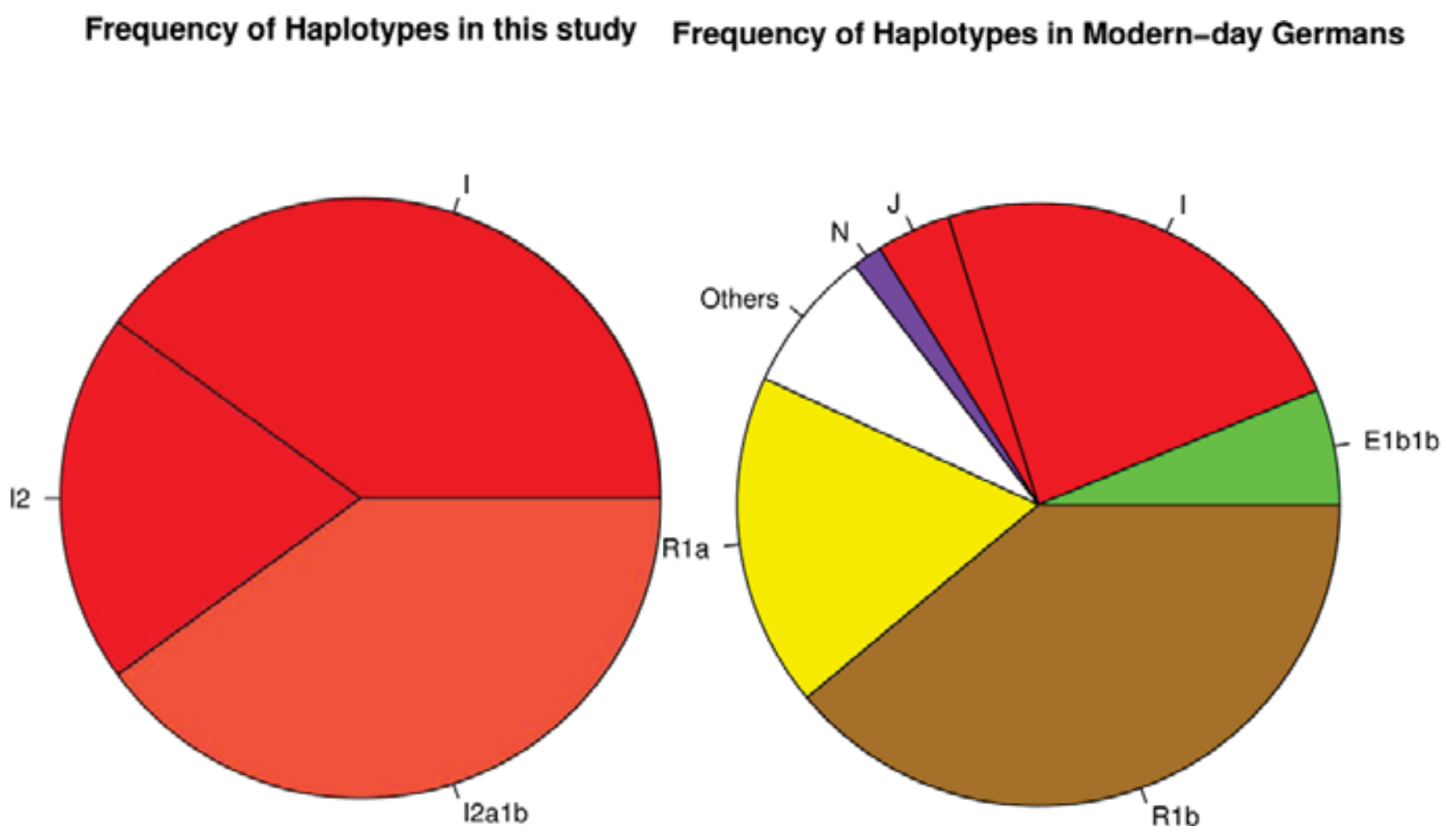

**Discussion**

We have found that Loschbour and all four Motala males whose haplogroups we could determine belong to Y-haplogroup I , a haplogroup that today, is found almost exclusively in Europe at a much lower frequency than it occurred around 8,000 years ago[7]. Its sister clade (haplogroup J) is hypothesized to have had a Near Eastern origin[8]. It has been suggested that haplogroup I was common in pre-agricultural Europeans[9], and our study confirms this directly as it documents its presence in two European hunter-gatherer groups from the period immediately antedating the Neolithic transition.

We cannot, at present, determine when Y chromosome haplogroup I entered Europe, although its occurrence in two Mesolithic European hunter-gatherer populations (Loschbour and Motala) and its near absence outside of Europe today suggest an old origin.

It is tempting to speculate that haplogroup I might be the dominant European Y chromosome haplogroup in Palaeolithic Europe, that is, the male counterpart of maternally inherited mitochondrial haplogroup U (SI4). Y chromosome haplogroup I[10] as well as mitochondrial haplogroup U, have also been identified in Neolithic Europeans, and are found throughout Europe in present-day populations. Thus, both maternally- and paternally-inherited genetic components of present-day Europeans may reflect a history of admixture: a genetic contribution from both the hunter-gatherers and early farmers of Europe. Y chromosome haplogroup I is scarce in the Near East today, with only sporadic occurrences of this haplogroup in the North Caucasus (~3% in frequency)[11], consistent with limited gene flow from Europe into this area. This finding is also consistent with the near absence of haplogroup U in the Near East and our findings from the autosomal data.

The present-day frequency of haplogroup I in Europe is variable, with local maxima in Scandinavia[2] and the western Balkans, which might reflect more recent expansions. Our finding that Loschbour, a Mesolithic west European, was M423+ is hard to reconcile with a previous suggestion[12] that this lineage diffused during the Neolithic from south-eastern Europe.

The absence of Y-haplogroup R1b in our two sample locations is striking given that it is, at present, the major west European lineage. Importantly, however, it has not yet been found in ancient European contexts prior to a Bell Beaker burial from Germany (2,800-2,000BC)[13], while the related R1a lineage has a first known occurrence in a Corded Ware burial from Germany (2,600BC)[14]. This casts doubt on early suggestions associating these haplogroups with Paleolithic Europeans[15], and is more consistent with their Neolithic entry into Europe at least in the case of R1b[16,17]. More research is needed to document the time and place of their earliest occurrence in Europe. Interestingly, the Mal'ta boy (MA1) belonged[18] to haplogroup R* and we tentatively suggest that some haplogroup R bearers may be responsible for the wider dissemination of Ancestral North Eurasian ancestry into Europe, as their haplogroup Q relatives may have plausibly done into the Americas[18].

This work provides a first glimpse into the pre-Neolithic Y chromosomes of Europe. Unlike the La Braña male[19], a Mesolithic Iberian whose C-V20 chromosome is extremely rare in present-day Europe[20], the Y-chromosomes of Loschbour and the Motala males appear to belong to haplogroups that persist in a substantial fraction of present Europeans. Despite the fact that our sample is limited to two locations and five male individuals, the results in this section are consistent with haplogroup I representing a major pre-Neolithic European clade, and hint at subsequent events during and after the Neolithic transition as important contributors to the Y chromosomal variation of living Europeans.

# Supplementary Information 6

## Neanderthal ancestry estimates in the ancient genomes

Nick Patterson* and David Reich

* To whom correspondence should be addressed (nickp@broadinstitute.org)

It is possible to obtain an unbiased estimate of Neanderthal ancestry proportion in a non-African population $\hat{\alpha}$ using an $f_4$-ratio or $S$-statistic[1-3]:

$$\hat{\alpha} = \frac{f_4(Altai,\ Denisova; Test,\ Yoruba)}{f_4(Altai,\ Denisova; Vindija,\ Yoruba)} \quad \text{(S6.1)}$$

Here, Altai is a high coverage (52×) Neanderthal genome sequence[4] and Denisova is a high coverage sequence[5] from another archaic human population (31×), both from Denisova Cave in the Altai Mountains of southern Siberia. Vindija is low coverage (1.3×) Neanderthal data from a mixture of three Neanderthal individuals from Vindija Cave in Croatia[1].

Intuitively, the $f_4$-statistic in the numerator measures the rate at which a Test modern human shares more alleles with the Altai Neanderthal than with the Denisova hominin, using a modern human population without appreciable Neanderthal ancestry (Yoruba) as a baseline. If Test is from a modern human population that has negligible Neanderthal ancestry, the Test and Yoruba populations will share alleles with the archaic samples at the same rate, and the expected value is 0. On the other hand, if Test has a fraction $\alpha$ of Neanderthal ancestry, it will share alleles with Altai at a higher rate than with Denisova. Thus, the expected value is $\alpha f_4(Altai,\ Denisova;\ Neanderthal,\ Yoruba)$. Since the denominator is the same $f_4$-statistic (using Vindija to represent Neanderthals), the expected value of the ratio is $\alpha$.

Table S6.1 reports estimates of Neanderthal that emerge for each ancient sample analyzed in this study. Some of the standard errors are large due to the limited amount of data available for the samples. The inferred values are all consistent with each other and the approximately the approximately 2% Neanderthal ancestry estimated for present-day humans[1-3].

***Table S6.1.*** *Estimates of Neanderthal ancestry for ancient samples*

| | Estimate | Std. Err. from a Block Jackknife |
|---|---|---|
| AG2 | 1.7% | 0.5% |
| MA1 | 1.6% | 0.3% |
| LaBrana | 2.2% | 0.3% |
| Loschbour | 2.1% | 0.3% |
| Motala1 | 2.2% | 0.5% |
| Motala2 | 1.8% | 0.6% |
| Motala3 | 1.9% | 0.4% |
| Motala4 | 1.2% | 0.7% |
| Motala6 | 2.1% | 1.1% |
| Motala9 | 0.8% | 1.8% |
| Motala12 | 1.9% | 0.3% |
| Motala_merge | 2.0% | 0.3% |
| Skoglund_hunter | 3.8% | 1.0% |
| Skoglund_farmer | 3.8% | 1.8% |
| Stuttgart | 1.8% | 0.3% |

# Supplementary Information 7

## Analysis of segmental duplications and copy number variants

Peter H. Sudmant* and Evan E. Eichler

* To whom correspondence should be addressed (psudmant@gmail.com)

We constructed read-depth based copy number maps for the Loschbour, Stuttgart and Motala12 individuals, and co-analyzed them with whole genome sequence data from the archaic Denisova[1] and the archaic Altai Neandertal genome[2], as well as 25 deeply sequenced present-day human genomes that we have described previously[3].

These maps consist of windowed copy number estimates across the genome 500 unmasked base-pairs wide and sliding by intervals of 100 unmasked base-pairs.

We first assessed the quality of each genome in regions putatively free of copy number variation[3] which allowed us to quantify our ability to accurately determine a diploid copy number state for each 500 bp window encompassed in these loci. As read-depth based copy number estimates are often affected by GC-associated sequencing biases we assessed our accuracy as a function of genomic GC% (Figure S7.1) and cumulatively across all regions examined. This is a fairly strict test as to actually call a copy number variant the aggregate signal of many windows is taken into account. All genomes with the exception of the low coverage Motala12 individual demonstrate a high fraction of correctly determined sites (>85%) with higher concordance in individuals sequenced to higher coverage, such as the Neandertal and Denisova genomes.

***Figure S7.1: Quality control and copy number calling.*** *(a) The fraction of incorrectly correctly determined diploid loci is plotted as a function of GC content. (b) The total fraction of correctly determined diploid loci in each individual assessed in this study.*

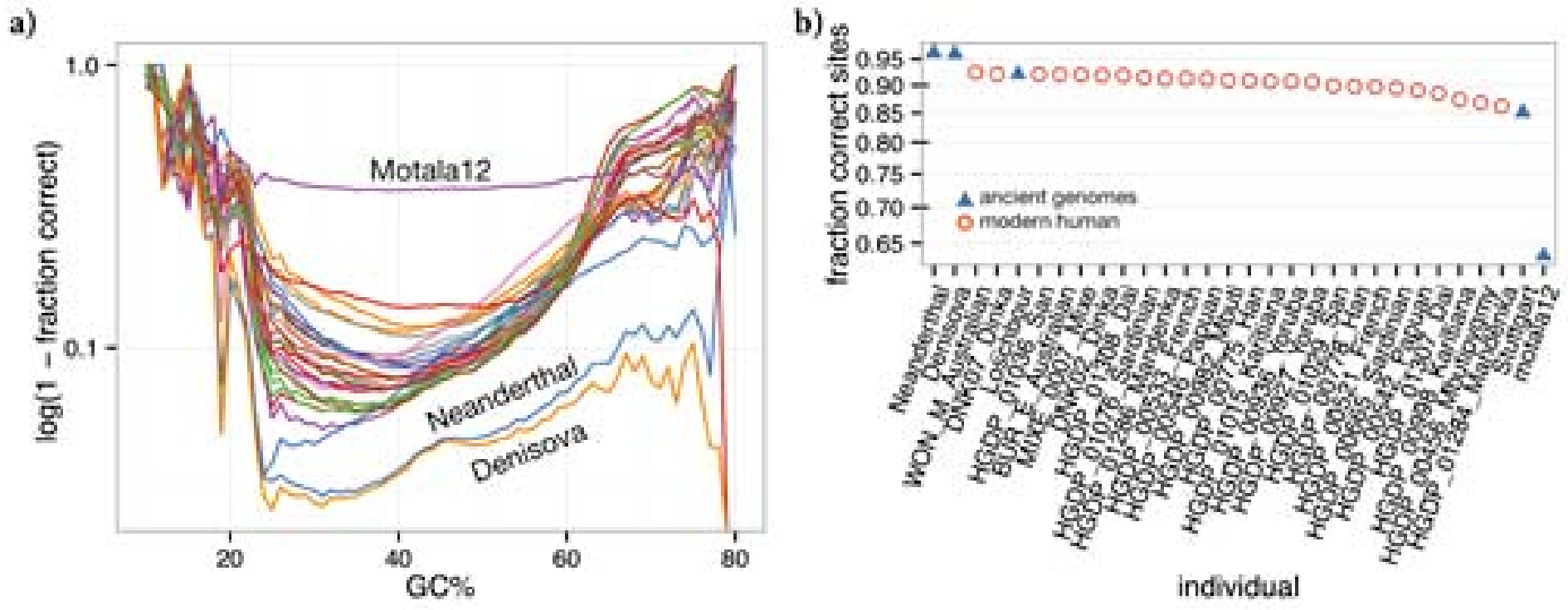

We next performed a genome-wide scan for copy number variants using digital array comparative genomic hybridization (dCGH)[4]. Briefly, for all-pairwise combinations of genomes, we calculate the $\log_2$-ratio across copy number windows. We then segment each of these $\log_2$-ratio maps using a scale-space filtering based technique[4]. We then compute the significance of the putative copy number variants determined by the segmentation using a modified T-statistic to account for the autocorrelation of the underlying data. Putative CNV calls amongst individual pairs of genomes are finally merged by calculating the reciprocal overlap between all overlapping calls and merging overlapping calls with cophenetic distance $\leq 0.85$. We restricted our analysis to calls with a log-likelihood of $\leq$-6.

The Motala12 individual was excluded from this initial scan due to its lower coverage. We identified 3,846 putative copy number variants, 2,094 of which intersected segmental duplications. For segmentally duplicated CNVs in the Stuttgart and Loschbour individuals, the copy and position of segmental duplications is within the range of present-day humans.

We focused on two biologically relevant loci in these individuals for further discussion. The first region is the *CCL3L1* locus on 17q12 (Figure S7.2). The *CCL3L1* gene encodes for a chemokine involved in immune and inflammatory processes. The copy number of *CCL3L1* varies widely among humans and is stratified between European and non-European populations (Figure S7.3). While European populations exhibit fewer copies of *CCL3L1* (a median of 2 copies), the Stuttgart individual has only a single copy of the locus, a state shared by only ~1.5% of individuals (as assessed from 1000 Genomes Project populations).

***Figure S7.2. A copy number heat-map of the 17q13 locus. (a)*** *The Stuttgart individual exhibits a deletion of the locus encompassing the chemokine genes CCL3L1, CCL3L3, CCL4L1 and CCL4L2. Deletions of these genes occurs in ~1.5% of 1000 genomes individuals. (b) Distribution of CCL3L1 copy number in the 1000 genomes Phase I project and the analyzed archaic genomes. The Stuttgart genome exhibits a single copy of CCL3L1.*

*(a)*

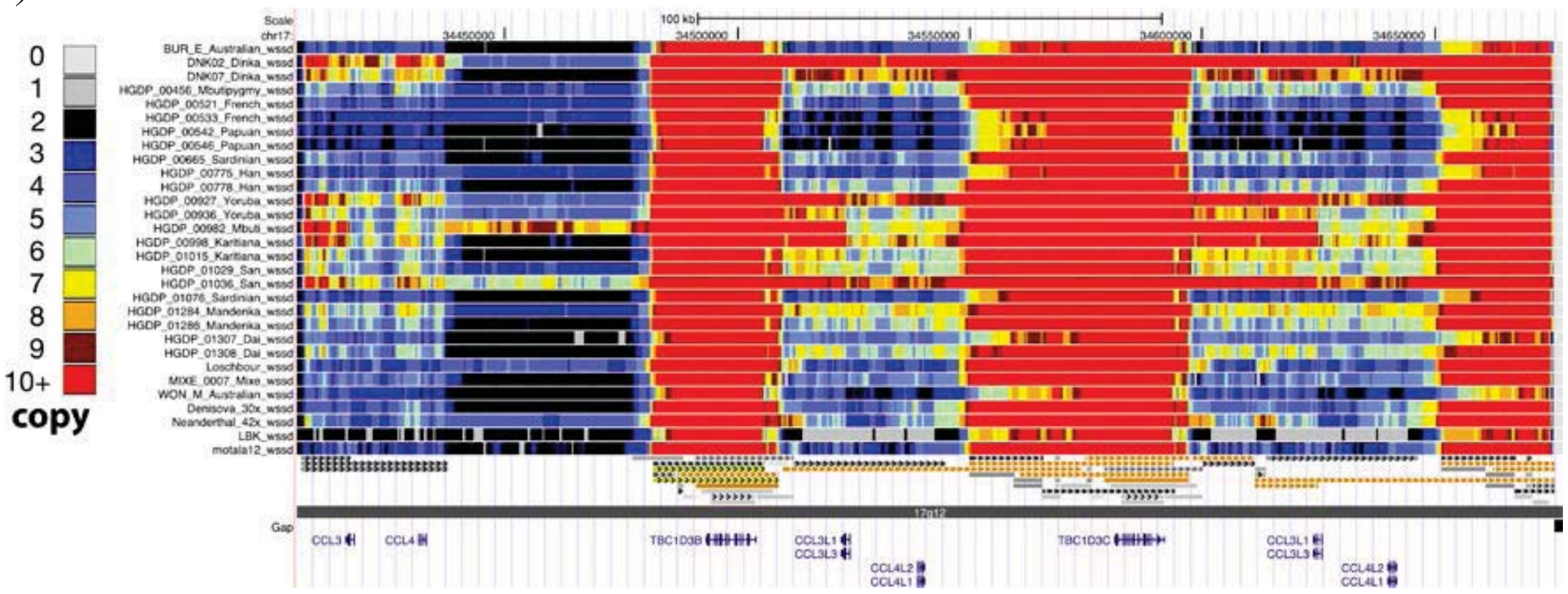

*(b)*

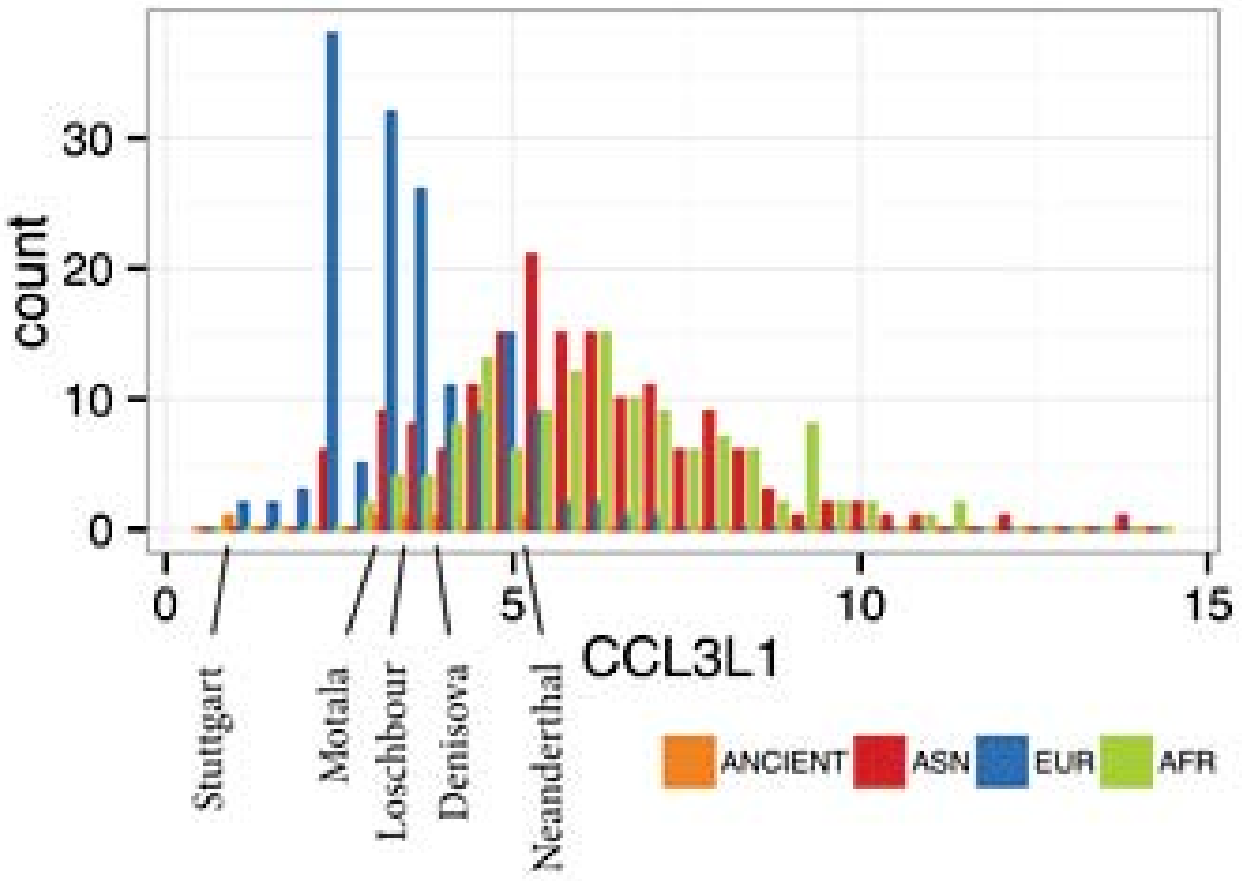

A second potentially interesting locus is the amylase gene (*AMY1*), which is has also recently expanded in human populations, potentially as a result of adaptations to start rich diets [5]. We have recently reported that the Denisova and Neandertal genomes have the ancestral state of two copies of amylase. We find that Motala12, Loschbour and Stuttgart have 6, 13, and 16 copies of *AMY1* respectively. As this is well within the range of current European populations it suggests that amylase copy number expanded in *Homo sapiens* before the advent of agriculture. Further sequencing of early modern humans will help to refine the picture of the emergence of extra amylase copies in homo-sapiens.

***Figure S7.3. Distribution of AMY1 copy number.*** *Both Stuttgart and Loschbour have high copy numbers of AMY1, while Motala12 has a low copy number.*

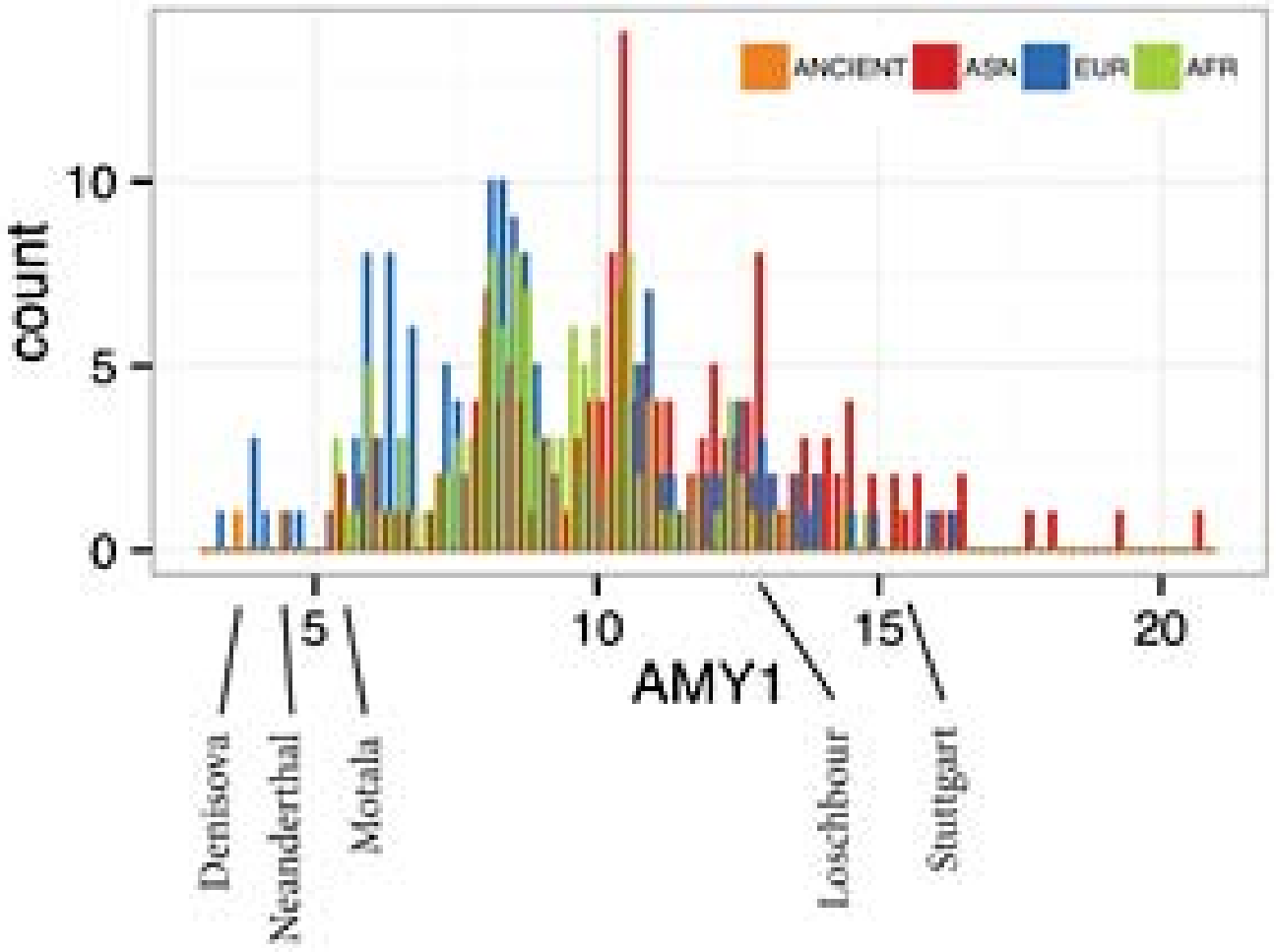

We identified 1,556 non-segmentally duplicated CNVs among the individuals assessed and genotyped these (Supplementary Online Table 1**).** These include 76 deletions and 168 duplications in the Stuttgart individual and 68 deletions and 104 duplications in the Loschbour individual. These loci include loss of 4 olfactory genes and exon intersecting homozygous deletions of the *LCE3C* and *LCE3B* genes in Stuttgart. In the Loschbour individual we identify the loss of 2 olfactory genes, the same homozygous *LCE3C* and *LCE3B* intersecting deletion, and a heterozygous deletion of the first exon of one isoform of *SLC25A24.* No Loschbour or Stuttgart specific events were identified, consistent with these individuals having variation within the range of present-day humans.

# Supplementary Information 8

## Phenotypic inference

Karola Kirsanow*

* To whom correspondence should be addressed (kirsanow@uni-mainz.de)

### Introduction

We assessed three ancient modern humans (the Loschbour forager, the Motala12 forager, and the Stuttgart farmer) at a panel of SNPs and multi-allelic markers having well-validated phenotypic effects in present-day humans. Many of the markers in our panel have also been affected by natural selection relatively recently in human prehistory.

Walsh et al. [1-4], among others [5-7], demonstrated that it is possible to predict human eye, hair, and skin color phenotypes with accuracy using a small number of DNA variants. Here, we predicted pigmentation phenotypes of the three ancient modern humans using two models [2,3,6-8] that have been validated in present-day populations [1,4,7-9] as well as on skeletal remains [10]. We used these models together with additional SNP and haplotype data to infer the most likely iris, hair and skin pigmentation for the Loschbour, Motala12, and Stuttgart individuals.

We also analyzed Loschbour and Stuttgart at 35 single nucleotide polymorphisms (SNPs) known from genome-wide association studies (GWAS) to be reproducibly associated with susceptibility to the Metabolic Syndrome (MetS) and compared the results of two different diabetes-risk score models incorporating 24 of these SNPs [11,12]. MetS-related SNPs have evidence of being under recent selection[13,14], possibly because of pressures related to changes in diet and climate associated with human migration and the adoption of agriculture.

Finally, we analyzed the Loschbour, Motala12, and Stuttgart samples at a panel of sites having well-described phenotypic effects that have been identified as targeted by selection in recent human prehistory [15-19].

### Methods

We analyzed DNA polymorphism data stored in the VCF format [20] using the VCFtools software package (http://vcftoools.sourceforge.net/). For the Loschbour and Stuttgart individuals, we included data from sites not flagged as LowQuality, with genotype quality (GQ) of ≥30, and SNP quality (QUAL) of ≥50. We chose to assess Motala12 because this individual was sequenced at higher average coverage than the other Motala samples. However, the coverage of the Motala12 individual (2.4×) was appreciably lower than that of the Loschbour (22×) and Stuttgart (19×) individuals, and we therefore altered our genotyping methodology to account for the limitations imposed by lower coverage at our sites of interest. Specifically, we included sites having at least 2× coverage which passed visual inspection of the local alignment using samtools tview (http://samtools.sourceforge.net)[21].

We carried out five sets of phenotypic analyses:

(1) We assessed the genotypes of the Loschbour forager and Stuttgart farmer at pigmentation SNPs included in the 8-plex and the Hirisplex pigmentation phenotype prediction models system (Table S8.1). We assigned probabilities to hair and eye color phenotypes using the enhanced version 1.0 Hirisplex Microsoft Excel macro[8] (Table S8.2) .

(2) We assessed the genotypes of the Loschbour and Motala foragers and the Stuttgart farmer at a panel of SNPs in the *HERC2/OCA2* and *SLC24A5* genes comprising several pigmentation-related haplotypes (Tables S8.3, S8.4, S8.5).

(3) We assessed the genotypes of the Loschbour and Stuttgart individuals at a panel of SNPs associated with risk for Metabolic Syndrome (Table S8.6) and that form the basis for two type 2 diabetes (T2D) risk score models [11,12]. We computed weighted genotype risk scores using the methods described in Meigs (2008) and Cornelis (2009). We additionally genotyped the three ancient modern humans (Table S8.7) at 5 SNPs in the *SLC16A11* gene forming a haplotype associated with T2D risk in which the risk haplotype appears to derive from Neanderthal introgression.

(4) We assessed the genotypes of the Loschbour, Motala12, and Stuttgart individuals at a panel of SNPs with evidence for recent natural selection, including several known to show high allele frequency differentiation between European and East Asian populations (Table S8.8).

(5) We assessed the genotypes of the Loschbour, Motala12, and Stuttgart individuals at 8 SNPS in the *NAT2* gene in order to determine the acetylation phenotype of the three ancient modern humans (Table S8.9).

We caution that the pigmentation phenotype models and metabolic syndrome risk scoring models are not independent. In particular, seven of the eight markers in the 8plex pigmentation model are also included in the Hirisplex model, and four metabolic syndrome-associated SNPs are shared between the two diabetes risk score models.

**Results**

*Pigmentation*

For hair color, the integrated results of the genotype-based pigmentation models indicate that there is at least a 98% probability that both the Stuttgart and Loschbour individuals had dark (brown or black) hair. The Hirisplex model assigns the highest probability to black hair color for both individuals (Table S8.2).

The results of the 8-plex skin pigmentation model were inconclusive for both the Loschbour and Stuttgart individuals. However, the Loschbour and Stuttgart genotypes at rs1426654 in *SLC24A5* indicate that the Stuttgart individual may have had lighter skin than the Loschbour hunter-gatherer. The Loschbour individual is homozygous for the rs1426654 ancestral allele, while Stuttgart is homozygous for the derived skin-lightening allele [22,23]. This allele has the single greatest effect on skin pigmentation of the SNPs identified to date in present-day populations[22-24].

For eye color, the single most significant determinant is the rs12913832 SNP in the *HERC2* gene. The genotype at this site excludes the possibility that the Stuttgart farmer had blue eyes. Positive iris color determinations are less secure. The Loschbour forager is homozygous for the derived allele at rs12913832, indicating that this individual is likely to have had blue (61% probability) or intermediate iris color (17% probability). It has been suggested that this mutation arose within the last 6,000 to 10,000 years, and thus the Loschbour individual would have been a relatively early carrier [25].

It should be noted that while these predictive models have been well-validated in present-day European populations, it is possible that heretofore undetected variation in pigmentation genes may have contributed to phenotypic variation in ancient modern humans. Any such variation would not be captured by this analysis.

The number of sites without coverage in the Motala12 sample prevented the inclusion of this individual in the model-based phenotypic inference. However, some inferences can be made from high-impact single pigmentation SNPs and short haplotypes. The Motala12 forager, like the Stuttgart farmer, carries at least one copy of the derived rs1426654 pigmentation-lightening allele, and may thus have had lighter skin pigmentation than the Loschbour forager. We typed the three ancient modern humans at 7 SNPs forming three short haplotypes associated with eye color in present-day worldwide populations (Table S8.3)[26]. The observed reads in the Motala12 forager, like the Loschbour forager match the blue-eye-associated allele at all 7 SNPs. However, this includes two SNPs (rs7495174 and rs1291382) at only 1× coverage. Motala12 carries the blue-eye haplotype at the two BEH3 SNPs, which are in LD with the causal SNP, rs1291382, in present-day Europeans (but not outside of Europe)[26]. However, the Stuttgart farmer is also homozygous for the two blue-associated BEH3 SNPs, despite being homozygous for the ancestral allele at rs1291382. Stronger support for the inference of non-brown eyes for Motala12 is the observation of the derived allele at rs1129038, a site at almost complete LD with rs1291382 in present-day populations[26].

In order to determine the haplotypes carried by the ancient modern humans at major pigmentation loci and compare these ancient 'superalleles' with haplotypes in present-day populations, we genotyped all three ancient modern humans at a number of SNPS in the *OCA2*, *HERC2*, and *SLC24A5* genes:

1. We evaluated the three ancient modern humans at a panel of 13 SNPs in the *OCA2*/*HERC2* region comprising the haplotype shared by 97% of blue-eyed individuals in a present-day study population[25] (Table S8.4).

2. We determined the genotypes of the Loschbour, Stuttgart, and Motala12 individuals at a panel of 16 SNPs in the *SLC24A5* gene comprising the *SLC25A5* haplotype observed in most present-day humans carrying the derived rs1426654 (A111T) allele [27](Table S8.5).

We find that the Loschbour forager is homozygous for the h-1 *HERC2*/*OCA2* haplotype observed in 97% of blue-eyed individuals in a present-day study population from Turkey, Jordan, and Denmark[25]. Due to a combination of missing and heterozygous sites in our unphased genotype data, the haplotypes of the Motala12 forager and Stuttgart farmer could not be conclusively determined. The identification of h-1 in one of the earliest reported carriers of the rs1291382 derived allele supports the inference that h-1 is the founder haplotype for the blue-eye mutation[25].

Examining the *SLC24A5* region, we find that the Stuttgart farmer is homozygous for the C11 haplotype found in 97% of all modern carriers of the derived rs1426654 pigmentation-lightening allele[27]. The A111T mutation is estimated to have arisen at ~22-28 kya[28], with the selective sweep favoring its rise beginning ~19kya (under a dominant model) or ~11kya (under an additive model)[29]. The Loschbour forager does not carry the derived rs1426654 allele. The Motala12 forager, like the Stuttgart farmer, is homozygous for the C11 haplotype. Although three of the SNPs defining C11 were genotyped at 1× coverage in the Motala12 sample, C11 is the only haplotype matching the possible patterns of variation.

*Metabolic Syndrome Risk Score*

Complex human disease phenotypes are less amenable to genotype-based prediction than externally visible characteristics such as pigmentation. The diabetes risk scoring systems developed to date thus do not have strong predictive power at the population level [11]. Nevertheless, we used these scoring systems to begin to characterize the metabolic genotypes of the Loschbour and Stuttgart ancient modern humans in comparison with the average present-day non-diabetic genotype.

We evaluated the Loschbour and Stuttgart individuals using two different type 2 diabetes (T2D) risk score models (Motala12 could not be included in this analysis because of inadequate coverage at crucial SNPs) (Table S8.6). We find that the two ancient modern humans display metabolic syndrome-associated allele spectra comparable to those observed in present-day Europeans.

The Meigs 2008 model indicates a higher T2D risk for the Loschbour individual relative to Stuttgart. The weighted genotype risk scores for both Loschbour and Stuttgart fall within the overlapping one standard deviation ranges of the present-day diabetic and non-diabetic ranges predicted by this model.

The Cornelis 2009 model predicts a roughly equal risk for both individuals. According to this model, the weighted genotype risk scores of both the Loschbour and Stuttgart individuals (10.7 and 10.6, respectively) are within the 95% CI of that of the median present-day non-diabetic individual (10.4).

Overall, the risk allele is the ancestral allele at 19 out of the 35 MetS-associated SNPs whose genotypes we evaluated. The Loschbour and Stuttgart individuals carried similar numbers of ancestral MetS-associated risk alleles (21 for Loschbour and 19 for Stuttgart), and derived MetS risk alleles (14 for Loschbour and 15 for Stuttgart). Moreover, the MetS risk scores of the ancient forager and farmer do not indicate any significant departures from the MetS risk score averages in present-day Europeans.

We also genotyped the three ancient modern humans at 5 SNPs in the *SLC16A11* gene comprising a haplotype associated with type 2 diabetes risk in a present-day Latin American population and for which the risk haplotype is believed to derive from Neanderthal introgression[30] (Table S8.7). None of the three ancient modern humans carried the risk haplotype.

*Other phenotypic characteristics*

We also assessed the Loschbour, Motala12, and Stuttgart individuals for their genotype at nine SNPs with well-validated phenotypic associations and evidence for recent positive selection (Table S8.8).

All three ancient modern humans are homozygous for the ancestral alleles at the *LCTa* and *LCTb* polymorphisms and as a result are predicted to have been unable to digest lactose as adults. The *LCTa* mutation has been estimated to have first experienced positive selection between 6,256 and 8,683 years ago in central Europe [31]. Thus, although the allele is associated with the spread of the LBK culture, it is likely to have been uncommon in early LBK populations, consistent with our results.

The heterozygous state of both the Stuttgart and Loschbour individuals at a SNP in the *AGT* gene suggests that they may have had a slightly increased risk of hypertension (Motala12 could not be genotyped at this locus). The risk allele in the *AGT* gene is an ancestral allele. The derived protective allele is estimated to have arisen 22,500-44,500 years ago [16].

All three ancient modern humans were homozygous for a derived allele at rs2740574 in *CYP3A4*, which is thought to confer protection from certain forms of cancer and is also possibly associated with protection from rickets [32]. Loschbour and Stuttgart are also homozygous for the derived allele at rs776746 in *CYP3A5*, which is estimated to have arisen ~75,000 years ago [33], and which affects drug metabolism (Motala12 could not be genotyped at this locus).

We additionally evaluated the three ancient modern humans for their genotypes at SNPs in *EDAR*, *ADH1B*, *ABCC1*, and *ALDH2* that are known to have high allele frequency differentiation between present-day Europeans and East Asians. All three individuals are homozygous for alleles associated with wet earwax (*ABCC1*) and non-shoveled incisors (*EDAR*), which are phenotypes known to occur at higher frequency in Europeans [34-36]. Both the Loschbour forager and the Stuttgart farmer carried the ancestral alleles at *ALDH2* or *ADH1Ba*, two loci associated with alcohol metabolism which are known to have been under recent positive selection in East Asian populations[18,19,37] . The derived alleles at these SNPs are associated with slower alcohol metabolism and reduced alcohol consumption.The Motala12 forager could not be genotyped at *ALDH2* or *ADH1Ba*, but was homozygous for the ancestral allele at *ADH1Bb*. The Stuttgart individual was also homozygous for the ancestral allele at a third alcohol-metabolism locus under recent positive selection, *ADH1Bb(Arg48His)*[37], (the Loschbour and Motala12 individuals could not be conclusively genotyped at this locus).

Finally, we assessed the genotypes of the three ancient modern humans at 8 SNPs in the *NAT2* gene in order to determine the acetylation phenotype (rapid>intermediate>slow) of each individual. *NAT2* is involved in the metabolism of a wide variety of xeniobiotics, including a number of carcinogens, and there is evidence from present-day populations for selection favoring the slow-acetylator phenotype, possibly related to dietary changes accompanying the transition to agriculture. [38,39] We inferred acetylation status using three partially independent methods: the 4 SNP panel proposed by Hein *et al.*[40]*;* the NAT2pred online tool (NAT2pred.rit.albany.edu)[41], which consists of the 4 SNP panel plus 3 additional SNPs; and a tag SNP which is in strong LD with a 7 SNP panel in present-day Europeans[42].The three inference methods we employed agreed that the Stuttgart farmer was most likely to have been a slow acetylator, the Loschbour forager was an intermediate acetylator, and the Motala12 forager was a rapid acetylator. The observation of a slow-acetylator phenotype in the sample from an early agriculturalist population supports the inference that selection on the *NAT2* region may be related to the adoption of farming.

***Table S8.1. Loshbour, Stuttgart and Motala12 genotypes for SNPs associated with pigmentation***

| | | **8plex SNPs** | | |
|---|---|---|---|---|
| **SNP** | ***Gene*** | **Loschbour** | **Stuttgart** | **Motala12** |
| rs1291382 | *HERC2* | G/G | A/A | G/G (1x) |
| rs1545397 | *OCA2* | A/A | A/A | ° |
| rs16891982 | *SLC45A2* | C/C | C/C | ° |
| rs885479 | *MC1R* | G/G | G/G | G/G (3x) |
| rs1426654 | *SLC24A5* | G/G | A/A | A/A (3x) |
| rs12896399 | *SLC24A4* | G/G* | T/T | ° |
| rs6119471 | *ASIP* | C/C | C/C | C/C (3x) |
| rs12203592 | *IRF4* | T/T | C/C | T/T (3x) |
| | | **Hirisplex SNPs** | | |
| **SNP** | **Gene** | **Loschbour** | **Stuttgart** | **Motala12** |
| n29insa | *MC1R* | C/C | C/C | C/C (5x) |
| rs11547464 | *MC1R* | G/G | G/G | ° |
| rs885479 | *MC1R* | G/G | G/G | G/G (3x) |
| rs1805008 | *MC1R* | C/C | C/C | C/C (2x) |
| rs1805005 | *MC1R* | G/G | G/G | G/G (6x) |
| rs1805006 | *MC1R* | C/C | C/C | C/C (6x) |
| rs1805007 | *MC1R* | C/C | C/C | ° |
| rs1805009 | *MCIR* | G/G | G/G | G/G (4x) |
| y152och | *MC1R* | C/C | C/C | ° |
| rs2228479 | *MC1R* | G/G* | G/G | G/G (4x) |
| rs1110400 | *MC1R* | T/T | T/T | ° |
| rs28777 | *SLC45A2* | C/A | C/C | A/A (4x) |
| rs16891982 | *SLC45A2* | C/C | C/C | ° |
| rs12821256 | *KITLG* | T/T | T/T | T/T (3x) |
| rs4959270 | *EXOC2* | A/A | C/C | A/A (1x) |
| rs12203592 | *IRF4* | T/T | C/C | T/T (3x) |
| rs1042602 | *TYR* | C/C | C/A | C/C (3x) |
| rs1800407 | *OCA2* | C/C | C/C* | C/C (1x) |
| rs2402130 | *SLC24A4* | G/A | A/A | A/A (1x) |
| rs12913832 | *OCA2/HERC2* | G/G | A/A | G/G (1x) |
| rs2378249 | *PIGU/ASIP* | A/A | A/A | A/A (1x) |
| rs12896399 | *SLC24A4* | G/G* | T/T | ° |
| rs1393350 | *TYR* | G/G | G/G | G/G (4x) |
| rs683 | *TYRP1* | A/A | A/A | A/A (1x) |

*Coverage at each position is given in parentheses for the Motala12 sample. *These SNPs had genotype quality between 20 and 30, but passed other quality filters. °These SNPs could not be genotyped.*

***Table S8.2. Hirisplex model probability scores for pigmentation.***

| | Loschbour | Stuttgart |
|---|---|---|
| | | |
| **HAIR** | **Probability** | **Probability** |
| **Brown** | 0.413 | 0.220 |
| **Red** | 0 | 0 |
| **Black** | 0.579 | 0.774 |
| **Blond** | 0.008 | 0.005 |
| | | |
| **HAIR SHADE** | **Probability** | **Probability** |
| **Light** | 0.022 | 0.006 |
| **Dark** | 0.978 | 0.994 |
| | | |
| **EYE** | **Probability** | **Probability** |
| **Blue** | 0.613 | 0 |
| **Intermediate** | 0.166 | 0.004 |
| **Brown** | 0.222 | 0.996 |

***Table S8.3. OCA2/HERC2 haplotypes observed in the Loschbour, Motala, and Stuttgart individuals***

| Haplotype | Blue-eye allele | SNP | Loschbour | Stuttgart | Motala12 |
|---|---|---|---|---|---|
| BEH1 | A | rs4778138 | A/A | A/G | A/A (3x) |
| BEH1 | C | rs4778241 | C/C | A/C | C/C (5x) |
| BEH1 | A | rs7495174 | A/A* | A/A | A/A (1x) |
| BEH2 | T | rs1129038 | T/T* | C/C | T/T (3x) |
| BEH2 | G | rs1291382 | G/G | A/A | G/G (1x) |
| BEH3 | C | rs916977 | C/C | C/C | C/C (2x) |
| BEH3 | T | rs1667394 | T/T | T/T | T/T (4x) |

*Genotypes of the Loschbour, Stuttgart, and Motala12 individuals at the sites comprising three haplotypes associated with blue eyes in modern populations*[26]*. Coverage at each position is given in parentheses for the Motala12 sample.*These sites had genotype quality between 20 and 30 but passed other quality filters.*

***Table S8.4. 13-SNP OCA2/HERC2 genotypes of the Loschbour, Motala, and Stuttgart individuals***

| SNP | Loschbour | Stuttgart | Motala12 |
|---|---|---|---|
| rs4778241 | C/C | A/C | C/C (5x) |
| rs1129038 | T/T* | C/C | T/T (3x) |
| rs12593929 | A/A | A/A | A/A (2x) |
| rs12913832 | G/G | A/A | G/G (1x) |
| rs7183877 | C/C | C/C | ° |
| rs3935591 | C/C | C/C | T/C (7x) |
| rs7170852 | A/A | A/A | A/A (7x) |
| rs2238289 | A/A | A/A | A/A (3x) |
| rs3940272 | G/G* | ° | ° |
| rs8028689 | T/T | T/T | T/T (2x) |
| rs2240203 | T/T | T/T | T/T (5x) |
| rs11631797 | G/G | G/G* | ° |
| rs916977 | CC | CC | C/C (2x) |
| **Haplotype** | **h-1** | * | * |

*Genotypes of the Loschbour, Stuttgart, and Motala12 individuals at the 13 sites comprising the haplotype found at high frequency in present blue-eyed individuals, along with haplotype asignment[25]. Coverage at each position is given in parentheses for the Motala12 sample.* These SNPs had genotype quality <30 but passed other quality filters; °the individual could not be genotyped at this locus.*

***Table S8.5. 16-SNP SLC25A5 genotypes of the Loschbour, Motala12, and Stuttgart individuals***

| SNP | Loschbour | Stuttgart | Motala12 |
|---|---|---|---|
| rs1834640 | A/G | A/A | A/A(2x) |
| rs2675345 | A/G* | A/A | A/A(3x) |
| rs2469592 | A/G | A/A | A/A(4x) |
| rs2470101 | T/C | T/T | T/T(5x) |
| rs938505 | C/T | C/C | C/C(2x) |
| rs2433354 | C/T | C/C | C/C(1x) |
| rs2459391 | A/G | A/A | A/A(5x) |
| rs2433356 | A/G | G/G | G/G(4x) |
| rs2675347 | A/G | A/A | A/A(1x) |
| rs2675348 | A/G | A/A | A/A(2x) |
| rs1426654 | G/G | A/A | A/A(3x) |
| rs2470102 | A/G* | A/A | A/A(5x) |
| rs16960631 | A/A | A/A | A/A(3x) |
| rs2675349 | A/G | A/A | A/A(1x) |
| rs3817315 | C/T | C/C | C/C(2x) |
| rs7163587 | T/C | C/C | C/C(4x) |
| **Haplotype** | * | **C11** | **C11** |

*Genotypes of the Loschbour, Stuttgart, and Motala12 individuals at the 16 sites comprising the SLC25A5 haplotype observed in most modern humans carrying the derived rs1426654 (A111T) allele, along with haplotype assignment (SLC25A5 haplotype assignment was not possible for the Loschbour forager) [27]. Coverage at each position is given in parentheses for the Motala12 sample.*These SNPs had genotype quality <30 .*

***Table S8.6. Metabolic syndrome SNPs assessed in Loschbour and Stuttgart, by risk score model.***

| **Metabolic syndrome associated SNPs** | | | |
|---|---|---|---|
| **SNP** | **Gene** | **Loschbour** | **Stuttgart** |
| rs7923837 | *HHEX* | G/G | A/A |
| rs5015480 | *HHEX/IDE* | C/C | T/T |
| rs3802678 | *GBF1* | A/A | A/T |
| rs6235 | *PCSK1* | C/C | G/G |
| rs7756992 | *CDKAL1* | A/G | A/G |
| rs6446482 | *WFS1* | C/G | C/G |
| rs11037909 | *EXT2* | T/C | T/C |
| rs6698181 | *PKN2* | T/T | C/T |
| rs17044137 | *FLJ39370* | T/A | T/A |
| rs12255372 | *TCF7L2* | G/G | G/G |
| rs7480010 | *LOC387761* | A/A | A/A |
| rs11634397 | *ZFAND6* | A/G | G/G |
| rs10946398 | *CDKAL1* | A/C | C/C |
| rs8050136 | *FTO* | A/A | C/A |
| **Meigs 2008** | | | |
| **SNP** | **Gene** | **Loschbour** | **Stuttgart** |
| rs7901695° | *TCF7L2* | T/T | ° |
| rs7903146° | *TCF7L2* | C/C | C/C |
| rs1470579 | *IGF2BP2* | A/C | A/A |
| rs10811661 | *CDKN2A/B* | T/C | T/T |
| rs864745 | *JAZF1* | T/C | C/C |
| rs5219 | *KCNJ11* | * | T/C |
| rs5215* | *KCNJ11* | C/T | C/T |
| rs12779790 | *CDC123/CAMK1D* | A/G | A/A |
| rs7578597 | *THADA* | T/T | T/T |
| rs7754840 | *CDKAL1* | G/C | C/C |
| rs7961581 | *TSPAN8/LGR5* | T/T | C/T |
| rs4607103 | *ADAMTS9* | C/C | C/C |
| rs1111875 | *HHEX* | C/C | T/T |
| rs10923931 | *NOTCH2* | G/T | G/T |
| rs13266634 | *SLC30A8* | C/C | C/C |
| rs1153188 | *DCD* | T/T | T/A |
| rs1801282 | *PPARG* | C/C | C/C |
| rs9472138 | *VEGFA* | C/C | C/C |
| rs10490072 | *BCL11A* | T/C | T/T |
| rs689 | *INS* | A/T | A/T |
| **Weighted genotype risk score** | | **118.0** | **101.6** |
| **Cornelis 2009** | | | |
| **SNP** | **Gene** | **Loschbour** | **Stuttgart** |
| rs564398 | *CDKN2A/B* | C/T | T/T |
| rs10010131 | *WFS1* | A/G | A/G |
| rs7754840 | *CDKAL1* | G/C | C/C |
| rs4402960 | *IGF2BP2* | G/T | G/G |
| rs1801282 | *PPARG* | C/C | C/C |
| rs5219 | *KCNJ11* | * | T/C |
| rs5215* | *KCNJ11* | C/T | C/T |
| rs1111875 | *HHEX* | C/C | T/T |
| rs13266634 | *SLC30A8* | C/C | C/C |
| rs10811661 | *CDKN2A/B* | T/C | T/T |
| rs7901695 | *TCF7L2* | T/T | T/T° |
| Rs7903146° | *TCF7L2* | C/C | C/C |
| **Weighted genotype risk score** | | **10.6** | **10.7** |

**For the purpose of computing the Weighted Genotype Risk Score, we use rs5215 as a proxy for rs5219, which failed to pass the quality filter for the Loschbour sample. These two SNPs are in strong LD ($r^2$=0.90)[43]*

*in present-day populations. °rs7903146 was used as a proxy for rs7901695, which for the Stuttgart individual failed to pass the quality filter. The two SNPs are in strong LD ($r^2$=0.98) [44] in present-day populations.*

***Table S8.7. SLC16A11 genotypes of the Loschbour, Stuttgart, and Motala12 individuals***

| SNP | Loschbour | Stuttgart | Motala12 |
|---|---|---|---|
| rs75493593 (P443T) | G/G* | G/G | G/G (2x) |
| rs75418188 (G340S) | C/C* | C/C* | C/C (2x) |
| rs13342232 (L187L) | A/A | A/A | A/A (3x) |
| rs13342692 (D127G) | T/T | T/T | T/T (7x) |
| rs117767867 (V113I) | C/C | C/C | C/C (3x) |

*Genotypes of the Loschbour, Stuttgart, and Motala12 individuals at 5 SNPs in the SLC16A11 gene comprising a haplotype associated with type 2 diabetes risk in a modern Latin American population[30]. Coverage at each position is given in parentheses for the Motala12 sample.*These sites had genotype quality between 20 and 30, but passed other quality filters.*

***Table S8.8. Loschbour, Stuttgart, and Motala12 genotypes for SNPs known to be under selection in modern humans***

| SNP | Gene | Loschbour | Stuttgart | Motala12 |
|---|---|---|---|---|
| rs182549 | *LCTb* | C/C | C/C | C/C(3x) |
| rs4988235 | *MCM6/LCTa* | G/G | G/G | G/G(2x) |
| rs699 | *AGT* | A/G | A/G | ° |
| rs4590952 | *KITLG* | A/G | G/G | G/G(2x) |
| rs2740574 | *CYP3A4* | T/T | T/T | T/T(3x) |
| rs776746 | *CYP3A5* | C/C | C/C | C/C(1x) |
| rs3827760 | *EDAR* | A/A | A/A | A/A(3x) |
| rs17822931 | *ABCC1* | C/C | C/C | C/C (3x) |
| rs671 | *ALDH2* | G/G | G/G | GG(1x) |
| rs3811801 | *ADH1Ba* | G/G | G/G | G/G(2x) |
| rs1229984 | *ADH1Bb* | ° | C/C | C/C(1x) |

*Coverage at each position is given in parentheses for the Motala12 sample. °The individual could not be genotyped at this locus.*

***Table S8.9. NAT2 genotypes and inferred acetylation status of the Loschbour, Stuttgart, and Motala12 individuals***

| SNP | Loschbour | Stuttgart | Motala12 |
|---|---|---|---|
| rs1801279 (191G>A) | G/G | G/G | GG (5x) |
| rs1801280 (341T>C) | T/C | C/C | TT (3x) |
| rs1799930 (590G>A) | G/G | G/A° | GG (4x) |
| rs1799931 (857G>A) | G/G | G/G* | GG (1x) |
| rs1495741(tag) | G/A | A/A | GG (3x) |
| rs1041983(282C>T) | C/C | C/C | CC (4x) |
| rs1799929(481C>T) | C/T° | T/T | CT (2x) |
| rs1208(803A<G) | G/A | G/G | AA (2x) |
| **Acetylation status** | **Intermediate** | **Slow** | **Rapid** |

*Genotypes of the Loschbour, Stuttgart, and Motala12 individuals at 8 SNPs in the NAT2 gene associated with acetylation status (rapid>intermediate>slow)[40-42,45] Coverage at each position is given in parentheses for the Motala12 sample.. °These sites had Qual <30 *This site had genotype quality between 20 and 30 but passed other quality filters.*

# Supplementary Information 9

## Affymetrix Human Origins genotyping dataset and ADMIXTURE analysis

Iosif Lazaridis, Nick Patterson, Susanne Nordenfelt, Nadin Rohland, George Ayodo, Hamza A. Babiker, Graciela Bailliet, Elena Balanovska, Oleg Balanovsky, Ramiro Barrantes, Gabriel Bedoya, Haim Ben-Ami, Judit Bene, Fouad Berrada, Claudio M. Bravi, Francesca Brisighelli, George B.J. Busby, Francesco Cali, Mikhail Churnosov, David E.C. Cole, Daniel Corach, Larissa Damba, George van Driem, Stanislav Dryomov, Jean-Michel Dugoujon, Sardana A. Fedorova, Irene Gallego Romero, Marina Gubina, Michael Hammer, Brenna Henn, Tor Hervig, Ugur Hodoglugil, Aashish R. Jha, Sena Karachanak-Yankova, Rita Khusainova, Elza Khusnutdinova, Rick Kittles, Toomas Kivisild, William Klitz, Vaidutis Kučinskas, Alena Kushniarevich, Leila Laredj, Sergey Litvinov, Theologos Loukidis, Robert W. Mahley, Béla Melegh, Ene Metspalu, Julio Molina, Joanna Mountain, Klemetti Näkkäläjärvi, Desislava Nesheva, Thomas Nyambo, Ludmila Osipova, Svante Pääbo, Jüri Parik, Fedor Platonov, Olga L. Posukh, Valentino Romano, Francisco Rothhammer, Igor Rudan, Ruslan Ruizbakiev, Hovhannes Sahakyan, Antti Sajantila, Antonio Salas, Elena B. Starikovskaya, Ayele Tarekegn, Draga Toncheva, Shahlo Turdikulova, Ingrida Uktveryte, Olga Utevska, René Vasquez, Mercedes Villena, Mikhail Voevoda, Cheryl Winkler, Levon Yepiskoposyan, Pierre Zalloua, Tatijana Zemunik, Cristian Capelli, Mark G. Thomas, Andres Ruiz-Linares, Sarah A. Tishkoff, Lalji Singh, Kumarasamy Thangaraj, Richard Villems, David Comas, Rem Sukernik, Mait Metspalu, David Reich*

* To whom correspondence should be addressed (reich@genetics.med.harvard.edu)

**Overview of the Human Origins dataset**

We begin by describing the Affymetrix Human Origins dataset of single nucleotide polymorphism (SNP) genotyped in diverse present-day humans.

Briefly, the Affymetrix Human Origins SNP array consists of 14 panels of SNPs for which the ascertainment is well understood[1,2]. The oligonucleotide probes on the array target of 627,421 SNPs of which 620,744 are on the autosomes, 4,331 are on chromosome X, 2,089 are on chromosome Y and 257 are on the mitochondrial DNA.

Genotypes of present-day humans on the array have already been reported in three studies:

- Patterson et al. 2012[2] is the original technical description of the array (File S1 of that study). The study also released genotyping data from 944 samples from the CEPH / Human Genome Diversity Panel from diverse worldwide populations (ftp://ftp.cephb.fr/hgdp_supp10/).
- Pickrell et al. 2012[3] and Pickrell et al. 2014[4] together presented genotyping data of an additional 212 individuals from southern and eastern African populations.

Here we report data from many additional populations, filling in sampling gaps in the dataset especially in West Eurasia, and also adding in sampling from other world regions.

We combined the genotypes from all samples into a single file. We then carried out a comprehensive curation of the data to identify a list of SNPs that appeared to perform reliably in genotyping, and to identify a list of samples that were not close relatives of others in the dataset or outliers relative to others from their own populations.

**SNP filtering of the Human Origins Dataset**

The genotyping was performed in seven batches over the course of several years. We were concerned that differences in the experimental or bioinformatic processing across batches might cause systematic differences in genotyping results for each batch that have nothing to do with population history. Moreover, a subset of samples were from whole genome amplified (WGA) material rather than from genomic DNA extracted from blood and saliva, and we were concerned that these samples might have systematic differences from the other samples. Populations comprised of samples derived from WGA material are identified with the suffix "_WGA" in the dataset that we are making available.

To curate the data, we began by computing the following statistics on each SNP:

(1) Genotyping concordance over 69 samples from the West African Yoruba (YRI) that overlapped between this study and low coverage sequencing data from the 1000 Genomes Project[5].

(2) Genotyping concordance rate over 25 samples from diverse populations that overlapped between this study and high coverage genome sequences reported in ref.[6].

(3) Completeness of genotyping (restricting to males on chromosome Y). This was performed for:
    (a) All samples except WGA
    (b) Just WGA
    (c) By genotyping batch excluding WGA (1, 2, 3, 4, 5, 6, 7)

(4) Alternate allele frequency of a pool of West Africans and a pool of West Eurasians in each batch.

(5) Homozygous, Heterozygous, and Variant genotype counts for a pool of West Africans and a pool of Europeans over all batches but excluding WGA samples. We include only females from chromosome X SNPs so that all genotypes are diploid.

(6) Male and female frequencies for a pool of West Africans and a pool of Europeans over all batches excluding WGA samples.

We first computed the following statistics to filter out potentially problematic SNPs.

• "Maxconcordance" – Maximum concordance with either the 69 1000 Genomes Project[5] or 25 deeply sequenced[6] samples. If a site has missing data in a sequencing dataset the concordance is reported as 0 for that dataset.

• "Completeness" – Completeness percentage of genotyping across the WGA samples.

• "Mincompleteness" – Minimum completeness percentage for the SNP across 7 genotyping batches.

• "WGAcompleteness" – Completeness for SNP in the WGA data.

Table S9.1 gives the fraction of autosomal SNPs that would be retained after applying different thresholds (for all but the WGAcompleteness metric).

***Table S9.1: Fraction of SNPs retained with different concordance and completeness***

| **Threshold:** | **50%** | **80%** | **85%** | **90%** | **95%** | **96%** | **97%** | **97.5%** | **98%** | **98.5%** | **99%** | **99.5%** |
|---|---|---|---|---|---|---|---|---|---|---|---|---|
| **Maxconcordance** | 0.997 | 0.996 | 0.996 | 0.996 | 0.994 | 0.992 | 0.991 | **0.990** | 0.976 | 0.974 | 0.963 | 0.919 |
| **Completeness** | 1.000 | 1.000 | 1.000 | 0.999 | **0.995** | 0.990 | 0.978 | 0.968 | 0.954 | 0.931 | 0.881 | 0.639 |
| **Mincompleteness** | 1.000 | 0.992 | 0.987 | **0.975** | 0.933 | 0.907 | 0.872 | 0.853 | 0.772 | 0.619 | 0.308 | 0.060 |

Note: We highlight in bold the thresholds we use for our main dataset.

We also pooled all non-whole genome amplified (non-WGA) West Eurasians and all non-WGA West Africans. This gave us counts of the three possible genotypes for each SNP: homozygous reference, heterozygous, and homozygous derived. We computed Hardy-Weinberg-like statistics for all SNPs, looking for a deficiency of heterozygous sites as might be expected from poor allele calling:

$$HW = \sum_{i=1}^{3} \frac{(obs_i - exp_i)^2}{exp_i + 1} \tag{S9.1}$$

We conservatively added 1 to the denominator terms to deflate the statistics in the context of low expected values. This resulted in a West African HW statistic and a West Eurasian HW statistic. We imposed thresholds for significance based on a $\chi^2$ distribution with 1 degree of freedom.

We next computed empirical derived allele frequencies for many different sample sets for each SNP *i*. We performed 21 different pairwise comparisons:

15 All West African pairwise comparisons for batches 1-6

3 All West Eurasian pairwise comparisons for batches 1, 6 and 7

1 All non-WGA male West Africans vs. all non-WGA female West Africans

1 All non-WGA male West Eurasians vs. all non-WGA female West Eurasians

1 All non-WGA West Eurasians vs. allele frequencies from randomly drawn reads from 107 YRI West Africans from the 1000 Genomes Project[5], computed as in Prüfer et al.[6]

Consider two allele frequencies $a_i$ and $b_i$ for sample sets $A$ and $B$ respectively, in a subset of the genome (either all the autosomes, or just chromosome X) with $n$ SNPs. Further define $\mu_i=(a_i+b_i)/2$ as the mean of these frequencies. Then we can compute the following statistic that is approximately $\chi^2$ distributed with 1 degree of freedom. In the denominator, we normalize by the mean of the numerator over all SNPs. This is a form of "genomic control" that normalizes by the mean of this over-dispersed chi-square distribution, so that the statistic is well described by a $\chi^2$ distribution with 1 d.f.

$$Stat_i = \left[\frac{(a_i - b_i)^2}{\mu_i(1-\mu_i)}\right] \Big/ \left[\frac{1}{n}\sum_{i=1}^{n} \frac{(a_i - b_i)^2}{\mu_i(1-\mu_i)}\right] \tag{S9.1}$$

In practice, we carried out our filtering by computing the maximum statistic "$\text{MaxStat}_i$" over 23 of the approximately $\chi^2$ statistics that we analyzed (2 Hardy-Weinberg and 21 frequency comparisons). We then only accepted SNPs with "$\text{MaxStat}_i$" less than a specified threshold.

The threshold we use for our main analysis is 20. For this threshold, we empirically found that we removed almost no SNPs from the bulk of the distribution that was symmetrically spread around the $y=x$ axis (as might be expected from the fact that it corresponds to a nominal P-value of $\sim 10^{-5}$, on the order of 1 divided by the number of SNPs in the dataset). However, this threshhold did remove a population of SNPs that had frequency of 0% in one population and were polymorphic in the other, which are clear genotyping failures suggesting that the filter is valuable.

***Table S9.2. Summary of SNP filters used***

| | Maxconc-ordance* | Comp-leteness | Mincomp-leteness | WGA comp-leteness | Max of 23 $\chi^2$ stats | SNPs removed | SNPs retained |
|---|---|---|---|---|---|---|---|
| Main dataset | >0.975 | >0.95 | >0.9 | None | >20 | 25,131 | 602,290 |
| With WGA data | >0.995 | >0.99 | >0.95 | >0.99 | >9 | 185,132 | 442,289 |

Note: We only remove SNPs on chromosomes 1-23. Users who wish to analyze the Y chromosome and mtDNA data should do so at their own discretion and need to design their own customized filtering.

Table S9.2 shows the filters we chose for the dataset. For the analyses reported in this study, we restrict to non-WGA samples with the exception of the Saami individual which we force in because of its high interest, and use thresholds that strike a balance between retaining a large fraction of SNPs while removing extreme outliers. For users who wish to analyze the data from WGA samples which have a higher missing data rate than the other samples and where the missing data is concentrated disproportionally at particular SNPs, we recommend imposing the stronger thresholds.

In our paper we use 594,924 SNPs for all analyses; these are autosomal SNPs from the 602,290 SNPs indicated in Table S9.2, from which we further removed 1,449 when merging with the ancient samples and requiring a homologous chimpanzee allele, diallelic SNPs and a valid genetic distance. For genetic distance, we used the linkage disequilibrium-based map that includes chromosome X and which is available on the 1000 Genomes Project website at
http://ftp.1000genomes.ebi.ac.uk/vol1/ftp/technical/working/20110106_recombination_hotspots/.

**Filtering of samples**
A total of 2,722 samples were successfully genotyped.

A total of 2,395 samples remained after outlier removal. For outlier removal, we manually curated the data using ADMIXTURE[7] and EIGENSOFT[8,9] to identify samples that were visual outliers compared with samples from their own populations. We also identified samples that were apparently closely relatives of others in the dataset. In the dataset that we release, the "verbose" population IDs for these individuals are prefixed by the string "Ignore_", so that users who wish to analyze these samples are still able to analyzed the data.

A total of 2,345 individuals were used in our analysis dataset, after further omitting all samples that were genotyped from whole genome amplified (WGA) material, with the exception of the Saami.

Table S9.3 summarizes the distribution of the samples and populations in the dataset according to broad geographic region. Table S9.4 presents detailed information on each of the populations. Online Table 1, a tab-delimited text file, provides detailed information on each of the individual samples.

We used two types of naming schemes for populations. For the "simple" naming scheme that we used for most analyses in this study, we used short names if possible and tended to lump populations from within the same region (e.g. within England, within Spain, within the Ukraine, and within Turkey). This resulted in 241 populations with at least 1 sample each in our analysis dataset. For the "verbose" naming scheme, we sometimes used longer names and did not lump populations within each region. This results in 203 populations with at least 1 sample each in our analysis dataset (Table S9.3).

***Table S9.3. Breakdown of genotyped samples by world region***

| | **Before curation** | | | **After curation** | | |
|---|---|---|---|---|---|---|
| **World Region** | **Number of Samples** | **Number of populations (verbose)** | **Number of populations (simple)** | **Number of Samples** | **Number of populations (verbose)** | **Number of populations (simple)** |
| Africa | 675 | 63 | 55 | 535 | 57 | 49 |
| America | 231 | 27 | 24 | 213 | 27 | 24 |
| Central Asia / Siberia | 323 | 29 | 23 | 266 | 28 | 23 |
| East Asia | 251 | 22 | 22 | 243 | 22 | 22 |
| Oceania | 39 | 3 | 3 | 27 | 3 | 3 |
| South Asia | 329 | 22 | 22 | 280 | 22 | 22 |
| West Eurasia | 874 | 89 | 67 | 781 | 82 | 60 |
| Total | 2,722 | 255 | 216 | 2,345 | 241 | 203 |

Note: The categorization by region is not based on genetic data, explaining why the number of populations and samples classified as "West Eurasian" by the ADMIXTURE analysis below does not match that in the table.

A total of 2,244 samples (of the 2,722) are in a version of the dataset we have made freely available at http://genetics.med.harvard.edu/reichlab/Reich_Lab/Datasets.html. The remaining 478 samples have more restrictive procedures for data access, and users who wish to access the data will need to send the corresponding author (DR) a signed letter containing the text shown in Box S9.1.

***Box S9.1. Text that needs to be included in a letter to access the data not posted publicly***

I affirm that
(a) I will not distribute the data outside my collaboration,
(b) I will not post it publicly,
(c) I will make no attempt to connect the genetic data to personal identifiers for the samples,
(d) I will use the data only for studies of population history,
(e) I will not use the data for any selection studies,
(f) I will not use the data for any medically or disease related analyses.
(g) I will not use the data for any commercial purposes

Note: Please send a PDF of a signed letter with this text to David Reich (reich@genetics.med.harvard.edu)

In summary, with this paper we are releasing genotyping data corresponding to two sets of samples
2,243 samples (1,935 after curation) that are fully publicly available
2,722 samples (2,345 after curation) for researchers who send a signed PDF letter.

For both of these sample sets, we include a "verbose.ind" list of populations that includes verbose sample identifiers which correspond to the 255 populations listed in Table S9.4 and plotted in Figure 1A. We also release a "simple.ind" list of sample identifier which corresponds to the merged groups of 216 populations and simpler names used for most of the analyses in the study.

***Table S9.4. List of populations genotyped on the Human Origins array and record of curation***

| Simple Population ID | Verbose Population ID | Region | Country | Lat-itude | Lon-gi-tude | Sam-ples | Passed QC | Contributor |
|---|---|---|---|---|---|---|---|---|
| Ain_Touta_WGA | Ain_Touta_WGA | Africa | Algeria | 35.4 | 5.9 | 3 | 0 | Mark G. Thomas / Leila Laredj |
| Algerian | Algerian | Africa | Algeria | 36.8 | 3.0 | 7 | 7 | David Comas |
| Mozabite | Mozabite | Africa | Algeria | 32.0 | 3.0 | 27 | 21 | Patterson et al. 2012 |
| BantuSA | Bantu_SA_Ovambo | Africa | Angola | -19.0 | 18.0 | 1 | 1 | Patterson et al. 2012 |
| Gana | Gana | Africa | Botswana | -21.7 | 23.4 | 9 | 8 | Pickrell et al. 2012 and 2014 |
| Gui | Gui | Africa | Botswana | -21.5 | 23.3 | 11 | 7 | Pickrell et al. 2012 and 2014 |
| Hoan | Hoan | Africa | Botswana | -24.0 | 23.4 | 7 | 7 | Pickrell et al. 2012 and 2014 |
| Ju_hoan_South | Ju_hoan_South | Africa | Botswana | -21.2 | 20.7 | 9 | 6 | Pickrell et al. 2012 and 2014 |
| Kgalagadi | Kgalagadi | Africa | Botswana | -24.8 | 21.8 | 5 | 5 | Pickrell et al. 2012 and 2014 |
| Khwe | Khwe | Africa | Botswana | -18.4 | 21.9 | 10 | 8 | Pickrell et al. 2012 and 2014 |
| Naro | Naro | Africa | Botswana | -22.0 | 21.6 | 10 | 8 | Pickrell et al. 2012 and 2014 |
| Shua | Shua | Africa | Botswana | -20.6 | 25.3 | 10 | 9 | Pickrell et al. 2012 and 2014 |
| Taa_East | Taa_East | Africa | Botswana | -24.2 | 22.8 | 8 | 7 | Pickrell et al. 2012 and 2014 |
| Taa_North | Taa_North | Africa | Botswana | -23.0 | 22.4 | 11 | 9 | Pickrell et al. 2012 and 2014 |
| Taa_West | Taa_West | Africa | Botswana | -23.6 | 20.3 | 17 | 16 | Pickrell et al. 2012 and 2014 |
| Tshwa | Tshwa | Africa | Botswana | -21.0 | 25.9 | 9 | 5 | Pickrell et al. 2012 and 2014 |
| Tswana | Tswana | Africa | Botswana | -24.1 | 25.4 | 5 | 5 | Pickrell et al. 2012 and 2014 |
| BantuSA | Bantu_SA_Herero | Africa | BotswanaOrNamibia | -22.0 | 19.0 | 2 | 2 | Patterson et al. 2012 |
| BantuSA | Bantu_SA_Tswana | Africa | BotswanaOrNamibia | -28.0 | 24.0 | 2 | 2 | Patterson et al. 2012 |
| Biaka | BiakaPygmy | Africa | CentralAfricanRepublic | 4.0 | 17.0 | 23 | 20 | Patterson et al. 2012 |
| Mbuti | MbutiPygmy | Africa | Congo | 1.0 | 29.0 | 14 | 10 | Patterson et al. 2012 |
| Egyptian | Egyptian_Comas | Africa | Egypt | 31.0 | 31.2 | 14 | 11 | David Comas / Pierre Zalloua |
| Egyptian | Egyptian_Metspalu | Africa | Egypt | 30.2 | 31.2 | 8 | 7 | Mait Metspalu / Richard Villems / Leila Laredj / Ene Metspalu |
| Afar_WGA | Afar_WGA | Africa | Ethiopia | 11.8 | 41.4 | 5 | 0 | Mark G. Thomas / Ayele Tarekegn |
| Ethiopian_Jew | Ethiopian_Jew | Africa | Ethiopia | 9.0 | 38.7 | 7 | 7 | The National Laboratory for the Genetics of Israeli Populations |
| Oromo | Oromo | Africa | Ethiopia | 9.0 | 36.5 | 5 | 4 | Anna Di Rienzo* / Cynthia Beall* / Amha Gebremedhin* |
| Gambian | Gambian_GWD | Africa | Gambia | 13.4 | 16.7 | 6 | 6 | Coriell Cell Repositories |
| BantuKenya | BantuKenya | Africa | Kenya | -3.0 | 37.0 | 11 | 6 | Patterson et al. 2012 |
| Kikuyu | Kikuyu | Africa | Kenya | -0.4 | 36.9 | 4 | 4 | George Ayodo |
| Luhya | Luhya_Kenya_LWK | Africa | Kenya | 1.3 | 36.8 | 8 | 8 | Coriell Cell Repositories |
| Luo | Luo | Africa | Kenya | -0.1 | 34.3 | 9 | 8 | George Ayodo |
| Masai | Masai_Ayodo | Africa | Kenya | -1.1 | 35.9 | 3 | 2 | George Ayodo |
| Masai | Masai_Kinyawa_MKK | Africa | Kenya | -1.5 | 35.2 | 10 | 10 | Coriell Cell Repositories |
| Somali | Somali | Africa | Kenya | 5.6 | 48.3 | 13 | 13 | George Ayodo |
| BantuSA | Bantu_SA_S_Sotho | Africa | Lesotho | -29.0 | 29.0 | 1 | 1 | Patterson et al. 2012 |
| Libyan_Jew | Libyan_Jew | Africa | Libya | 32.9 | 13.2 | 9 | 9 | The National Laboratory for the Genetics of Israeli Populations |
| Burbur_WGA | Burbur_WGA | Africa | Morocco | 33.5 | 5.1 | 5 | 0 | Mark G. Thomas / Fouad Berrada |
| Moroccan | Moroccan | Africa | Morocco | 32.3 | -6.4 | 7 | 0 | David Comas |
| Moroccan_Jew | Moroccan_Jew | Africa | Morocco | 34.0 | -6.8 | 7 | 6 | The National Laboratory for the Genetics of Israeli Populations |
| Damara | Damara | Africa | Namibia | -19.8 | 16.2 | 13 | 12 | Pickrell et al. 2012 and 2014 |
| Haiom | Haiom | Africa | Namibia | -19.3 | 17.0 | 9 | 7 | Pickrell et al. 2012 and 2014 |
| Himba | Himba | Africa | Namibia | -19.1 | 14.1 | 5 | 4 | Pickrell et al. 2012 and 2014 |
| Ju_hoan_North | Ju_hoan_North | Africa | Namibia | -18.9 | 21.5 | 24 | 22 | Patterson et al. 2012 |
| Nama | Nama | Africa | Namibia | -24.3 | 17.3 | 18 | 16 | Pickrell et al. 2012 and 2014 |
| Wambo | Wambo | Africa | Namibia | -17.7 | 18.1 | 5 | 5 | Pickrell et al. 2012 and 2014 |
| Xuun | Xuun | Africa | Namibia | -18.7 | 19.7 | 15 | 13 | Pickrell et al. 2012 and 2014 |
| Esan | Esan_Nigeria_ESN | Africa | Nigeria | 6.5 | 6.0 | 8 | 8 | Coriell Cell Repositories |
| Yoruba | Yoruba | Africa | Nigeria | 7.4 | 3.9 | 108 | 70 | Coriell Cell Repositories |
| Mandenka | Mandenka | Africa | Senegal | 12.0 | -12.0 | 22 | 17 | Patterson et al. 2012 |
| Mende | Mende_Sierra_Leone_MSL | Africa | SierraLeone | 8.5 | -13.2 | 8 | 8 | Coriell Cell Repositories |
| Khomani | Khomani | Africa | South_Africa | -27.8 | 21.1 | 12 | 11 | Brenna Henna |
| BantuSA | Bantu_SA_Pedi | Africa | SouthAfrica | -29.0 | 30.0 | 1 | 1 | Patterson et al. 2012 |
| BantuSA | Bantu_SA_Zulu | Africa | SouthAfrica | -28.0 | 31.0 | 1 | 1 | Patterson et al. 2012 |
| Dinka | Dinka | Africa | Sudan | 8.8 | 27.4 | 9 | 7 | Michael Hammer |
| Shaigi_WGA | Shaigi_WGA | Africa | Sudan | 15.6 | 32.5 | 3 | 0 | Mark G. Thomas / Hamza A. Babiker |
| Datog | Datog | Africa | Tanzania | -3.3 | 35.7 | 3 | 3 | Brenna Henna / Joanna Mountain |
| Hadza | Hadza | Africa | Tanzania | -3.8 | 35.3 | 20 | 17 | Sarah A. Tishkoff / Thomas Nyambo |
| Hadza | Hadza_Henn | Africa | Tanzania | -3.6 | 35.1 | 8 | 5 | Brenna Henna / Joanna Mountain |
| Hadza_WGA | Hadza_Henn_WGA | Africa | Tanzania | -3.6 | 35.1 | 1 | 0 | Brenna Henna / Joanna Mountain |
| Sandawe | Sandawe | Africa | Tanzania | -5.5 | 35.5 | 28 | 22 | Sarah A. Tishkoff / Thomas Nyambo |
| Tunisian | Tunisian | Africa | Tunisia | 36.8 | 10.2 | 8 | 8 | David Comas |
| Tunisian_Jew | Tunisian_Jew | Africa | Tunisia | 36.8 | 10.2 | 7 | 7 | The National Laboratory for the Genetics of Israeli Populations |
| Saharawi | Saharawi | Africa | WesternSahara(Morocco) | 27.3 | -8.9 | 7 | 6 | David Comas |
| Chane | Chane | America | Argentina | -25.0 | -60.0 | 1 | 1 | Andres Ruiz-Linares / Claudio M. Bravi / Graciela Bailliet / Daniel Corach |
| Guarani | Guarani | America | Argentina | -27.5 | -59.0 | 5 | 5 | Andres Ruiz-Linares / Claudio M. Bravi / Graciela Bailliet / Daniel Corach |
| Aymara | Aymara | America | Bolivia | -16.5 | -68.2 | 6 | 5 | Andres Ruiz-Linares / Francisco Rothhammer / Jean-Michel Dugoujon / René Vasquez / Mercedes Villena |
| Bolivian | Bolivian_Cochabamba | America | Bolivia | -17.4 | -66.2 | 1 | 1 | Antonio Salas |
| Bolivian | Bolivian_LaPaz | America | Bolivia | -16.5 | -68.2 | 3 | 3 | Antonio Salas |
| Bolivian | Bolivian_Pando | America | Bolivia | -11.2 | -67.2 | 3 | 3 | Antonio Salas |
| Quechua | Quechua_RuizLinares | America | Bolivia | -20.0 | -66.0 | 2 | 2 | Andres Ruiz-Linares / Jean-Michel Dugoujon / René Vasquez / Mercedes Villena |
| Karitiana | Karitiana | America | Brazil | -10.0 | -63.0 | 14 | 12 | Patterson et al. 2012 |
| Surui | Surui | America | Brazil | -11.0 | -62.0 | 8 | 8 | Patterson et al. 2012 |
| Algonquin | Algonquin | America | Canada | 48.4 | -71.1 | 9 | 9 | Damian Labuda* |
| Chipewyan | Chipewyan | America | Canada | 59.6 | -107.3 | 32 | 30 | Damian Labuda* |
| Cree | Cree | America | Canada | 50.3 | -102.5 | 13 | 13 | Damian Labuda* |
| Ojibwa | Ojibwa | America | Canada | 46.5 | -81.0 | 28 | 19 | David E. C. Cole / Damian Labuda* |
| Chilote | Chilote | America | Chile | -42.5 | -73.9 | 4 | 4 | Andres Ruiz-Linares / Francisco Rothhammer |
| Inga | Inga | America | Colombia | 1.0 | -77.0 | 2 | 2 | Andres Ruiz-Linares / Gabriel Bedoya |
| Piapoco | Piapoco | America | Colombia | 3.0 | -68.0 | 5 | 4 | Patterson et al. 2012 |
| Ticuna | Ticuna | America | Colombia | -3.8 | -70.0 | 1 | 1 | Andres Ruiz-Linares / Gabriel Bedoya |
| Wayuu | Wayuu | America | Colombia | 11.0 | -73.0 | 1 | 1 | Andres Ruiz-Linares / Gabriel Bedoya |
| Cabecar | Cabecar | America | Costa Rica | 9.5 | -84.0 | 6 | 6 | Andres Ruiz-Linares / Ramiro Barrantes |
| Kaqchikel | Kaqchikel | America | Guatemala | 15.0 | -91.0 | 5 | 5 | Andres Ruiz-Linares / Julio Molina |
| Mayan | Mayan | America | Mexico | 19.0 | -91.0 | 21 | 18 | Patterson et al. 2012 |
| Mixe | Mixe | America | Mexico | 17.0 | 96.6 | 10 | 10 | William Klitz / Cheryl Winkler |
| Mixtec | Mixtec | America | Mexico | 16.7 | -97.2 | 10 | 10 | William Klitz / Cheryl Winkler |
| Pima | Pima | America | Mexico | 29.0 | -108.0 | 14 | 14 | Patterson et al. 2012 |
| Zapotec | Zapotec | America | Mexico | 17.0 | -96.5 | 10 | 10 | William Klitz / Cheryl Winkler |
| Quechua | Quechua_Coriell | America | Peru | -13.5 | -72.0 | 5 | 5 | Coriell Cell Repositories |
| AA | AA_Denver | America | USA | 39.7 | -105.0 | 12 | 12 | Rick Kittles |
| Mongola | Mongola | CentralAsiaSiberia | China | 45.0 | 111.0 | 11 | 6 | Patterson et al. 2012 |
| Kyrgyz | Kyrgyz | CentralAsiaSiberia | Kyrgyzstan | 42.9 | 74.6 | 10 | 9 | Robert W. Mahley / Ugur Hodoglugil |
| Aleut | Aleut | CentralAsiaSiberia | Russia | 53.6 | 160.8 | 7 | 7 | Rem Sukernik / Stanislav Dryomov |
| Altaian | Altaian | CentralAsiaSiberia | Russia | 51.9 | 86.0 | 7 | 7 | Mait Metspalu / Richard Villems / Leila Laredj / Ene Metspalu / Olga L. Posukh |
| Chukchi | Chukchi | CentralAsiaSiberia | Russia | 69.5 | 168.8 | 24 | 20 | Rem Sukernik / Stanislav Dryomov |
| Chukchi | Chukchi_Reindeer | CentralAsiaSiberia | Russia | 64.4 | 173.9 | 1 | 1 | Rem Sukernik / Stanislav Dryomov |
| Chukchi | Chukchi_Sir | CentralAsiaSiberia | Russia | 64.4 | 173.9 | 2 | 2 | Rem Sukernik / Stanislav Dryomov |
| Dolgan | Dolgan | CentralAsiaSiberia | Russia | 73.0 | 115.4 | 4 | 3 | Mait Metspalu / Richard Villems / Leila Laredj / Ene Metspalu / Sardana A. Fedorova / Fedor Platonov |
| Eskimo | Eskimo_Chaplin | CentralAsiaSiberia | Russia | 64.5 | 172.9 | 5 | 4 | Rem Sukernik / Stanislav Dryomov |
| Eskimo | Eskimo_Naukan | CentralAsiaSiberia | Russia | 66.0 | 169.7 | 20 | 13 | Rem Sukernik / Stanislav Dryomov |
| Eskimo | Eskimo_Sireniki | CentralAsiaSiberia | Russia | 64.4 | 173.9 | 5 | 5 | Rem Sukernik / Stanislav Dryomov |
| Even | Even | CentralAsiaSiberia | Russia | 57.5 | 135.9 | 10 | 10 | Rem Sukernik / Stanislav Dryomov |
| Itelmen | Itelmen | CentralAsiaSiberia | Russia | 57.2 | 156.9 | 7 | 6 | Rem Sukernik / Stanislav Dryomov |
| Kalmyk | Kalmyk | CentralAsiaSiberia | Russia | 46.2 | 45.3 | 10 | 10 | Mait Metspalu / Richard Villems / Leila Laredj / Ene Metspalu / Elza Khusnutdinova / Rita Khusainova / Sergey Litvinov |
| Koryak | Koryak | CentralAsiaSiberia | Russia | 58.1 | 159.0 | 13 | 9 | Rem Sukernik / Stanislav Dryomov |
| Mansi | Mansi | CentralAsiaSiberia | Russia | 62.5 | 63.3 | 8 | 8 | Rem Sukernik / Stanislav Dryomov |
| Nganasan | Nganasan | CentralAsiaSiberia | Russia | 71.1 | 96.1 | 14 | 11 | Rem Sukernik / Elena B. Starikovskaya |
| Selkup | Selkup | CentralAsiaSiberia | Russia | 65.5 | 82.3 | 10 | 10 | Mait Metspalu / Richard Villems / Leila Laredj / Ene Metspalu / Ludmila Osipova |
| Tlingit | Tlingit | CentralAsiaSiberia | Russia | 54.7 | 164.5 | 5 | 4 | Rem Sukernik / Stanislav Dryomov |
| Tubalar | Tubalar | CentralAsiaSiberia | Russia | 51.1 | 87.0 | 31 | 22 | Rem Sukernik / Stanislav Dryomov |
| Tuvinian | Tuvinian | CentralAsiaSiberia | Russia | 50.3 | 95.2 | 10 | 10 | Mait Metspalu / Richard Villems / Leila Laredj / Ene Metspalu / Larissa Damba / Mikhail Voevoda / Marina Gubina |
| Ulchi | Ulchi | CentralAsiaSiberia | Russia | 52.2 | 140.4 | 33 | 25 | Rem Sukernik / Stanislav Dryomov |
| Yakut | Yakut | CentralAsiaSiberia | Russia | 63.0 | 129.5 | 25 | 20 | Patterson et al. 2012 |
| Yakut | Yakut_Metspalu | CentralAsiaSiberia | Russia | n/a | n/a | 1 | 0 | Mait Metspalu / Richard Villems / Leila Laredj / Ene Metspalu / Sardana A. Fedorova / Fedor Platonov |
| Yukagir | Yukagir_Forest | CentralAsiaSiberia | Russia | 65.5 | 151.0 | 5 | 5 | Rem Sukernik / Stanislav Dryomov |
| Yukagir | Yukagir_Tundra | CentralAsiaSiberia | Russia | 68.6 | 153.0 | 20 | 14 | Rem Sukernik / Stanislav Dryomov |
| Tajik_Pomiri | Tajik_Pomiri | CentralAsiaSiberia | Tajikistan | 37.4 | 71.7 | 8 | 8 | Mait Metspalu / Richard Villems / Leila Laredj / Ene Metspalu / Oleg Balanovsky / Elena Balanovska |
| Turkmen | Turkmen | CentralAsiaSiberia | Uzbekistan | 42.5 | 59.6 | 7 | 7 | Mait Metspalu / Richard Villems / Leila Laredj / Ene Metspalu / Oleg Balanovsky / Elena Balanovska / Shahlo Turdikulova |
| Uzbek | Uzbek | CentralAsiaSiberia | Uzbekistan | 41.3 | 69.2 | 10 | 10 | Mait Metspalu / Richard Villems / Leila Laredj / Ene Metspalu / Elza Khusnutdinova / Rita Khusainova / Sergey Litvinov |
| Cambodian | Cambodian | EastAsia | Cambodia | 12.0 | 105.0 | 10 | 8 | Patterson et al. 2012 |
| Dai | Dai | EastAsia | China | 21.0 | 100.0 | 10 | 10 | Patterson et al. 2012 |
| Daur | Daur | EastAsia | China | 48.5 | 124.0 | 9 | 9 | Patterson et al. 2012 |
| Han | Han | EastAsia | China | 32.3 | 114.0 | 35 | 33 | Patterson et al. 2012 |
| Han_NChina | Han_NChina | EastAsia | China | 32.3 | 114.0 | 10 | 10 | Patterson et al. 2012 |
| Hezhen | Hezhen | EastAsia | China | 47.5 | 133.5 | 9 | 8 | Patterson et al. 2012 |
| Lahu | Lahu | EastAsia | China | 22.0 | 100.0 | 8 | 8 | Patterson et al. 2012 |
| Miao | Miao | EastAsia | China | 28.0 | 109.0 | 10 | 10 | Patterson et al. 2012 |
| Naxi | Naxi | EastAsia | China | 26.0 | 100.0 | 9 | 9 | Patterson et al. 2012 |
| Oroqen | Oroqen | EastAsia | China | 50.4 | 126.5 | 9 | 9 | Patterson et al. 2012 |
| She | She | EastAsia | China | 27.0 | 119.0 | 10 | 10 | Patterson et al. 2012 |
| Tu | Tu | EastAsia | China | 36.0 | 101.0 | 10 | 10 | Patterson et al. 2012 |
| Tujia | Tujia | EastAsia | China | 29.0 | 109.0 | 10 | 10 | Patterson et al. 2012 |
| Uygur | Uygur | EastAsia | China | 44.0 | 81.0 | 10 | 10 | Patterson et al. 2012 |
| Xibo | Xibo | EastAsia | China | 43.5 | 81.5 | 9 | 7 | Patterson et al. 2012 |
| Yi | Yi | EastAsia | China | 28.0 | 103.0 | 10 | 10 | Patterson et al. 2012 |
| Japanese | Japanese | EastAsia | Japan | 38.0 | 138.0 | 29 | 29 | Patterson et al. 2012 |
| Korean | Korean | EastAsia | Korea | 37.6 | 127.0 | 6 | 6 | Coriell Cell Repositories |
| Ami | Ami_Coriell | EastAsia | Taiwan | 22.8 | 121.2 | 10 | 10 | Coriell Cell Repositories |
| Atayal | Atayal_Coriell | EastAsia | Taiwan | 24.6 | 121.3 | 10 | 9 | Coriell Cell Repositories |
| Thai | Thai | EastAsia | Thailand | 13.8 | 100.5 | 10 | 10 | European Collection of Cell Cultures |
| Kinh | Kinh_Vietnam_KHV | EastAsia | Vietnam | 21.0 | 105.9 | 8 | 8 | Coriell Cell Repositories |
| Australian | Australian_ECCAC | Oceania | Australia | -13.0 | 143.0 | 9 | 3 | European Collection of Cell Cultures |
| Bougainville | Bougainville | Oceania | PapuaNewGuinea | -6.0 | 155.0 | 12 | 10 | Patterson et al. 2012 |
| Papuan | Papuan | Oceania | PapuaNewGuinea | -4.0 | 143.0 | 18 | 14 | Patterson et al. 2012 |
| Bengali | Bengali_Bangladesh_BEB | SouthAsia | Bangladesh | 23.7 | 90.4 | 8 | 7 | Coriell Cell Repositories |
| Cochin_Jew | Cochin_Jew | SouthAsia | India | 10.0 | 76.3 | 5 | 5 | The National Laboratory for the Genetics of Israeli Populations |
| GujaratiA | GujaratiA_GIH | SouthAsia | India | 23.2 | 72.7 | 5 | 5 | Coriell Cell Repositories |
| GujaratiB | GujaratiB_GIH | SouthAsia | India | 23.2 | 72.7 | 5 | 5 | Coriell Cell Repositories |
| GujaratiC | GujaratiC_GIH | SouthAsia | India | 23.2 | 72.7 | 5 | 5 | Coriell Cell Repositories |
| GujaratiD | GujaratiD_GIH | SouthAsia | India | 23.2 | 72.7 | 5 | 5 | Coriell Cell Repositories |
| Kharia | Kharia | SouthAsia | India | 25.8 | 82.7 | 15 | 12 | Lalji Singh / Kumarasamy Thangaraj |
| Lodhi | Lodhi | SouthAsia | India | 25.5 | 78.6 | 14 | 13 | Lalji Singh / Kumarasamy Thangaraj |
| Mala | Mala | SouthAsia | India | 18.7 | 78.2 | 15 | 13 | Lalji Singh / Kumarasamy Thangaraj |
| Onge | Onge | SouthAsia | India | 10.8 | 92.5 | 17 | 11 | Lalji Singh / Kumarasamy Thangaraj |
| Tiwari | Tiwari | SouthAsia | India | 21.9 | 83.4 | 15 | 15 | Lalji Singh / Kumarasamy Thangaraj |
| Vishwabrahmin | Vishwabrahmin | SouthAsia | India | 16.3 | 80.5 | 15 | 13 | Lalji Singh / Kumarasamy Thangaraj |
| Kusunda | Kusunda | SouthAsia | Nepal | 28.1 | 82.5 | 10 | 10 | Aashish R. Jha / George van Driem / Irene Gallego Romero / Toomas Kivisild |
| Balochi | Balochi | SouthAsia | Pakistan | 30.5 | 66.5 | 24 | 20 | Patterson et al. 2012 |
| Brahui | Brahui | SouthAsia | Pakistan | 30.5 | 66.5 | 24 | 21 | Patterson et al. 2012 |
| Burusho | Burusho | SouthAsia | Pakistan | 36.5 | 74.0 | 25 | 23 | Patterson et al. 2012 |
| Hazara | Hazara | SouthAsia | Pakistan | 33.5 | 70.0 | 22 | 14 | Patterson et al. 2012 |
| Kalash | Kalash | SouthAsia | Pakistan | 36.0 | 71.5 | 19 | 18 | Patterson et al. 2012 |
| Makrani | Makrani | SouthAsia | Pakistan | 26.0 | 64.0 | 25 | 20 | Patterson et al. 2012 |
| Pathan | Pathan | SouthAsia | Pakistan | 33.5 | 70.5 | 24 | 19 | Patterson et al. 2012 |
| Punjabi | Punjabi_Lahore_PJL | SouthAsia | Pakistan | 31.5 | 74.3 | 8 | 8 | Coriell Cell Repositories |
| Sindhi | Sindhi | SouthAsia | Pakistan | 25.5 | 69.0 | 24 | 18 | Patterson et al. 2012 |
| Abkhasian | Abkhasian | WestEurasia | Abkhazia | 43.0 | 41.0 | 9 | 9 | Mait Metspalu / Richard Villems / Leila Laredj / Ene Metspalu / Elza Khusnutdinova / Rita Khusainova / Sergey Litvinov |
| Albanian | Albanian | WestEurasia | Albania | 41.3 | 19.8 | 6 | 6 | David Comas |
| Armenian | Armenian | WestEurasia | Armenia | 40.2 | 44.5 | 10 | 10 | Mait Metspalu / Richard Villems / Leila Laredj / Ene Metspalu / Levon Yepiskoposyan / Hovhannes Sahakyan |
| Armenian_WGA | Armenian_WGA | WestEurasia | Armenia | 40.2 | 44.5 | 3 | 0 | Mark G. Thomas / Levon Yepiskoposyan |
| Assyrian_WGA | Assyrian_WGA | WestEurasia | Armenia | 40.3 | 44.6 | 5 | 0 | Mark G. Thomas / Levon Yepiskoposyan |
| Kurd_WGA | Kurd_WGA | WestEurasia | Armenia | 40.7 | 44.3 | 2 | 0 | Mark G. Thomas / Levon Yepiskoposyan |
| Baku_WGA | Baku_WGA | WestEurasia | Azerbaijan | 40.4 | 49.9 | 3 | 0 | Mark G. Thomas / Ruslan Ruizbakiev |
| Belarusian | Belarusian | WestEurasia | Belarus | 53.9 | 28.0 | 10 | 10 | Mait Metspalu / Richard Villems / Leila Laredj / Ene Metspalu / Alena Kushniarevich |
| Bulgarian | Bulgarian | WestEurasia | Bulgaria | 42.2 | 24.7 | 10 | 10 | Mait Metspalu / Richard Villems / Leila Laredj / Ene Metspalu / Draga Toncheva / Sena Karachanak-Yankova / Mari Nelis* |
| Croatian | Croatian | WestEurasia | Croatia | 43.5 | 16.4 | 10 | 10 | Cristian Capelli / George B. J. Busby / Igor Rudan / Tatijana Zemunik |
| Cypriot | Cypriot | WestEurasia | Cyprus | 35.1 | 33.4 | 8 | 8 | David Comas / Pierre Zalloua |
| Czech | Czech | WestEurasia | Czechoslovia(pre1989) | 50.1 | 14.4 | 10 | 10 | Coriell Cell Repositories |
| English | English_Cornwall_GBR | WestEurasia | England | 50.3 | -4.9 | 5 | 5 | Coriell Cell Repositories |
| English | English_Kent_GBR | WestEurasia | England | 51.2 | 0.7 | 5 | 5 | Coriell Cell Repositories |
| Scottish | Scottish_Argyll_Bute_GBR | WestEurasia | England | 56.0 | -3.9 | 4 | 4 | Coriell Cell Repositories |
| Estonian | Estonian | WestEurasia | Estonia | 58.5 | 24.9 | 10 | 10 | Mait Metspalu / Richard Villems / Leila Laredj / Ene Metspalu / Jüri Parik |
| Finnish | Finnish_FIN | WestEurasia | Finland | 60.2 | 24.9 | 8 | 7 | Coriell Cell Repositories |
| Saami_WGA | Saami_WGA | WestEurasia | Finland | 68.4 | 23.6 | 1 | 1 | Svante Pääbo / Antti Sajantila / Klemetti Näkkäläjärvi |
| Basque | Basque_French | WestEurasia | France | 43.0 | 0.0 | 22 | 20 | Patterson et al. 2012 |
| French | French | WestEurasia | France | 46.0 | 2.0 | 29 | 25 | Patterson et al. 2012 |
| French_South | French_South | WestEurasia | France | 43.4 | -0.6 | 7 | 7 | David Comas |
| Georgian | Georgian_Megrels | WestEurasia | Georgia | 42.5 | 41.9 | 10 | 10 | Mait Metspalu / Richard Villems / Leila Laredj / Ene Metspalu / Elza Khusnutdinova / Rita Khusainova / Sergey Litvinov |
| Georgian_Jew | Georgian_Jew | WestEurasia | Georgia | 41.7 | 44.8 | 9 | 7 | The National Laboratory for the Genetics of Israeli Populations |
| Georgian_WGA | Georgian_WGA | WestEurasia | Georgia | 41.7 | 44.8 | 2 | 0 | Mark G. Thomas / Haim Ben-Ami |
| Greek | Greek_Comas | WestEurasia | Greece | 40.6 | 22.9 | 14 | 14 | David Comas |
| Greek | Greek_Coriell | WestEurasia | Greece | 38.0 | 23.7 | 8 | 6 | Coriell Cell Repositories |
| Greek_WGA | Greek_WGA | WestEurasia | Greece | 37.9 | 23.7 | 18 | 0 | Mark G. Thomas / Theologos Loukidis |
| Hungarian | Hungarian_Coriell | WestEurasia | Hungary | 47.5 | 19.1 | 10 | 10 | Coriell Cell Repositories |
| Hungarian | Hungarian_Metspalu | WestEurasia | Hungary | 47.5 | 19.1 | 10 | 10 | Mait Metspalu / Richard Villems / Leila Laredj / Ene Metspalu / Béla Melegh / Judit Bene |
| Icelandic | Icelandic | WestEurasia | Iceland | 64.1 | -21.9 | 12 | 12 | Coriell Cell Repositories |
| Iranian | Iranian | WestEurasia | Iran | 35.6 | 51.5 | 9 | 8 | Mait Metspalu / Richard Villems / Leila Laredj / Ene Metspalu |
| Iranian_Jew | Iranian_Jew | WestEurasia | Iran | 35.7 | 51.4 | 10 | 9 | The National Laboratory for the Genetics of Israeli Populations |
| Iraqi_Jew | Iraqi_Jew | WestEurasia | Iraq | 33.3 | 44.4 | 9 | 6 | The National Laboratory for the Genetics of Israeli Populations |
| Druze | Druze | WestEurasia | Israel(Carmel) | 32.0 | 35.0 | 42 | 39 | Patterson et al. 2012 |
| Palestinian | Palestinian | WestEurasia | Israel(Central) | 32.0 | 35.0 | 45 | 38 | Patterson et al. 2012 |
| BedouinA | BedouinA | WestEurasia | Israel(Negev) | 31.0 | 35.0 | 25 | 25 | Patterson et al. 2012 |
| BedouinB | BedouinB | WestEurasia | Israel(Negev) | 31.0 | 35.0 | 21 | 19 | Patterson et al. 2012 |
| Italian_South | Italian_South | WestEurasia | Italy | 39.4 | 15.5 | 1 | 1 | Cristian Capelli / George B. J. Busby / Francesca Brisighelli |
| Sicilian | Italian_EastSicilian | WestEurasia | Italy | 37.1 | 15.3 | 5 | 5 | Cristian Capelli / George B. J. Busby / Francesco Cali / Valentino Romano |
| Sicilian | Italian_WestSicilian | WestEurasia | Italy | 38.0 | 12.5 | 6 | 6 | Cristian Capelli / George B. J. Busby / Francesca Brisighelli |
| Bergamo | Italian_Bergamo | WestEurasia | Italy(Bergamo) | 46.0 | 10.0 | 13 | 12 | Patterson et al. 2012 |
| Sardinian | Sardinian | WestEurasia | Italy(Sardinia) | 40.0 | 9.0 | 29 | 27 | Patterson et al. 2012 |
| Tuscan | Italian_Tuscan | WestEurasia | Italy(Tuscany) | 43.0 | 11.0 | 8 | 8 | Patterson et al. 2012 |
| Jordanian | Jordanian | WestEurasia | Jordan | 32.1 | 35.9 | 10 | 9 | Mait Metspalu / Richard Villems / Leila Laredj / Ene Metspalu |
| Lebanese | Lebanese | WestEurasia | Lebanon | 33.8 | 35.6 | 8 | 8 | Mait Metspalu / Richard Villems / Leila Laredj / Ene Metspalu |
| Lithuanian | Lithuanian | WestEurasia | Lithuania | 54.9 | 23.9 | 10 | 10 | Mait Metspalu / Richard Villems / Leila Laredj / Ene Metspalu / Vaidutis Kučinskas / Ingrida Uktveryte |
| Maltese | Maltese | WestEurasia | Malta | 35.9 | 14.4 | 8 | 8 | David Comas / Pierre Zalloua |
| Norwegian | Norwegian | WestEurasia | Norway | 60.4 | 5.4 | 11 | 11 | Cristian Capelli / George B. J. Busby / Tor Hervig |
| Orcadian | Orcadian | WestEurasia | OrkneyIslands | 59.0 | -3.0 | 13 | 13 | Patterson et al. 2012 |
| Ashkenazi_Jew | Ashkenazi_Jew | WestEurasia | Poland | 52.2 | 21.0 | 9 | 7 | The National Laboratory for the Genetics of Israeli Populations |
| Balkar | Balkar | WestEurasia | Russia | 43.5 | 43.6 | 10 | 10 | Mait Metspalu / Richard Villems / Leila Laredj / Ene Metspalu / Elza Khusnutdinova / Rita Khusainova / Sergey Litvinov |
| Chechen | Chechen | WestEurasia | Russia | 43.3 | 45.7 | 9 | 9 | Mait Metspalu / Richard Villems / Leila Laredj / Ene Metspalu / Elza Khusnutdinova / Rita Khusainova / Sergey Litvinov |
| Chuvash | Chuvash | WestEurasia | Russia | 56.1 | 47.3 | 10 | 10 | Mait Metspalu / Richard Villems / Leila Laredj / Ene Metspalu / Elza Khusnutdinova / Rita Khusainova / Sergey Litvinov |
| Kumyk | Kumyk | WestEurasia | Russia | 43.3 | 46.6 | 9 | 8 | Mait Metspalu / Richard Villems / Leila Laredj / Ene Metspalu / Elza Khusnutdinova / Rita Khusainova / Sergey Litvinov |
| Lezgin | Lezgin | WestEurasia | Russia | 42.1 | 48.2 | 10 | 9 | Mait Metspalu / Richard Villems / Leila Laredj / Ene Metspalu / Elza Khusnutdinova / Rita Khusainova / Sergey Litvinov |
| Mordovian | Mordovian | WestEurasia | Russia | 54.2 | 45.2 | 10 | 10 | Mait Metspalu / Richard Villems / Leila Laredj / Ene Metspalu / Elza Khusnutdinova / Rita Khusainova / Sergey Litvinov |
| Nogai | Nogai | WestEurasia | Russia | 44.4 | 41.9 | 9 | 9 | Mait Metspalu / Richard Villems / Leila Laredj / Ene Metspalu / Elza Khusnutdinova / Rita Khusainova / Sergey Litvinov |
| North_Ossetian | North_Ossetian | WestEurasia | Russia | 43.0 | 44.7 | 10 | 10 | Mait Metspalu / Richard Villems / Leila Laredj / Ene Metspalu / Elza Khusnutdinova / Rita Khusainova / Sergey Litvinov |
| Russian | Russian | WestEurasia | Russia | 61.0 | 40.0 | 23 | 22 | Patterson et al. 2012 |
| Adygei | Adygei | WestEurasia | Russia(Caucasus) | 44.0 | 39.0 | 25 | 17 | Coriell Cell Repositories |
| Saudi | Saudi | WestEurasia | Saudi_Arabia | 18.5 | 42.5 | 10 | 8 | Mait Metspalu / Richard Villems / Leila Laredj / Ene Metspalu |
| Basque | Basque_Spanish | WestEurasia | Spain | 43.1 | -2.1 | 10 | 9 | David Comas |
| Canary_Islanders | Spanish_Canarias_IBS | WestEurasia | Spain | 28.1 | -15.4 | 2 | 2 | Coriell Cell Repositories |
| Spanish | Spanish_Andalucia_IBS | WestEurasia | Spain | 37.4 | -6.0 | 4 | 4 | Coriell Cell Repositories |
| Spanish | Spanish_Aragon_IBS | WestEurasia | Spain | 41.0 | -1.0 | 6 | 6 | Coriell Cell Repositories |
| Spanish | Spanish_Baleares_IBS | WestEurasia | Spain | 39.5 | 3.0 | 4 | 4 | Coriell Cell Repositories |
| Spanish | Spanish_Cantabria_IBS | WestEurasia | Spain | 43.3 | -4.0 | 5 | 5 | Coriell Cell Repositories |
| Spanish | Spanish_Castilla_la_Mancha_IBS | WestEurasia | Spain | 39.9 | -4.0 | 5 | 5 | Coriell Cell Repositories |
| Spanish | Spanish_Castilla_y_Leon_IBS | WestEurasia | Spain | 41.4 | -4.5 | 5 | 5 | Coriell Cell Repositories |
| Spanish | Spanish_Cataluna_IBS | WestEurasia | Spain | 41.8 | 1.5 | 5 | 5 | Coriell Cell Repositories |
| Spanish | Spanish_Extremadura_IBS | WestEurasia | Spain | 39.0 | -6.0 | 5 | 5 | Coriell Cell Repositories |
| Spanish | Spanish_Galicia_IBS | WestEurasia | Spain | 42.5 | -8.1 | 5 | 5 | Coriell Cell Repositories |
| Spanish | Spanish_Murcia_IBS | WestEurasia | Spain | 38.0 | -1.1 | 5 | 4 | Coriell Cell Repositories |
| Spanish | Spanish_Valencia_IBS | WestEurasia | Spain | 39.5 | -0.4 | 5 | 5 | Coriell Cell Repositories |
| Spanish_North | Spanish_Pais_Vasco_IBS | WestEurasia | Spain | 42.8 | -2.7 | 5 | 5 | Coriell Cell Repositories |
| Syrian | Syrian | WestEurasia | Syrian | 35.1 | 36.9 | 8 | 8 | Mait Metspalu / Richard Villems / Leila Laredj / Ene Metspalu |
| Turkish | Turkish | WestEurasia | Turkey | 39.6 | 28.5 | 4 | 4 | David Comas / Pierre Zalloua |
| Turkish | Turkish_Adana | WestEurasia | Turkey | 37.0 | 35.3 | 10 | 10 | Robert W. Mahley / Ugur Hodoglugil |
| Turkish | Turkish_Aydin | WestEurasia | Turkey | 37.9 | 27.8 | 10 | 7 | Robert W. Mahley / Ugur Hodoglugil |
| Turkish | Turkish_Balikesir | WestEurasia | Turkey | 39.4 | 27.5 | 10 | 6 | Robert W. Mahley / Ugur Hodoglugil |
| Turkish | Turkish_Istanbul | WestEurasia | Turkey | 41.0 | 29.0 | 10 | 10 | Robert W. Mahley / Ugur Hodoglugil |
| Turkish | Turkish_Kayseri | WestEurasia | Turkey | 38.7 | 35.5 | 10 | 10 | Robert W. Mahley / Ugur Hodoglugil |
| Turkish | Turkish_Trabzon | WestEurasia | Turkey | 41.0 | 39.7 | 10 | 9 | Robert W. Mahley / Ugur Hodoglugil |
| Turkish_Jew | Turkish_Jew | WestEurasia | Turkey | 41.0 | 29.0 | 9 | 8 | The National Laboratory for the Genetics of Israeli Populations |
| Ukrainian | Ukrainian_East | WestEurasia | Ukraine | 50.3 | 31.6 | 6 | 6 | Mait Metspalu / Richard Villems / Leila Laredj / Ene Metspalu / Oleg Balanovsky / Elena Balanovska / Mikhail Churnosov |
| Ukrainian | Ukrainian_West | WestEurasia | Ukraine | 49.9 | 24.0 | 3 | 3 | Mait Metspalu / Richard Villems / Leila Laredj / Ene Metspalu / Oleg Balanovsky / Elena Balanovska / Mikhail Churnosov / Olga Utevska |
| Uzbek_WGA | Uzbek_WGA | WestEurasia | Uzbekistan | 41.3 | 69.3 | 1 | 0 | Mark G. Thomas / Ruslan Ruizbakiev |
| Yemen | Yemen | WestEurasia | Yemen | 14.0 | 44.6 | 7 | 6 | Mait Metspalu / Richard Villems / Leila Laredj / Ene Metspalu |
| Yemenite_Jew | Yemenite_Jew | WestEurasia | Yemen | 15.4 | 44.2 | 8 | 8 | The National Laboratory for the Genetics of Israeli Populations |

We note that in practice, for each of these two sets of samples we are releasing 14 genotyping datasets: 14 genotype files and SNP lists. The file that the majority of researchers are likely to wish to use is the "allsnps" dataset that includes all SNPs. However, we also release separately SNP datasets for each of the 14 panels for researchers who wish to take advantage of the uniform ascertainment.

**Merging in of ancient samples and whole genome sequences**

We next merged in 22 samples into the genotyping dataset whose data were obtained by sequencing (either the human reference genome sequence "Href", several primates, archaic genome sequences (Neandertals and Denisovans), or ancient modern humans (Table S9.5).

Many of the analyses reported in this study include ancient samples that had too-low sequencing coverage to permit confident diploid genotype calls. To analyze these samples in conjunction with genotyping data from the Affymetrix Human Origins array, we picked a single allele at random for each individual from each site in the genome for which there was a high quality sequence. That allele was then used to represent that individual at that nucleotide position (the individual was treated as homozygous there). This procedure has the effect of (artifactually) inferring a high level of genetic drift on the lineage specific to the individual. However, it is not expected to induce correlations in drift with other samples, and thus it is not expected to bias inferences about population relationships.

The ancient DNA sequences to which this procedure was applied are listed in Table S9.5, and discussed in more detail below:

(1) *Motala:* The number of Human Origins array SNPs for which there was sequencing coverage after this procedure was 352,966 for Motala_merge and 411,453 for Motala12.

(2) *Swedish farmers and hunter-gathers.* BAM files mapped to *hg19* were downloaded from ref.[10] for one Swedish Neolithic farmer (Gök4 in ref. 9; Skoglund_farmer in this paper), and three Swedish Neolithic hunter-gatherers (Ajv52, Ajv70 and Ire8; combined as Skoglund_HG in this paper). The number of SNPs with coverage after this procedure was 4,548 for Skoglund_farmer and 18,261 for Skoglund_HG.

(3) *Iceman:* The *hg18*-mapped genotype calls for this individual were downloaded from the VCF file reported by ref.[11]. liftOver (http://genome.ucsc.edu) was used to convert the coordinates to *hg19*. There was coverage on 518,229 Human Origins array SNPs after this procedure.

(4) *LaBrana:* BAM files mapped to *hg19* were downloaded from ref[12] and merged for the La Braña 1 individual. The total number of SNPs for which there was sequencing coverage was 549,671. We did not include much lower-quality data from a previous study[13] of La Braña 1 and a second La Braña 2 individual as we found that these two individuals roughly clustered with each other and with Loschbour but the newly published La Braña 1 data[12] gave us two orders of magnitude more SNPs for analysis.

(5) *Upper Paleolithic Siberians*: BAM files mapped to *hg19* were downloaded from ref[14]. The number of SNPs for which there was sequencing coverage after this procedure was 427,211 for MA1 and 92,486 for AG2.

We also included a Saqqaq Paleo-Eskimo[15] in the dataset which we do not analyze in our paper as it is not relevant to European origins. We used the genotyping files for this sample as raw reads were not available and called the single allele when the sample had a homozygous reported state at a site or one of the two alleles with 50% probability if it was heterozygous. The *hg18* coordinates were lifted to *hg19* using liftOver (http://genome.ucsc.edu).

***Table S9.5: Sequence data merged into the genotyping dataset.***

| Sample ID | Gender | "Population" Name |
|---|---|---|
| Href | M | hg19_reference_sequence |
| Chimp | M | Primate_Chimp |
| Gorilla | M | Primate_Gorilla |
| Orang | M | Primate_Orangutan |
| Macaque | M | Primate_Macaque |
| Marmoset | M | Primate_Marmoset |
| Vindija_light | F | Ancient_Neandertal |
| Mez1 | F | Neandertal_Mezmaiskaya |
| Altai | F | Neandertal_Altai |
| Denisova_light | F | Ancient_Denisova |
| Denisova | F | Denisovan |
| Loschbour | M | Luxembourg_Mesolithic |
| LBK380 | F | GermanStuttgart_LBK |
| Otzi | M | Tyrolean_Iceman |
| Saqqaq | M | Greenland_Saqqaq |
| MA1 | M | Siberian_Upper_Paleolithic |
| AG2 | M | Siberian_Ice_Age |
| Skoglund_HG | M | Swedish_HunterGatherer |
| Skoglund_farmer | F | Swedish_Farmer |
| Motala_merge | M | Swedish_Motala_Merge |
| Motala12 | M | Swedish_Motala |
| LaBrana | M | LaBrana |

**ADMIXTURE analysis**
We carried out model-based clustering analysis using ADMIXTURE[7] 1.23 on the dataset, combining the present-day humans with Loschbour, Stuttgart, Motala12, Motala_merge, and LaBrana.

ADMIXTURE is a commonly used method for investigating admixture proportions in human populations, although its interpretation in terms of history is not straightforward. ADMIXTURE searches for a specified number (K) "ancestral populations" described by allele frequencies, and then assigns each study sample coefficients of ancestry from these reconstructed populations. While ADMIXTURE is a commonly used method for revealing genetic structure in a set of samples, it is blind to the genesis of the structure. Two populations showing shared membership to a particular cluster might do so due to either shared ancestry or recent admixture. Nevertheless, ADMIXTURE is useful as an exploratory tool in analyses of genetic structure, and in the context of the present paper we use it mainly to (i) identify a set of West Eurasian populations for further analysis, and (ii) to identify a set of non-West Eurasian populations from the rest of the world to be used as references for our methods of ancestry estimation. This analysis also serves as an exploration of populations included in the Affymetrix Human Origins Array dataset made available with this paper.

We used PLINK[16] 1.07 to thin the original dataset of 594,924 autosomal SNPS to remove SNPs in strong linkage disequilibrium, employing a window of 200 SNPs advanced by 25 SNPs and an $r^2$ threshold of 0.4 (--indep-pairwise 200 25 0.4). A total of 291,184 SNPs remained for analysis after this procedure. We ran ADMIXTURE with default 5-fold cross-validation (--cv=5), varying the number of ancestral populations between K=2 and K=20 in 100 replicates.

***Figure S9.1: Cross-validation error of ADMIXTURE analysis.*** *Cross-validation error decreases as* K *increases and appears to plateau above* K=16.

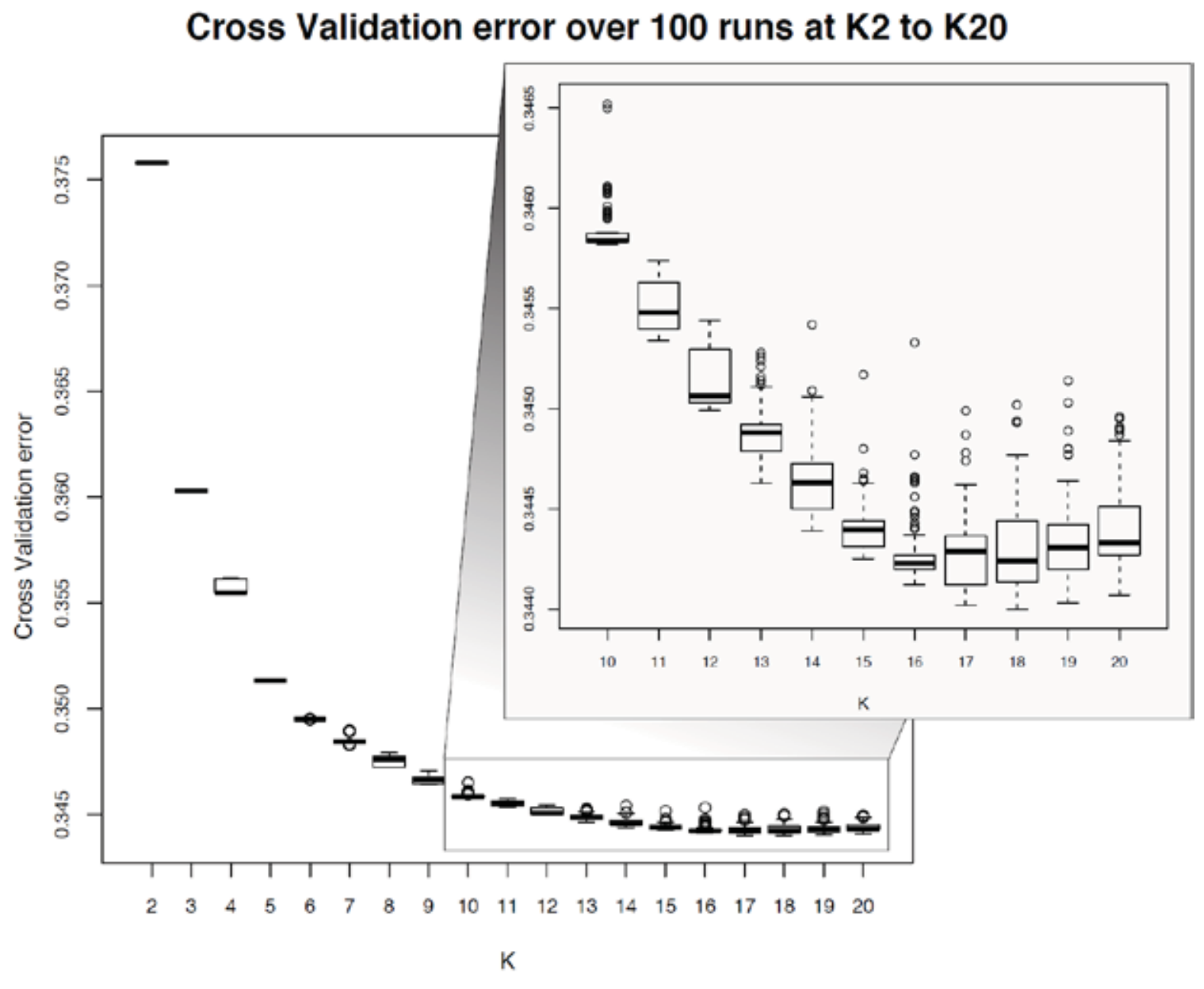

The cross-validation (CV) procedure in ADMIXTURE is designed to help choose the best *K* as one for which the model has the best predictive accuracy and thus lowest CV error. We observe that CV error drops as *K* increases while no real differentiation observable above *K*=16 (Figure S9.1). We note however that for our purposes the quest for the best *K* is not entirely relevant as we are interested in the hierarchical nature of the genetic structure revealed by ADMIXTURE in successive models with increasing number of "ancestral populations" (increasing K).

***Figure S9.2: The comparison of Log Likelihood (LL) scores and CLUMPP scores of individual ADMIXTURE runs against the run that yielded the highest LL score at each K.*** *Over all models (K) all ADMIXTURE runs that reached a LL score within 600 units of the LL score of the best run, yielded identical results in terms of inferred "ancestry" proportions (CLUMPP score >0.99).*

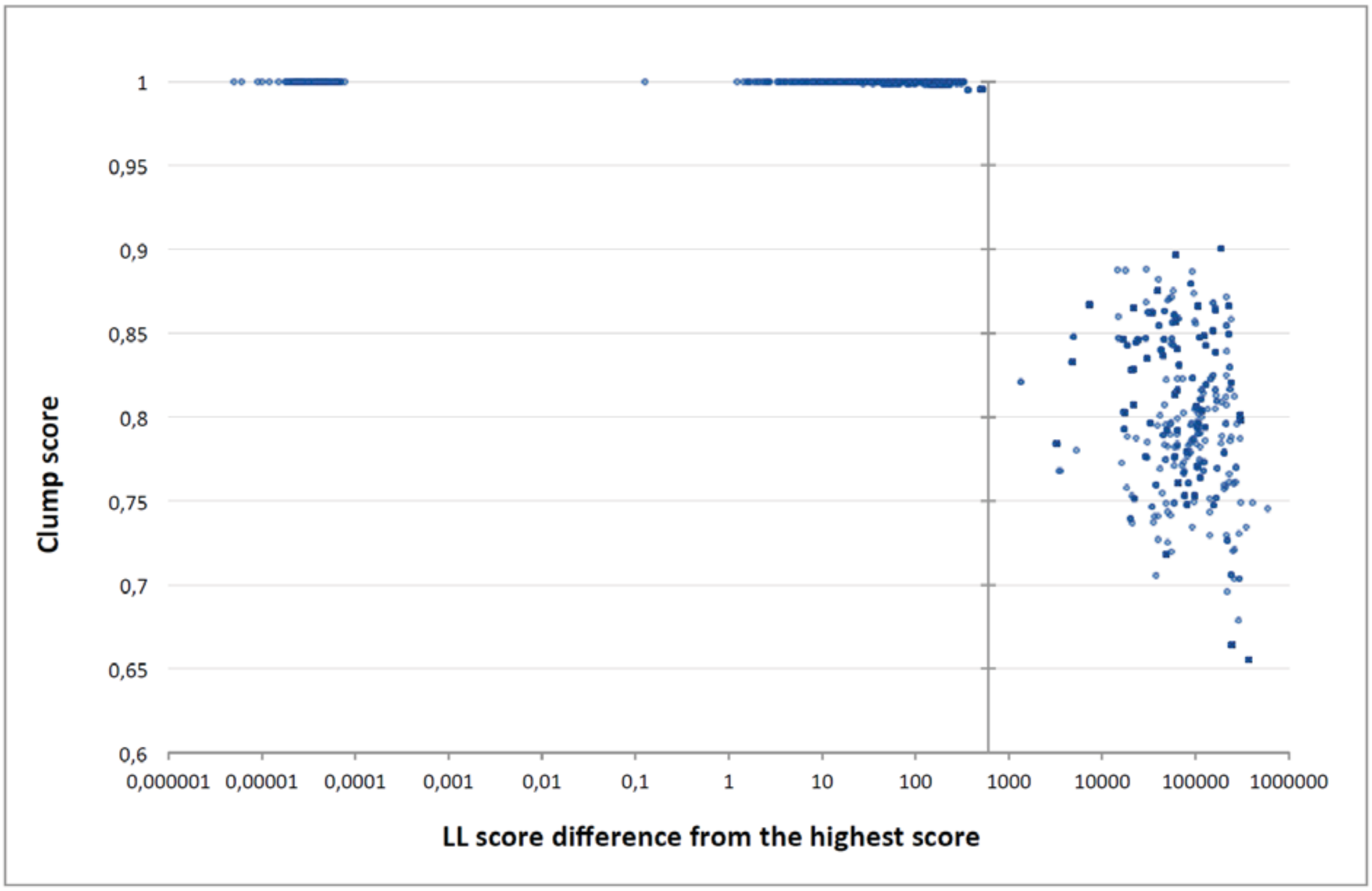

In order to examine the convergence of different runs at each K we examined the Log Likelihood (LL) scores. We assume that the global LL maximum for a given model (K) was reached if 10% of runs that reach the highest LL scores vary minimally in LL scores and yield identical ancestry proportions for the studied samples. The latter was inspected using CLUMPP[17]. We found that up to K=17 at least 20% of runs (20 runs) that reached the highest LL-scores also produced identical results (CLUMPP score >0.99 as compared to the run that yielded the highest LL score). Thus we have reasonable confidence for ADMIXTURE results for K=2 to K=17. Nevertheless, we present results for all models in the ADMIXTURE analysis in Extended Data Figure 3. For plotting we use for each K the run with the highest LL – but note again that at least 20% of runs at each K between K=2 to K=17 that reached the highest LL scores also yielded identical results. Thus any of those 20% could have been used for plotting at each K. We also note that the variation in LL scores within the top 10% or 20% of runs which all end up with identical results according to CLUMPP, is about an order of magnitude higher (around 600 LL units; Figure S9.2) than what we have observed using data from the conventionally ascertained genotyping arrays.

We also assessed convergence at each K by noting the proportion of replicates at each K that reached the "best" solution, defined by examining the set of K populations that maximized the K ancestral populations (Figure S9.3), breaking ties by taking the population with highest sample size. For example, at K=2 and K=3 we observed that 100% of the replicates converged on the solutions (Han, Ju_hoan_North) and (Basque, Ju_hoan_North, Karitiana) respectively, while at K=4 the best solution was (Ami Ju_hoan_North Karitiana Sardinian), supported by 64 replicates, while an alternative solution (Basque Ju_hoan_North Karitiana Yoruba) was supported by the remaining 36.

***Figure S9.3: Proportion of replicates that arrived at the best solution for each K.*** *More than 20% of replicates arrived at the best solution up to K=17.*

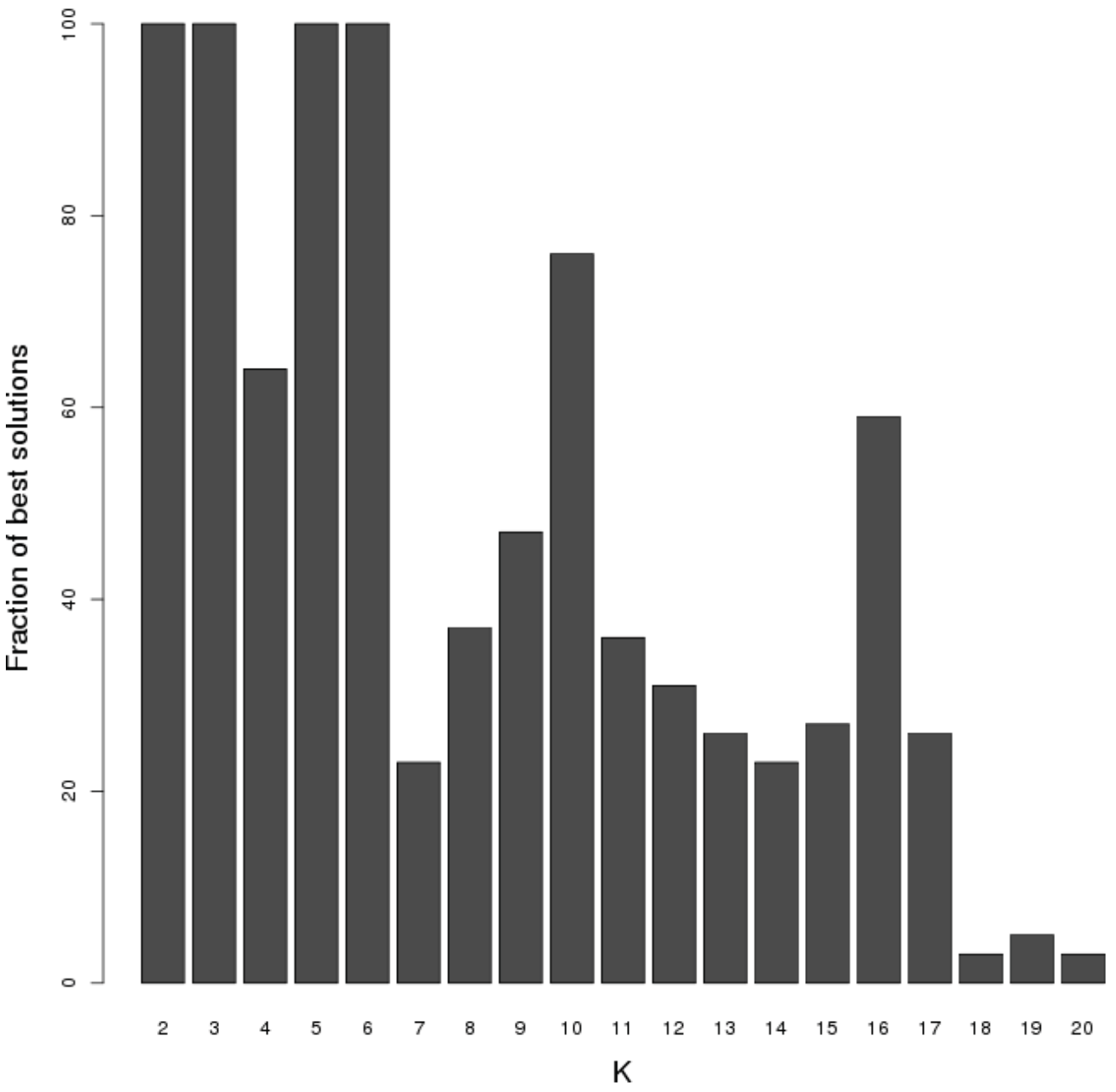

The results of the ADMIXTURE analysis can be found in Extended Data Figure 3.

We observe the following:

K=2 separates African from non-African populations.

K=3 reveals a West Eurasian ancestry component. The ancient samples appear to be mostly West Eurasian in their ancestry, although the hunter gatherers are also inferred to have greater or lesser extents of an eastern non-African (ENA) component lacking in Stuttgart. This is consistent with the positive $f_4$*(ENA, Chimp; Hunter Gatherer, Stuttgart)* statistic reported in SI12, which we interpret there as showing that ENA populations are closer to European Hunter-Gatherers than to Stuttgart.

K=4 breaks the ENA component down into one maximized in Native American populations like the Karitiana and one characterizing the East Asian populations and maximized in the Ami from Taiwan. This analysis further suggests that the ENA affinity of Hunter-Gatherers is related to the Karitiana.

K=5 breaks the African component into an African hunter-gatherer ancestry maximized in Bushmen such as the Ju_hoan_North and an African farmer component maximized in the Yoruba.

K=6 reveals a south Eurasian component maximized in Papuans, which is also represented in South Asians. MA1 shows some affinity to this component, in contrast to more recent European hunter gatherers who continue to mainly show ties to Native Americans.

K=7 reveals a Northeast Siberian component, which is maximized in the Itelmen.

The ancient European hunter-gatherers who previously had shown minor ancestry in Native American and East Asian component now replace the latter for the Northeast Siberian one. This is consistent with contemporary (eastern) Europeans who if they do show membership in East Eurasian components, show largely the same Northeast Siberian component. It is of course impossible to tell from this analysis if this is due to shared ancestry or admixture.

***Figure S9.4: A West Eurasian component first appears at* K=3 *and we use it to identify a set of "West Eurasian" populations for most analyses in our paper.***

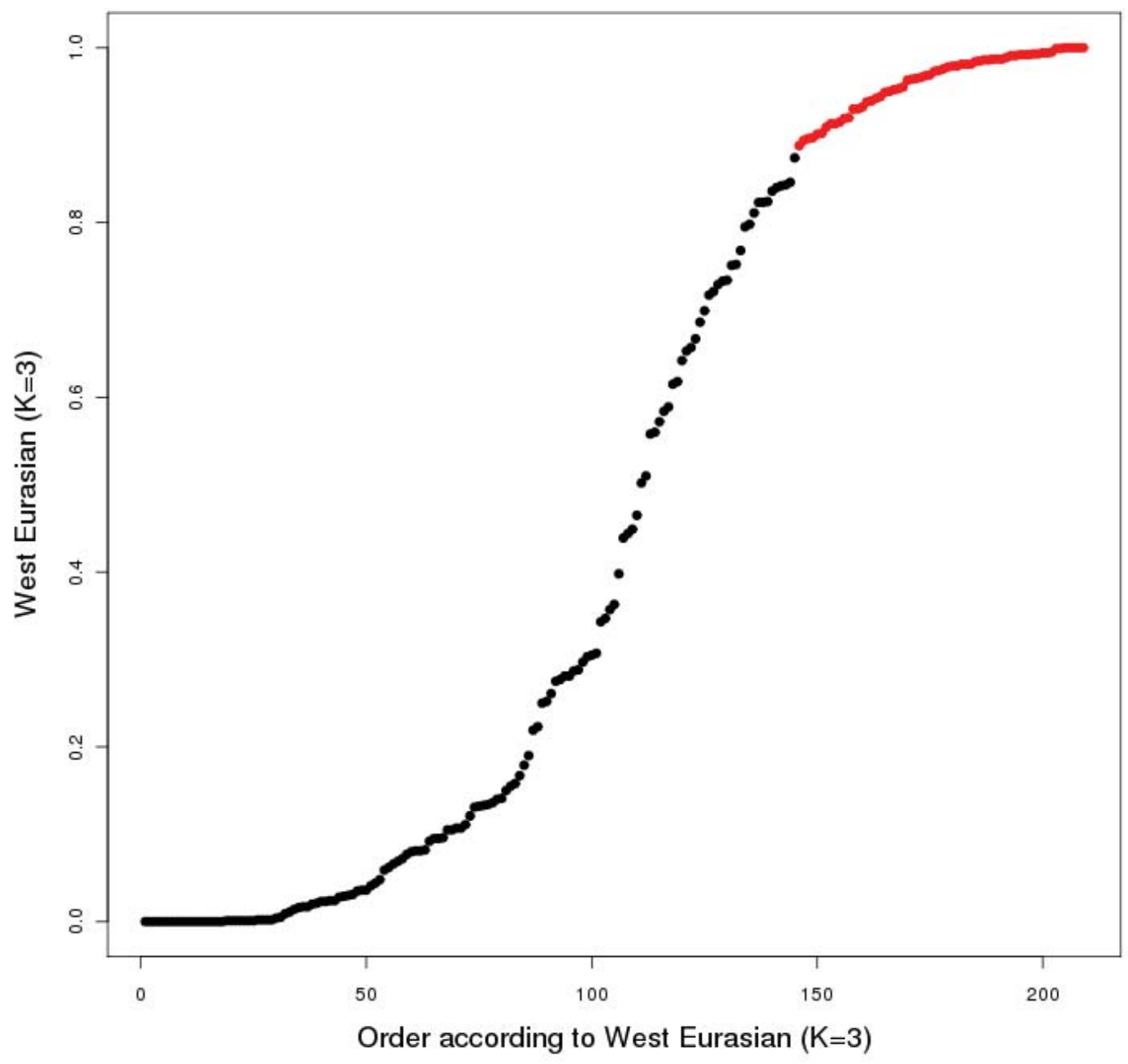

K=8 reveals a South Asian component maximized in the Mala. This is separated from the earlier south Eurasian component, and MA1 derives approximately one third of its ancestry to this new component, rather than from the Papuan maximized Oceanian component that also results from this split. We note that the ancestry proportions in ancient samples like MA1 are more likely explained by shared ancestry than admixture. This is more likely to explain the nearly three-way distribution of South Asian, West Eurasian and Native American (plus Northeast Siberian) ancestry proportions in MA1, than three-way admixture of established populations.

K=9 reveals a split within West Eurasia, with one component maximized in Loschbour and Northern European contemporary populations (both NE and NW) and one maximized in BedouinB. From the ancient samples, only Stuttgart shows mixed membership in these two components, consistent with the hypothesis that Early European Farmers represented a mixture of West European Hunter Gatherers and Near Eastern farmers. Membership in the Near Eastern component is prevalent in most of Southern Europe, consistent with the hypothesis that Europeans have inherited some Near Eastern ancestry via early farmers. However, Northwestern Europeans show minimal membership to this cluster while Northeastern populations like Estonians lack it altogether, in agreement with the ordering of EEF ancestry inferred using formal methods (Fig. 2B and Extended Data Table 3).

Most West Eurasian populations appear to be made up of the Loschbour and BedouinB-related component but populations from Northeastern Europe continue to possess partial ancestry from the Northeast Siberian-related component that is maximized in the Itelmen.

K=10 reveals a component maximized in the East African Hadza.

K=11, an Onge-maximized component appears. Until K=11 Onge was composed of approximately 50% South Asian, 30% East Asian and 20% Papuan components. Several South Asian populations show varying memberships in East Asian and to a much lesser extent also in the Papuan components. It is again impossible to determine from an ADMIXTURE analysis alone if this is due to shared ancestry or admixture. However, the differentiation of the Onge is likely best explained by random genetic drift mediated differentiation of allele frequencies in the small population. The Mala-maximized component disappears and a new GujaratiD-maximized component makes up most of the ancestry of South Asian populations.

K=12 A Chipewyan maximized component appears and is spread roughly around the Bering Strait. Interestingly, the Native American-like ancestry in MA1 largely resolves to this component.

K=13 shows the appearance of a Mbuti-maximized African Pygmy-related component.

K=14 A component maximized in the Chukchi appears.

K=15 shows the appearance of a component, maximized in the Kalash, that becomes the most predominant signal in Indus Valley and Caucasus populations. It is also prominent in the rest of South Asia, Central Asia, Near East and in diminishing strength in Europe. It is absent in Sardinians, Basques, and all ancient Europeans, although it is present in MA1. This component also does not appear in North and East Africa (except for Egyptians and Tunisians) where other West Eurasian admixture is observed. This is consistent with MA1 related population having contributed some ancestry to present-day Europeans not accounted for by West Eurasian Hunter Gatherers and Early European Farmers. The presence of this component in the Near East contrasts with its absence in Stuttgart, consistent with the widely shared negative $f_3$*(Near East; Stuttgart, MA1)* statistics (Table 1) indicating that present-day Near Easterners have been affected by gene flow not present in early Near Eastern migrants into Europe. Interestingly, a study of present-day South Asian populations[18] also revealed the presence of a common ancestral population (labeled "k5" there) between northern parts of South Asia, the Caucasus and Central Asia and Europe (but not Sardinians). The absence of the similar component in ancient Europeans reinforces our idea that preent-day Europeans have ancestry from a third ancestral population after the Neolithic transition and that this may be related to the evidence for admixture between Ancestral North Indians (related to the "k5" population) and Ancestral South Indians that took place over the last ~4,000 years[19].

K=16 reveals a Masai-maximized East African component

K=17 The Native American specific component splits in two: one maximized in Karitiana and the other a Central American centered cluster maximized in the Pima.

K=18 The Ju_hoan_North-maximized component is split into one that continues to center on the Ju_hoan_North and one maximized for the Gui.

K=19 A component maximized in the BedouinB population and otherwise rather restricted to Near Eastern populations is joined by one maximized in Stuttgart and well-represented in Sardinians and other Mediterranean populations.

K=20 A component maximized and largely restricted to the Kalash population appears. A Georgian-maximized component appears that seems to encompass similar ancestry as the Kalash-maximized component at lower K.

We wish to avoid over-interpretation of the admixture proportions, but nonetheless highlight some patterns each of which is validated by *f*-statistic analyses reported in this study and previous studies:

1. The absence of a Near Eastern relatedness in all European hunter-gatherer groups but its presence in Stuttgart.

2. The clear affinity of MA1 to Native American populations but not to East Asian or present-day Siberian populations.

3. The occurrence of low levels of additional gene flows in west Eurasia from Africa (in parts of the Near East or southern Europe) or recent Siberia (in parts of Northeastern Europe or the Near East and Caucasus).

4. Evidence tying MA1 to Europe, the northern Near East and Caucasus, and south/central Asia.

For our main analyses, we identified a set of West Eurasian (European and Near Eastern) populations as those that had maximum membership of the West Eurasian ancestral population at K=3 (Fig. S9.4). Restricting to the 59 present-day populations ("simple" naming scheme), this is 777 individuals. This count differs from the 781 West Eurasian individuals reported in Table S9.3, as some populations that are geographically African (North African Jewish groups) cluster with West Eurasians in ADMIXTURE analysis, while some groups that are geographically West Eurasian have substantial African (e.g., Yemen) or East Eurasian (e.g., Nogai) ancestry. The list of 64 populations that are classified as West Eurasian in this way (include 5 ancient samples indicated in italics) is as follows:

**"West Eurasian" set:** Abkhasian, Adygei, Albanian, Armenian, Ashkenazi_Jew, Balkar, Basque, BedouinA, BedouinB, Belarusian, Bergamo, Bulgarian, Canary_Islanders, Chechen, Croatian, Cypriot, Czech, Druze, English, Estonian, Finnish, French, French_South, Georgian, Georgian_Jew, Greek, Hungarian, Icelandic, Iranian, Iranian_Jew, Iraqi_Jew, Italian_South, Jordanian, Kumyk, *LaBrana*, Lebanese, Lezgin, Libyan_Jew, Lithuanian, *Loschbour*, Maltese, Mordovian, Moroccan_Jew, *Motala12*, *Motala_merge*, North_Ossetian, Norwegian, Orcadian, Palestinian, Russian, Sardinian, Saudi, Scottish, Sicilian, Spanish, Spanish_North, *Stuttgart*, Syrian, Tunisian_Jew, Turkish, Turkish_Jew, Tuscan, Ukrainian, Yemenite_Jew

We also identified a set of 54 world populations that maximize an ADMIXTURE component at least once (across all 100 replicates and across K=2 to K=20):

**"World Foci" set:** AA, Algonquin, Ami, Atayal, Basque, BedouinB, Biaka, Bougainville, Brahui, Cabecar, Chipewyan, Chukchi, Damara, Datog, Dinka, Esan, Eskimo, Georgian, Gui, GujaratiD, Hadza, Han, Itelmen, Ju_hoan_North, Kalash, Karitiana, Kharia, Korean, Koryak, LaBrana, Lahu, Lodhi, Loschbour, MA1, Mala, Mandenka, Masai, Mbuti, Mozabite, Naxi, Nganasan, Onge, Papuan, Pima, Sandawe, Sardinian, She, Somali, Stuttgart, Surui, Tubalar, Ulchi, Vishwabrahmin, Yoruba

We wanted to identify a subset of these populations without evidence of West Eurasian admixture. Doing so using ADMIXTURE output is not straightforward. At low K, some populations (e.g., Papuans at K=3) show an admixture coefficient in the West Eurasian ancestral population that is spurious and is the result of their ancestry being "forced" into a low number of ancestral populations (West Eurasian, Native American/East Asian, and Sub-Saharan) not representative of their distinctive ancestry. Conversely, at higher K, some populations (e.g., Mozabite, Kalash, Hadza, Mala, or Chipewyan) appear to be near-completely descended from an ancestral population maximized in them, masking the West Eurasian-related admixture they possess and which has been detected using other methods[4,19-22].

We retained 37 populations with sample size of at least 10, and further removed populations from the "West Eurasian" set identified above, resulting in the following set of 33 populations:

AA, Ami, Biaka, Bougainville, Brahui, Chipewyan, Chukchi, Damara, Eskimo, Hadza, Han, Han_NChina, Ju_hoan_North, Kalash, Karitiana, Kharia, Lodhi, Mala, Mandenka, Masai, Mbuti, Mozabite, Nganasan, Onge, Papuan, Pima, Sandawe, She, Somali, Tubalar, Ulchi, Vishwabrahmin, Yoruba

As a first step we removed populations with any West Eurasian ancestry at $K$=11. This value was chosen because it is the lowest one in which all 100 replicates consistently show ancestral components maximized in BedouinB and Loschbour and we wanted to remove populations of either Near Eastern or European partial ancestry. This resulted in the following set of 15 populations:

Ami, Biaka, Bougainville, Chukchi, Eskimo, Han, Ju_hoan_North, Karitiana, Kharia, Mbuti, Onge, Papuan, She, Ulchi, Yoruba

We removed Han_NChina as a precaution due to recent evidence[20] that they possess some West Eurasian admixture and because we did not want to include two populations from the Han ethnic group. This set includes populations of East Asian (Ami, Han, She), Sub-Saharan (Biaka, Ju_hoan_North, Mbuti, Yoruba), Native American (Karitiana), Oceanian (Bougainville, Papuan), North Asian (Chukchi, Eskimo, Ulchi), and South Asian (Kharia, Onge) ancestry.

Eliminating populations with any West Eurasian admixture whatsoever is difficult in view of recent results that even Sub-Saharan Africans[6] possess a trace of such ancestry. Nonetheless the above-identified set appears to consist of individuals of overwhelmingly non-West Eurasian ancestry which we can use for our analyses as reference points for our analyses. As more ancient genomes from other parts of the world become available, it may be possible to use them instead, thus removing even the tiny effects that more recent gene flows may have contributed to distant human populations.

# Supplementary Information 10
## Principal Components Analysis

Iosif Lazaridis*, Nick Patterson and David Reich

* To whom correspondence should be addressed (lazaridis@genetics.med.harvard.edu)

We used *smartpca*[1] (version: 10210) from EIGENSOFT[2,3] 5.0.1 to carry out Principal Components Analysis (PCA) on the Human Origins dataset. We performed PCA on a subset of the individuals and then projected the remainder using the *lsqproject: YES* option which accounts for samples with substantial missing data (which is important for many ancient DNA samples). We did not perform any outlier removal iterations (*numoutlieriter: 0)*. We set all other options to the default. Significance was assessed with a Tracy-Widom test[1] using the *twstats* program from EIGENSOFT 5.0.1 (all the first few principal components plotted in this section were highly significant, P=0).

### Global PCA
We first used a subset of populations from around the world ("World Foci" list of SI9) to build a PCA of world variation. We show the ancient samples projected onto PC1 and PC2 (Fig. S10.1) and PC3 and PC4 (Fig. S10.2).

In PC1 vs. PC2 (Fig. S10.1, explaining 9.1% and 2.7% of variance) we observe that Early European farmers group with Sardinians but European hunter-gatherers and especially Ancient North Eurasians deviate towards present-day Eastern non-African populations. This pattern is consistent with our model (Fig. 2a) according to which hunter-gatherers share common genetic drift with Eastern non-Africans that is only partially shared by Early European Farmers who trace part of their ancestry to a "Basal Eurasian" population that diverged prior to the split of European hunter-gatherers from Eastern non-Africans.

We next turned to PC3 vs. PC4 (Fig. S10.2, explaining 2.1% and 1.5% of variance). PC4 maximally differentiates Native American from Oceanian populations. Both Western European Hunter-Gatherers and Early European Farmers (WHG or EEF) have similar values of this component but Ancient North Eurasians and, to a lesser degree, Scandinavian Hunter-Gatherers deviate towards higher values of PC4, specifically towards Native Americans. Our model (Fig. 2a) is consistent with this as Karitiana has partial ancestry from Ancient North Eurasians, whereas both West European Hunter-Gatherers and Early European Farmers do not. Note that the Basal Eurasian ancestry plays no role here as both the Basal Eurasian ancestry (which EEF have) and the "main" Eurasian ancestry (shared by WHG and eastern non-Africans) are symmetrically related to different eastern non-African groups. The Scandinavian Hunter-Gatherers, however, deviate towards Native Americans, consistent with the fitted model which derives part of their ancestry from Ancient North Eurasians (SI14).

***Figure S10.1: Projection of ancient samples onto "Global" PCA dimensions 1 and 2.*** *Early European Farmers overlap Sardinians; all ancient Europeans group with West Eurasian populations; European hunter-gatherers and to a greater degree Ancient North Eurasians deviate from West Eurasians in the direction of eastern non-African populations.*

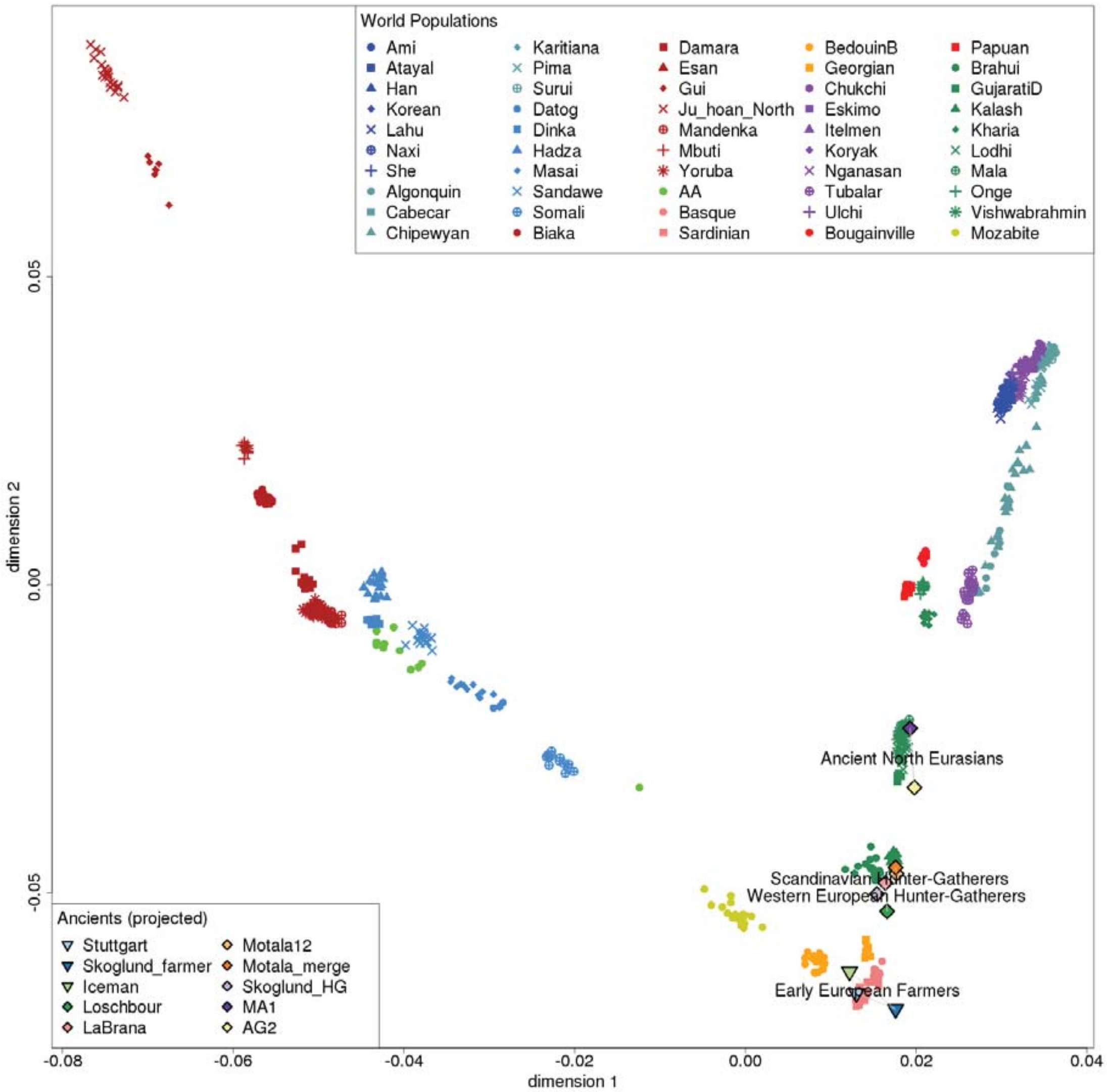

***Figure S10.2: Projection of ancient samples onto "Global" PCA dimensions 3 and 4.*** *PC4 differentiates eastern non-African groups with Native Americans occupying one end and Oceanians the other. Ancient North Eurasians, and (to a lesser degree) Scandinavian hunter-gatherers deviate from other ancient Europeans towards the Native American end of PC4.*

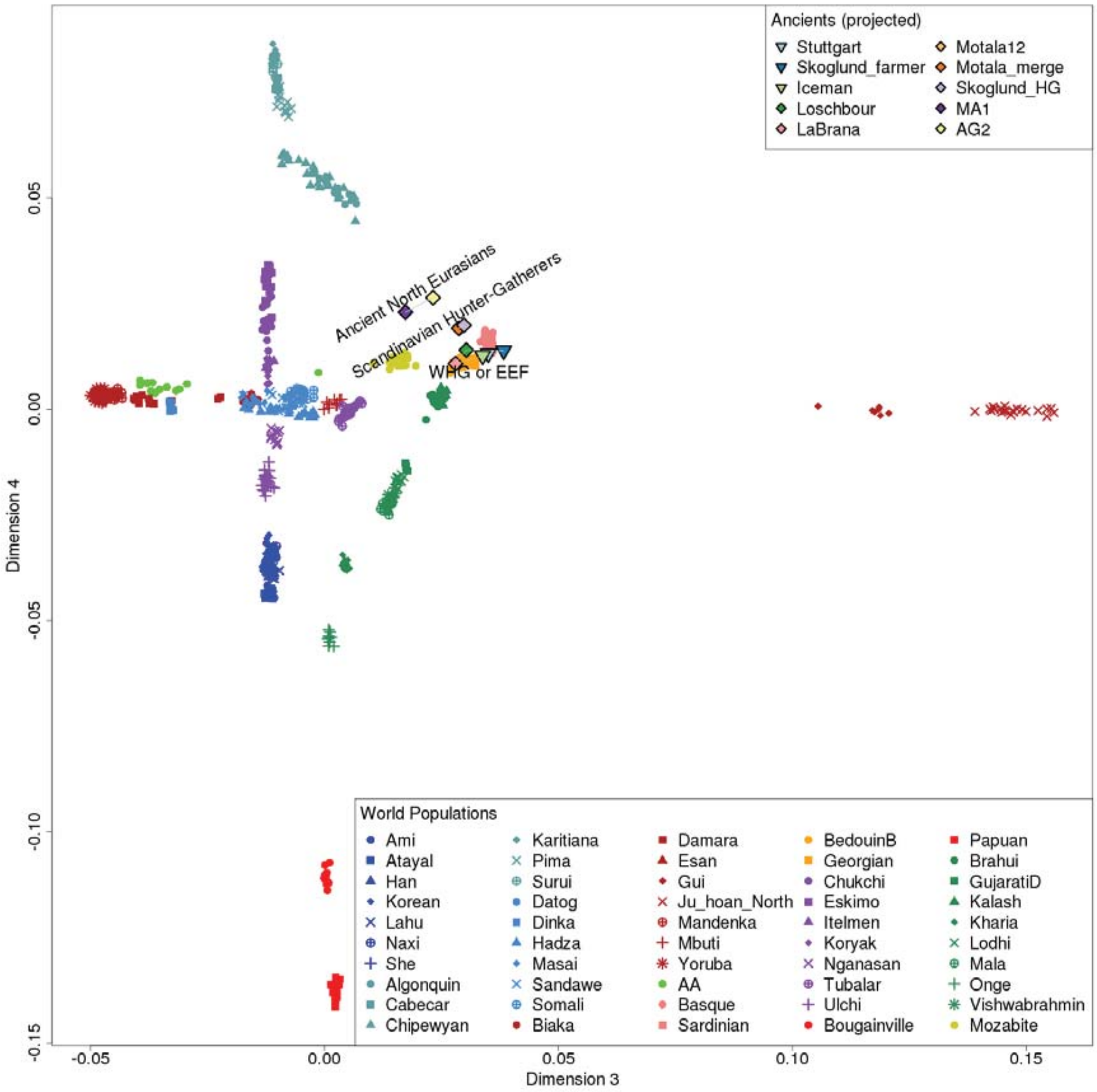

**West Eurasian PCA (Ancients Projected)**

Fig. 1B shows PC1 and PC2 of a PCA built on West Eurasian populations (explaining 0.9 and 0.4% of variance respectively) with the ancient samples projected. European and Near Eastern populations form parallel clines with a south-north orientation. The space between them is partially filled by Mediterranean and Jewish populations. Early European Farmers (EEF) including Stuttgart cluster with Mediterranean Europeans, especially Sardinians. Ancient North Eurasians (ANE) like MA1 project at the north of West Eurasian variation, between Europe and the Near East (AG2 is closer to Europe than the Near East but it is of poorer quality and contaminated[4]). European hunter-gatherers like Loschbour and Stuttgart project beyond present-day Europeans along PC1 and in the direction of Near Eastern-European differentiation. Loschbour groups with LaBrana, a Mesolithic Iberian hunter-gatherer[5], into a Western European Hunter-Gatherer (WHG) cluster. Motala12 and Motala_merge are similar to each other and to Skoglund_HG a merge of Neolithic hunter-gatherers from Sweden[6], thus forming a Scandinavian Hunter-Gatherer (SHG) cluster. The SHG appear intermediate between WHG and ANE, similar to the Global PCA and consistent with the evidence from $f_4$-statistics (SI14).

The PCA suggests that Europeans were formed by admixture involving Near Eastern and WHG populations (since they appear intermediate between the Near East and WHG) and also EEF and ANE populations. This impression is confirmed by the analysis of $f_3$-statistics (Table 1, Extended Data Table 1) which demonstrate that European populations have negative *$f_3$(European; Loschbour, Near East)* and *$f_3$(European; Stuttgart, MA1)* statistics and with our detailed modeling (SI14) which confirms that a model in which the Near Eastern ancestry is mediated via EEF and Europeans have additional WHG/ANE-related ancestry successfully fits most European populations and makes predictions consistent with those of a method with minimal modeling assumptions (SI17).

**West Eurasian PCA (Moderns Projected)**

An alternative method of visualizing the relationships of the ancient samples to present-day populations is to infer the PCs using ancient populations (Loschbour, Stuttgart, and MA1) and to project modern populations (Fig. S10.3). Dimension 1 differentiates MA1 from Europeans while Dimension 2 differentiates Loschbour from Stuttgart. Fig. S10.4 shows in magnification the central portion of the plot, that is, the projected present-day West Eurasians. While this is noticeably “noisier” than Fig. 1B as it is based on only three individuals, several patterns are clear: (i) Europeans deviate from Near Easterners along PC2 towards Loschbour, and (ii) in both Europe and the Near East there are clines with the most southern populations having the least proximity to MA1.

***Figure S10.3: Projection of West Eurasian populations onto the first two principal components inferred using Loschbour, Stuttgart, and MA1 (full version).***

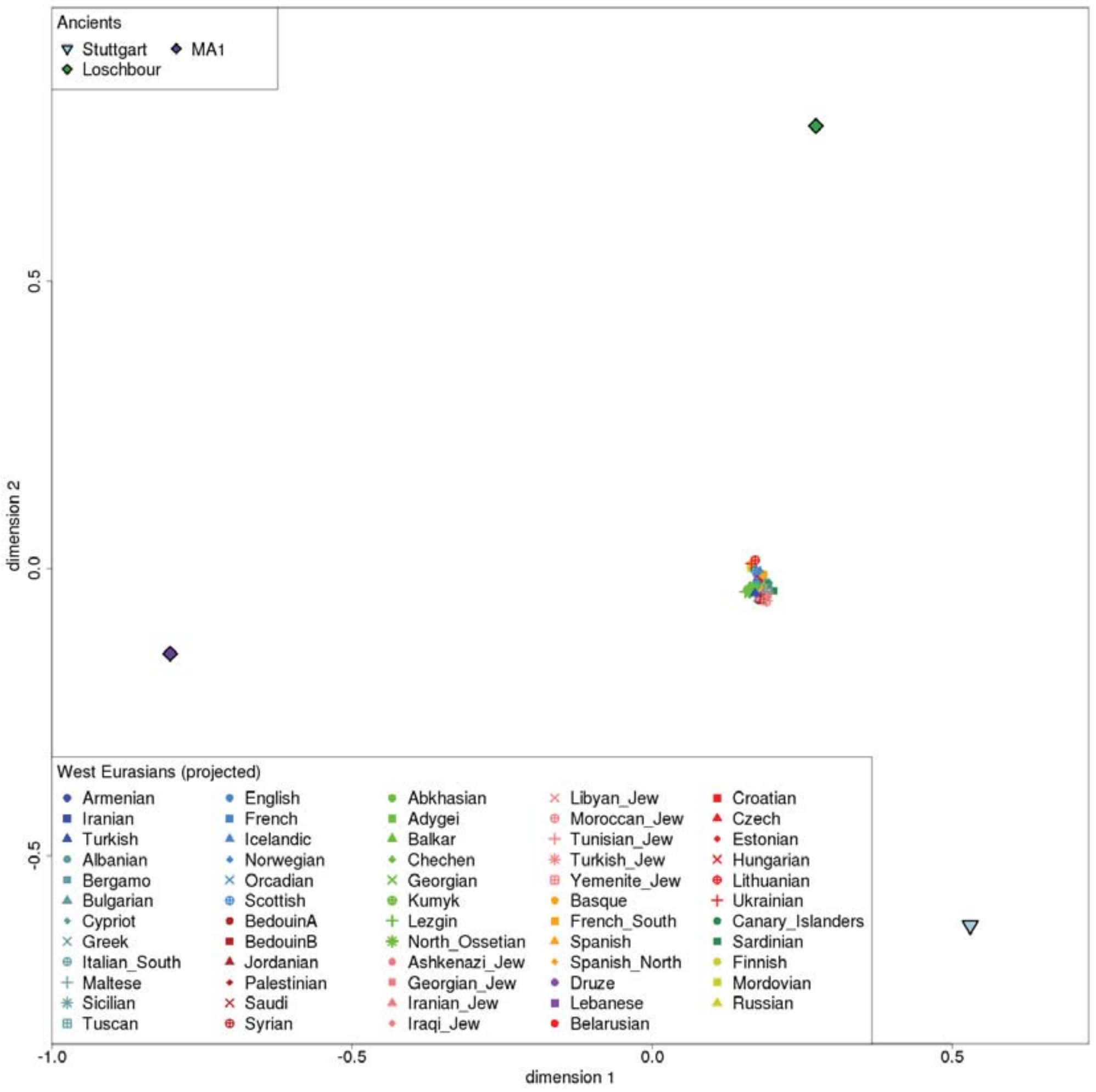

***Figure S10.4: Projection of West Eurasian populations onto the first two principal components inferred using Loschbour, Stuttgart, and MA1 (magnified version of Fig. S10.3 focusing on West Eurasian populations).***

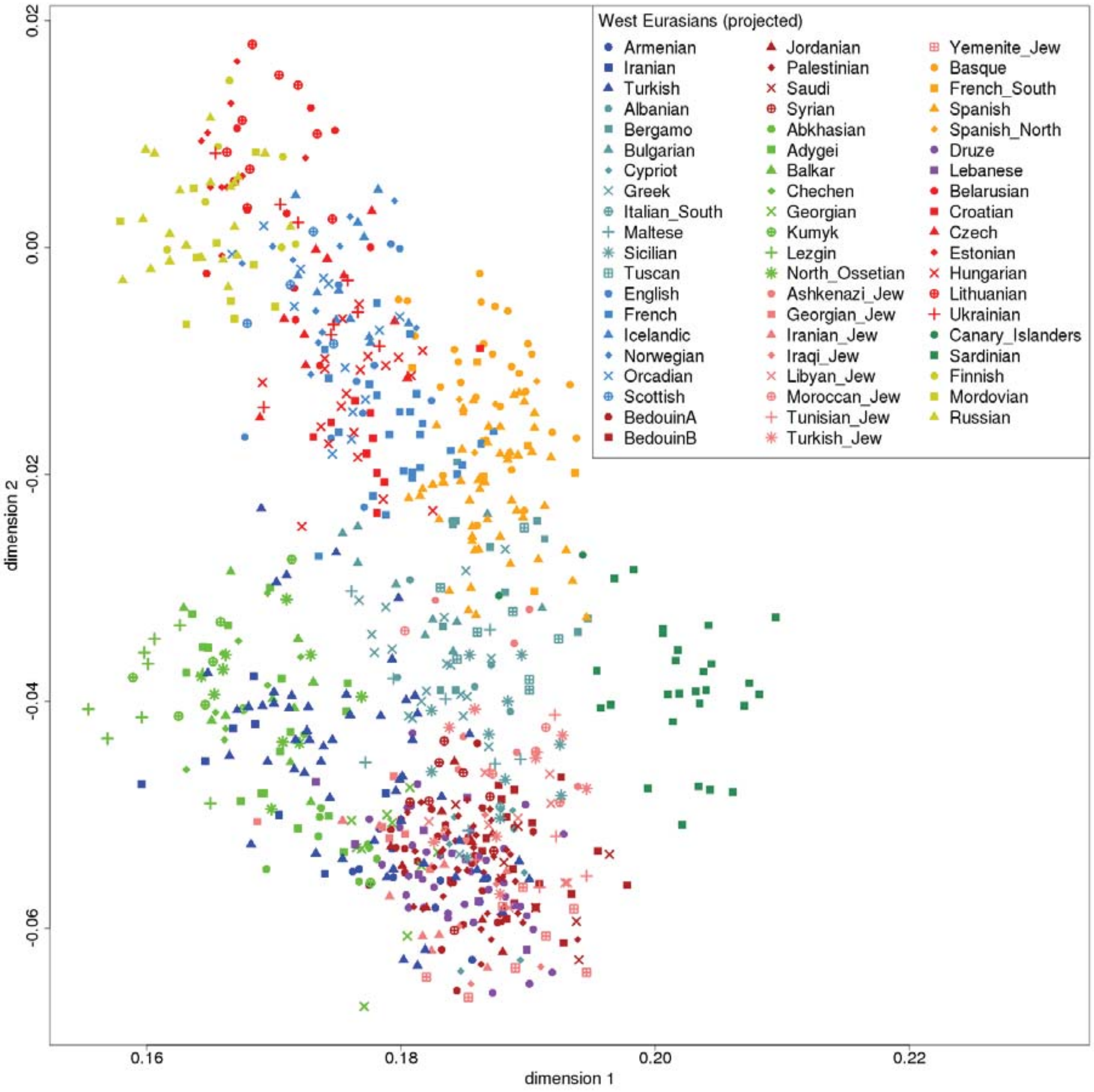

**European PCA (Ancients Projected)**

We also projected ancient samples onto the PCA inferred using only European populations (Fig. S10.5, PC1 and PC2 explaining 0.7 and 0.5% of variance). The absence of Near Eastern populations removes the European-Near Eastern differentiation (which dominates PC1 in Fig. 1B) and now the first dimension of the PCA corresponds to the main Sardinian-Northeast European cline of Europe.

Early European farmers and Ancient North Eurasians continue to occupy "southern" and "northern" ends of this cline as in Fig. 1B but European hunter-gatherers now occupy the northern end of the cline as well, and their position contrasts with that of Fig.1B. European hunter-gatherers project beyond Europeans in Fig. 1B (in the direction of European-Near Eastern differentiation) along PC1 as all present-day European groups have European hunter-gatherer ancestry that distinguishes them from Near Easterners; by contrast, in Fig. S10.5 the PC1 captures intra-European differentiation (south-northeast). We estimate (Fig. 2B, Extended Data Table 3) that variable WHG-related ancestry *above* *and beyond* that which is inherited via EEF contributes to intra-European differentiation and that modern Europeans share additional common drift with both Loschbour and MA1 (Extended Data Fig. 4). Thus, Europeans appear to be a mixture of one element related to Early European Farmers and one related to both MA1 and Loschbour, consistent with our model (Fig. 2A) that derives them from a "Hunter" population that had both Loschbour- and MA1-related ancestry (SI14).

***Figure S10.5: Projection of ancient samples onto the "European" PCA.***

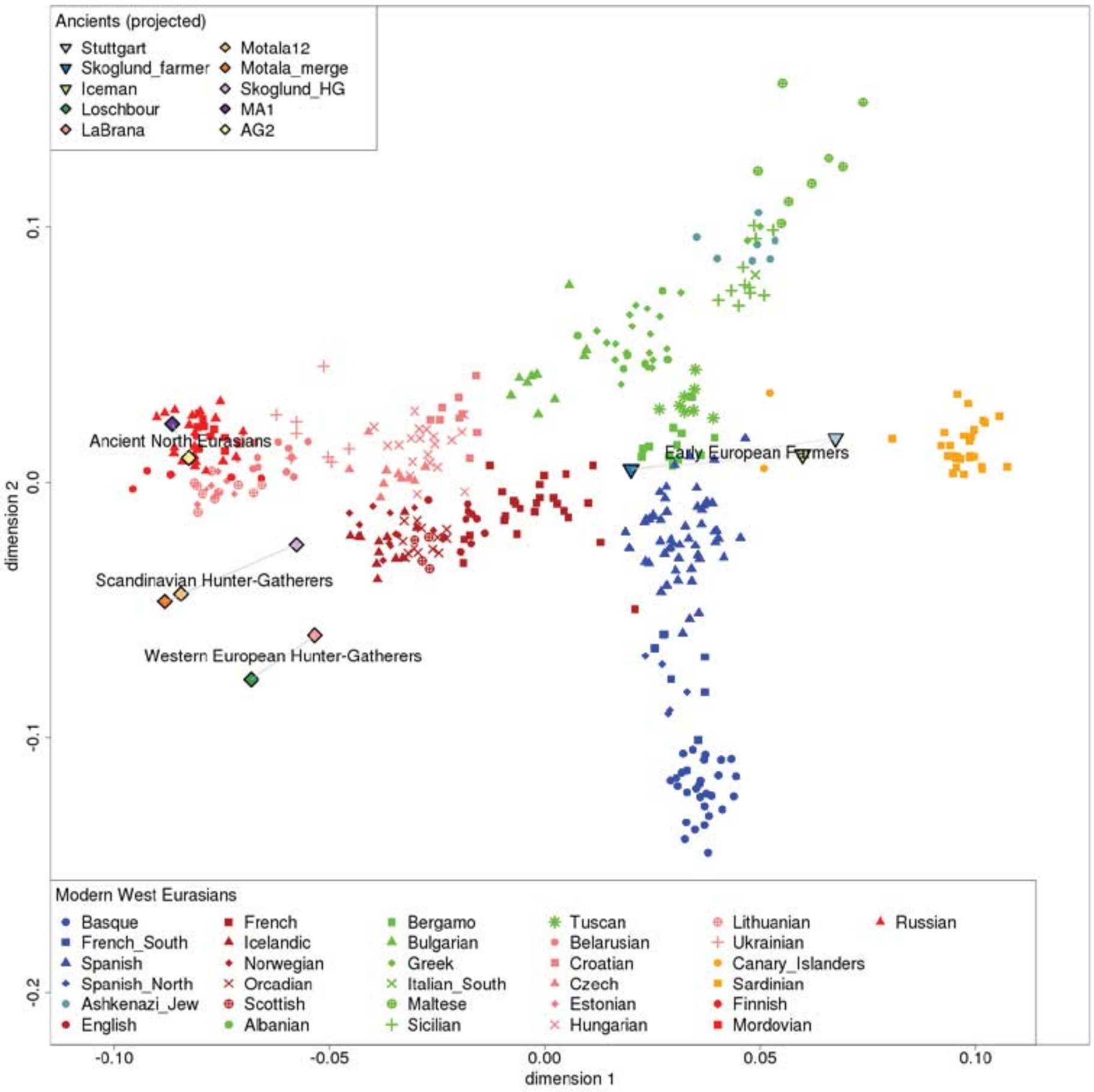

**Why do our PCAs of European samples correlate poorly to geographic maps of Europe?**

Previous studies have shown that remarkably, when one carries out PCA on a diverse set of European populations, a plot of the first two PCs can represent a map of Europe[7,8]. It is notable, however, that Fig. 1B does not resemble a map of Europe at all. Moreover, even when we remove Near Eastern populations from the input populations into the PCA—to make our study more similar to the previous studies that showed a correlation to a map of Europe and which almost entirely analyzed European populations—the resulting PCA does not appear to correlated well to a map of Europe (Fig. S10.5).

To investigate why a plot of the first two PCs in a PCA of diverse European samples differs in such a qualitatively clear way between our study and previous reports[7,8], we classified the European populations in our study accordance to the geographical scheme of Novembre et al. (2008)[8] (Table S10.1) and contrasted the fraction of individuals from each region in our Human Origins dataset to that used in that paper. We include in Table S10.1 Cypriot and Turkish populations which are not included in Fig. S10.5. While these samples do not appear to cluster with Europeans in any of our analyses, Cypriot and Turkish samples were used in Novembre et al. (2008), and so we include them for consistency in the comparison.

***Table S10.1: Assignment of Human Origins populations into geographic categories.***
*We use the same categorization as Novembre* et al. *(2008)*[8].

| Population | Sample Size | Category |
|---|---|---|
| Albanian | 6 | SE |
| Ashkenazi_Jew | 7 | J |
| Basque | 29 | SW |
| Belarusian | 10 | NE |
| Bergamo | 12 | S |
| Bulgarian | 10 | SE |
| Canary_Islanders | 2 | SW |
| Croatian | 10 | SE |
| Cypriot | 8 | ESE |
| Czech | 10 | E |
| English | 10 | NW |
| Estonian | 10 | NE |
| Finnish | 7 | NE |
| French | 25 | W |
| French_South | 7 | SW |
| Greek | 20 | SE |
| Hungarian | 20 | E |
| Icelandic | 12 | N |
| Italian_South | 1 | S |
| Lithuanian | 10 | NE |
| Maltese | 8 | S |
| Mordovian | 10 | NE |
| Norwegian | 11 | N |
| Orcadian | 13 | NW |
| Russian | 22 | NE |
| Sardinian | 27 | S |
| Scottish | 4 | NW |
| Sicilian | 11 | S |
| Spanish | 53 | SW |
| Spanish_North | 5 | SW |
| Tuscan | 8 | S |
| Turkish | 56 | ESE |
| Ukrainian | 9 | NE |

***Table S10.2: Number of sampled individuals per region***

| (this study) | | | Novembre et al. (2008) | | |
|---|---|---|---|---|---|
| Region | Sample Size | Fraction | Region | Sample Size | Fraction |
| C | 0 | 0.000 | C | 186 | 0.134 |
| E | 30 | 0.065 | E | 31 | 0.022 |
| ESE | 64 | 0.138 | ESE | 8 | 0.006 |
| J | 7 | 0.015 | J | 0 | 0.000 |
| N | 23 | 0.050 | N | 14 | 0.010 |
| NE | 78 | 0.168 | NE | 31 | 0.022 |
| NW | 27 | 0.058 | NW | 266 | 0.192 |
| S | 67 | 0.145 | S | 232 | 0.167 |
| SE | 46 | 0.099 | SE | 96 | 0.069 |
| SW | 96 | 0.207 | SW | 264 | 0.190 |
| W | 25 | 0.054 | W | 259 | 0.187 |

It is clear from the comparison of Table S10.2 that our study has a qualitatively very different distribution of samples compared to Novembre et al. (2008). In particular, in the combination of Central Europe (C), Western (W) and Northwestern (NW) Europe, our study has 11.2% of samples compared with 51.3% in Novembre et al. (2008). Conversely, in Eastern (E) and Northeastern Europe (NE) our study has 23.3% of samples compared with 4.5% in Novembre et al. (2008).

Our results highlight the fact that the correspondence between the map of Europe and the first two principal components depends on the composition of the included samples. Unfortunately, the lack of coverage of C and opposite balance of W/NW and E/NE in the two studies does not allow us to "mimic" the distribution of Novembre et al. (2008) and to test if the genetic-geographic correspondence might emerge in our dataset. An interesting avenue for future research is to study the conditions under which sample selection may lead to a correspondence between PCA and geography.

Fig. S10.5 suggests that the main axis of differentiation in Europe when the subcontinent is considered as a whole may tend to Northeastern Europe rather than SSE/NNW[8]. This is consistent with our analysis of ancestry proportions in European populations (Fig. 2B, Extended Data Table 3) which indicate a cline of reduced EEF (and increasing WHG) ancestry along that direction.

Importantly, the cline of ancestry proportions that we detect in this study does not depend on the relative sampling representation of different regions (as does PCA) as is evident when individual European populations are either fit to a formal model (SI14) or to an algebraic procedure (SI17).

# Supplementary Information 11

**All-pair $f_3$-statistics**

Iosif Lazaridis*, Nick Patterson and David Reich

* To whom correspondence should be addressed (lazaridis@genetics.med.harvard.edu)

The $f_3$-statistic[1,2] $f_3(Test, Ref_1, Ref_2)$ can be significantly negative if the *Test* population is a mixture of populations related to two reference populations $Ref_1$ and $Ref_2$. It is not necessary that the two reference populations be identical to the admixing ones. Fig. S11.1 shows a demonstration of a *Test* population produced by a mixture (in proportions $\alpha$ and $\beta=1-\alpha$) of two populations X and Y that are related to two reference populations $Ref_1$ and $Ref_2$.

***Figure S11.1:*** *Illustration of* f$_3$*-statistics (modified from Patterson* et al. *2012*[1]*).*

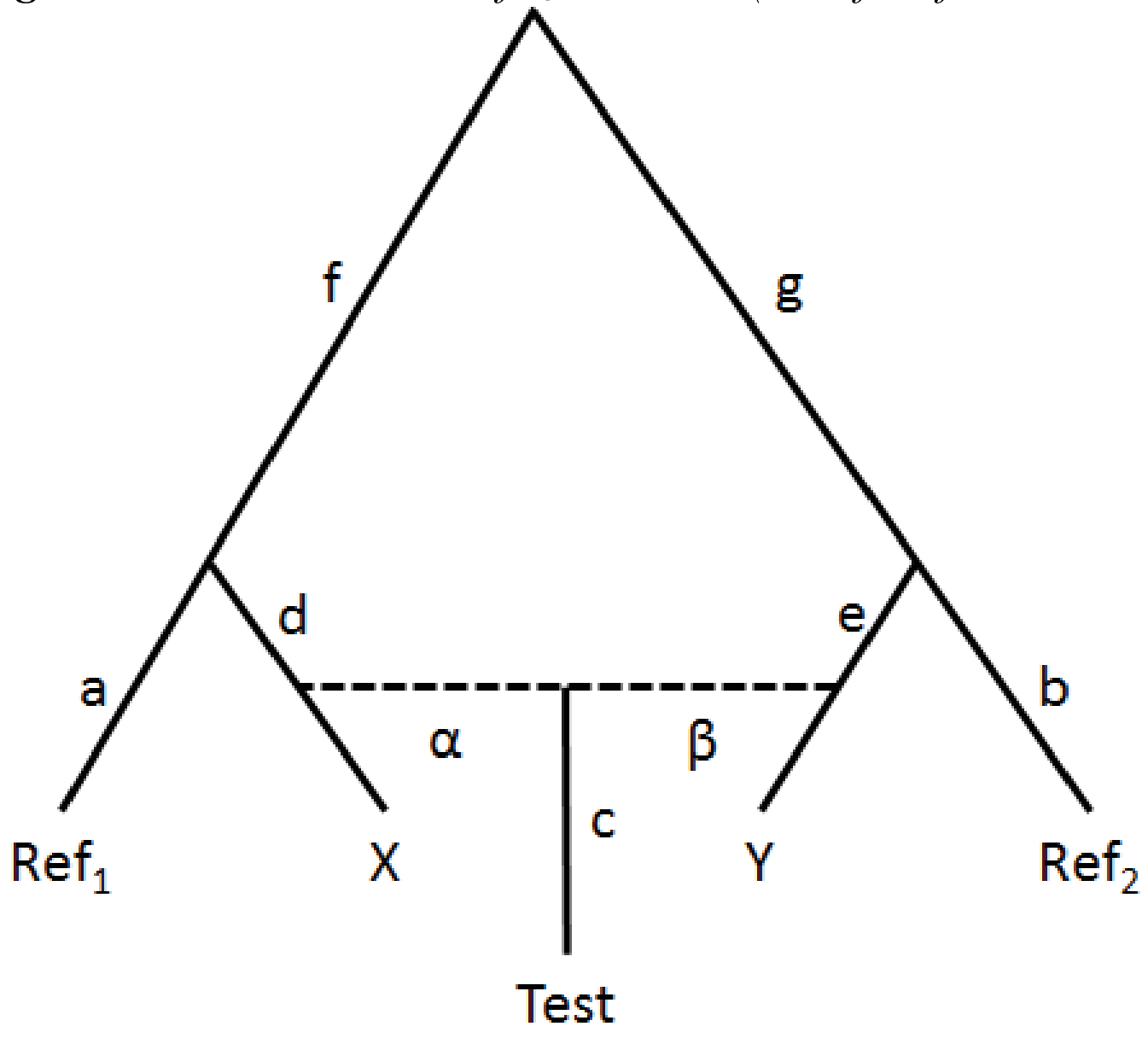

The value of $f_3(Test, Ref_1, Ref_2) = c+\alpha^2d+\beta^2e-\alpha\beta(g+f)$.The statistic is negative if the term $\alpha\beta(g+f)$ exceeds $c+\alpha^2d+\beta^2e$. Thus, if significant post-admixture drift ($c$) has occurred in a population, the statistic can be positive, and also if the true admixing populations (X and Y) have significant drift ($d$ and $e$) separating them from the reference populations. Note that the reference population-specific drifts ($a$ and $b$) do not feature in the expression of the statistic. Thus, the statistic is not always negative even if admixture did take place. It becomes negative only if post-admixture drift is not substantial and reasonable reference populations exist[1].

The ancient Europeans sequenced in our study are plausible candidates of being surrogates for at least some of the ancestral populations of Europeans, a relationship also suggested by the PCA of Fig. 1B. We did not, however, assume this *a priori*, but sought to discover (in an unsupervised manner), the answers to the following questions:

(i) Do diverse European populations show a signal of admixture?
(ii) If admixture took place, what are the sources of this admixture?
(iii) If admixture took place, are the ancient samples better surrogates for the admixing populations than any present-day populations?

For each West Eurasian population, we computed all 19,306 possible $f_3$-statistics for all pairs of references that included: 192 other present-day human populations from the Affymetrix Human Origins dataset (SI9) as well as the five ancient individuals: Loschbour and LaBrana (WHG), MA1 (ANE), Stuttgart (ANE), and Motala12 (SHG).

Question (i) is definitively answered in our analysis by our finding that when we search for pairs of populations that produces the lowest $f_3$-statistic for each target population in Europe, the lowest statistic is usually highly significantly negative even after correction for multiple hypothesis testing (Table 1). We note that we have previously reported widespread evidence of mixture via $f_3$-statistics in Europeans[1]. What is new here is the larger number of European populations analyzed, and the fact that with the ancient genomes in hand, we can show that some of the ancestral mixing populations are phylogenetically more closely related to the ancient genomes than to any present-day human populations in our dataset.

Question (ii) is addressed by the finding that all Europeans form their lowest statistic with the pairs EEF-ANE, EEF-WHG, or present-day Near East-WHG. The key observation here is that an ancient sample is always involved in producing the lowest $f_3$-statistic, even though the great majority of populations tested as references are present-day ones (Table 1). These pairings confirm the visual impression of Fig. 1B, which is that Europeans form a cline between between EEF and ANE and are intermediate between the Near East and WHG. Fig. 1B also suggests that Europeans are not a simple EEF-WHG mixture as, with the exception of Sardinians all others strongly deviate "northwards" along the European cline, i.e., in the same direction as ANE deviate from EEF, and indeed Sardinians are the only European population whose lowest statistic involved the EEF-WHG pairing.

We also examined the all-pair $f_3$-statistic to gain insight into whether a 2-way admixture model is adequate for European variation. Extended Data Table 1 shows that populations for which the lowest statistic involves EEF-ANE (e.g., Albanians) often also have a negative statistic involving Near East-WHG as well. Conversely, populations whose lowest statistic involves Near East-WHG (e.g., English) often also have a negative statistic involving EEF-ANE.

We calculate the difference between two $f_3$-statistics using a standard Block Jackknife[1,3]. In the case of Europeans (Extended Data Table 1) the two statistics are often within 3 standard errors of each other, so there is no strong evidence that Near East-WHG admixture or EEF-ANE admixture alone suffices to explain European admixture. In other words, if Europeans derive from a 2-way admixture, it is from ancestral populations that are not perfectly represented by the ancient genomes we sequenced, even if some of those ancient genomes are also shown by our analysis to be better surrogates for the admixing populations than any of the present-day genomes in our data.

These observations motivated us to explore models that involve all three ancient populations in the ancestry of Europeans. This is reported in SI14, where we formally develop a model that derives a test population as a 3-way mixture of EEF-WHG-ANE. In terms of that model, the Near East-WHG evidence of admixture corresponds to the admixture present in EEF who are a mixture of Near Eastern farmers and European hunter-gatherers (SI10), while the EEF-ANE and EEF-WHG statistics correspond to the admixture of Early European farmers with populations that harbored more WHG and ANE ancestry. Similar inferences are also suggested by $f_4$-statistics (Extended Data Fig. 4).

In Near Easterners the lowest statistic (Extended Data Table 1) always involved Stuttgart, providing evidence that Stuttgart has ancestry from the ancient Near East, or, alternatively that there has been massive migration of European farmers into the Near East. The latter scenario is implausible, however, given the known trajectory of farming dispersals in the opposite direction. Notably, the lowest statistic of the form EEF-WHG or WHG-Near East is virtually always much higher (more than 3 standard errors). This is not surprising as the PCA suggests no WHG ancestry in the Near East and neither do related $f_4$-statistics (Extended Data Fig. 4).

Finally, we checked the difference of the lowest $f_3$-statistic for European populations (which always always involves at least one ancient sample, Table 1) from the lowest $f_3$-statistic that involved only present-day populations.

Table S11.1 shows both the lowest $f_3$-statistics involving ancient populations and the lowest $f_3$-statistics involving only present-day populations, as well as a Z-score (Zdiff) of the difference between the two. It is clear that the ancient samples produce significantly lower $f_3$-statistics than any present-day population pair. Basques, for example, do not have a single negative $f_3$-statistic when only present-day populations are considered, but they have a Z=-10.3 significantly negative statistic involving the (Iraqi_Jew, Loschbour) pair. In a case where admixture was previously detected, e.g., the French[1] for the pairing (Sardinian, Karitiana), a much lower $f_3$-statistic ($Z_{diff}$=4.3 standard errors) is produced by the pairing (Stuttgart, MA1) involving two ancient samples.

The strongest decrease in the value of the $f_3$-statistics when we including ancient genomes is observed for Europeans, for which $Z_{diff}$>3 except for three cases:

(i) Ashkenazi Jews where the (Stuttgart, MA1) pairing produces $Z_{diff}$=2.3 lower statistic than the (Basque, Dinka) (the Dinka may reflect small recently gene flow from Africans[4]).

(ii) Maltese where the (Stuttgart, MA1) pairing is $Z_{diff}$=2 lower than the similar (Basque, Esan)

(iii) Russians where the (Loschbour, Chuckchi) pairing is $Z_{diff}$=2.7 lower than (Chuckchi, Sardinian).

For all three cases more recent admixture events involving non-European populations probably took place as discussed in SI14

For Near Easterners, decreases in the minimum $f_3$-statistics that are obtained by using ancient samples are often marginal and do not reach statistical significance. For example, the (Stuttgart, Esan) statistic of admixture in the Lebanese is not significantly lower than the (Sardinian, Wambo) one; both may reflect the same history of African admixture into the Levant. This is not surprising as the ancient genomes come from Europe and are less directly relevant ancestral populations (in the case of Stuttgart) for Near Easterners.

Our analysis of $f_3$-statistics strongly suggests that European populations today are likely all admixed, that the sources of admixture are related to the ancient individuals, and that the ancient individuals are better surrogates for the European ancestral populations than any present-day populations.

We anticipate that even better surrogates may be discovered as more ancient genomes are sequenced. However, our modeling analysis (SI14) already supports the idea that a 3-way mixture model using the available samples can successfully describe the ancestry of many Europeans in the context of world populations, within the limits of the resolution of our analyses.

**Table S11.1: Lowest $f_3$-statistics for West Eurasian populations across all reference populations and lowest $f_3$-statistics across only modern reference populations.**

| | All reference populations | | | | Only modern reference populations | | | | |
|---|---|---|---|---|---|---|---|---|---|
| Test | $Ref_1$ | $Ref_2$ | $f_3(Test; Ref_1, Ref_2)$ | Z | $Ref_1$ | $Ref_2$ | $f_3(Test; Ref_1, Ref_2)$ | Z | Zdiff |
| Abkhasian | Stuttgart | MA1 | -0.005268 | -2.929 | Georgian | Tuvinian | -0.002362 | -5.261 | 1.5 |
| Adygei | Piapoco | Stuttgart | -0.00733 | -5.92 | Georgian | Nganasan | -0.004125 | -8.916 | 2.5 |
| Albanian | Stuttgart | MA1 | -0.012083 | -7.027 | Sardinian | Surui | -0.005728 | -8.798 | 3.8 |
| Armenian | GujaratiC | Stuttgart | -0.006987 | -8.238 | GujaratiD | Sardinian | -0.003249 | -8.993 | 4.3 |
| Ashkenazi_Jew | Stuttgart | MA1 | -0.005731 | -3.403 | Basque | Dinka | -0.001987 | -3.784 | 2.3 |
| Balkar | Piapoco | Stuttgart | -0.01132 | -8.879 | Georgian | Nganasan | -0.008063 | -16.446 | 2.2 |
| Basque | Iraqi_Jew | Loschbour | -0.008332 | -10.293 | French_South | Spanish_North | 0.001546 | 5.624 | 11.9 |
| BedouinA | Esan | Stuttgart | -0.016223 | -18.196 | Sardinian | Wambo | -0.014322 | -39.459 | 1.9 |
| BedouinB | Esan | Stuttgart | 0.008926 | 7.825 | Mbuti | Sardinian | 0.010868 | 17.985 | 2.0 |
| Belarusian | Georgian | Loschbour | -0.013265 | -17.59 | Karitiana | Sardinian | -0.004456 | -7.685 | 9.7 |
| Bergamo | Stuttgart | MA1 | -0.010586 | -6.203 | Pima | Sardinian | -0.004502 | -10.818 | 3.4 |
| Bulgarian | Stuttgart | MA1 | -0.013021 | -8.2 | Aymara | Sardinian | -0.006795 | -14.869 | 3.7 |
| Chechen | Stuttgart | MA1 | -0.005568 | -3.183 | Karitiana | Sardinian | -0.002605 | -4.049 | 1.7 |
| Croatian | Stuttgart | MA1 | -0.011444 | -6.732 | Aymara | Sardinian | -0.005176 | -10.237 | 3.7 |
| Cypriot | Stuttgart | MA1 | -0.00568 | -3.229 | Sardinian | Taa_West | -0.002243 | -4.743 | 2.0 |
| Czech | Georgian | Loschbour | -0.013668 | -17.931 | Karitiana | Sardinian | -0.005439 | -10.393 | 8.9 |
| Druze | Stuttgart | MA1 | -0.002437 | -1.504 | Mbuti | Sardinian | -0.000524 | -1.374 | 1.3 |
| English | Iraqi_Jew | Loschbour | -0.012855 | -14.846 | Karitiana | Sardinian | -0.004063 | -7.548 | 8.6 |
| Estonian | Abkhasian | Loschbour | -0.012446 | -15.118 | Karitiana | Sardinian | -0.003061 | -5.039 | 9.1 |
| Finnish | Abkhasian | Loschbour | -0.010152 | -11.333 | Karitiana | Sardinian | -0.005529 | -8.085 | 4.1 |
| French | Stuttgart | MA1 | -0.013113 | -8.403 | Karitiana | Sardinian | -0.005939 | -15.053 | 4.3 |
| French_South | Iraqi_Jew | Loschbour | -0.009543 | -9.489 | Quechua | Sardinian | -0.00196 | -3.708 | 7.2 |
| Georgian | GujaratiC | Stuttgart | -0.003613 | -3.977 | Abkhasian | Tuscan | -0.00003 | -0.092 | 4.0 |
| Georgian_Jew | GujaratiC | Stuttgart | -0.000867 | -0.903 | GujaratiC | Sardinian | 0.002308 | 4.642 | 3.8 |
| Greek | Stuttgart | MA1 | -0.011782 | -7.396 | Guarani | Sardinian | -0.005114 | -13.613 | 4.0 |
| Hungarian | Stuttgart | MA1 | -0.013286 | -8.42 | Karitiana | Sardinian | -0.006569 | -15.329 | 4.1 |
| Icelandic | Abkhasian | Loschbour | -0.012141 | -15.62 | Karitiana | Sardinian | -0.002542 | -4.436 | 9.9 |
| Iranian | Piapoco | Stuttgart | -0.009425 | -7.195 | Cabecar | Yemenite_Jew | -0.006529 | -9.39 | 2.1 |
| Iranian_Jew | GujaratiC | Stuttgart | -0.001831 | -2.021 | Sardinian | Vishwabrahmin | 0.002541 | 5.701 | 4.9 |
| Iraqi_Jew | Vishwabrahmin | Stuttgart | -0.002572 | -2.582 | Naro | Sardinian | 0.000794 | 1.3 | 3.2 |
| Jordanian | Esan | Stuttgart | -0.01449 | -14.338 | Mbuti | Sardinian | -0.011667 | -23.716 | 2.9 |
| Kumyk | Piapoco | Stuttgart | -0.011068 | -8.165 | Karitiana | Sardinian | -0.00809 | -13.376 | 2.2 |
| Lebanese | Esan | Stuttgart | -0.010474 | -9.439 | Sardinian | Wambo | -0.0086 | -15.997 | 2.0 |
| Lezgin | Stuttgart | MA1 | -0.010038 | -5.968 | Karitiana | Sardinian | -0.003934 | -6.877 | 3.5 |
| Libyan_Jew | Esan | Stuttgart | -0.005063 | -4.403 | Mende | Sardinian | -0.002107 | -3.834 | 2.7 |
| Lithuanian | Abkhasian | Loschbour | -0.011918 | -14.854 | Sardinian | Surui | -0.000264 | -0.408 | 12.1 |
| Maltese | Stuttgart | MA1 | -0.008592 | -4.949 | Basque | Esan | -0.005311 | -10.767 | 2.0 |
| Mordovian | Abkhasian | Loschbour | -0.011454 | -14.373 | Sardinian | Surui | -0.008201 | -13.294 | 3.1 |
| Moroccan_Jew | Esan | Stuttgart | -0.006245 | -5.219 | Mandenka | Sardinian | -0.005205 | -8.68 | 0.8 |
| North_Ossetian | Piapoco | Stuttgart | -0.009273 | -7.176 | Georgian | Nganasan | -0.006987 | -13.623 | 1.6 |
| Norwegian | Georgian | Loschbour | -0.011951 | -14.836 | Aymara | Sardinian | -0.003877 | -7.133 | 8.9 |
| Orcadian | Armenian | Loschbour | -0.01016 | -13.388 | Karitiana | Sardinian | -0.00236 | -4.461 | 8.3 |
| Palestinian | Esan | Stuttgart | -0.011994 | -13.215 | Sardinian | Tswana | -0.009722 | -28.067 | 2.5 |
| Russian | Chukchi | Loschbour | -0.011865 | -11.345 | Chukchi | Sardinian | -0.008905 | -23.315 | 2.7 |
| Sardinian | Stuttgart | LaBrana | -0.00437 | -2.625 | Basque | Yemenite_Jew | 0.003776 | 14.999 | 4.8 |
| Saudi | Kgalagadi | Stuttgart | -0.004159 | -3.555 | Kgalagadi | Sardinian | -0.000862 | -1.427 | 2.6 |
| Scottish | Iraqi_Jew | Loschbour | -0.010343 | -8.275 | Karitiana | Sardinian | -0.00249 | -2.863 | 5.6 |
| Sicilian | Stuttgart | MA1 | -0.010803 | -6.483 | Basque | Esan | -0.004847 | -11.966 | 3.7 |
| Spanish | Iraqi_Jew | Loschbour | -0.012595 | -17.828 | Piapoco | Sardinian | -0.00543 | -15.438 | 8.8 |
| Spanish_North | Iraqi_Jew | Loschbour | -0.011152 | -9.87 | Karitiana | Sardinian | -0.001475 | -1.975 | 7.2 |
| Syrian | Esan | Stuttgart | -0.01013 | -8.712 | Mbuti | Sardinian | -0.008903 | -15.47 | 1.4 |
| Tunisian_Jew | Gambian | Stuttgart | -0.002606 | -2.025 | Mende | Sardinian | -0.000939 | -1.519 | 1.5 |
| Turkish | Piapoco | Stuttgart | -0.012922 | -11.324 | Chukchi | Sardinian | -0.009911 | -32.052 | 2.6 |
| Turkish_Jew | Stuttgart | MA1 | -0.007541 | -4.284 | Basque | Damara | -0.00516 | -11.407 | 1.2 |
| Tuscan | Stuttgart | MA1 | -0.01093 | -6.424 | Karitiana | Sardinian | -0.004631 | -8.334 | 3.5 |
| Ukrainian | Georgian | Loschbour | -0.013398 | -16.736 | Piapoco | Sardinian | -0.005217 | -9.327 | 9.0 |
| Yemenite_Jew | Esan | Stuttgart | -0.002704 | -2.36 | Sardinian | Tswana | 0.000483 | 0.799 | 3.2 |

# Supplementary Information 12

## Statistical evidence for at least three source populations for present-day Europeans

Nick Patterson*, Iosif Lazaridis and David Reich

* To whom correspondence should be addressed (nickp@broadinstitute.org)

**Overview**

In a previous study on Native American population history, we showed that it is possible to provide formal evidence for a minimum number of migrations into the ancestors of a test set of populations[1].

The method involves studying a matrix of $f_4$-statistics relating a set of test populations to a set of proposed outgroups.

To infer the minimum number of ancestral populations that must have mixed to form the test set of populations, the method exploits the fact that each of these ancestral mixing populations must have had a vector of $f_4$-statistics relating them to the outgroup populations.

Thus, the test populations today must be linear combinations of these ancestral $f_4$-statistic vectors.

By using linear algebra techniques to infer the minimum number of ancestral $f_4$-statistic vectors that are necessary (in linear combination) to explain the $f_4$-statistic vectors in all the test populations, we can infer a minimum on the number of migration events that must have occurred.

Concretely, we have a scenario where we have a set of "left" populations $L$ (proposed outgroups) and a set of "right" populations $R$ (test populations from a geographic region of interest, like Europe or the Americas) (Note S6 of ref. 1). We define:

$$X(l, r) = f_4(l_0, l; r_0, r) \quad \text{(S12.1)}$$

Here, $l_0$, $r_0$ are arbitrarily chosen "base" populations within the sets $L$ and $R$, and $l$, $r$ range over all choices of other populations in $L$ and $R$. The choice of "base" populations does not matter statistically (we obtain mathematically identical results for any choice of base population).

We showed in[1] that if $X(l, r)$ has rank $r$ and there were $n$ waves of immigration into $R$ with no back-migration from $R$ to $L$, then:

$$r+1 \leq n \quad \text{(S12.2)}$$

We used this to show that there were at least 3 waves of immigration into pre-Colombian America.

**Evidence for at least three source populations for most present-day Europeans**

To investigate whether a subset of European populations could be derived from $n$ waves of immigration, or equivalently that $X(l,r)$ has rank $n+1$, we with the following sets $L$ and $R$:

*L = {Stuttgart, Loschbour, MA1, Onge, Karitiana, Mbuti}*

*R = {Albanian, Basque, Belorussian, Bulgarian, Croatian, Czech, English, Estonian, French, French_South, Greek, Hungarian, Icelandic, Italian, Lithuanian, Norwegian, Orcadian, Pais_Vasco, Sardinian, Scottish, Spanish, Tuscan, Ukrainian}*

The set $L$ is chosen to match the populations used in SI14 for modeling, and includes a Sub-Saharan African group (Mbuti), two eastern non-Africans (Onge and Karitiana) that are differentially related to West Eurasians and MA1, and the three representatives of the ancestral populations inferred by our study (the Stuttgart individual representing EEF, the Loschbour individual representing WHG, and the MA1 individual representing ANE). The set $R$ includes all populations identified in both SI14 and SI17 as compatible with being derived from the same 3 ancestral populations, and excludes Sicilians, Maltese, Ashkenazi Jews, Finnish, Russians and Mordovians which have evidence of additional complex history.

From the $f_4$ statistics we can empirically estimate the matrix $X$ and test its consistency with a specified rank as described in ref. 1. For each possible rank $r$ we assume that $X$ has that rank (a null hypothesis) and test $X$ for rank $r+1$. In our previous study[1], we published a likelihood ratio test that yields a $\chi^2$ statistic to evaluate the consistency of this null hypothesis with the data[1]. In the tables below we present $r$, the number of degrees of freedom (d.o.f), the $\chi^2$ statistic value, and a P-value.

For the chosen $L$ and $R$ lists, we find that rank 2 is excluded, and hence at least 4 ancestral populations have contributed to the populations of $R$ (Table S12.1).

***Table S12.1: At least 4 ancestral populations for 23 European groups**. Rank 2 is excluded ($p<10^{-12}$), so rank 3, or at least 4 ancestral populations are inferred for European populations.*

| R | d.o.f. | $\chi^2$ | P-value |
|---|---|---|---|
| 0 | 26 | 2088.9 | $<10^{-12}$ |
| 1 | 24 | 740.8 | $<10^{-12}$ |
| 2 | 22 | 149.4 | $<10^{-12}$ |
| 3 | 20 | 30.4 | 0.063 |
| 4 | 18 | 15.1 | 0.654 |

The finding of at least 4 ancestral populations is seemingly at odds with our modeling approach which assumes 3 populations, so we sought to determine the cause of the added complexity.

We removed each of the populations of $R$ in turn and repeated the analysis over all 23 subsets. If the 4$^{th}$ ancestral population has largely affected only one of the populations in $R$, the evidence for four populations should disappear or greatly weaken when one of the affected population is removed.

We find that the P-value for rank 2 remains $<10^{-12}$ for 22 subsets, but for the subset $R$-{Spanish} it becomes 0.019, which is not significant after correcting for multiple hypothesis testing.

We next repeated the analysis of 253 subsets, removing all pairs of populations in turn. Again, for the vast majority of subsets the $P$-value for rank 2 remains $<10^{-12}$ but for all 22 pairs involving Spanish and another population, the P-value increases, ranging from 0.013-0.104, all non-significant.

We conclude that additional complexity exists in the Spanish population. It is possibly that this is due to the presence of low levels of Sub-Saharan ancestry in the Mediterranean[2] or of North African[3] admixture as has been reported previously. Such ancestry has also been suggested to occur at low levels in other European populations, and perhaps the Spanish stand out in our analysis because of their large sample size.

To shed more light on the additional source of ancestry that we detected in the Spanish we used ALDER[4], a method that uses admixture linkage disequilibrium to infer the time and extent of admixture. We used Mbuti, Yoruba, and Mozabite as African reference populations (Table S12.2).

This analysis confirm that gene flow from Sub-Saharan or North African populations has occurred in the Spanish sample.

***Table S12.2: Estimates of African admixture in Spanish population.*** *The Spanish population may harbor some African-related admixture representing a fourth wave of migration into Europe, but affecting Spain much more than the other groups.*

| | **African admixture (%)** | | **Time of African admixture (%)** | |
|---|---|---|---|---|
| **African reference** | **Lower bound** | **Std. error** | **Generations** | **Std. Error** |
| Mbuti | 0.7 | 0.1 | 66.2 | 9.7 |
| Yoruba | 1.5 | 0.2 | 65.5 | 9.7 |
| Mozabite | 12.6 | 2.0 | 73.7 | 10.4 |

**Adding outgroups to a minimal set of European populations**
A different approach is not to start with the full set of populations, but to choose a "small" $R$ as:

*R = {Belorussian, Bulgarian, Croatian, Czech, English, French, Hungarian, Icelandic, Norwegian, Orcadian, Sardinian, Scottish}*

This set of populations includes members of the main south-north European cline (Fig. 1B), and avoids most Mediterranean and Baltic populations where there may be more complex history involving Near Eastern, African, or East Eurasian ancestry.

We want to investigate whether this "simpler" subset of populations could be the result of admixture between only two ancestral populations. We had to "guess" a smaller set because of the combinatorial explosion of possible subsets of 23 populations (e.g., 1,352,078 possible subsets of 12 populations).

We first used a minimal set of proposed outgroup populations $L$:

*L = {MA1, Karitiana, Stuttgart, Loschbour}*

We find that rank 1 is excluded ($P < 10^{-12}$), and thus there must be at least 3 source populations related to the outgroups even for this restricted set of European populations (Table S12.3)

***Table S12.3. Test for L = {MA1, Karitiana, Stuttgart, Loschbour}***

| **R** | **d.o.f.** | $\chi^2$ | **P-value** |
|---|---|---|---|
| 0 | 13 | 1067 | $<10^{-12}$ |
| 1 | 11 | 121 | $<10^{-12}$ |
| 2 | 9 | 10.5 | 0.312 |

We next added Onge and Yoruba to $L$ (the Onge are an indigenous group from the Andaman Islands who have been genetically isolated for tens of thousands of years[5]). Again the data indicate at least 3 source populations, without significant evidence for more (Table S12.4).

***Table S12.4. Test for L = {MA1, Karitiana, Stuttgart, Loschbour, Onge, Yoruba}***

| **R** | **d.o.f.** | $\chi^2$ | **P-value** |
|---|---|---|---|
| 0 | 15 | 1504 | $<10^{-12}$ |
| 1 | 13 | 145 | $<10^{-12}$ |
| 2 | 11 | 17 | 0.114 |

A limitation of these methods is that they only work when there has been no back-migration from the populations related to the test set *R* into the ancestors of the outgroups *L*. In Native Americans, this seemed like a reasonable assumption, although even here there is evidence of back-migration from Native Americans into far northeastern Siberians (Naukan and Chukchi)[1].

For West Eurasians, the situation is potentially more problematic, as Europe and the Near East (and the Caucasus) have been far from isolated. Thus if enough Near East populations are introduced into *L* we can expect that the rank of *X* will increase if we have enough statistical power. In practice, however, such effects are mild. Specifically, we added each population *P* from the following list to the outgroup set *L* consisting of four populations.

> *P* = {*Abkhasian, Armenian, Ashkenazi_Jew, BedouinA, BedouinB, Chechen, Cypriot, Dinka, Druze, Georgian, Georgian_Jew, Han, Iranian, Iranian_Jew, Iraqi_Jew, Jordanian, Kalmyk, Lebanese, Libyan_Jew, Moroccan_Jew, Onge, Palestinian, Saudi, Syrian, Tunisian_Jew, Turkish, Turkish_Jew, Turkmen, Vishwabrahmin, Yemenite_Jew, Yoruba*}

For each population *P* in turn we computed the $\chi^2$ statistic (here with 12 d.o.f.) for the null that the rank of *X* is 2. The smallest P-value that we obtained was 0.024 for the 7 samples from *Turkish_Jew* population. On further exploration we obtained a P-value of 0.000048 (which is likely significant even after correcting for multiple hypothesis testing) by adding both *Yoruba* and *Turkish_Jew* to the 4 population *L* set and testing for consistency with rank 2. The underlying genetic history here is not clear to us. We conclude that the set *R* of European populations specified above cannot have arisen from a mixture of as few as 2 ancestral populations, but there is no strong evidence for more than 3 even when we add additional outgroup populations.

**Conclusion**

The strength of the approach in this section is that it formally tests for the number of ancestral components for all populations in *R* without assuming a model of population relationships.

Our results confirm that a large number of European populations cannot be derived from a mixture of just two ancestral populations. However, large subsets of populations are formally consistent with a mixture of at least three ancestral populations, without substantial evidence for a fourth ancestral population if the added complexity in the Spanish population is removed.

Finally we find that even for a much reduced set of European populations, at least three ancestral populations are inferred, and that this result is robust to addition of many non-European populations into the outgroup panel.

We anticipate that with larger population sample sizes additional minor inputs into Europe may be identified, further refining the history of European populations beyond the three ancestral populations identified by our study. However, these results increase our confidence that a model of three ancestral inputs can explain important features of the data.

# Supplementary Information 13

## Admixture proportions for Stuttgart

Iosif Lazaridis*, Nick Patterson and David Reich

* To whom correspondence should be addressed (lazaridis@genetics.med.harvard.edu)

A few lines of evidence suggest that the Stuttgart female harbors ancestry not only from Near Eastern farmers but also from pre-Neolithic European hunter-gatherers:

1. Her position in Fig. 1B, intermediate between the Near East and European hunter-gatherers.

2. The fact that the statistic $f_4$*(Stuttgart, X; Loschbour, Chimp)* is nearly always positive when *X* is a Near Eastern population (Table S13.1).

3. The results of ADMIXTURE analysis (SI 9), which show that when the West Eurasian ancestral population is split into European/Near Eastern sub-populations centered in Loschbour and southern Near Easterners respectively, Stuttgart is assigned ancestry from both.

***Table S13.1: Loschbour shares more genetic drift with Stuttgart than with Near Easterners.*** *This pattern is consistent with European hunter-gatherer admixture in Stuttgart.*

| **Population *X*** | $f_4$***(Stuttgart, X; Loschbour, Chimp)*** | **Z** |
|---|---|---|
| Kumyk | 0.00153 | 3.094 |
| Turkish_Jew | 0.00169 | 3.563 |
| Turkish | 0.00179 | 3.837 |
| Cypriot | 0.00191 | 3.904 |
| Abkhasian | 0.00199 | 4.151 |
| Georgian | 0.00200 | 4.155 |
| Moroccan_Jew | 0.00214 | 4.309 |
| Georgian_Jew | 0.00216 | 4.284 |
| Armenian | 0.00218 | 4.490 |
| Tunisian_Jew | 0.00257 | 5.169 |
| Iranian_Jew | 0.00276 | 5.672 |
| Druze | 0.00277 | 5.924 |
| Libyan_Jew | 0.00297 | 6.214 |
| Iraqi_Jew | 0.00305 | 6.066 |
| Iranian | 0.00311 | 6.290 |
| Lebanese | 0.00377 | 7.741 |
| Saudi | 0.00423 | 8.575 |
| Syrian | 0.00437 | 8.618 |
| Yemenite_Jew | 0.00458 | 9.100 |
| BedouinB | 0.00464 | 9.331 |
| Palestinian | 0.00474 | 10.183 |
| Jordanian | 0.00480 | 9.603 |
| BedouinA | 0.00618 | 12.951 |

Note: Only significant Z>3 statistics with X being any West Eurasian are shown (the complete set of these statistics for all West Eurasian populations is given in Extended Data Table 1).

A history of such admixture is plausible archaeologically, as the Linearbandkeramik postdates the earliest Neolithic of southeastern Europe, and there may have been opportunity for Near Eastern

Neolithic farmers to acquire a portion of European hunter-gatherer ancestry prior to the establishment of the central European Neolithic, either en route to central Europe (e.g., in the Balkans) or by mixing with the indigenous central European hunter-gatherers who they encountered.

A challenge in estimating mixture proportions for Stuttgart is that the two constituent elements contributing to it may not be well represented in our data. Present-day Near Eastern populations appear to have been strongly affected by events postdating movements of Neolithic migrants into Europe, as nearly all Near Eastern populations show negative $f_3$*(Near East; Stuttgart, X)* statistics where *X* is *MA1*, *Native American*, *South Asian*, or *African* (Table 1, Extended Data Table 1). As a result, it is risky to treat any individual Near Eastern population as an unmixed descendant of early Near Eastern farmers. Similarly, the ancient European hunter gatherer samples that we have sequenced (Loschbour and Motala12) are useful for these analyses, but it is not clear how closely they are related to the Mesolithic inhabitants of the Balkans and central Europe.

Recognizing the challenge posed by the lack of accurate surrogates for the ancestral populations, we hypothesized that Stuttgart is a mixture of an unknown hunter-gatherer population that forms a clade with Loschbour in proportion 1-$\alpha$ and an unknown Near Eastern population (NE) in proportion $\alpha$. While we do not know the exact NE population contributing ancestry to Stuttgart, we explored using BedouinB as a surrogate, as this is the population that appears at the southern end of the Near Eastern cline in Fig. 1B and has no evidence of eastern non-African ancestry by ADMIXTURE analysis (SI 9). A complication of using the BedouinB population for this purpose is that it has some African admixture, as indicated by the ADMIXTURE analysis (SI 9). We estimated a lower bound (4.2 ± 0.3%) on this admixture proportion using ALDER[1], using the Yoruba as a reference population. The advantage of this linkage-disequilibrium based method is that, unlike $f_4$-ratio estimation[2], no explicit model of population relationships is needed. We can also use the 5.1% estimate from ADMIXTURE K=3, or 7.3% from ADMIXTURE K=5 (SI 9). The two estimates differ because the Yoruba are inferred to have low levels of West Eurasian admixture at K=3, but to belong 100% to their own ancestral component at K=5. We did not use the K=4 value as in some ADMIXTURE replicates Yoruba formed their own component while in others they did not, whereas they formed their own component in all 100 replicates at K=5.

***Figure S13.1: $f_4$-ratio estimation of Near Eastern admixture in Stuttgart***

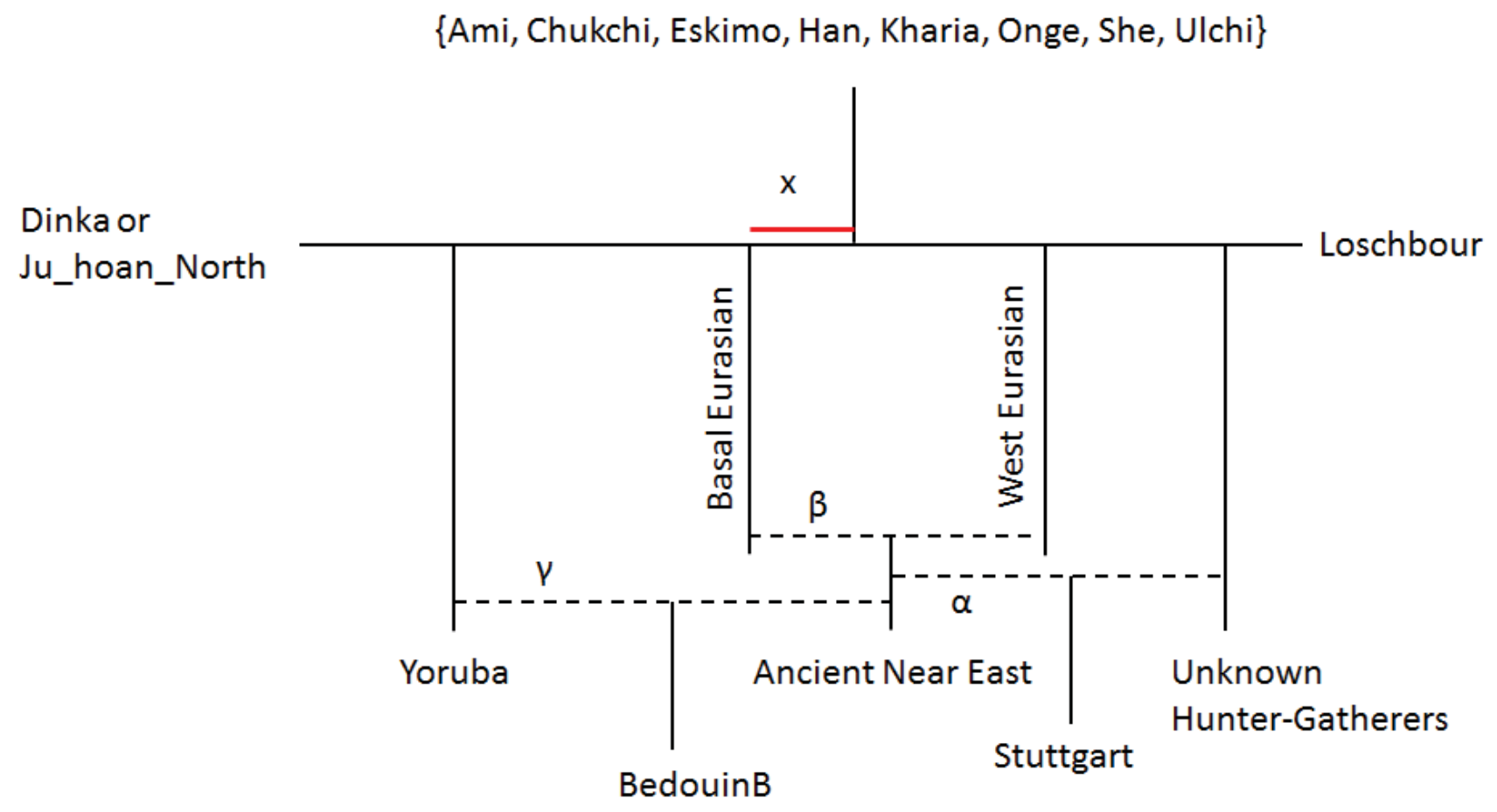

Consider Fig. S13.1 in which we model Stuttgart as a mixture of an unknown hunter-gatherer (UHG) population and Ancient Near East (NE) in proportions (1-$\alpha$, $\alpha$). From our modeling note (SI14), NE is plausibly a mixture of a West Eurasian element plus a basal Eurasian one, so let 1-$\beta$, $\beta$ be the mixture proportions of these two elements. We also assume the phylogenetic position of eastern non-African population *X*, alternatively using Ami, Chukchi, Eskimo, Han, Kharia, Onge, She, Ulchi, from the set of 15 populations identified in SI 9. We exclude Karitiana because of their substantial ANE ancestry, and Oceanians because of their Denisovan ancestry, neither of which conforms to the Fig. S13.1. model.

We can then write:

$$f_4(\textit{African outgroup, X; Loschbour, Stuttgart}) = -\alpha\beta x \qquad \text{(S13.1)}$$

where $x$ is the drift shared by most Eurasians but not basal Eurasians. We can also write:

$$f_4(\textit{African outgroup, X; Loschbour, NE}) = -\beta x \qquad \text{(S13.2)}$$

The ratio of the two yields the Near Eastern admixture of Stuttgart, $\alpha$. While $f_4$*(Outgroup, X; Loschbour, NE)* is unknown, we can estimate it via ancestry subtraction[3] as follows:

$$\begin{aligned} &f_4(\textit{African outgroup, X; Loschbour, BedouinB}) = \qquad \text{(S13.3)}\\ &= \gamma f_4(\textit{African outgroup, X; Loschbour, Yoruba}) + (1\text{-}\gamma) f_4(\textit{African outgroup, X; Loschbour, NE}) \end{aligned}$$

or, equivalently:

$$\begin{aligned} &f_4(\textit{African outgroup, X; Loschbour, NE}) = \qquad \text{(S13.4)}\\ &= [f_4(\textit{African outgroup, X; Loschbour, BedouinB})\\ &\quad - \gamma f_4(\textit{African outgroup, X; Loschbour, Yoruba})]/(1\text{-}\gamma) \end{aligned}$$

We choose Yoruba as a source of the African admixture as the source of the admixture in BedouinB appears to be African-farmer related (K=5, SI 9), and Yoruba are the population of African farmers with the highest sample size in the Human Origins dataset.

Shared common drift between "African outgroup" and Yoruba in the above equation complicates analysis, so we choose the "African outgroup" to be Dinka and Ju_hoan_North, two populations that do not appear to have recent common ancestry with West Africans and have very different histories.

We estimate $\gamma$=4.2%, or 5.1%, or 7.3%, as mentioned previously; these differ by only a few percent, but because they are used to subtract a portion of African ancestry from the BedouinB that is quite divergent from Eurasians, these small differences have substantial effects.

***Table S13.2: Near Eastern admixture estimates for Stuttgart***

| | **Outgroup=Dinka** | | | **Outgroup=Ju_hoan_North** | | |
|---|---|---|---|---|---|---|
| **African ancestry assumed in BedouinB:** | 4.2 | 5.1 | 7.3 | 4.2 | 5.1 | 7.3 |
| Ami | 0.667 | 0.727 | 0.941 | 0.662 | 0.729 | 0.981 |
| Chukchi | 0.700 | 0.752 | 0.925 | 0.697 | 0.755 | 0.954 |
| Eskimo | 0.721 | 0.774 | 0.950 | 0.720 | 0.778 | 0.979 |
| Han | 0.632 | 0.689 | 0.893 | 0.625 | 0.689 | 0.928 |
| Kharia | 0.549 | 0.608 | 0.835 | 0.535 | 0.601 | 0.873 |
| Onge | 0.665 | 0.717 | 0.896 | 0.660 | 0.718 | 0.927 |
| She | 0.684 | 0.744 | 0.958 | 0.680 | 0.748 | 1.001 |
| Ulchi | 0.706 | 0.760 | 0.945 | 0.703 | 0.764 | 0.978 |

The amount of Near Eastern admixture estimated for Stuttgart can be seen in Table S13.2 and ranges between 55-100% with estimates increasing as the amount of estimated African admixture in BedouinB increases. Estimates using Dinka or Ju_hoan_North as an African outgroup are similar. There are reasons to doubt both the lower estimates (near 55%), since ALDER provides only a lower bound on African ancestry, but also the higher estimates (near 100%) since there is direct evidence that Stuttgart has European hunter-gatherer ancestry (Fig. 1B and Table S13.1). Determining the precise levels of Near Eastern admixture in Stuttgart must await further ancient DNA studies from both Europe and the Near East, but we can at least reasonably suggest that most of the sample's ancestry was Near Eastern, consistent with the mtDNA evidence for the Linearbandkeramik, which demonstrated a strong Near Eastern influence[4-6].

# Supplementary Information 14
## Admixture Graph Modeling

Iosif Lazaridis*, Nick Patterson and David Reich

* To whom correspondence should be addressed (lazaridis@genetics.med.harvard.edu)

**Overview**
In this note we use $f_4$-statistics and ADMIXTUREGRAPH methodology[1] as implemented in the *qpGraph* software of ADMIXTOOLS[2] to investigate the relationships of Stuttgart and Loschbour to present-day human populations from Eurasia, Oceania, and the Americas. The ADMIXTUREGRAPH software allows us to test models relating a number of populations that may also contain admixture edges. ADMIXTUREGRAPH estimates model parameters (genetic drift along branches and admixture proportions) and assesses model fit by comparing fitted and estimated *f*-statistics using a block jackknife[3] and reporting outliers when these differ by more than 3 standard errors[2]. Our procedure does not attempt to devise a definitive model for the deep prehistoric relationships of present-day humans – which is certainly far more complex than the models we identify here – but rather to develop a working model that fits the *f*-statistics for many past and present populations within the limits of our resolution, and that can serve as a null hypothesis for further study of human history. In this section, we first compare ancient individuals against each other to better understand their inter-relationships and then compare them against a wide range of eastern non-African populations identified in SI9 as having no evidence of recent West Eurasian admixture. These comparisons using $f_4$-statistics identify features of the genetic data that a successful model must address. We then show how models with either no admixture events or one admixture event cannot fit the data, and identify a parsimonious model (the only one we could identify with two admixture events) that fits the data successfully, is robust to the addition of additional ancient samples, has a structure that is similar to those produced by different graph fitting methods (SI15, SI16), and makes mixture proportion estimates that are consistent (Extended Data Table 3) with those of a methodology described in SI17 that makes minimal modeling assumptions.

**Relationships of the ancient genomes to each other**
We begin by investigating some simple relationships using $f_4$-statistics which will inform the more detailed models we will later investigate. We report only statistics with $|Z|\geq 2$ in the tables that follow.

We first report (Table S14.1) statistics of the form $f_4(Ancient_1, Chimp; Ancient_2, Ancient_3)$ for the ancient samples: Loschbour, Stuttgart, Motala12, MA1, and LaBrana. Such statistics determine whether ($Ancient_2$, $Ancient_3$) are consistent with being a clade relative to $Ancient_1$. If they are not a clade, the statistic shows whether $Ancient_1$ is more closely related to $Ancient_2$ (in which case it is positive), or $Ancient_3$ (in which case it is negative).

We summarize our findings for each ancient sample:

LaBrana is closer to all ancient Europeans than to MA1, it is closer to both Loschbour and Motala12 than to Stuttgart, and it is closer to Loschbour than to Motala12. Thus, in order of increasing detail, LaBrana is identified as a European, a European hunter-gatherer, and a "West European" hunter-gatherer most related to Loschbour.

Loschbour behaves similarly to LaBrana, in the sense that it is closer to ancient Europeans than to MA1, closer to both LaBrana and Motala12 than to Stuttgart, and closer to LaBrana than to Motala12. These results suggest that Loschbour and LaBrana are relatively close relatives, consistent with the visual impression of them clustering in the PCA of Fig. 1B.

***Table S14.1: Relationships between ancient Eurasians***

| **Ancient$_1$** | **Ancient$_2$** | **Ancient$_3$** | ***$f_4$(Ancient$_1$, Chimp; Ancient$_2$, Ancient$_3$)*** | **Z** |
|---|---|---|---|---|
| LaBrana | Stuttgart | MA1 | 0.001744 | 2.246 |
| LaBrana | Loschbour | Stuttgart | 0.018555 | 25.288 |
| LaBrana | Loschbour | MA1 | 0.02018 | 24.356 |
| LaBrana | Loschbour | Motala12 | 0.008498 | 11.324 |
| LaBrana | Motala12 | Stuttgart | 0.010002 | 13.606 |
| LaBrana | Motala12 | MA1 | 0.011512 | 14.426 |
| Loschbour | LaBrana | Stuttgart | 0.017057 | 23.082 |
| Loschbour | LaBrana | MA1 | 0.019669 | 23.297 |
| Loschbour | LaBrana | Motala12 | 0.004742 | 5.962 |
| Loschbour | Stuttgart | MA1 | 0.002536 | 3.174 |
| Loschbour | Motala12 | Stuttgart | 0.012104 | 15.940 |
| Loschbour | Motala12 | MA1 | 0.014795 | 17.991 |
| MA1 | LaBrana | Stuttgart | 0.003929 | 5.612 |
| MA1 | Loschbour | Stuttgart | 0.004455 | 6.222 |
| MA1 | Motala12 | LaBrana | 0.004326 | 5.712 |
| MA1 | Motala12 | Stuttgart | 0.008463 | 11.263 |
| MA1 | Motala12 | Loschbour | 0.003815 | 5.276 |
| Motala12 | LaBrana | Stuttgart | 0.008594 | 11.369 |
| Motala12 | LaBrana | MA1 | 0.007186 | 8.256 |
| Motala12 | Loschbour | LaBrana | 0.003756 | 5.382 |
| Motala12 | Loschbour | Stuttgart | 0.012391 | 16.721 |
| Motala12 | Loschbour | MA1 | 0.01098 | 12.505 |
| Stuttgart | LaBrana | MA1 | 0.005673 | 7.749 |
| Stuttgart | Loschbour | LaBrana | 0.001499 | 2.538 |
| Stuttgart | Loschbour | MA1 | 0.006991 | 9.708 |
| Stuttgart | Motala12 | LaBrana | 0.001408 | 2.125 |
| Stuttgart | Motala12 | MA1 | 0.007008 | 9.190 |

MA1 is closer to all European hunter-gatherers than to Stuttgart and it is closer to Motala12 than to both Loschbour and LaBrana. Notice the asymmetry with the results of the previous paragraph: MA1 is closer to Loschbour than to Stuttgart, but Loschbour is closer to Stuttgart than to MA1. Thus, (Loschbour, Stuttgart) cannot be a "European clade" relative to MA1: this violates the fact that MA1 is closer to Loschbour than to Stuttgart; similarly, (Loschbour, MA1) cannot be a "Eurasian Hunter-Gatherer" clade relative to Stuttgart: this violates the fact that Loschbour is closer to Stuttgart than to MA1.

Motala12 is closer to both Loschbour and LaBrana than to MA1 and closer to both Loschbour and LaBrana than to Stuttgart. However, the statistic *$f_4$(Motala12, Chimp; Stuttgart, MA1)* is not positive (as is the case if we substitute Loschbour or LaBrana for Motala12), but instead is non-significantly negative -0.001455 ($Z=-1.718$). Together with the results of the preceding paragraph, this suggests a history of gene flow between Motala12 and MA1.

Stuttgart is uniformly closer to European hunter-gatherers than to MA1; this is expected given the evidence of European hunter-gatherer ancestry (SI13) in this early Neolithic European. There is also a hint from the data that Stuttgart is closer to Loschbour than to LaBrana; however, this particular statistic does not reach a "highly significant" $|Z|>3$ and we could not confirm it on the basis of whole genome transversion polymorphisms (Extended Data Table 3, Z=1.79).

**Summary of results from the *$f_4$*-statistic analysis relating ancient Eurasians**

The most salient findings from the survey of $f_4$-statistics involving only ancient Eurasians are:

1. The ancient Europeans (Loschbour, LaBrana, Motala12, and Stuttgart) share more alleles with each other than with MA1, with the only exception being that Motala12 is not more similar to Stuttgart than to MA1.

2. MA1 is more similar to Motala12 than to other European hunter-gatherers and more similar to European hunter-gatherers than to Stuttgart.
3. Loschbour and LaBrana are consistent with being a clade to the limits of our resolution.

We next analyzed how the ancient samples related to a set of non-West Eurasian populations identified by ADMIXTURE analysis (SI9). For each non-West Eurasian geographical region we computed statistics of the form *$f_4$(Ancient$_1$, Ancient$_2$; non-West Eurasian, Chimp)* and *$f_4$(Ancient, Chimp; non-West Eurasian$_1$, non-West Eurasian$_2$)*. These statistics test, respectively, whether two ancient individuals form a clade with respect to a non-West Eurasian population and whether two non-West Eurasian groups form a clade with respect to an ancient Eurasian individual.

**Relationship of ancient samples to South Asian populations without West Eurasian admixture**

We first consider the relationship of ancient samples to Onge (indigenous Little Andaman Islanders[4]), an island population from the Bay of Bengal that is distantly related to Ancestral South Indians[1]. We also test relationships to the Kharia, an Austroasiatic-speaking population from India that does not appear to be part of the Indian Cline of varying West Eurasian-related Ancient North Indian ancestry—in particular, it seems to have little or no West Eurasian admixture unlike the Indo-European and Dravidian speaking populations in India—and instead appears to have some East Asian-related admixture[1].

***Table S14.2: Onge and Kharia are closer to Eurasian hunter-gatherers than to Stuttgart.***

| **South Asian** | **Ancient$_1$** | **Ancient$_2$** | ***$f_4$(South Asian, Chimp; Ancient$_1$, Ancient$_2$)*** | **Z** |
|---|---|---|---|---|
| Kharia | Loschbour | Stuttgart | 0.001154 | 2.564 |
| Kharia | MA1 | Stuttgart | 0.001447 | 2.563 |
| Kharia | Motala12 | Stuttgart | 0.0012 | 2.365 |
| Onge | LaBrana | Stuttgart | 0.001528 | 2.797 |
| Onge | Loschbour | Stuttgart | 0.00191 | 3.452 |
| Onge | MA1 | Stuttgart | 0.001842 | 2.987 |
| Onge | Motala12 | Stuttgart | 0.002105 | 3.660 |

The results of Table S14.2 provide suggestive evidence that Onge and Kharia share more common ancestry with ancient Eurasian hunter-gatherers than with Stuttgart. All statistics involving two hunter-gatherer populations have $|Z|<1$, so ancient Eurasian hunter-gatherers are approximately symmetrically related to Onge and Kharia, and they are more closely related to them than is Stuttgart.

**Relationship of ancient samples to East Asians**

We next consider the relationship of ancient samples to East Asians using the set (Ami, Han, She). East Asians are more closely related to all hunter-gatherers than to Stuttgart, but there are no significant differences between hunter-gatherers (all such statistics have $|Z|<1.1$) (Table S14.3).

***Table S14.3: East Asians are more closely related to ancient hunter-gatherers than to Stuttgart.***

| **East Asian** | **Ancient$_1$** | **Ancient$_2$** | ***$f_4$(East Asian, Chimp; Ancient$_1$, Ancient$_2$)*** | **Z** |
|---|---|---|---|---|
| Ami | LaBrana | Stuttgart | 0.001453 | 2.841 |
| Ami | Loschbour | Stuttgart | 0.001745 | 3.424 |
| Ami | MA1 | Stuttgart | 0.001751 | 2.884 |
| Ami | Motala12 | Stuttgart | 0.001495 | 2.713 |
| Han | LaBrana | Stuttgart | 0.001464 | 2.973 |
| Han | Loschbour | Stuttgart | 0.001634 | 3.275 |
| Han | MA1 | Stuttgart | 0.001548 | 2.634 |
| Han | Motala12 | Stuttgart | 0.001626 | 3.022 |
| She | LaBrana | Stuttgart | 0.001449 | 2.856 |
| She | Loschbour | Stuttgart | 0.001814 | 3.538 |
| She | MA1 | Stuttgart | 0.001719 | 2.824 |
| She | Motala12 | Stuttgart | 0.001731 | 3.137 |

We also found no statistics of the form $f_4$(*Ancient, Chimp; East Asian$_1$, East Asian$_2$*) (all |Z|<1.6). Thus, there is no evidence of differential relatedness of East Asians to ancient West Eurasians and Siberians.

**Relationship of ancient samples to North Asians**
We next consider the relationship of ancient samples to North Asian populations (Chukchi, Eskimo, Ulchi). North Asians are more closely related to the ancient hunter-gatherers than to Stuttgart and they are also more closely related to MA1 than to the European hunter-gatherers (Table S14.4). This suggests that North Asian groups have some ancestry related to MA1.

***Table S14.4: North Asians are more closely related to ancient hunter-gatherers than to Stuttgart, and more closely related to MA1 than to European hunter-gatherers***

| North Asian | Ancient$_1$ | Ancient$_2$ | $f_4$(*North Asian, Chimp; Ancient$_1$, Ancient$_2$*) | Z |
|---|---|---|---|---|
| Chukchi | LaBrana | Stuttgart | 0.002113 | 4.251 |
| Chukchi | Loschbour | Stuttgart | 0.002358 | 4.639 |
| Chukchi | MA1 | LaBrana | 0.00244 | 4.346 |
| Chukchi | MA1 | Stuttgart | 0.004756 | 8.044 |
| Chukchi | MA1 | Loschbour | 0.002388 | 3.978 |
| Chukchi | MA1 | Motala12 | 0.001176 | 2.072 |
| Chukchi | Motala12 | LaBrana | 0.001368 | 2.626 |
| Chukchi | Motala12 | Stuttgart | 0.003631 | 6.624 |
| Chukchi | Motala12 | Loschbour | 0.001152 | 2.140 |
| Eskimo | LaBrana | Stuttgart | 0.002091 | 4.135 |
| Eskimo | Loschbour | Stuttgart | 0.002481 | 4.818 |
| Eskimo | MA1 | LaBrana | 0.003206 | 5.643 |
| Eskimo | MA1 | Stuttgart | 0.005367 | 9.003 |
| Eskimo | MA1 | Loschbour | 0.002884 | 4.787 |
| Eskimo | MA1 | Motala12 | 0.00167 | 2.888 |
| Eskimo | Motala12 | LaBrana | 0.0016 | 3.040 |
| Eskimo | Motala12 | Stuttgart | 0.003775 | 6.884 |
| Eskimo | Motala12 | Loschbour | 0.001167 | 2.165 |
| Ulchi | LaBrana | Stuttgart | 0.001664 | 3.429 |
| Ulchi | Loschbour | Stuttgart | 0.002206 | 4.420 |
| Ulchi | MA1 | Stuttgart | 0.002429 | 4.091 |
| Ulchi | Motala12 | Stuttgart | 0.002335 | 4.272 |

**Relationship of ancient samples to Oceanians**
We considered the relationship of the ancient samples to Oceanians using the set (Papuan, Bougainville). The statistics in Table S14.5 border on |Z|=3 and are suggestive that hunter-gatherer groups share more genetic drift with Oceanian populations than with Stuttgart. All statistics involving two ancient hunter-gatherers are non-significant with |Z|<1.1.

***Table S14.5: Oceanian populations are genetically closer to hunter-gatherers than to Stuttgart.***

| Oceanian | Ancient$_1$ | Ancient$_2$ | $f_4$(*Oceanian, Chimp; Ancient$_1$, Ancient$_2$*) | Z |
|---|---|---|---|---|
| Bougainville | LaBrana | Stuttgart | 0.001127 | 2.182 |
| Bougainville | Loschbour | Stuttgart | 0.001566 | 2.951 |
| Bougainville | MA1 | Stuttgart | 0.001491 | 2.337 |
| Bougainville | Motala12 | Stuttgart | 0.001724 | 3.119 |
| Papuan | LaBrana | Stuttgart | 0.001183 | 2.256 |
| Papuan | Loschbour | Stuttgart | 0.001364 | 2.599 |
| Papuan | MA1 | Stuttgart | 0.00141 | 2.165 |
| Papuan | Motala12 | Stuttgart | 0.001755 | 3.181 |

Statistics of the form $f_4$(*Ancient, Chimp; Bougainville, Papuan*) (not shown) are all positive (|Z|>2) but do not suggest gene flow between Bougainville and west Eurasia, as they are affected by differential Denisovan admixture into the two Oceanian groups[5]. We conclude that Oceanian populations are genetically closer to Eurasian hunter-gatherers than to Stuttgart.

**Relationship of ancient samples to Native Americans**

We explored the relationship of ancient samples to Native Americans without post-Colombian European admixture (Cabecar, Karitiana, Mixe, Piapoco, Surui) in Table S14.6.

Native American populations are more closely related to hunter-gatherers than to Stuttgart, but also more closely related to MA1 than to European hunter-gatherers. This recapitulates the recently reported evidence of gene flow involving MA1 and the ancestors of Native Americans[6]. In this paper we use the Karitiana as a recently unadmixed population[7] with the largest sample size in the Human Origins dataset to investigate more ancient gene flow between the Americas and Eurasia.

***Table S14.6: Native American populations are more closely related to ancient hunter-gatherers than to Stuttgart, and are more closely related to MA1 than to European hunter-gatherers.***

| **Nat. Am.** | **Ancient$_1$** | **Ancient$_2$** | $f_4$***(Nat. Am., Chimp; Ancient$_1$, Ancient$_2$)*** | **Z** |
|---|---|---|---|---|
| Cabecar | LaBrana | Stuttgart | 0.001765 | 3.001 |
| Cabecar | Loschbour | Stuttgart | 0.00256 | 4.161 |
| Cabecar | MA1 | LaBrana | 0.004402 | 6.434 |
| Cabecar | MA1 | Stuttgart | 0.006242 | 8.825 |
| Cabecar | MA1 | Loschbour | 0.003709 | 5.331 |
| Cabecar | MA1 | Motala12 | 0.002088 | 3.093 |
| Cabecar | Motala12 | LaBrana | 0.002268 | 3.654 |
| Cabecar | Motala12 | Stuttgart | 0.004088 | 6.668 |
| Cabecar | Motala12 | Loschbour | 0.001417 | 2.268 |
| Karitiana | LaBrana | Stuttgart | 0.002298 | 3.923 |
| Karitiana | Loschbour | Stuttgart | 0.002813 | 4.861 |
| Karitiana | MA1 | LaBrana | 0.005195 | 8.170 |
| Karitiana | MA1 | Stuttgart | 0.007701 | 11.423 |
| Karitiana | MA1 | Loschbour | 0.004746 | 7.056 |
| Karitiana | MA1 | Motala12 | 0.003135 | 4.815 |
| Karitiana | Motala12 | LaBrana | 0.002135 | 3.542 |
| Karitiana | Motala12 | Stuttgart | 0.004477 | 7.413 |
| Karitiana | Motala12 | Loschbour | 0.001629 | 2.648 |
| Mixe | LaBrana | Stuttgart | 0.002287 | 4.199 |
| Mixe | Loschbour | Stuttgart | 0.002497 | 4.507 |
| Mixe | MA1 | LaBrana | 0.004544 | 7.263 |
| Mixe | MA1 | Stuttgart | 0.006886 | 10.869 |
| Mixe | MA1 | Loschbour | 0.004386 | 6.718 |
| Mixe | MA1 | Motala12 | 0.002626 | 4.306 |
| Mixe | Motala12 | LaBrana | 0.001787 | 3.036 |
| Mixe | Motala12 | Stuttgart | 0.004208 | 7.218 |
| Mixe | Motala12 | Loschbour | 0.00159 | 2.748 |
| Piapoco | LaBrana | Stuttgart | 0.002652 | 4.681 |
| Piapoco | Loschbour | Stuttgart | 0.002976 | 5.129 |
| Piapoco | MA1 | LaBrana | 0.004497 | 6.978 |
| Piapoco | MA1 | Stuttgart | 0.007275 | 11.246 |
| Piapoco | MA1 | Loschbour | 0.004136 | 6.286 |
| Piapoco | MA1 | Motala12 | 0.002692 | 4.248 |
| Piapoco | Motala12 | LaBrana | 0.00165 | 2.755 |
| Piapoco | Motala12 | Stuttgart | 0.004363 | 7.304 |
| Piapoco | Motala12 | Loschbour | 0.001293 | 2.206 |
| Surui | LaBrana | Stuttgart | 0.001848 | 3.107 |
| Surui | Loschbour | LaBrana | 0.001144 | 2.108 |
| Surui | Loschbour | Stuttgart | 0.002905 | 4.763 |
| Surui | MA1 | LaBrana | 0.005033 | 7.704 |
| Surui | MA1 | Stuttgart | 0.006936 | 10.041 |
| Surui | MA1 | Loschbour | 0.00385 | 5.559 |
| Surui | MA1 | Motala12 | 0.002698 | 4.121 |
| Surui | Motala12 | LaBrana | 0.002278 | 3.800 |
| Surui | Motala12 | Stuttgart | 0.004099 | 6.582 |

***Table S14.7: Ancient Eurasians are closest to Karitiana and most distant to Papuans.***

| **Ancient** | **$ENA_1$** | **$ENA_2$** | **$f_4(Ancient, Chimp; ENA_1, ENA_2)$** | **Z** |
|---|---|---|---|---|
| LaBrana | Ami | Onge | 0.000684 | 1.736 |
| LaBrana | Ami | Papuan | 0.004295 | 8.820 |
| LaBrana | Karitiana | Ami | 0.002518 | 6.191 |
| LaBrana | Karitiana | Onge | 0.003201 | 6.657 |
| LaBrana | Karitiana | Papuan | 0.006812 | 12.532 |
| LaBrana | Karitiana | Ulchi | 0.002077 | 5.506 |
| LaBrana | Onge | Papuan | 0.003611 | 7.554 |
| LaBrana | Ulchi | Ami | 0.000440 | 1.809 |
| LaBrana | Ulchi | Onge | 0.001124 | 2.949 |
| LaBrana | Ulchi | Papuan | 0.004735 | 10.210 |
| Loschbour | Ami | Onge | 0.000511 | 1.255 |
| Loschbour | Ami | Papuan | 0.004513 | 8.929 |
| Loschbour | Karitiana | Ami | 0.002677 | 6.297 |
| Loschbour | Karitiana | Onge | 0.003188 | 6.406 |
| Loschbour | Karitiana | Papuan | 0.007189 | 12.361 |
| Loschbour | Karitiana | Ulchi | 0.002042 | 5.300 |
| Loschbour | Onge | Papuan | 0.004001 | 7.762 |
| Loschbour | Ulchi | Ami | 0.000635 | 2.573 |
| Loschbour | Ulchi | Onge | 0.001146 | 2.898 |
| Loschbour | Ulchi | Papuan | 0.005147 | 10.313 |
| MA1 | Ami | Onge | 0.000578 | 1.402 |
| MA1 | Ami | Papuan | 0.004339 | 8.594 |
| MA1 | Karitiana | Ami | 0.007689 | 17.253 |
| MA1 | Karitiana | Onge | 0.008267 | 15.464 |
| MA1 | Karitiana | Papuan | 0.012028 | 20.574 |
| MA1 | Karitiana | Ulchi | 0.006702 | 16.498 |
| MA1 | Onge | Papuan | 0.003760 | 7.506 |
| MA1 | Ulchi | Ami | 0.000987 | 3.924 |
| MA1 | Ulchi | Onge | 0.001565 | 4.009 |
| MA1 | Ulchi | Papuan | 0.005326 | 10.904 |
| Motala12 | Ami | Onge | 0.000291 | 0.751 |
| Motala12 | Ami | Papuan | 0.003891 | 7.934 |
| Motala12 | Karitiana | Ami | 0.004562 | 10.811 |
| Motala12 | Karitiana | Onge | 0.004853 | 9.984 |
| Motala12 | Karitiana | Papuan | 0.008454 | 15.097 |
| Motala12 | Karitiana | Ulchi | 0.003546 | 9.155 |
| Motala12 | Onge | Papuan | 0.003600 | 7.259 |
| Motala12 | Ulchi | Ami | 0.001017 | 4.160 |
| Motala12 | Ulchi | Onge | 0.001308 | 3.525 |
| Motala12 | Ulchi | Papuan | 0.004908 | 10.386 |
| Stuttgart | Ami | Onge | 0.000757 | 2.049 |
| Stuttgart | Ami | Papuan | 0.004146 | 8.613 |
| Stuttgart | Karitiana | Ami | 0.001619 | 4.127 |
| Stuttgart | Karitiana | Onge | 0.002376 | 5.113 |
| Stuttgart | Karitiana | Papuan | 0.005765 | 10.838 |
| Stuttgart | Karitiana | Ulchi | 0.001471 | 4.084 |
| Stuttgart | Onge | Papuan | 0.003388 | 6.843 |
| Stuttgart | Ulchi | Ami | 0.000148 | 0.670 |
| Stuttgart | Ulchi | Onge | 0.000906 | 2.551 |
| Stuttgart | Ulchi | Papuan | 0.004294 | 9.164 |

**Relationship of ancient samples to eastern non-Africans**

We finally explore the relationship of ancient samples to all eastern non-Africans (ENA) together using the set (Onge, Papuan, Ami, Ulchi, Karitiana) which includes representatives from all major ENA populations. In Table S14.7 we show all $f_4$-statistics of the form $f_4(Ancient, Chimp, ENA_1, ENA_2)$ and not only those with $|Z|\geq 2$ as in previous tables. Papuans appear to be more distant from ancient

Eurasians than are any other ENA population, consistent with their additional admixture from archaic Denisovans. Comparisons involving (Onge, Ami) show a slightly closer relationship of ancient Eurasians to Ami than to Onge, but barely reach significance and we do not view this evidence as compelling. Karitiana, on the other hand, appear generally closer to present-day west Eurasians than to all other ENA populations, while the Siberian Ulchi appear intermediate in their closeness, between Karitiana and Onge/Ami.

In what follows, we develop models for West Eurasia that take into account Karitiana and Onge, forcing us to account for both the evidence of a specific link between MA1 and Native Americans, and also for the more general evidence of a link between eastern non-Africans and ancient Eurasian hunter-gatherers. These two populations serve as a "sanity check" for model development, ensuring that our reconstruction of deep population relationships related to the ancestral populations of present-day Europeans are also consistent with non-West Eurasian outgroups. We include only Karitiana and Onge as our focus is not in fully modeling eastern non-African origins which are likely to be also complex. However, the results of the model developed using just these two outgroups are consistent with those of a method that uses many more eastern non-African outgroups (SI17), as the two methods produce consistent admixture estimates for present-day Europeans (SI17, Extended Data Table 2).

**Summary of $f_4$-statistics on Eastern non-Africans**

Our survey of $f_4$-statistics serves to identify features of the relationships between different populations that must be accounted for in a successful model. We itemize the most pertinent observations:

1. Ancient Eurasians (Europeans and MA1) are genetically closer to Karitiana than to North Asians, intermediately related to Onge and East Asians, and least related to Papuans
2. Hunter-gatherers do not differ in their relationships to eastern non-Africans from East, South Asia and Oceania, while the Karitiana and North Asians are clearly more related to MA1 than to European hunter-gatherers. MA1 is more closely related to Karitiana than to North Asians.
3. Eastern non-Africans are all more closely related to ancient hunter gatherers than to Stuttgart

We confirm these key findings using only transversions on both the Human Origins dataset (Extended Data Table 2) and using whole genome sequences[8], thus showing that our results are not artifacts of SNP ascertainment bias (Extended Data Table 2). We refer to items #1, #2 and #3 above in what follows as we explore the space of possible models relating the populations

***Figure S14.1: A (failed) model with no admixture.***

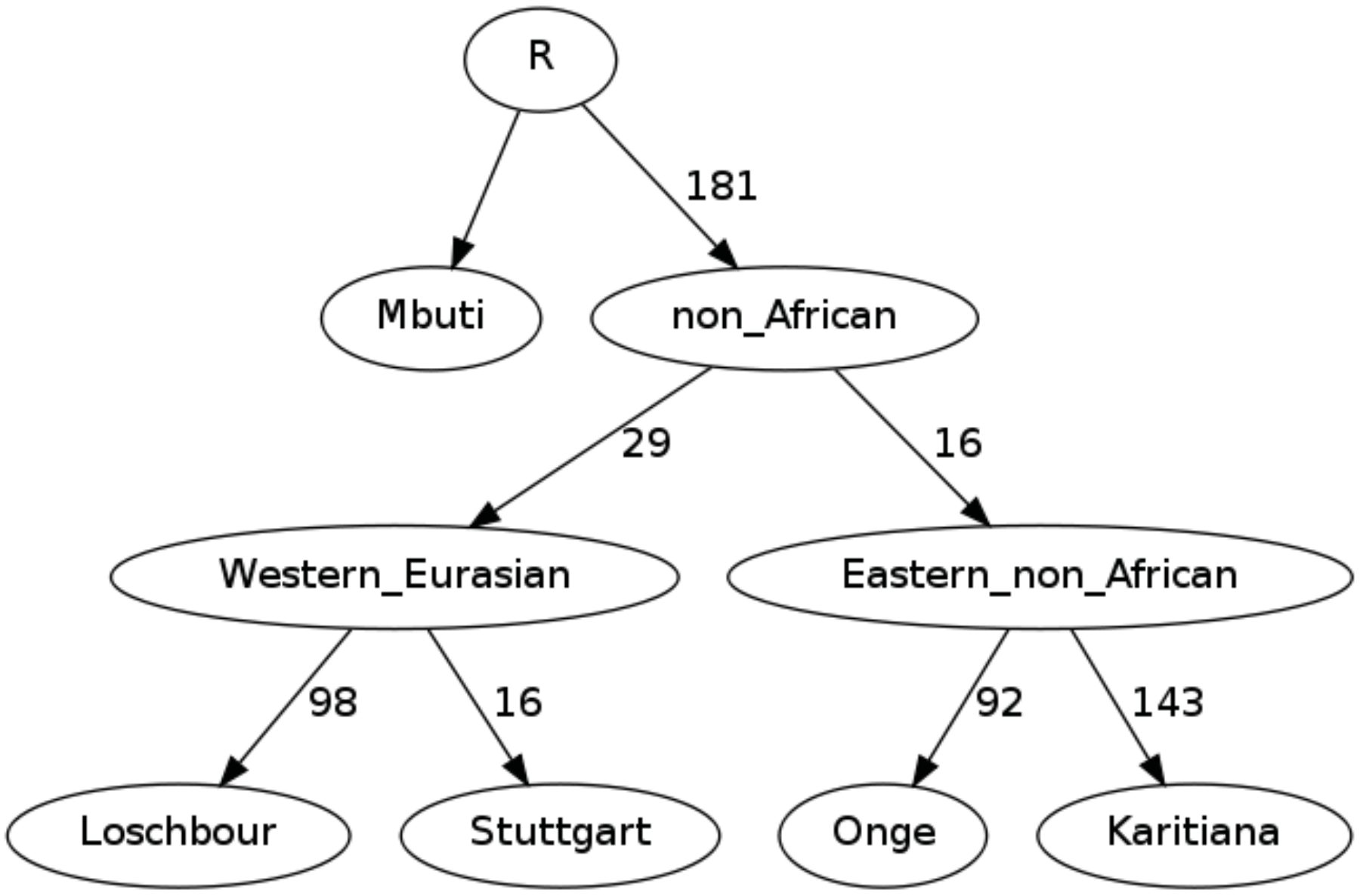

### A tree model fails

We begin with a simple model that we fit unsuccessfully with ADMIXTUREGRAPH (Fig. S14.1). This model corresponds to the "no admixture" maximum likelihood tree inferred by TreeMix analysis for these populations (SI16) and groups the ancient European samples (Loschbour and Stuttgart) and eastern non-African populations (Karitiana and Onge) as two separate clades with no gene flow between them. The tree of Figure S14.1 fails to fit, as it predicts that Stuttgart is equally related to Onge and Karitiana (contradicting item #1), and it predicts that Stuttgart and Loschbour are equally related to Karitiana (contradicting item #3). Note that drifts along edges are multiplied by 1000 in this and following figures.

***Figure S14.2: Failed models with one admixture edge.*** *We show the best fits, but caution that all of these models are poor fits in the sense of f-statistics more than 3 standard errors from the data.*

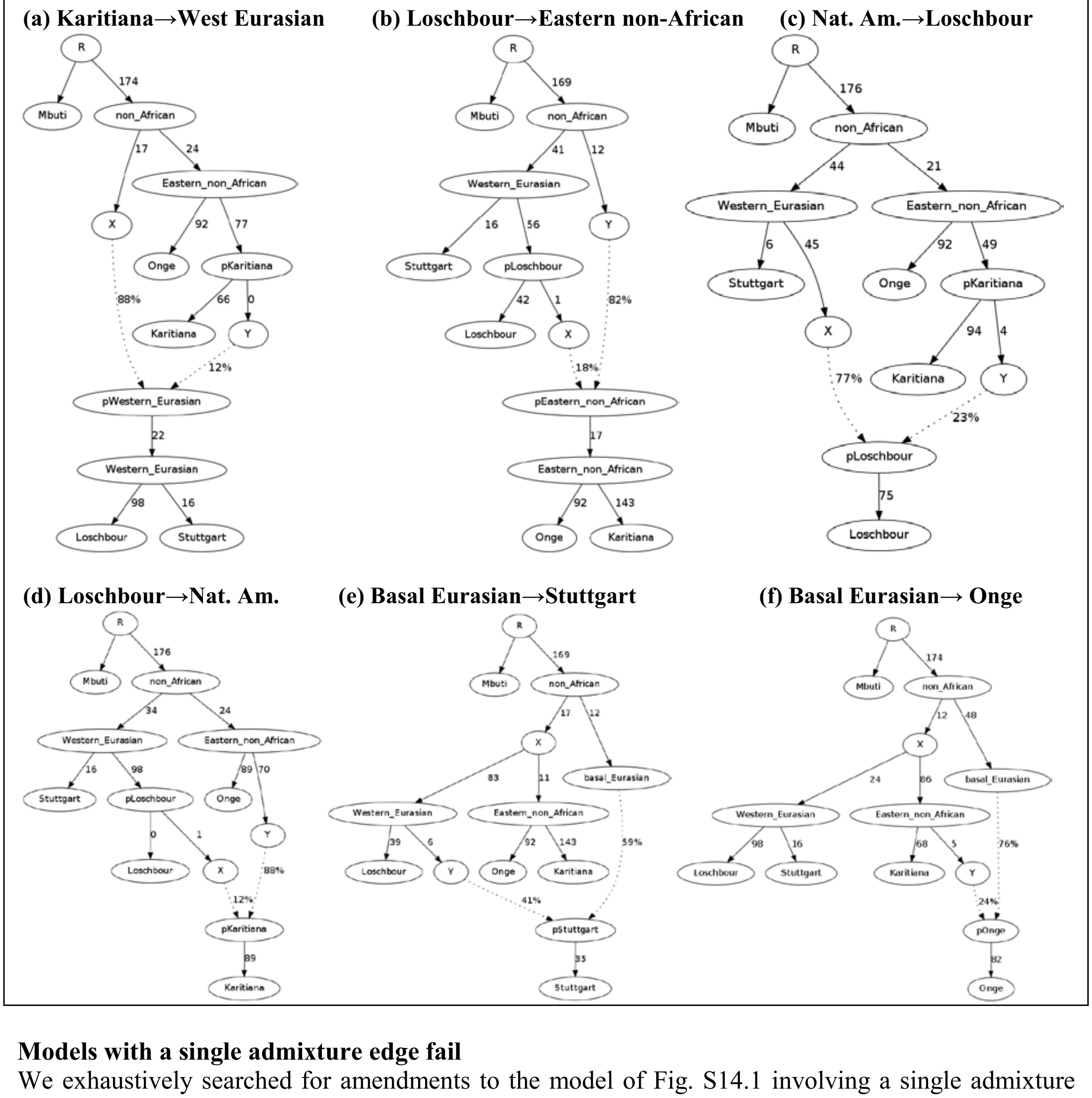

### Models with a single admixture edge fail

We exhaustively searched for amendments to the model of Fig. S14.1 involving a single admixture edge, but find that they all fail to account for the observed $f_4$-statistics and the asymmetries between both Stuttgart/Loschbour and Onge/Karitiana.

*A single admixture event between Eastern non-Africa and West Eurasia fails*

We attempted to amend the model by adding one admixture event between the West Eurasian and Eastern non-African subtrees, but this fails:

1. Admixture into Western Eurasian from the Karitiana branch (Fig. S14.2a) fails, because it predicts that Stuttgart and Loschbour are equally related to Onge (contradicting item #3).
2. Admixture into Eastern non-African from Loschbour (Fig. S14.2b) fails, because it predicts that Stuttgart are equally related to Karitiana and Onge (contradicting item #1).
3. Admixture into Loschbour from the Karitiana branch (Fig. S14.2c) fails, because it predicts that Stuttgart is equally related to Onge and Karitiana (contradicting item #1).
4. Admixture into Karitiana from Loschbour (Fig. S14.2d) fails, because it predicts that Onge are equally related to Stuttgart and Loschbour (contradicting item #3).

We further considered scenarios of early admixture between West Eurasians and Eastern non-Africans (i.e., emanating from before the Loschbour/Stuttgart and Onge/Karitiana split) but found that this does not help as it preserves the topological form of Fig. S14.1.

We also considered Eastern non-African admixture into Stuttgart or conversely West Eurasian admixture into Onge, but these break the symmetry in the wrong direction, making fits worse. Thus, a single admixture between Eastern non-Africa and Western Eurasia is insufficient to explain the data.

We also considered a scenario of "Basal Eurasian" admixture into either Stuttgart or Onge (Fig. S14.2e and f respectively). This is admixture from a source that branched off before the divergence of West Eurasians and eastern non-Africans. By adding this type of admixture into Stuttgart we explain the observed greater Loschbour proximity to eastern non-Africans (#3), but not the observed greater proximity of Stuttgart to Karitiana than to Onge (#1). Conversely, by adding this admixture into Onge we explain the observed greater Karitiana proximity to West Eurasians (#1), but not the observed greater proximity of eastern non-Africans to Loschbour than to Stuttgart (#3). Basal Eurasian admixture to either Loschbour or Karitiana break the symmetry in the wrong direction, implying that Karitiana should be closer to Stuttgart than to Loschbour or that Loschbour should be closer to Onge than to Karitiana respectively.

To summarize, models with one admixture edge cannot resolve the observed asymmetries, motivating a search for a model with at least two admixture edges that can fit.

**Successful models with two admixture edges**

The idea of basal Eurasian is nonetheless attractive, so we pursued it further.

A single such admixture event into Stuttgart (as in Fig. S14.2e) would fully explain #3, i.e., that all eastern non-Africans are more closely related to hunter-gatherers than to Stuttgart. Such an idea is also archaeologically plausible on account of the Near Eastern related admixture that we have detected in Stuttgart. The Near East was the staging point for the peopling of Eurasia by anatomically modern humans. As a result, it is entirely plausible that it harbored deep Eurasian ancestry which did not participate in the initial peopling of Eurasia, but was much later brought into Europe by Near Eastern farmers. More speculatively, some basal Eurasian admixture in the Near East may reflect the early presence of anatomically modern humans[9] in the Levant, or the populations responsible for the appearance of the Nubian Complex in Arabia[10], both of which date much earlier than the widespread dissemination of modern humans across Eurasia. Finally, it could reflect continuing more recent gene flows between the Near East and nearby Africa after the initial out-of-Africa dispersal, perhaps associated with the spread of Y-chromosome haplogroup E subclades from eastern Africa[11,12] into the Near East, which appeared at least 7,000 years ago in Neolithic Europe[13], or the detection of African skeletal morphology in Epipaleolithic Natufians from Israel[14].

Equally archeologically plausible is basal Eurasian admixture in Onge (Fig. S14.2f), which would partially explain #1. The Onge are a southern Eurasian population, and a scenario of a "southern route" peopling of Eurasia (of which the Onge are plausible partial descendants) might have resulted

***Figure S14.3: Successful models with all admixture in one part of Eurasia***

***(Left) all admixture in West Eurasia*** ***(Right) all admixture in eastern non-Africa***

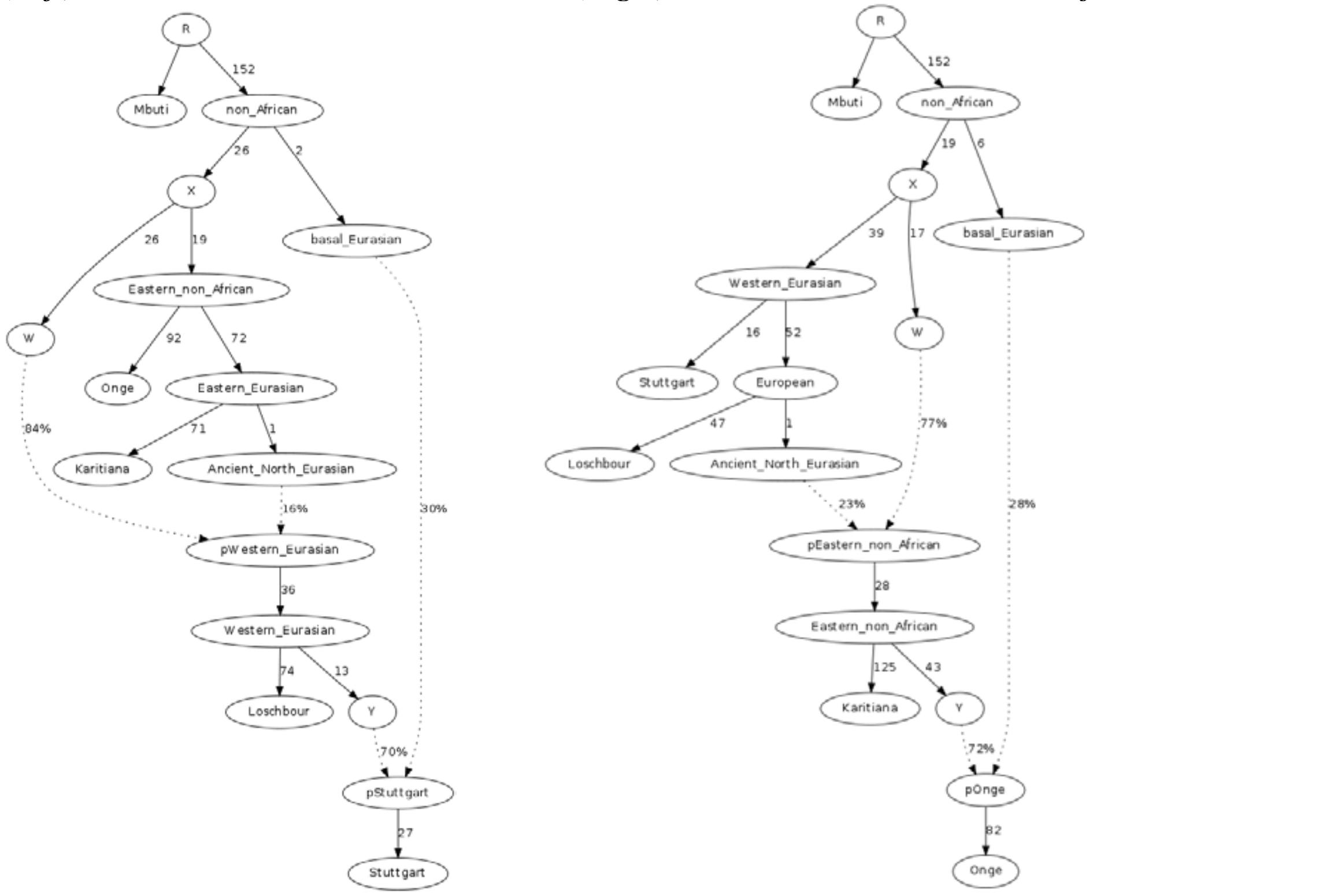

in them having deep Eurasian ancestry, similar to a model proposed for the early peopling of Australia by anatomically modern humans[15]. Such ancestry would cause them to share less genetic drift with West Eurasians than with the Karitiana.

As shown in Fig. S14.2, basal Eurasian admixture into either Stuttgart or Onge fails to explain the data. However, we can combine it with gene flow between West Eurasians and eastern non-Africans and thereby obtain a successful model.

Fig. S14.3 shows scenarios that fit the data involving basal Eurasian admixture. If Stuttgart harbors basal Eurasian admixture (left), then the affinity of Loschbour to eastern non-Africans is maintained, but the greater proximity of Karitiana than Onge to west Eurasians is not. We can amend our model by proposing gene flow from Karitiana into the ancestors of West Eurasians. Note that this admixture must go to the ancestor of West Eurasians, because both Stuttgart and Loschbour are genetically closer to Karitiana than to Onge (#2). The situation is symmetrical if Onge has basal Eurasian admixture (right), in which case the affinity of west Eurasians to Karitiana is maintained, but the greater proximity of Loschbour to eastern non-Africans (#3) is not. This problem can be fixed by proposing admixture from relatives of Loschbour into the ancestor of eastern non-Africans. In both the models of Fig. S14.3, all admixture takes place either in west Eurasia (left), or eastern non-Africa (right), with the other populations not being admixed.

Fig. S14.4 proposes a second set of possibilities, also involving basal Eurasian admixture. If Stuttgart has basal Eurasian admixture (left), then the greater proximity of Karitiana than Onge to West Eurasians could be explained by gene flow from West Eurasians into Native American ancestors; this could originate either in the Loschbour branch (left-top), or from a basal West Eurasian lineage (left-bottom). Symmetrically, basal Eurasian admixture into Onge (right) can be combined with eastern non-African gene flow into Loschbour from Native Americans (right-top) or eastern non-Africans (right-bottom).

***Figure S14.4: Successful models combining basal admixture with a second gene flow event***

**(Left Top) Basal→Stuttgart / Loschbour→Karitiana** **(Right Top) Basal→Onge / Karitiana→Loschbour**

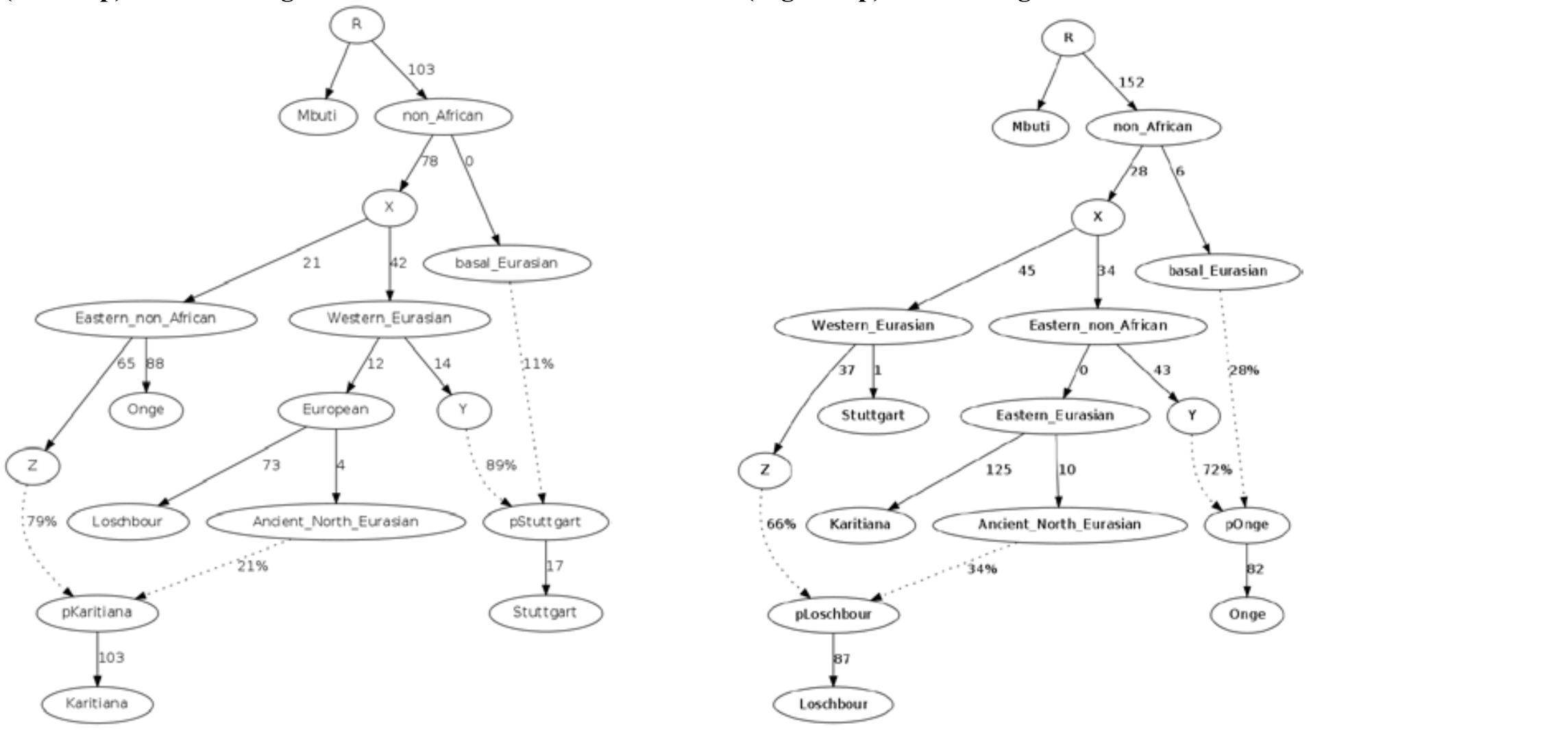

**(Left Bottom) Basal→Stuttgart / West Eurasian→Karitiana** **(Right Bottom) Basal→Onge / Eastern non-African→Loschbour**

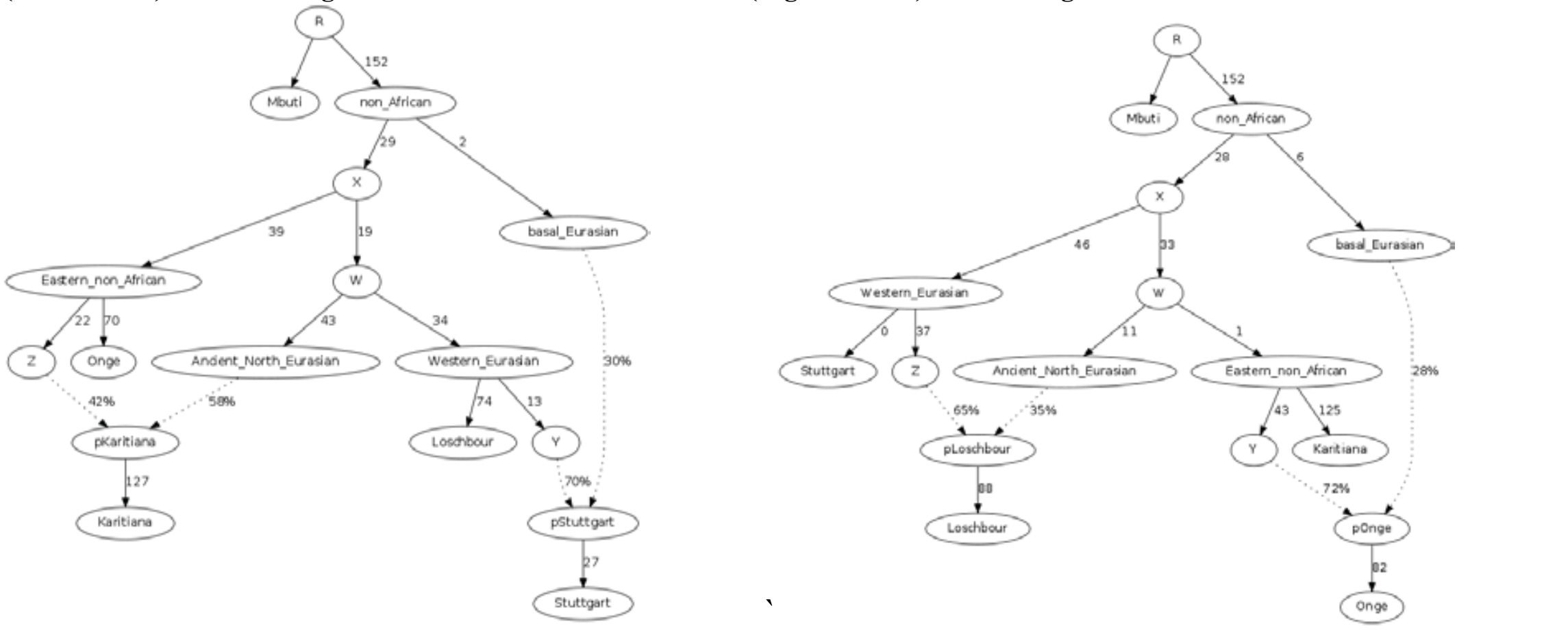

In Fig. S14.5 we propose two successful models (without basal Eurasian admixture) which invoke variable admixture in either direction across Eurasia. These models propose two admixture events for the set of considered populations, but make Karitiana and Onge (left) and Loschbour and Stuttgart (right) be composed of the same ancestral elements but in different proportions.

We have thus identified a total of eight models (Figs S14.3, S14.4 and S14.5) each with two admixture events, that are all consistent with the *f*-statistics for the four populations and yet make quite different predictions about the prehistory of Eurasia. We note that even more complex models could be devised (with more than two admixture events) that would be equally consistent, but may be unparsimonious for a set of only four populations. For the time being, we conclude that very simple models (with one admixture event) fail, while a plethora of consistent models exist for slightly more complex models (with two admixture events).

**MA1 as representative of Ancient North Eurasians**

A possible way to constrain the choice of model is to attempt to fit additional populations into their structure. MA1 is an Upper Paleolithic Siberian with demonstrated genetic links to both Europe and Native Americans[6] and thus is a powerful sample for constraining possible historical scenarios. It is potentially a "missing link": a representative of a population mediating gene flow between east and west across Eurasia, so we consider whether it could be incorporated into the models of Figs S14.3 to S14.5 without breaking them. We summarize the results for the eight models in Table S14.8.

***Figure S14.5: Two successful models without basal Eurasian admixture.***

**(Left) variable Karitiana-related admixture into Loschbour and Stuttgart.** **(Right) variable Loschbour-related admixture into Karitiana and Onge**

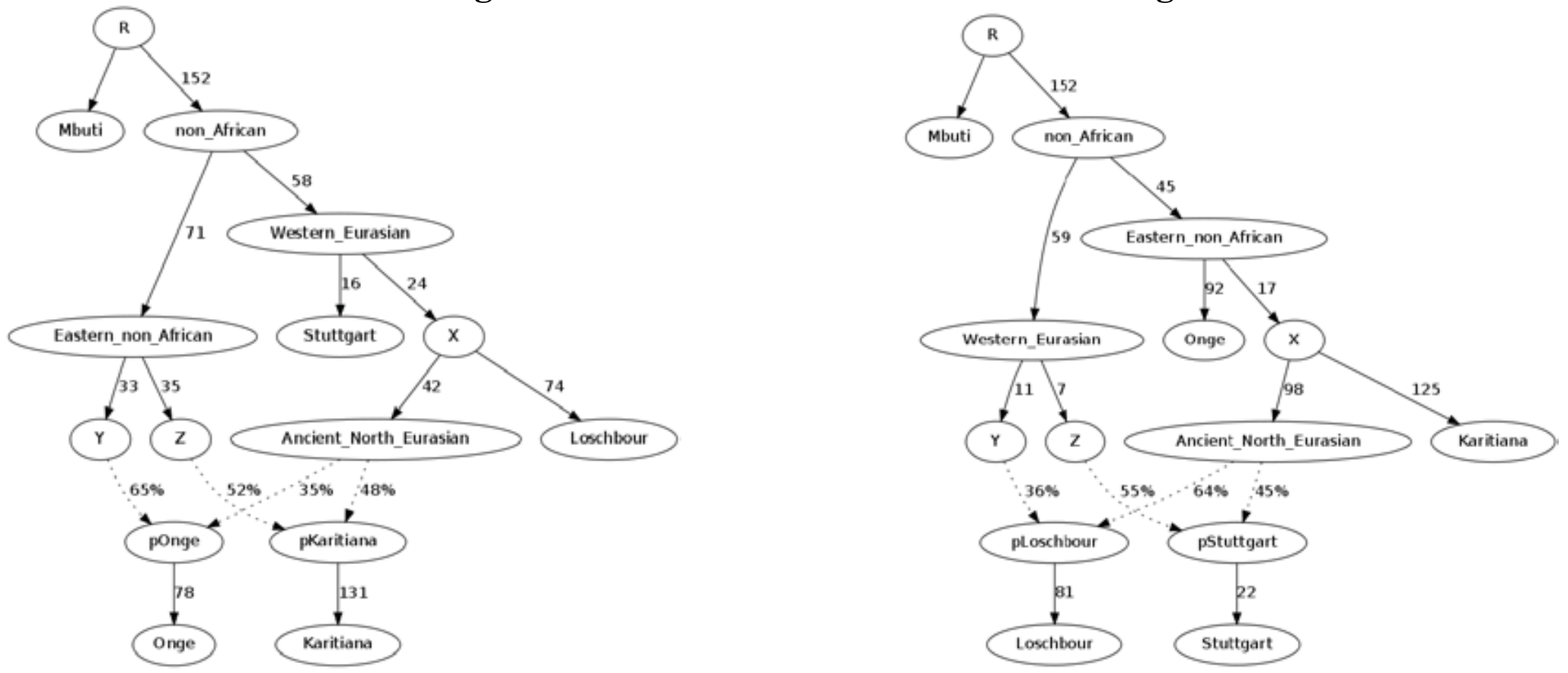

We find that the model in Figure S14.4 (left-bottom) is the only model that fits MA1 without any additional admixture events, specifically by specifying that MA1 is a clade with the Ancient_North_Eurasian node mediating gene flow between West Eurasia and Native Americans.

***Table S14.8: Attempting to fit MA1 into the structure of models of Figures S14.3, S14.4 and S14.5.***

| **Fig.** | **Admixture event 1** | **Admixture event 2** | **Violation** | **Z** |
|---|---|---|---|---|
| S14.3L | Basal→Stuttgart | Karitiana→West Eurasian | *$f_2$(Onge, MA1)* | 5.1 |
| S14.3R | Basal→Onge | Loschbour→East non- | *$f_2$(Loschbour, Stuttgart)* | -6.2 |
| S14.4LT | Basal→Stuttgart | Loschbour→Karitiana | *$f_3$(MA1; Loschbour,* | 8.0 |
| S14.4RT | Basal→Onge | Karitiana→Loschbour | *$f_4$(Onge, Stuttgart; Kar.,* | 10.5 |
| S14.4LB | Basal→Stuttgart | West Eurasian→Karitiana | ✓ | |
| S14.4RB | Basal→Onge | East non-African→ Losch. | *$f_4$(Onge, MA1; Losch.,* | 3.8 |
| S14.5L | Loschbour→Karitiana | Loschbour→Onge | *$f_2$(Loschbour, Stuttgart)* | -6.0 |
| S14.5R | Karitiana→Loschbour | Karitiana→Stuttgart | *$f_2$(Onge, MA1)* | 5.2 |

For the remaining models we list the *f*-statistics that are most discrepant between empirically estimated and fitted parameters together with their Z-score for deviation from expectation; for example model S14.4 (right-top) makes Karitiana and MA1 sister clades, so we fit zero for the violating statistic, but we in fact observe a positive value with Z=10.5.

We conclude that (i) gene flow into the Karitiana originated from a basal West Eurasian population and (ii) Neolithic farmers such as Stuttgart had admixture from a Basal Eurasian population is consistent with the evidence. The model of Figure S14.4 (left-bottom) including MA1 is shown in Figure S14.6. We should caution that this population is termed "Basal Eurasian" on account of its phylogenetic position in the model (basal to all Eurasian groups and contributing only to Stuttgart), but its geographical distribution in the past is unknown. Fitted parameters, however, indicate that Basal Eurasians share with other Eurasians most of the (Mbuti→non_African) genetic drift associated with "Out-of-Africa" populations.

***Figure S14.6: A successful model involving Stuttgart, Loschbour, MA1, Onge and Karitiana.*** *The high genetic drift in the MA1-specific branch is an artifact of the low coverage (about 1x) of this sample, which means that many sites that are in fact heterozygous appear as homozygous. However, this is not expected to affect inferences of the relationships between MA1 and the other samples.*

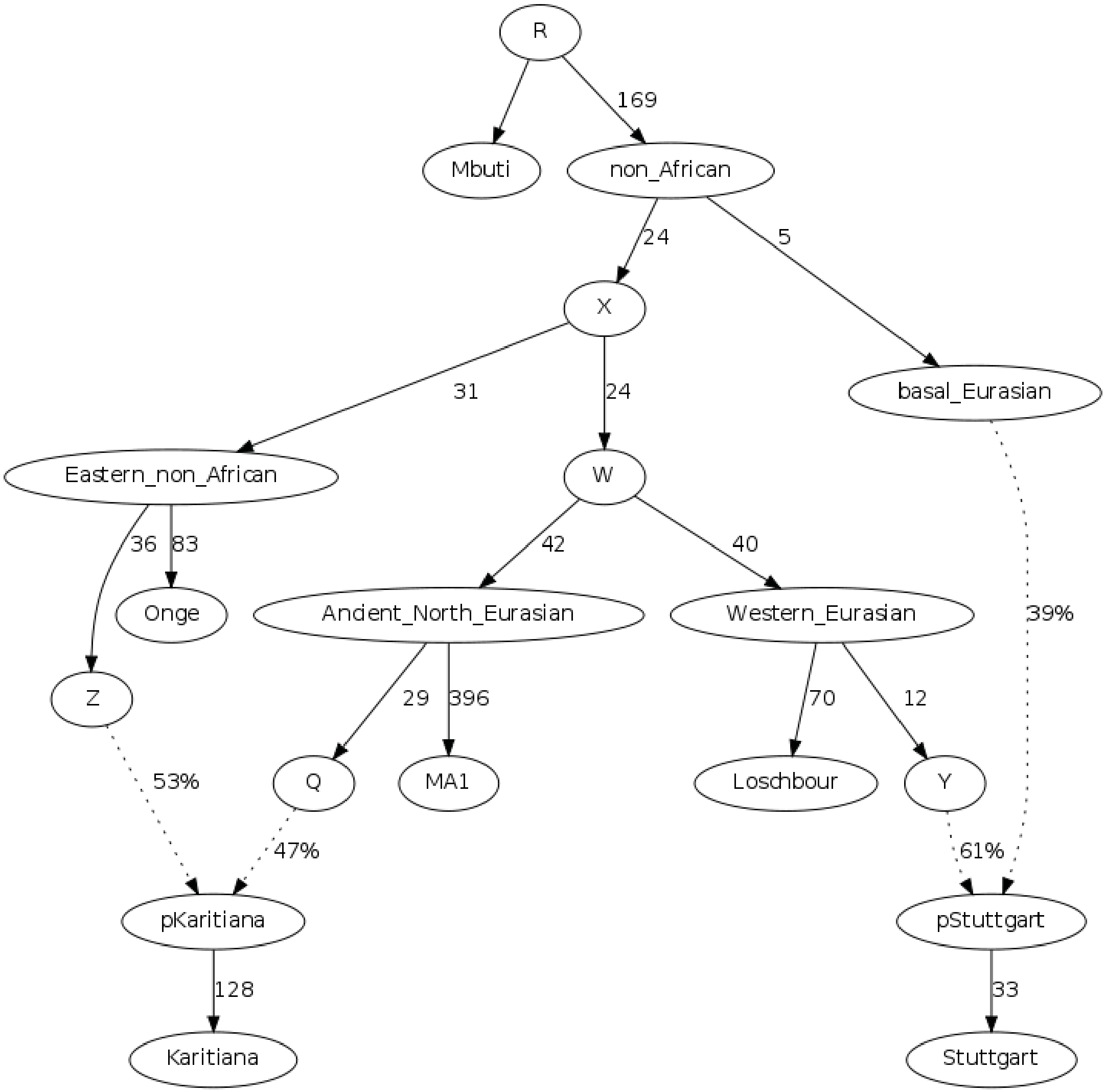

**No evidence of Basal East Asian admixture in MA1**

The model of Fig. S14.6 proposes that MA1 is unadmixed, but it was argued[6] that MA1 may have basal East Asian (basal eastern non-African in our terminology) admixture based on the evidence that MA1 shares more alleles than Sardinians with either Oceanians or East Asians. This was a reasonable suggestion because of the sample's provenance, but statistics of the form $f_4$*(ENA, Chimp; Loschbour, MA1)* appear symmetric for any eastern-non African (ENA) population from the set (Ami, Atayal, Han, Naxi, She, Papuan, Bougainville, Onge) with $|Z|<0.3$. If MA1 had more basal East Asian admixture than Loschbour, these statistics should be negative. It is possible that both Loschbour and MA1 experienced eastern non-African gene flow, but it is not parsimonious (under the model of eastern non-African gene flow) that two samples from widely separated geographical locations (Western Europe and Siberia) and times (8-24 thousand years ago) would experience such gene flow in amounts that precisely cancel themselves out to produce perfectly symmetric statistics of the given form.

Our model provides a simpler alternative explanation for the asymmetry between MA1/Sardinians with respect to ENA: not through admixture into MA1 but instead through basal Eurasian admixture into Neolithic farmers. This scenario accounts for both the fact that ENA share more alleles with MA1 than with Stuttgart (because Stuttgart has basal Eurasian admixture), and for the fact that Loschbour and MA1 are symmetrically related to ENA (because they both lack Neolithic Near Eastern ancestry).

The model of Fig. S14.6 was developed as a solution to our observations on the relationships of ancient Eurasians to eastern non-African groups, but it also specifies relationships among ancient Eurasians themselves as indicated by statistics of the form $f_4(Ancient_1, Chimp; Ancient_2, Ancient_3)$ at the beginning of this SI (Table S14.1). The fact that Loschbour and Stuttgart are closer to each other than either one is to MA1 is accounted for by their descent from the "Western Eurasian" node, thus sharing the W→Western_Eurasian drift. The fact that MA1 is closer to Loschbour than to Stuttgart is accounted for by the fact that Stuttgart has a proportion of "Basal Eurasian" ancestry while Loschbour and MA1 share the non_African→X→W drift for the entirety of their ancestry.

While drift lengths do not correspond to time units (as drift accumulation along branches depends on population sizes), the chronological date of MA1 (~24 thousand years) represents a minimum for the separation (node W) of Ancient North Eurasians from European hunter-gatherers. A study of European samples before that time may better define this date, as a sample along the X→W branch would be symmetrically related to (MA1, Loschbour) and would thus provide another minimum date. The Tianyuan sample[16] from China clusters with eastern non-Africans at the exclusion of Europeans in an analysis of capture data from chromosome 21, suggesting that the split of eastern non-Africans from Eurasian hunter-gatherers (node X) occurred >40 thousand years ago. The split of basal Eurasians from other Eurasians (node non_African) must then be older than 40 thousand years ago. There is uncertainty about the human autosomal mutation rate with implications about the African/non-African divergence[17,18]; the resolution of this question may provide an upper bound for the split of basal Eurasians from other non-Africans.

As suggested previously for Basal Eurasians, we caution against a too literal reading of terminology, as the spatial and temporal distribution of the populations associated with the nodes of the model are still incompletely known. For example, the category "West Eurasian" could be enlarged to include "Ancient North Eurasians" as these share common drift (X→W) with Europeans: this may have plausibly occurred in West Eurasia prior to an eastward migration of the ancestors of MA1[6]. In a different, geographical, sense the category "West Eurasian" could be transferred to the "basal Eurasian" element instead, as it is the only one whose presence we can detect only in West Eurasia, while the common ancestry of both MA1 and Loschbour with eastern non-Africans (drift non_African→X) raises the alternative possibility of an eastern sojourn of their ancestors and a temporal priority of "basal Eurasians" in western Eurasia. Ancient DNA from earlier Eurasians may better resolve these questions of terminology and interpretation.

**Fitting LaBrana as a clade with Loschbour and the Iceman as a clade with Stuttgart**

We identified a model of deep Eurasian inter-relationships by successfully fitting MA1 to a range of models compatible with Loschbour and Stuttgart, and observing that the structure of Fig. S14.6 was the only one that could accommodate MA1 without modification.

As a test of the robustness of the model of Fig. S14.6, we next attempted to add more ancient samples.We did this for LaBrana, the Iceman, and Motala12 which were the highest coverage ancient samples. AG2 clusters with MA1 as observed in its initial publication[6] and as we observe (Fig. 1B), but its interpretation is made difficult by the presence of contamination, likel. The other samples included in Fig. 1B from a Swedish study[19] have much lower coverage and while we observe that they cluster in expected ways in PCA (Fig. 1B), we make no formal claim about their relationship to the higher quality ancient genomes. Of the three good quality genomes not included in the model of Fig.

S14.6, we find that LaBrana could be fit as a sister group of Loschbour and the Iceman could be fit as a sister group of Stuttgart, and indeed both could be fit simultaneously; we show the fitted model in Fig. S14.7.

This confirms the visual impression of clustering of Fig. 1B and motivates us to use the high quality diploid genomes from Loschbour and Stuttgart for the remainder of this SI.

***Figure S14.7: LaBrana and the Iceman can be fitted as sister groups of Loschbour and Stuttgart.***

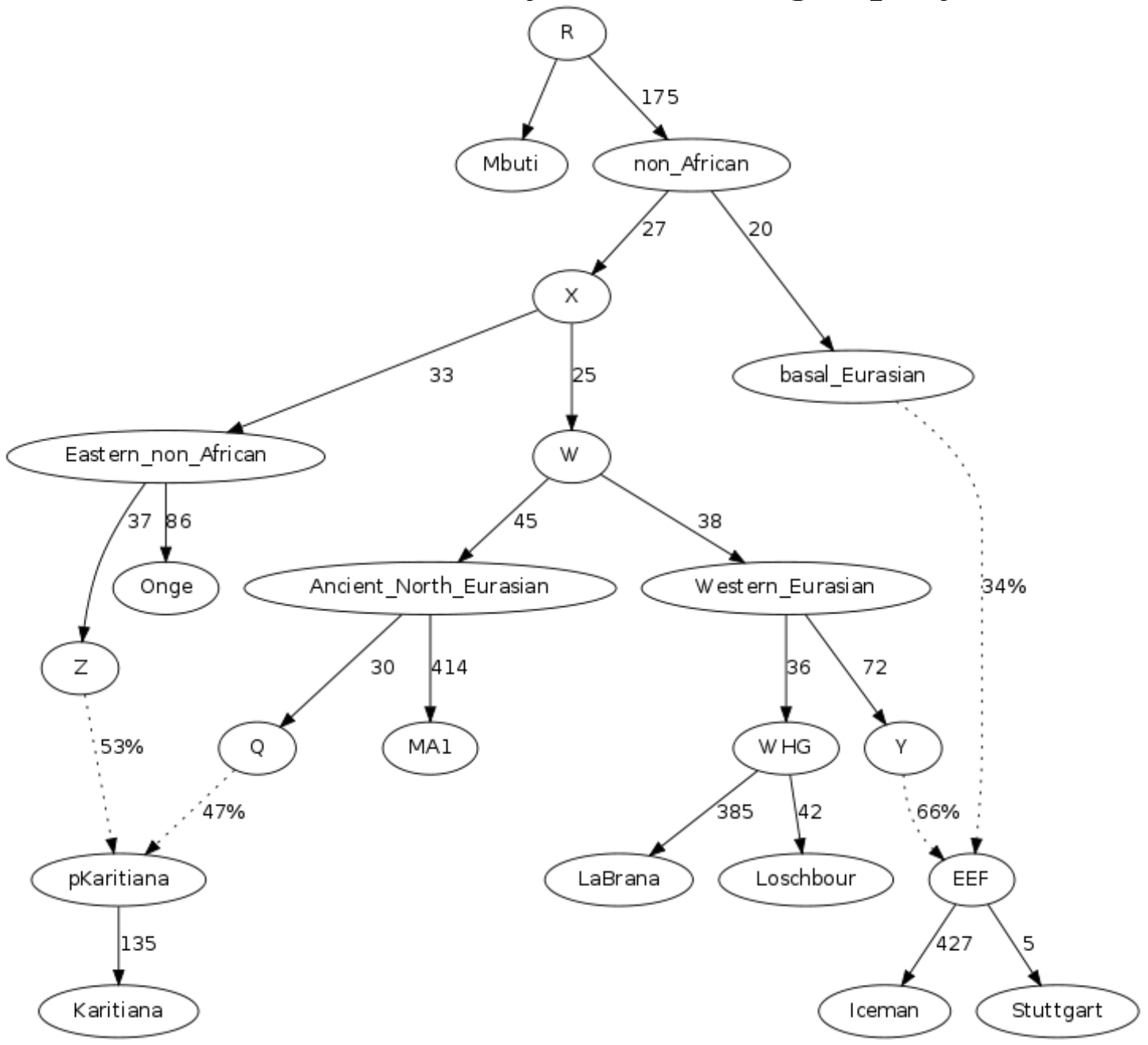

***Figure S14.8: Motala12 can be fit as a mixture of Loschbour and MA1***

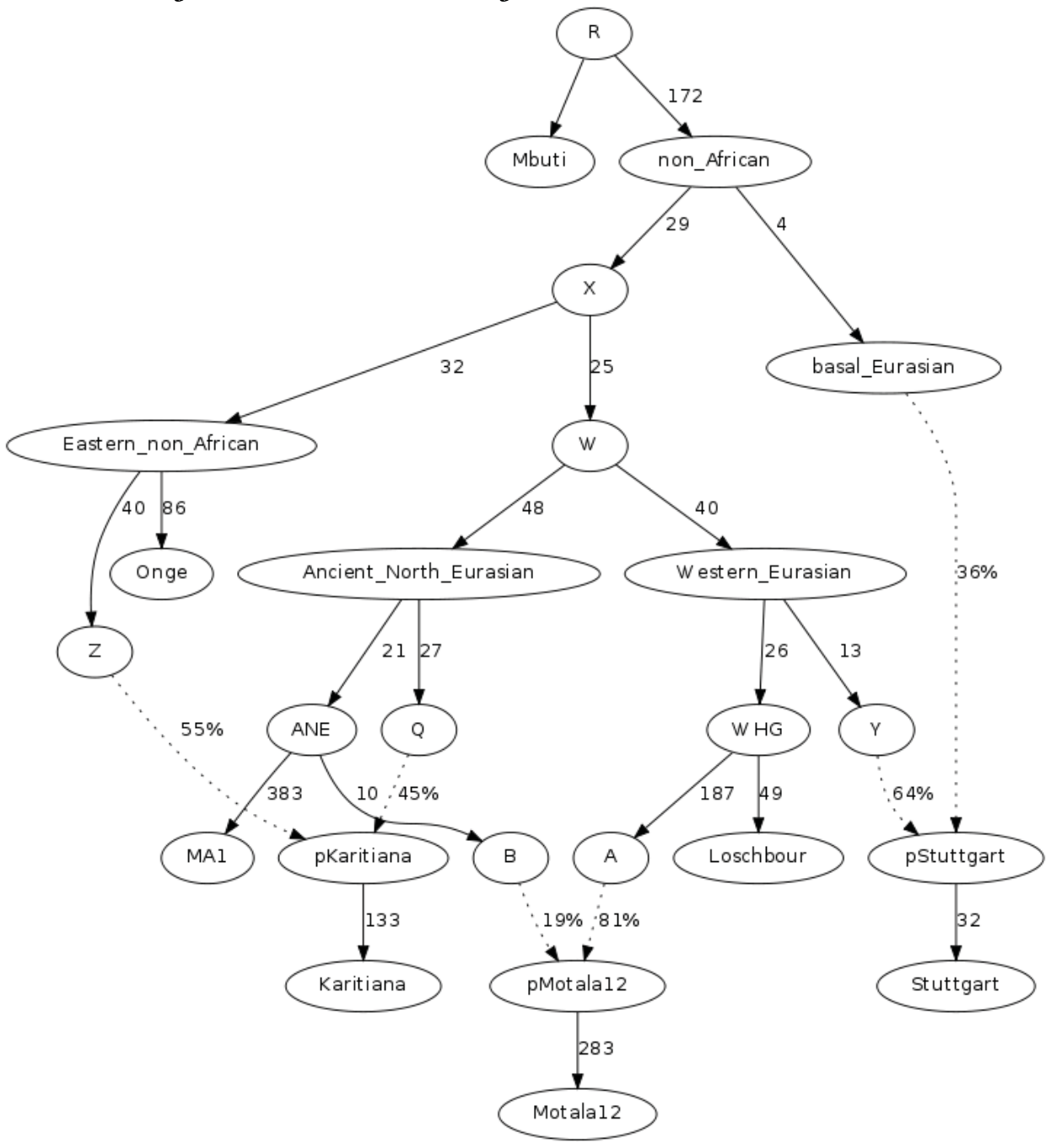

**Motala12 is not a clade with Loschbour as it has MA1-related admixture**

We next attempted to fit Motala12 as a clade with Loschbour in the topology of Fig. S14.6, but were unable to do so, because $f_4$*(Loschbour, Motala12; Stuttgart, MA1)* is significantly positive (Z=5.6). A possible explanation for this is that the European hunter-gatherers who admixed with Near Eastern farmers to form Stuttgart were more like Loschbour than like Motala12. However, the statistic $f_4$*(Motala12, Loschbour; MA1, Mbuti)* is also significantly positive (Z=5.2), as is the statistic $f_4$*(Motala12, Loschbour; MA1, Chimp)* (Table S14.1), and this suggests that MA1 and Motala12 share more alleles than MA1 and Loschbour. Scandinavian hunter-gatherers can be fit, if they are modeled as a mixture of Loschbour and MA1 (Fig. S14.8). This scenario is consistent with the above statistics, Motala12's intermediate geographical position between Western Europe and Siberia, and their intermediate position between West European hunter-gatherers and Ancient North Eurasians (Fig. 1B).

We next attempt to fit West Eurasian populations as simple clades, 2-way mixtures, and finally 3-way mixtures.

**No present-day West Eurasians form a clade with either Loschbour or Stuttgart**

We attempted to fit each individual West Eurasian population in turn as simple clades with Loschbour or Stuttgart. We did not expect this to be possible on the basis of Fig. 1B which shows that none of them cluster with the ancient samples, except possibly Sardinians. However $f_3$-statistics indicate widespread admixture appear in nearly all West Eurasians (SI11) and we show in SI12 that at least 3 source populations are needed for present-day Europeans. Consistent with this evidence, we find that no West Eurasian populations form clades with either Loschbour or Stuttgart, suggesting that these ancient individuals belonged to populations that no longer exist in unadmixed form.

***Figure S14.9: A successful 2-way mixture for Sardinians on the Fig. S14.6 scaffold.*** *They fit as a mix of Loschbour and Stuttgart-related "Hunter" and "Farmer" populations in proportions 21/79%.*

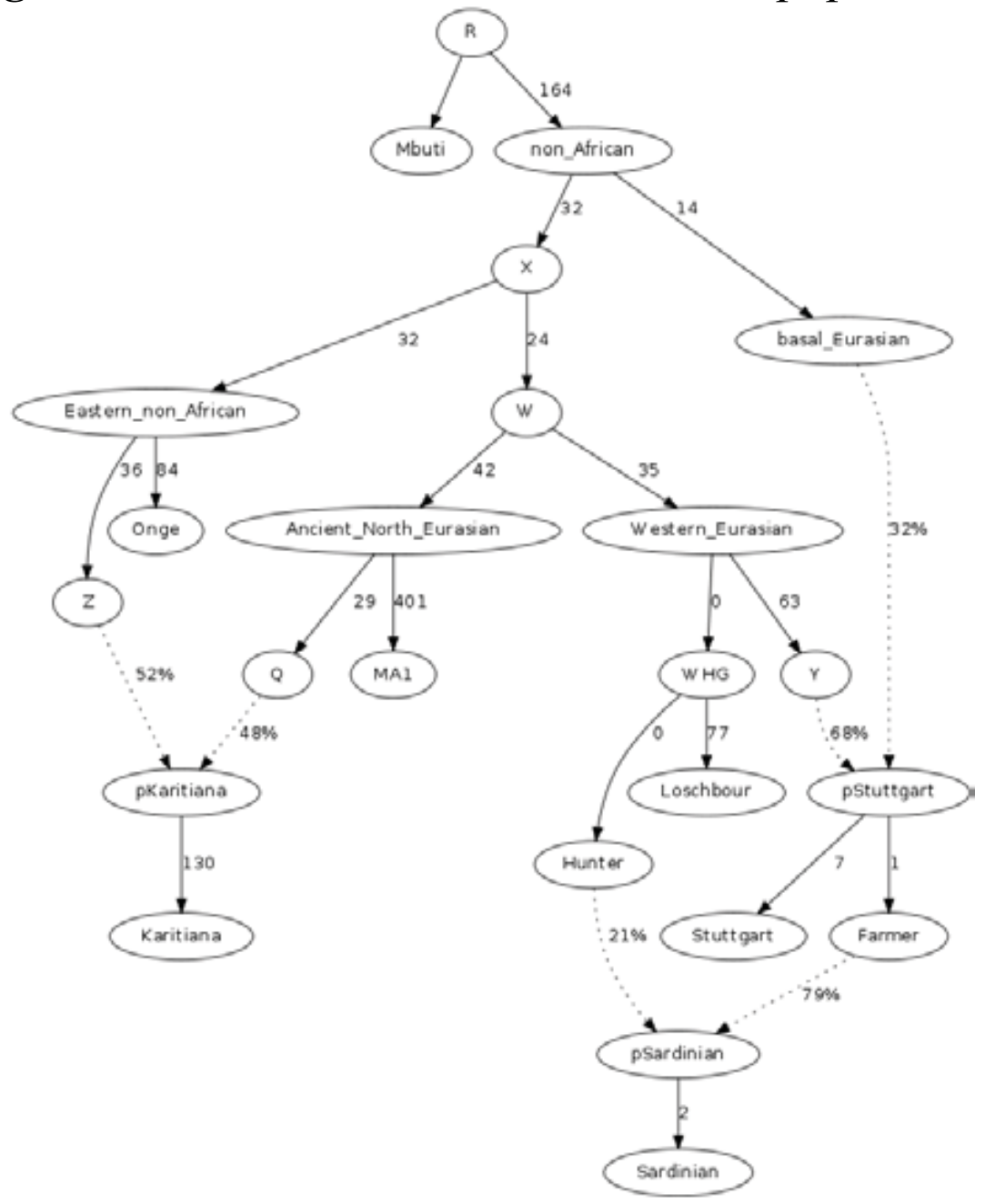

**Most Europeans are not a 2-way mixture of Loschbour and Stuttgart**

We observed that when we attempted to fit Europeans as a clade with Stuttgart, the violated $f_4$-statistics included $f_4$*(European, Stuttgart; Loschbour, Mbuti)* whose estimated values are positive. This is also indicated by Extended Data Fig. 4 which indicates that Loschbour shares more alleles with present-day Europeans than with Stuttgart, so we next attempted to fit individual West Eurasian populations as a 2-way mixture of Loschbour and Stuttgart, representing Early European farmers and West European Hunter Gatherers.

Fig. 1B suggests that such a fit may not be possible, as most Europeans form a cline that cannot be explained by such a mixture; this is also suggested by the lowest $f_3$-statistics (Table 1, Extended Data Table 1) involving any pair of reference populations which involve the (EEF, WHG) pair only for Sardinians. This suggests that at least some Europeans may be consistent with having been formed by a simple 2-way mixture of populations related to Stuttgart and Loschbour. We thus fit each West Eurasian population into the topology of Fig. S14.6. Only Basques, Spanish_North, and Sardinians, can be fit successfully with this model. Fig. S14.9 shows a successful fit, which suggests that Sardinians, despite their well-documented isolation, were more admixed with indigenous Europeans than the first farmers of northern Europe represented by Stuttgart (21% western European hunter-gatherer and 79% Early European Farmer).

Most European populations cannot be fit as this type of 2-way mixture and, intuitively, this is due to their tendency (Fig. 1B) towards Ancient North Eurasians that is not modeled by such a mixture. Indeed, when we examined the set of $f_4$-statistics exceeding $|Z|>3$ for European populations, MA1 was involved for all populations who did not fit the model structure of Fig. S14.9, ranging from Bergamo (fitted *$f_4$(Loschbour, MA1; Stuttgart, Bergamo)* = -0.002162, Z=3.04 standard errors lower than the estimated value of 0.003951) to Mordovians (fitted *f4(Stuttgart, Mordovian; MA1, Mordovian)* = 0.000886, Z=7.4 standard errors higher than the estimated value of -0.010302).

***Figure S14.10: A successful 2-way mixture for Sardinians on the Fig. S14.7 scaffold.*** *They fit as a mix of Motala12 and Stuttgart-related "Hunter" and "Farmer" populations in proportions 12/88%.*

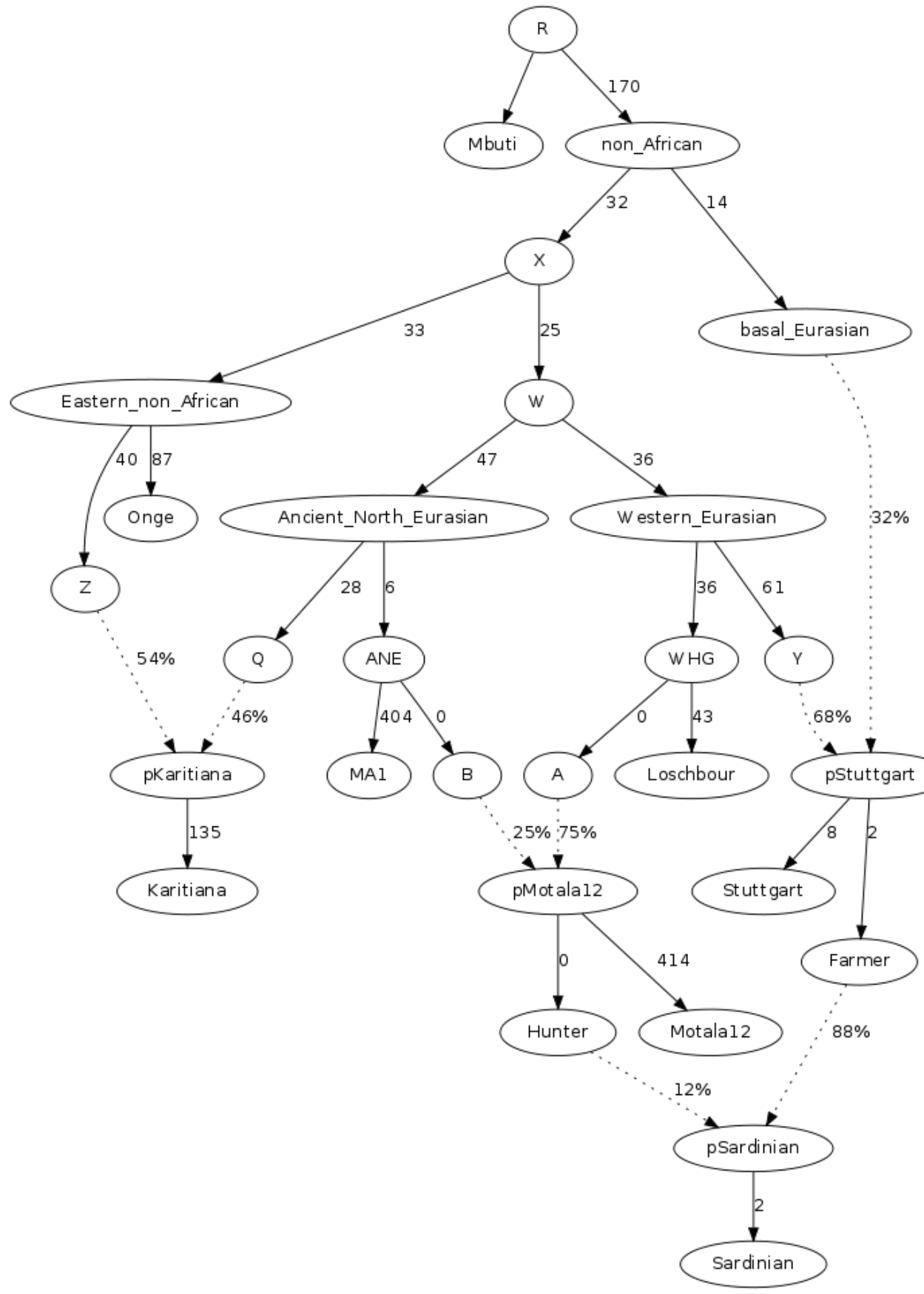

**Most Europeans are not a 2-way mixture of Motala12 and Stuttgart**

The fact that a Stuttgart/Loschbour mixture did not preserve the relationship of European populations to MA1 motivated us to try modeling them as a Stuttgart/Motala12 mixture, given the evidence that Motala12 has some MA1-related admixture. Fig. 1B suggests that this may not be enough to explain the data, since, despite being intermediate between Loschbour and MA1, Scandinavian hunter gatherers are still fairly close to Western European ones in PCA. Additionally, Motala12 does not feature at all in the all-pairs $f_3$-statistics documenting admixture in West Eurasians in SI11. We thus fit individual European populations into the topology of Fig. S14.8, but, only Basque, French_South, and Sardinian could be accommodated. We show a successful fit for Sardinians in Fig. S14.10. We do not propose that southwestern Europeans were formed by a mixture of Early European Farmers and Scandinavian hunter-gatherers, but the fact that they can be fit as such indicates that Scandinavian hunter-gatherers were close enough to their West European relatives so that they can serve as a proxy for them.

***Figure S14.11: The ratio $f_4$(X, Stuttgart; Karitiana, Chimp) / $f_4$(X, Stuttgart, MA1, Chimp) is <1 for different European populations.*** *This suggests that MA1 is a better surrogate for Ancient North Eurasians than is Karitiana. The bars indicate ± 1 standard error.*

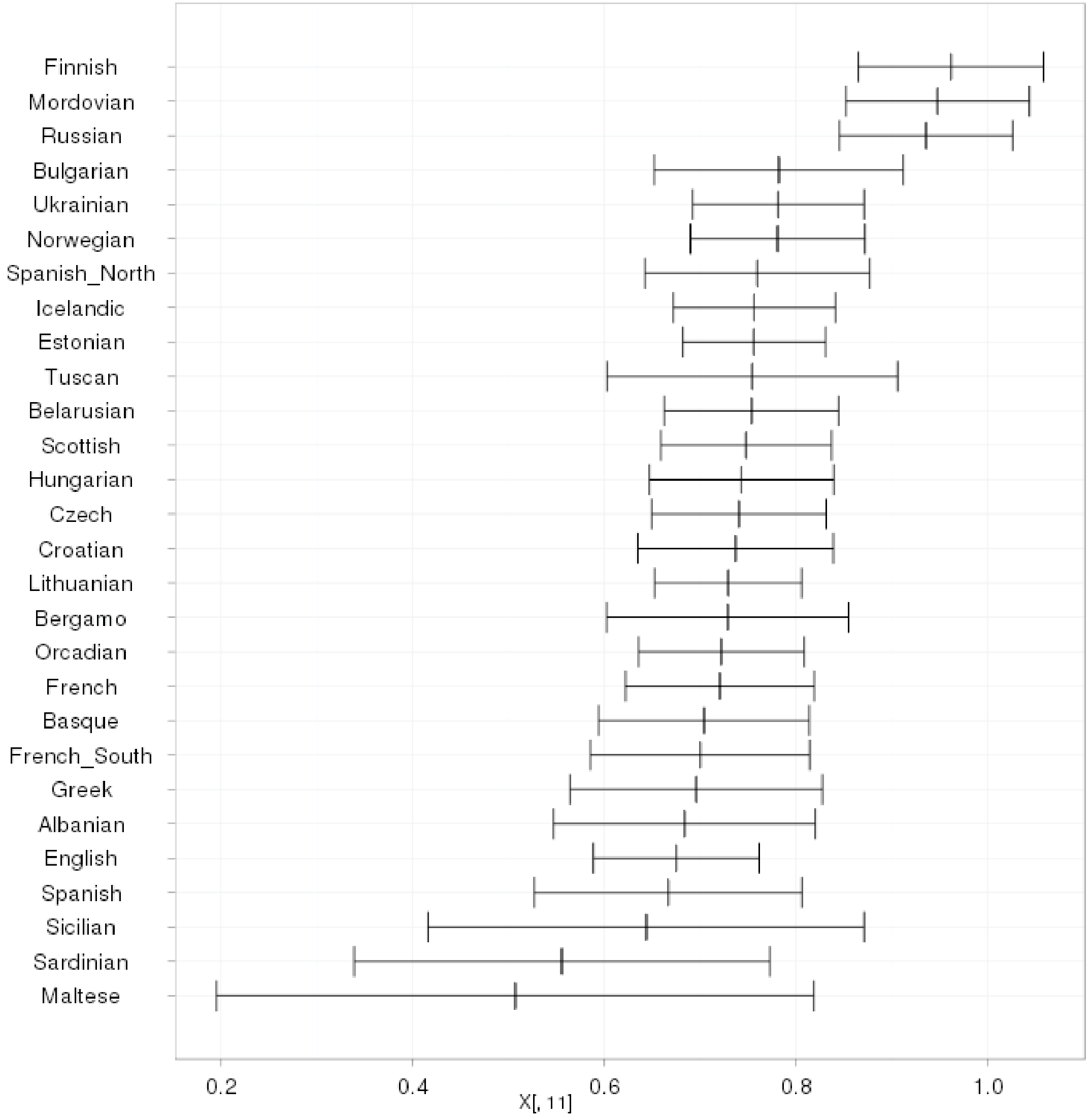

**Europeans can be fit as a 3-way mixture of Loschbour, Stuttgart, and MA1**

We inspected the statistics that precluded European populations from fitting both the Loschbour/Stuttgart (Fig. S14.9) and Motala12/Stuttgart (Fig. S14.10) models, and we noticed that these often involved either Karitiana or MA1. We plot the ratio of $f_4(X, Stuttgart; Karitiana, Chimp)$ / $f_4(X, Stuttgart, MA1, Chimp)$ in Fig. S14.11 for different European populations.

The related statistics $f_4(X, BedouinB; Karitiana, Chimp)$ and $f_4(X, BedouinB; MA1, Chimp)$ are plotted in Extended Data Fig. 5. By using BedouinB instead of Stuttgart, we can also plot Stuttgart in the space of these statistics. Europeans uniformly share more drift with MA1 than with Karitiana, and form a cline in this space with slope >1. Karitiana, because of its Ancient North Eurasian ancestry was crucial in detecting the presence of such ancestry in Europeans[2,6] but can now be replaced in the study of this ancestry by a better proxy for this ancestry (MA1), as also discussed in SI11. We hope that in the future additional representatives of the Ancient North Eurasian population may be studied, with either higher sequencing coverage or a closer genetic relationship to the ANE population admixing into Europe.

Motivated by these observations, we modeled Europeans to be not only a mix of Stuttgart and one of the available ancient samples (Loschbour or Motala12), but also of a "Hunter" population whose amount of MA1-related ancestry was allowed to be variable across Europeans, reflecting a hypothesis that ANE, WHG and EEF ancestry may have mixed in different proportions across Europe. Unlike Fig. S14.9 where zero MA1-related ancestry is assumed in Europeans, and Fig. S14.10 where "Hunter" is constrained to be a sister group of Scandinavian hunter-gatherers, we fit a model in which "Hunter" would only be constrained to be a mixture of Loschbour- and MA1-related ancestry. Fig. S14.12 shows the successful model structure, and Table S14.9 the inferred admixture proportions.

A total of 26 European populations fit this model, and we are encouraged by the fact that none of the Near Eastern populations fit. Thus, the model fitting correctly detects that they cannot be derived as a mixture of the exact same three ancestral populations as Europeans (they lack the European hunter-gatherer ancestry that EEF have in part (SI13) and that the WHG have in full).

It is evident that southern European populations have a greater affinity to early European farmers, and northern European populations to Western European hunter gatherers, consistent with the analysis of a Swedish Funnelbeaker farmer[19] (Skoglund_farmer in Fig. 1B) who resembled southern Europeans, and two Iberian Mesolithic hunter-gatherers[20] (represented by the higher-quality LaBrana genome[21] in Fig. 1B) who resembled Northern Europeans. Our analysis supports the view that ancestry from the two groups is variable across Europe, and suggests that a third element related to Upper Paleolithic Siberians, which in our analysis is best represented by MA1, contributed to present Europeans.

An interesting feature of these proportions is that they contrast the Basques to their Iberian neighbors, with nearly a third of their ancestry coming from WHG; this reflects the same genetic patterns as Fig. 1B which shows the Basques to the left off their Iberian neighbors, and European hunter gatherers projected in the same direction. Basques appear to possess a geographically local maximum of European hunter-gatherer ancestry.

***Figure S14.12: A successful 3-way fit for French, a population that cannot fit as a 2-way mixture.*** *Estimated mixture proportions are 45/55% "Hunter"/"Farmer", or 55/31/14% EEF/WHG/ANE.*

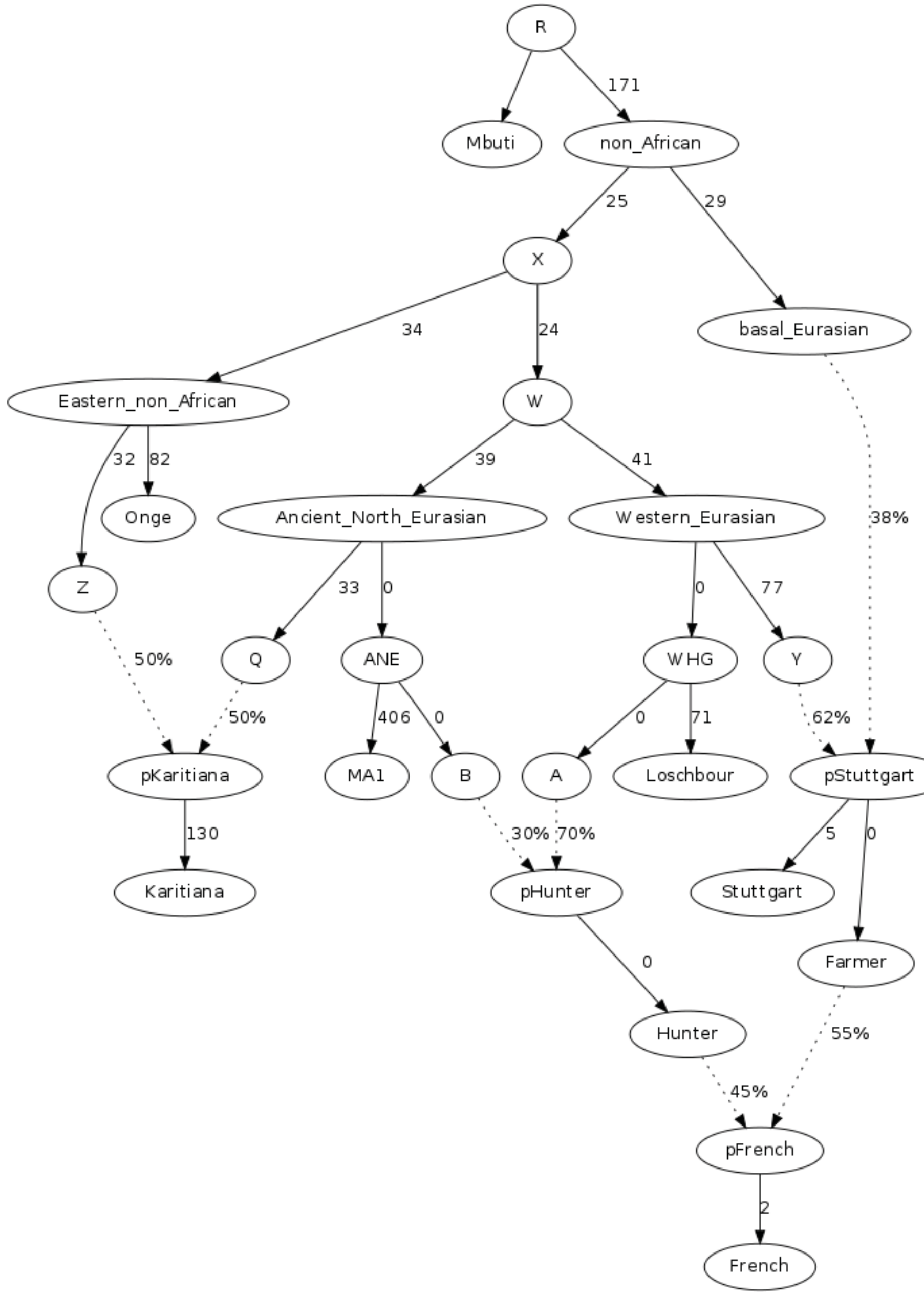

***Table S14.9: Admixture proportions for West Eurasian fit as a 3-way mixture of Early European Farmers (EEF), West European Hunter-Gatherers (WHG) and Ancient North Eurasians (ANE).*** *(These proportions are also shown in Extended Data Table 2.)*

| | **EEF** | **WHG** | **ANE** |
|---|---|---|---|
| Albanian | 0.781 | 0.092 | 0.127 |
| Ashkenazi_Jew | 0.931 | 0.000 | 0.069 |
| Basque | 0.593 | 0.293 | 0.114 |
| Belarusian | 0.418 | 0.431 | 0.151 |
| Bergamo | 0.715 | 0.177 | 0.108 |
| Bulgarian | 0.712 | 0.147 | 0.141 |
| Croatian | 0.561 | 0.293 | 0.145 |
| Czech | 0.495 | 0.338 | 0.167 |
| English | 0.495 | 0.364 | 0.141 |
| Estonian | 0.322 | 0.495 | 0.183 |
| French | 0.554 | 0.311 | 0.135 |
| French_South | 0.675 | 0.195 | 0.130 |
| Greek | 0.792 | 0.058 | 0.151 |
| Hungarian | 0.558 | 0.264 | 0.179 |
| Icelandic | 0.394 | 0.456 | 0.150 |
| Lithuanian | 0.364 | 0.464 | 0.172 |
| Maltese | 0.932 | 0.000 | 0.068 |
| Norwegian | 0.411 | 0.428 | 0.161 |
| Orcadian | 0.457 | 0.385 | 0.158 |
| Sardinian | 0.817 | 0.175 | 0.008 |
| Scottish | 0.390 | 0.428 | 0.182 |
| Sicilian | 0.903 | 0.000 | 0.097 |
| Spanish | 0.809 | 0.068 | 0.123 |
| Spanish_North | 0.713 | 0.125 | 0.163 |
| Tuscan | 0.746 | 0.136 | 0.118 |
| Ukrainian | 0.462 | 0.387 | 0.151 |

The model fit in Fig. S14.12 is for the French population, but for each of the 26 successfully fit populations, the internal structure of the tree may be different. In Fig. S14.13 we present the range of parameter estimates. Some of these appear quite stable, achieving very similar values regardless of which individual population is fit, while others are less so, with the extreme being the amount of WHG ancestry in "Hunter", ranging from 0 to 95.7%. In that particular case, it was Ashkenazi Jews, Maltese and Sicilians for whom the value was 0, and Sardinians who had the highest 95.7% value.

***Figure S14.13: Range of parameter estimates of Fig. S14.10 model for successfully fit populations***

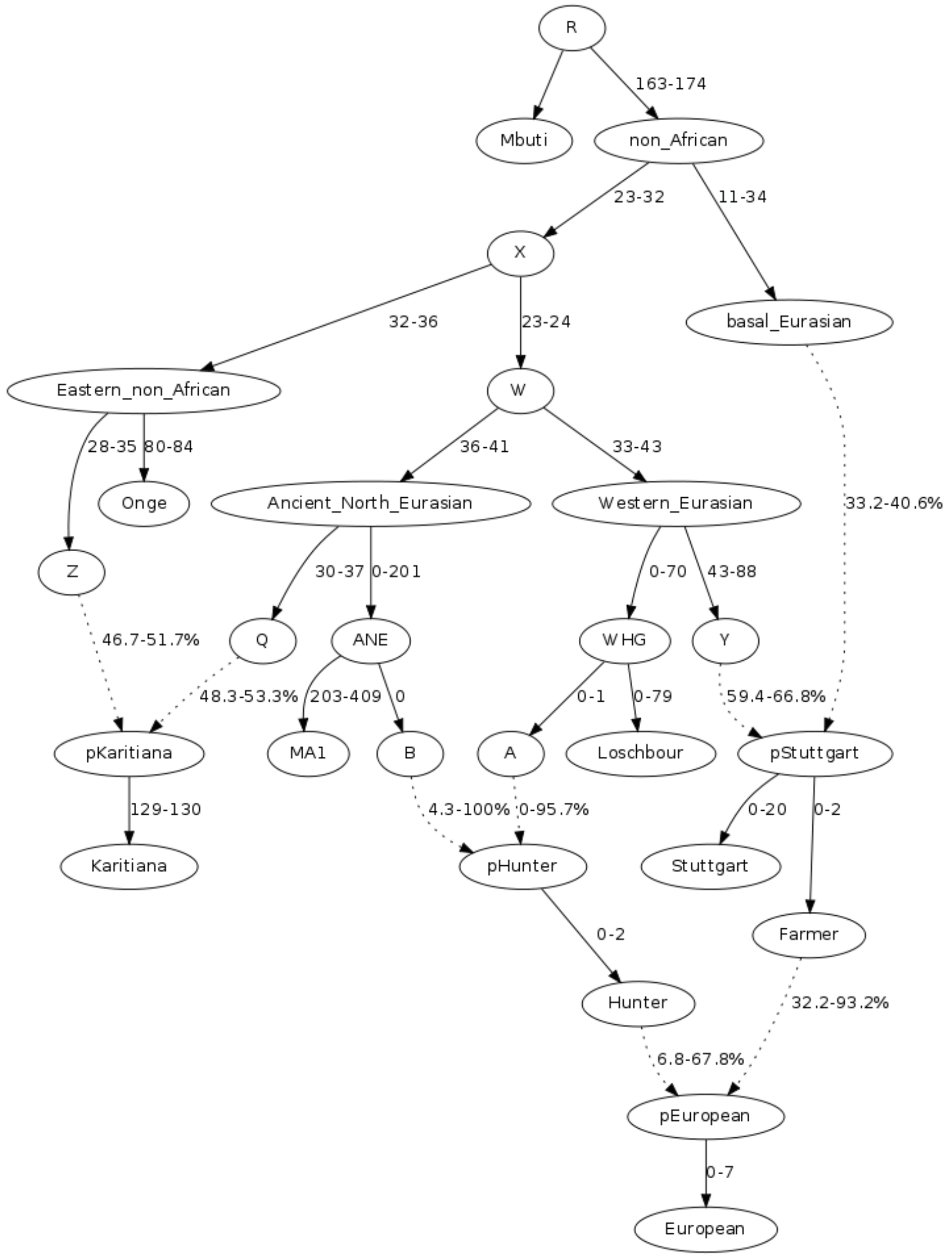

**Most European populations have both WHG and ANE ancestry**

We were interested in the possibility that a single "Hunter" population could have been the conduit for the ANE ancestry in most European populations. To determine whether this was plausible, we explicitly checked which West Eurasian populations could be fit as a mixture of Stuttgart and a "Hunter" population with $x$% WHG and 100-$x$% ANE ancestry. We fit the model of Figures S14.12 and S14.13 again for the populations of Table S14.9, but this time did not allow the proportions of WHG and ANE to vary freely but rather "locked" them in 5% increments, from (0, 100), (5,95), …, (100,0), thus exploring the whole range of possible mixtures for "Hunter".

Fig. S14.14 shows the range of values of $x$ that were compatible with each population. While a wide range of values is consistent with each population, with the exception of some populations which are consistent with no WHG ancestry (Albanian, Ashkenazi_Jew, Greek, Maltese, Sicilian), and some consistent with no ANE ancestry (Basque, French_South, Bergamo, Spanish_North, Sardinian), most Europeans can only be fit as having both WHG and ANE ancestry. Moreover, even for populations compatible with no WHG or no ANE ancestry, the best fit (Table S14.9) includes some such ancestry. For example, Basques are compatible with having no ANE ancestry, but according to Table S14.9, the best fit has 0.293 WHG and 0.114 ANE ancestry, for an $x$ ratio of 72%, that is, an intermediate value within the range indicated in Fig. S14.13

***Figure S14.14: WHG/(WHG+ANE) ratio (%) feasible ranges (in pink  for populations of Table S14.10.*** *Most European populations have both WHG and ANE ancestry, while a few are compatible with having no WHG ancestry (ratio 0%) and a few are compatible with having no ANE ancestry (ratio 100%).*

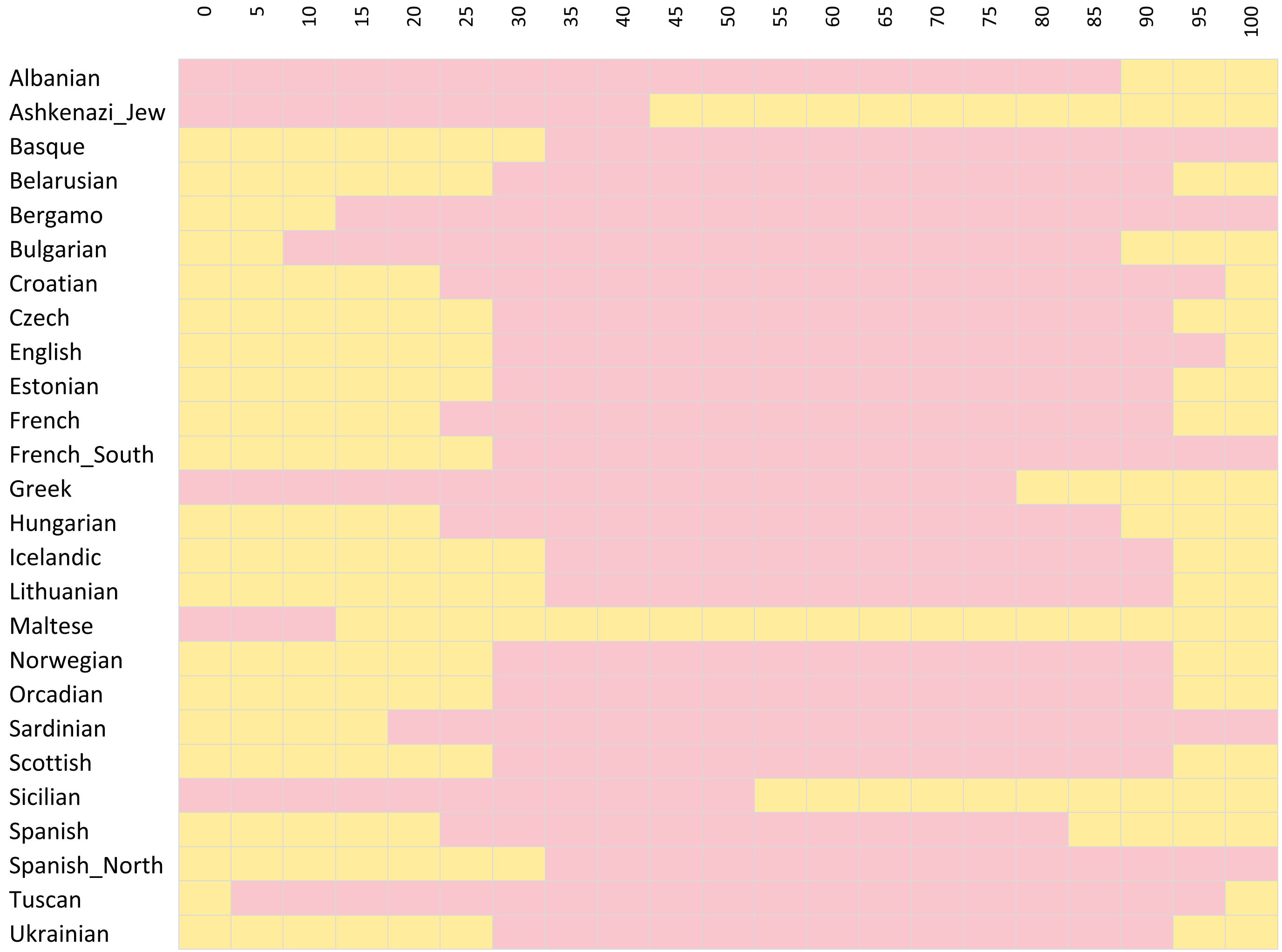

**Pairs of European populations consistent with descent from the same "Farmers" and "Hunters"**
Fig. S14.14 suggests that a large number of European populations can be successfully fit over a wide range of the WHG/(WHG+ANE) ratio. While this could in principle be consistent with their descending from the same "Farmer" and "Hunter" populations, this hypothesis may not in fact be a fit to the data because the internal tree parameters inferred for two populations may differ.

To explore this issue further, an idea is to attempt to fit two European populations *A* and *B* simultaneously as independent mixtures of "Farmer" and "Hunter". This has the advantage of forcing the tree to accommodate both *A* and *B*, and can thus determine whether a common tree can fit both. However, this simple modeling ignores the post-admixture histories of *A* and *B*, which may be complex and involve gene flow between them. It is unrealistic to model European populations as independent mixtures of "Farmer" and "Hunter" in the context of the major gene flows that must have occurred within Europe itself since the advent of agriculture.

To address this problem, we modified ADMIXTUREGRAPH. As discussed in Patterson et al. (2012)[9] a basis for *f*-statistics involving populations $(A_0, A_1, ..., A_n)$ is found from $f_3(A_0; A_i, A_j), f_2(A_0, A_i)$ $0 < i < j$. We think of $A_0$ as a base population. Suppose *A* and *B* are 2 populations whose descendants have a complex recent history such as two European populations descended from the "Farmer" and "Hunter", above. ADMIXTUREGRAPH calculates an empirical covariance matrix for the *f*-statistics involving the base point $A_0$. Our modification is simply to add a large constant (we chose 10,000) to the variance term for $f_3(A_0, A, B)$. This has the effect that ADMIXTUREGRAPH regards *f*-statistics involving both *A* and *B* as essentially uninformative, which has precisely the desired effect. This has the advantage of fitting a tree structure for both *A* and *B* simultaneously while avoiding the interactions between *A* and *B* that might reflect details of their more recent common history.

In Fig. S14.15 we show populations pairs that are consistent with descent from identical "Farmer" and "Hunter" populations. Sicilians, Ashkenazi Jews, and Maltese are only compatible with each other and not with any other populations, consistent with Fig. S14.14 and Table S14.9 which show them to be have less or even no WHG ancestry in contrast to other populations. Greeks are compatible with their geographical neighbors in the Balkans (Albanians and Bulgarians) and Italy (Bergamo and Tuscans). Basques and Spanish_North are incompatible with several populations from Mediterranean and Southeastern Europe. Mediterranean and Southeastern Europeans such as Spanish, Albanians, Bulgarians, Bergamo, Tuscans, Croatians, and Hungarians are compatible with each other

Importantly, this analysis confirms that a large number of European populations are consistent with descent from identical "Farmer" and "Hunter" populations. Overall, 202 of the 325 possible pairs for the 26 populations resulted in graph fits with no outlier $f_4$-statistics. We conclude that a substantial fraction of modern European populations are consistent with having inherited ancestry from the three EEF/WHG/ANE groups via only two proximate ancestral populations. This inference is not inconsistent with that of SI12 that at least three sources are needed for present-day Europeans, as that analysis considers a large set of European populations as a whole, whereas the analysis in this section only considers population pairs. The analysis of SI12 documents 3-way admixture for present-day Europeans while that of the current section indicates which pairs of European populations have similar WHG/(WHG+ANE) ratios and can thus be fit as a mixture of a "Farmer" and a "Hunter" population.

In Fig. S14.16 we plot the WHG/(WHG+ANE) ratio over all 202 compatible pairs. It is clear that the bulk of the distribution is in the 60-80% interval, with a visible peak around 71-74%. This suggests that for many Europeans, "Hunter" was a population of predominantly WHG-related ancestry but with a substantial ANE-related component.

*Figure S14.15: Population pairs marked in pink are consistent with common descent from identical "Farmer" and "Hunter" populations.*

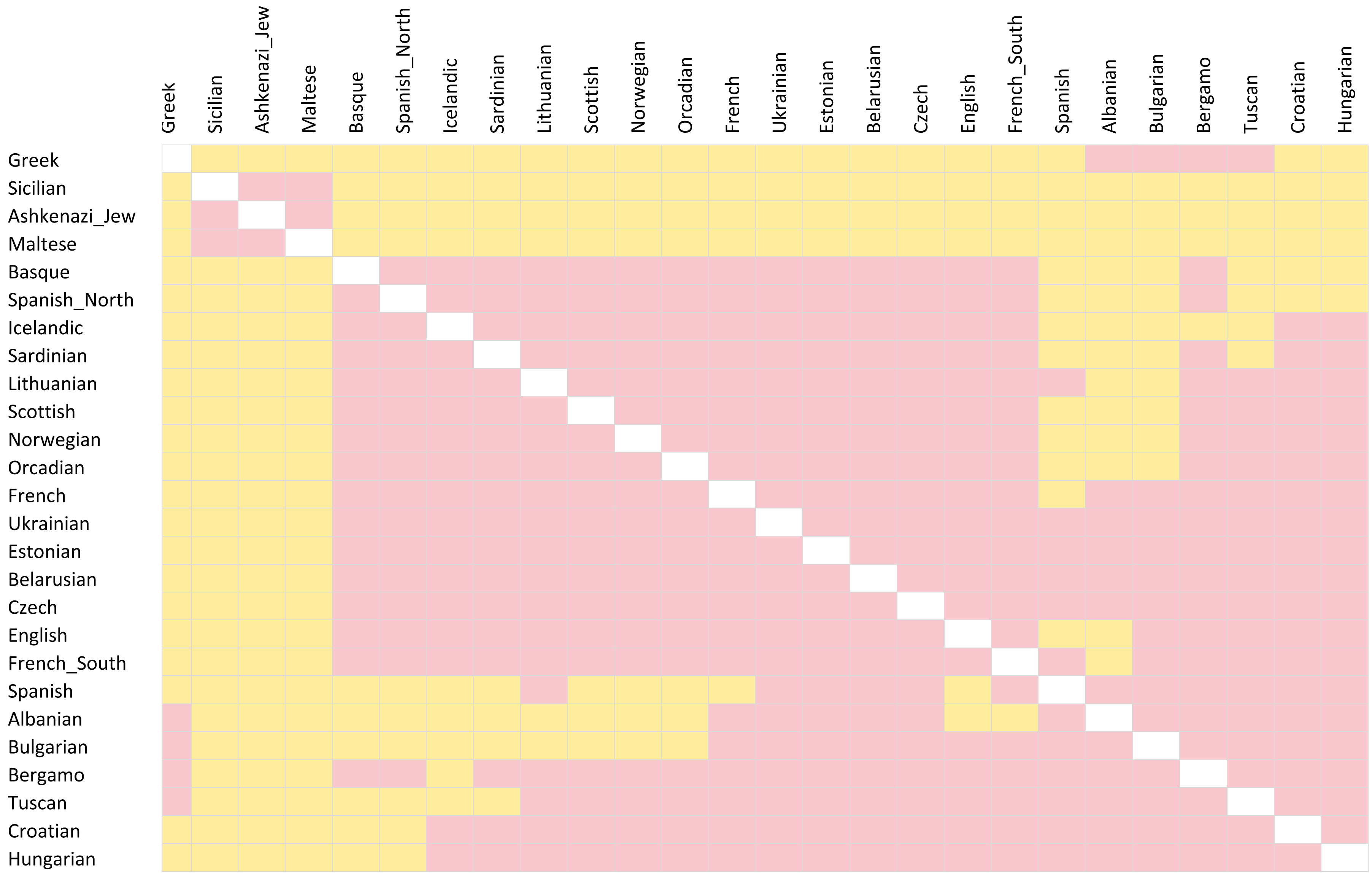

***Figure S14.16: Distribution of WHG/(WHG+ANE) ratio for population pairs that can be successfully fit as descendants of identical "Farmer" and "Hunter" populations (Fig. S14.14).***

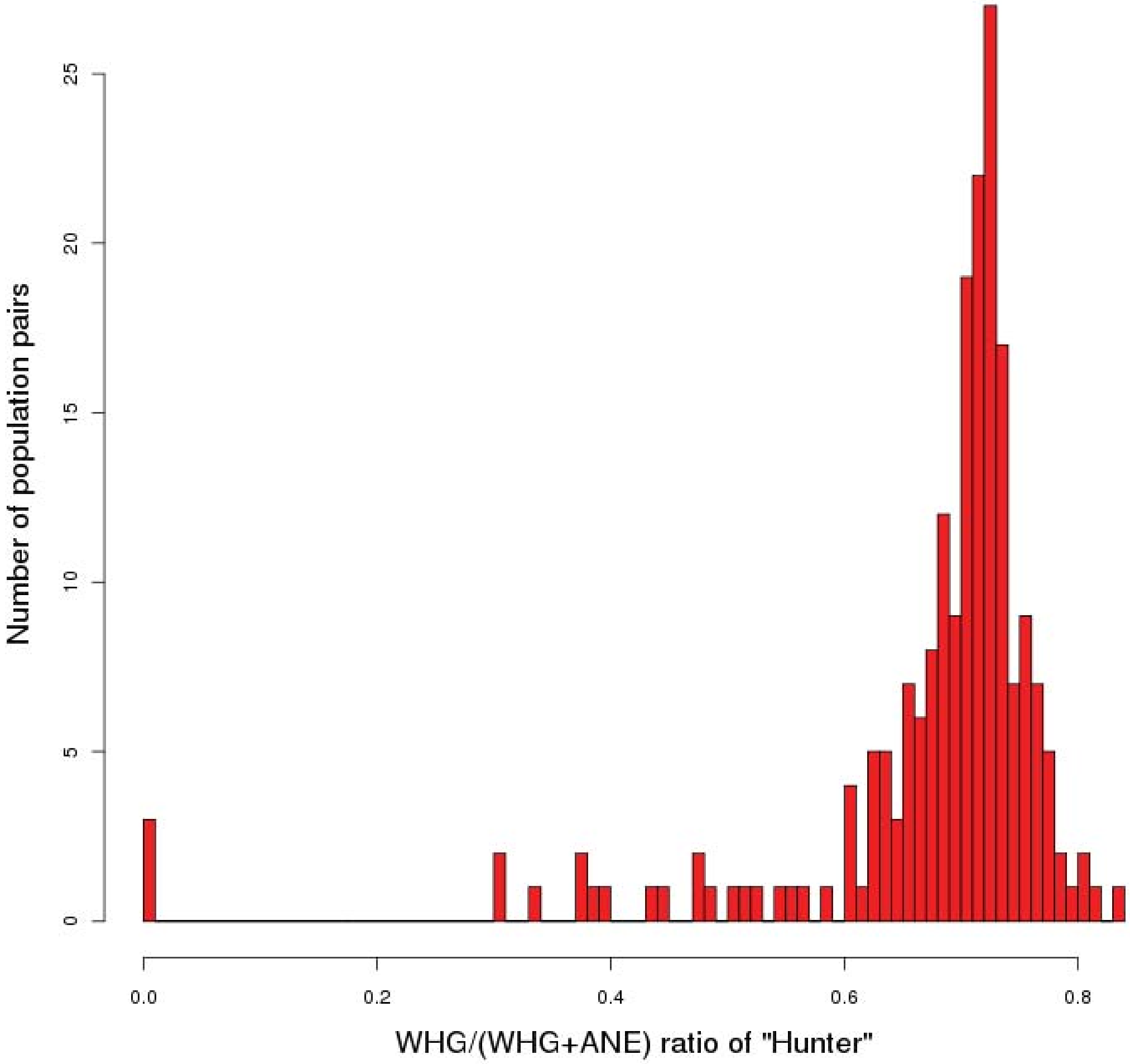

**Almost all pairs of European populations consistent with descent from the same ""EEF", "WHG", and "ANE" populations**

We repeated the joint fitting of population pairs, but allowed each population in a pair to descend from a different "Hunter" population, i.e., with a variable WHG/(WHG+ANE) ratio. Almost all population pairs were now successful (264 of 325, Fig. S14.17), with the exception of Ashkenazi Jews, Maltese, and Sicilians who could often not be fit with other populations. It appears that these populations have Near Eastern ancestry that is not well-modeled by the 3-population model. This is consistent with their position in Fig. 1B, and the results of analysis of SI 17 which do not explicitly model deep population history.

We estimated averaged admixture proportions for 23 populations (excluding Ashkenazi Jews, Sicilians, and Maltese) who appear in Fig. S14.17 to be consistent with descent from identical EEF, WHG, and ANE populations. Whereas the proportions of Table S14.9 were derived from individual fits of the populations, those of Table S14.10 represent the average, for each population, over all compatible population pairs. The proportions of Table S14.10 are the ones plotted in Fig. 2B.

*Figure S14.17: Population pairs marked in pink are consistent with common descent from identical “EEF”, “WHG”, and “ANE” populations.*

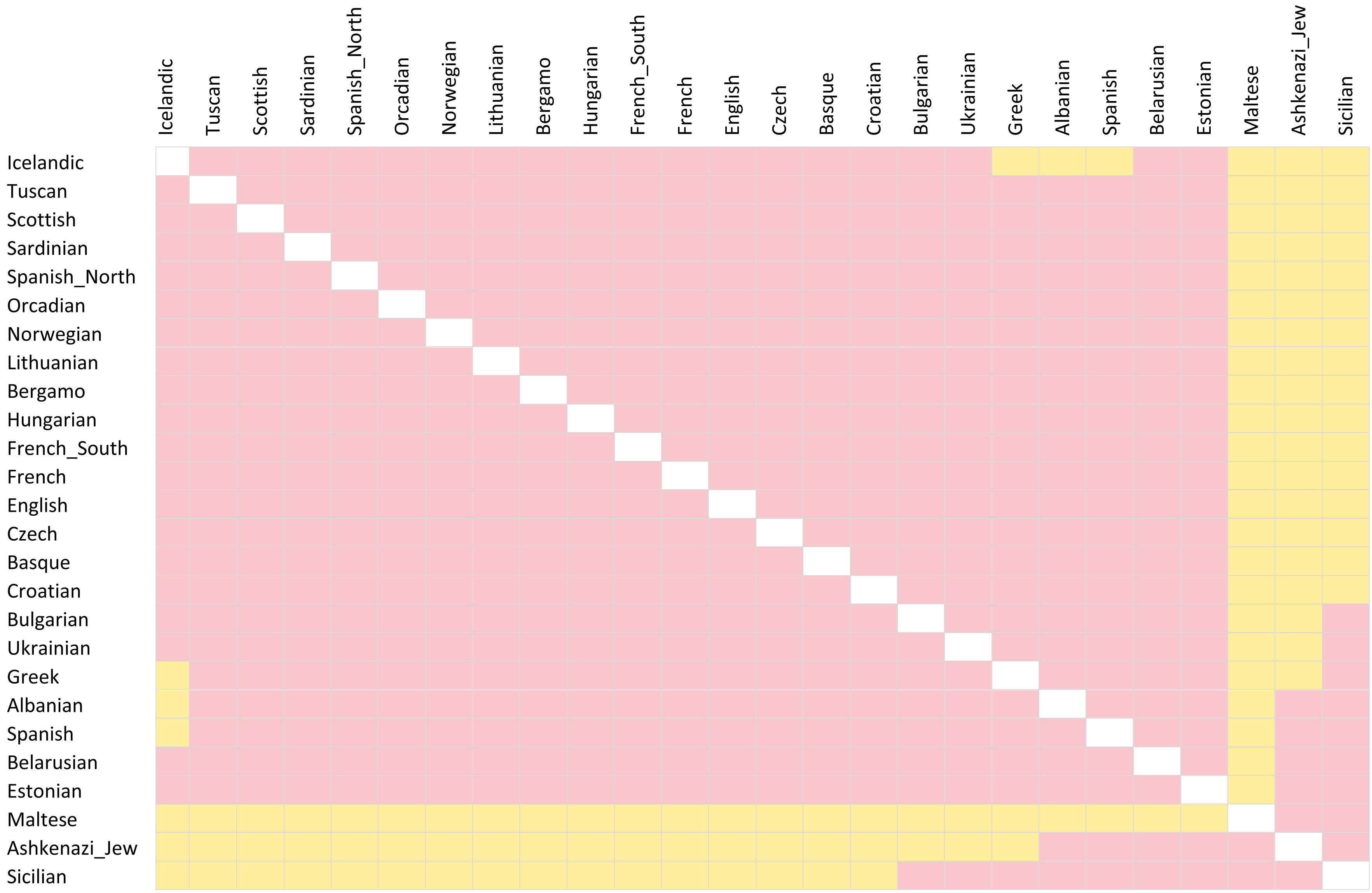

***Table S14.10: Averaged admixture proportions for European populations.*** *Each proportion represents the mean over all fits with compatible populations; the range of the successful fits is also shown. (These proportions are also included with other mixture estimates in Extended Data Table 2).*

| | **EEF** | | **WHG** | | **ANE** | |
|---|---|---|---|---|---|---|
| | Mean | Range | Mean | Range | Mean | Range |
| Albanian | 0.781 | 0.772-0.819 | 0.082 | 0.032-0.098 | 0.137 | 0.129-0.158 |
| Basque | 0.569 | 0.527-0.616 | 0.335 | 0.255-0.392 | 0.096 | 0.076-0.129 |
| Belarusian | 0.426 | 0.397-0.464 | 0.408 | 0.338-0.443 | 0.167 | 0.150-0.199 |
| Bergamo | 0.721 | 0.704-0.793 | 0.163 | 0.061-0.189 | 0.117 | 0.104-0.147 |
| Bulgarian | 0.718 | 0.707-0.778 | 0.132 | 0.047-0.151 | 0.151 | 0.138-0.175 |
| Croatian | 0.564 | 0.548-0.586 | 0.285 | 0.242-0.310 | 0.151 | 0.137-0.172 |
| Czech | 0.489 | 0.460-0.531 | 0.348 | 0.273-0.382 | 0.163 | 0.145-0.196 |
| English | 0.503 | 0.476-0.536 | 0.353 | 0.296-0.382 | 0.144 | 0.130-0.169 |
| Estonian | 0.323 | 0.293-0.345 | 0.49 | 0.451-0.520 | 0.187 | 0.172-0.205 |
| French | 0.563 | 0.537-0.601 | 0.297 | 0.230-0.328 | 0.140 | 0.126-0.169 |
| French_South | 0.636 | 0.589-0.738 | 0.256 | 0.111-0.323 | 0.108 | 0.088-0.151 |
| Greek | 0.791 | 0.780-0.816 | 0.048 | 0.019-0.060 | 0.161 | 0.150-0.171 |
| Hungarian | 0.548 | 0.520-0.590 | 0.279 | 0.199-0.313 | 0.174 | 0.156-0.210 |
| Icelandic | 0.409 | 0.386-0.424 | 0.448 | 0.409-0.473 | 0.143 | 0.126-0.170 |
| Lithuanian | 0.352 | 0.327-0.384 | 0.488 | 0.433-0.527 | 0.160 | 0.135-0.184 |
| Norwegian | 0.417 | 0.388-0.438 | 0.423 | 0.383-0.450 | 0.160 | 0.140-0.181 |
| Orcadian | 0.465 | 0.439-0.493 | 0.378 | 0.329-0.403 | 0.157 | 0.140-0.179 |
| Sardinian | 0.818 | 0.791-0.874 | 0.141 | 0.058-0.182 | 0.041 | 0.026-0.068 |
| Scottish | 0.408 | 0.387-0.424 | 0.421 | 0.384-0.448 | 0.171 | 0.149-0.201 |
| Spanish | 0.759 | 0.736-0.804 | 0.126 | 0.066-0.170 | 0.115 | 0.091-0.151 |
| Spanish_North | 0.612 | 0.561-0.660 | 0.292 | 0.214-0.365 | 0.096 | 0.072-0.126 |
| Tuscan | 0.751 | 0.737-0.806 | 0.123 | 0.047-0.145 | 0.126 | 0.114-0.150 |
| Ukrainian | 0.463 | 0.445-0.491 | 0.376 | 0.322-0.399 | 0.160 | 0.148-0.187 |

**$f_4$-ratio based estimation of Early European Farmer ancestry**

The proportions of Table S14.10 are based on model fits using ADMIXTUREGRAPH, which simultaneously optimizes *f*-statistics over several populations. This may make the estimates more robust, but is also based on assuming that the model we fit is accurate in all its detail. We also confirmed these estimates using a simpler approach applied to the proposed graph.

Consider Fig. S14.18. In this model which we have argued above is a fit to the data for many European populations to within the limits of our resolution, a European population has $\alpha\beta$ of its ancestry from Basal_Eurasian and Stuttgart has $\beta$ of its ancestry from EEF. It is then the case that:

$$f_4(Mbuti, Onge; Loschbour, European) = -\alpha\beta x$$
$$f_4(Mbuti, Onge; Loschbour, Stuttgart) = -\beta x$$

This exploits the fact that the paths Mbuti→Onge and Loschbour→European or Loschbour→Stuttgart intersect only over the segment non_African→X whose drift length is $x$. We can then apply $f_4$-ratio estimation in a straightforward way by dividing the two[1,2]. We show in Table S14.11 the estimates we obtain as well as their differences from those of Table S14.10.

The $f_4$-ratio estimates differ from those of ADMIXTUREGRAPH by no more than 1.3 standard errors. The mean and standard deviation over all populations is 0.047±0.506. Thus, an $f_4$-ratio estimation of this proportion over the proposed model is consistent with the optimization-based estimate.

***Figure S14.18: The fact that a European population has* a fraction αβ *of Basal Eurasian ancestry and Stuttgart has* β *such ancestry, allows for an estimate of EEF ancestry via an* $f_4$*-ratio.***

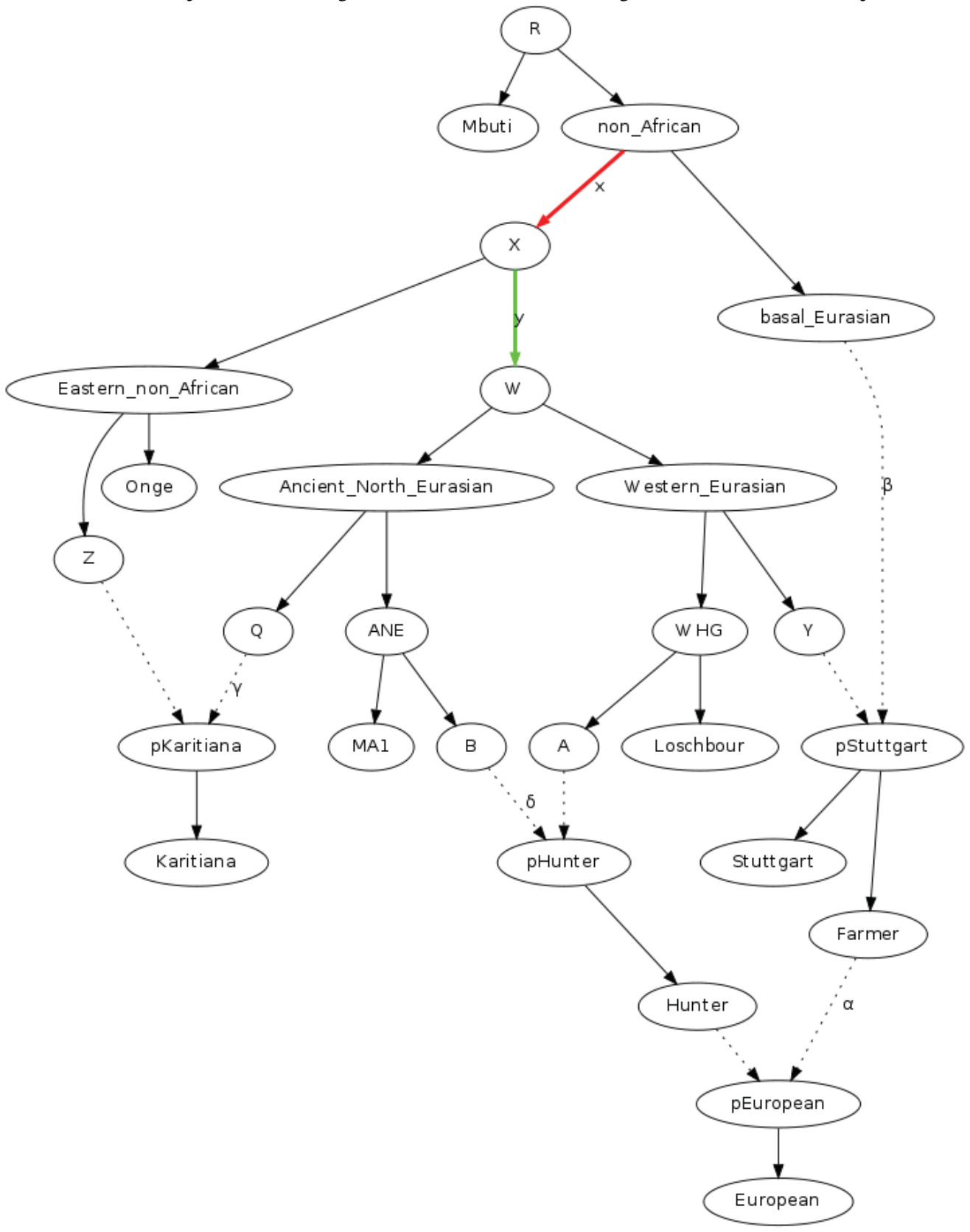

***Table S14.11: $f_4$-ratio based estimates of EEF ancestry are consistent with ADMIXTUREGRAPH***

| | **$f_4$-ratio estimate** | **std. error** | **ADMIXTUREGRAPH − $f_4$-ratio (Z-score)** |
|---|---|---|---|
| Albanian | 0.628 | 0.116 | 1.320 |
| Basque | 0.621 | 0.104 | -0.494 |
| Belarusian | 0.392 | 0.116 | 0.293 |
| Bergamo | 0.699 | 0.112 | 0.192 |
| Bulgarian | 0.686 | 0.115 | 0.275 |
| Croatian | 0.502 | 0.113 | 0.548 |
| Czech | 0.501 | 0.111 | -0.110 |
| English | 0.511 | 0.110 | -0.075 |
| Estonian | 0.286 | 0.128 | 0.284 |
| French | 0.577 | 0.105 | -0.136 |
| French_South | 0.700 | 0.119 | -0.539 |
| Greek | 0.685 | 0.113 | 0.939 |
| Hungarian | 0.555 | 0.106 | -0.071 |
| Icelandic | 0.398 | 0.116 | 0.095 |
| Lithuanian | 0.374 | 0.121 | -0.180 |
| Norwegian | 0.395 | 0.115 | 0.194 |
| Orcadian | 0.497 | 0.110 | -0.293 |
| Sardinian | 0.903 | 0.132 | -0.646 |
| Scottish | 0.332 | 0.135 | 0.567 |
| Spanish | 0.838 | 0.121 | -0.648 |
| Spanish_North | 0.690 | 0.123 | -0.639 |
| Tuscan | 0.773 | 0.125 | -0.179 |
| Ukrainian | 0.419 | 0.115 | 0.382 |

**$f_4$-ratio estimate of Basal Eurasian admixture in Stuttgart**
A different parameter that can be estimated via an $f_4$-ratio is the amount of basal_Eurasian admixture into Stuttgart. Consider the edge $X \rightarrow W$ with drift length $y$ in Fig. S14.18.

We can estimate $y$ directly by the following statistic:

$$y = f_4(Mbuti, MA1; Onge, Loschbour) \quad \text{(S14.1)}$$

But also:

$$\beta y = f_4(Stuttgart, Loschbour; Onge MA1) \quad \text{(S14.2)}$$

Taking the ratio we estimate $\beta$=0.44 ± 0.10. The fitted values of $\beta$ are within 1 standard error of this estimate (Fig. S14.13). These results suggest that the hypothesized Basal Eurasian ancestry would need to have made a major contribution to ancient Near Eastern populations to explain our data. The amount of Basal Eurasian admixture in the ancient Near East is uncertain, as the lack of an unadmixed Near Eastern reference makes the amount of Near Eastern admixture into Stuttgart uncertain (SI13). However, we can confidently say that it must have been higher than the estimated value for Stuttgart.

**$f_4$-ratio estimate of Ancient North Eurasian admixture in Karitiana and North Asians**
A different parameter that can be estimated via an $f_4$-ratio is the amount of Ancient North Eurasian admixture into Karitiana. Consider again the edge $X \rightarrow W$ with drift length $y$ in Fig. S14.18. We write:

$$\gamma\beta y = f_4(Stuttgart, Loschbour; Onge Karitiana) \quad \text{(S14.3)}$$

We already estimated an expression for $\beta y$ in Equation (S14.2). Taking the ratio we estimate $\gamma$=0.413 ± 0.176. The fitted values of $\beta$ are within 1 standard error of this estimate (Fig. S14.13) and our estimate is in good agreement with the TreeMix[22] estimate[6] using whole genome data.

We also verified that three North Asian populations (Ulchi, Eskimo, Chukchi) that are more closely related to MA1 than to European hunter-gatherers (Table S14.4) could also be fit into our model in place of Karitiana, suggesting that they, too can be expressed as a mixture of Ancient North Eurasians (ANE) and an eastern non-African (ENA) component.

However, as MA1 is closer to Karitiana than to North Asians (Table S14.7), we expect that the balance of the two components would be different in Karitiana and North Asians; this is supported by the $f_4$-ratio estimation of ANE ancestry for these populations (Table S14.12).

| ***Table S14.12: Ancient North Eurasian ancestry in Karitiana and North Asians*** **Population** | **Ancient North Eurasian ancestry** | **std. error** |
|---|---|---|
| Karitiana | 0.413 | 0.176 |
| Chukchi | 0.199 | 0.148 |
| Ulchi | 0.129 | 0.145 |
| Eskimo | 0.244 | 0.149 |

**East Eurasian gene flow into far Northeastern European populations**
Three European populations failed to successfully fit the model of Fig. S14.12, and we list them in Table S14.13 together with the most significantly differing $f$-statistics.

These three far northeastern European populations share more alleles with Karitiana/Onge than is predicted by the model (both Onge and Karitiana-related statistics are violated for all three). This is consistent with the ADMIXTURE analysis (SI9), which suggests that they possess a Siberian ancestral component not shared with other Europeans. It is also consistent with the results of Fig. S14.11, which show that these three populations share more alleles with Karitiana relative to other Europeans. A possible explanation for this is distinct gene flow from Siberia, perhaps related to the migration of Y-haplogroup N from east Asia into west Eurasia[23,24], as this lineage is present in the northeast and rare elsewhere in Europe.

***Table S14.13: European populations that cannot be fit as a 3-way mix of EEF, WHG, and ANE***

| Population | Violated statistic | fitted | estimated | Z |
|---|---|---|---|---|
| Finnish | Karitiana, MA1; Loschbour, Finnish | 0.002025 | -0.003984 | -3.161 |
| Mordovian | Karitiana, MA1; Loschbour, Mordovian | 0.002050 | -0.004990 | -3.790 |
| Russian | Karitiana, MA1; Loschbour, Russian | 0.001947 | -0.004214 | -3.398 |

In Extended Data Fig. 7, we plot the statistics $f_4$*(Test, BedouinB; Han, Mbuti)* and $f_4$*(Test, BedouinB; MA1, Mbuti).* We use BedouinB here so that we can also plot Stuttgart in the same figure. Populations that fit the model of Fig. S14.12 form a cline from Stuttgart in the south, to Lithuanians and Estonians in the north, but the three populations violating our model (Table S14.13) are clearly to the right, sharing relatively more alleles with the Han. We also add the single Saami individual from our dataset and the Chuvash on this plot. These two additional European groups deviate from the main European cline even more strongly, in the same direction as the Finnish, Mordovian, and Russian.

While we find no evidence that the Han have West Eurasian admixture (SI9), it is still possible in principle that they share some unknown common ancestry with Northeast Europeans. We also computed the statistic" $f_3$*(Mbuti; Ami, Test)* statistic for West Eurasian populations. This "outgroup $f_3$-statistic"[6] measures the amount of common genetic drift shared between Ami (a Taiwanese aboriginal population that seems extremely unlikely to have historical connections with Northeastern Europeans in particular). We plot the results in Figure S14.19, with the three northeastern European populations appearing as clear outliers (red).

***Figure S14.19: Northeast Europeans share more drift with Ami than other Europeans***

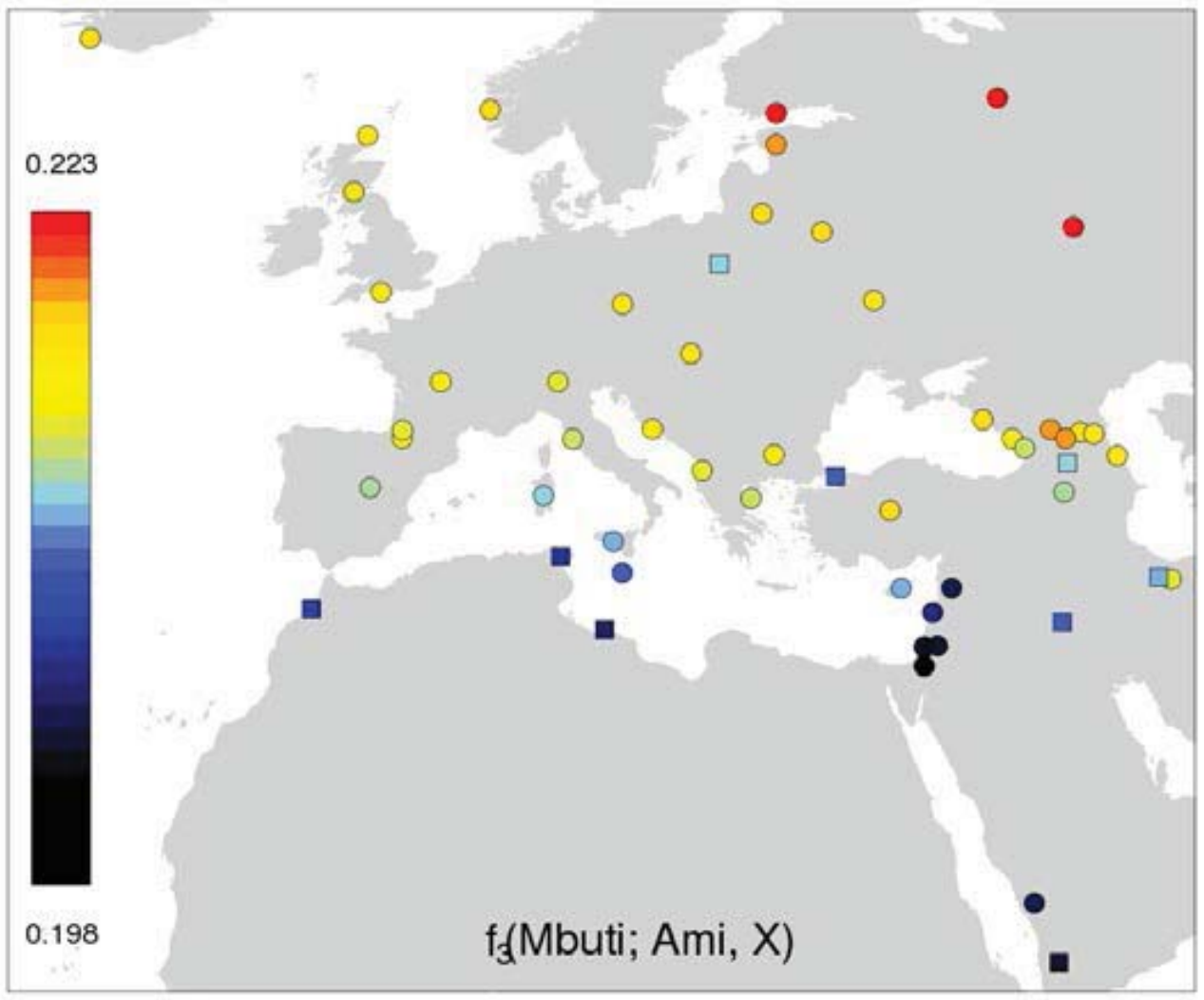

Note that the fact that Europe has higher values of this statistic than the Near East does not indicate East Eurasian admixture across Europe as this statistic is also expected to be reduced by the presence of either Basal Eurasian or African admixture.

Finally, we used ALDER[25] to investigate whether a linkage disequilibrium signal of recent admixture exists in Northeastern Europe (Table S14.14) using the Han as a reference and found a significant (Z>3) curve for three populations (for Finnish, Z=1.27, while we could not use this method on a single Saami individual). The estimated dates are 52-70 generations ago, or 1,500-2,000 years assuming 29 years per generation[26].

***Table S14.14: Chuvash, Mordovians & Russians have LD evidence for recent East Asian mixture.***

| | East Eurasian admixture (%) | | Time | |
|---|---|---|---|---|
| | Lower bound (%) | std. error | Generations | std. error |
| Chuvash | 11.7 | 1.7 | 62.5 | 11.0 |
| Mordovian | 6.7 | 1.3 | 69.8 | 16.9 |
| Russian | 5.7 | 0.4 | 52.3 | 5.8 |

The most straightforward explanation for these combined observations is that Northeastern Europeans possess some ancestry from an eastern Eurasian population, although more complicated explanations involving a population that affected both Northeastern Europe and eastern non-Africans are also possible. It is clear that the genetic landscape of Siberia has changed since the time of MA1 (~24,000 years ago), as this would explain both the fact that present-day Siberians share less drift with MA1 than both Europeans and Native Americans[11], that "First Americans" like the Karitiana already possessed east Eurasian admixture, and also that later waves of migration into the Americas also share additional common drift with Han Chinese than the wave of "First Americans", analogous to the pattern we observe in far northeast Eurasians[13].

Modern Siberians especially from the most eastern part of Siberia continue to have Ancient North Eurasian ancestry (Table S14.12) to a lesser extent than the Karitiana. Northeastern Europe may also have received genetic input from a later period of the Siberian gene pool in which (unlike the time of MA1), the eastern Eurasian influence was present. More ancient DNA research in both Northeastern Europe and Siberia should be useful for clarifying the historical events that explain these patterns.

**High levels of Ancient North Eurasian ancestry in the Northeast Caucasus**

Finally, we turned to the Near East and Caucasus to explore the implications of our model for admixture events there. We note (Table 1, Extended Data Table 1, SI11) that Near Easterners all have their lowest $f_3$-statistics involving Stuttgart, consistent with the hypothesis that Stuttgart possesses a substantial proportion of ancient Near Eastern ancestry. However, different populations appear to have their strongest signal of admixture involving pairings of Stuttgart with Africans, South Asians, Native Americans or MA1. Together with the evidence of Fig. 1B, this points to Near Eastern and Caucasian populations having a common ancestry related to Stuttgart, which is, however, modified by different influences related to many world populations. Unlike Europe, where several ancient DNA samples now exist, including the ones sequenced for the present study, no ancient human genomes exist for the Near East, making reconstructions of its past even more difficult.

We intersected the set of Near Eastern populations without substantial (<1%) African admixture as inferred by ADMIXTURE K=6 (SI 9) with those whose most significant $f_3$-statistic involved the pairing (Stuttgart, MA1) (Table 1). Five populations met these criteria: Abkhasian, Chechen, Cypriot, Druze, Lezgin. We modified the model of Fig. S14.12 to model these populations as a mixture of a Near Eastern population that also contributed to Stuttgart and an MA1-related ANE population (but no WHG ancestry) (Fig. S14.20). All five populations fit successfully, and we report their admixture proportions in Table S14.15.

An interesting detail of Fig. S14.20 is that the Near_East is modeled as a mixture of basal_Eurasian and a node Y which forms a clade with Loschbour. Present-day Near Eastern populations are indeed more closely related to European hunter-gatherers than to MA1 despite having some MA1-related ancestry. This can be easily seen in Extended Data Fig. 6C where the range of the statistic $f_4$*(Test, Chimp; MA1, Loschbour)* is *negative* for all West Eurasian populations including all Near Eastern ones, suggesting that they share more drift with Loschbour than with MA1 (the statistic is Z<-4 for all West Eurasian populations except the Lezgin where it is Z=-3.6). If we attempt to fit Near Eastern populations as either “pure Basal Eurasian” or “Basal Eurasian” plus an element predating the WHG/ANE split such models fail as they do not explain such statistics.

***Figure S14.20: A model for Near Eastern populations with Ancient North Eurasian admixture. Stuttgart is a mixture of Near_East and a sister group of Loschbour (UHG: Unknown Hunter-Gatherers); A Test population (shown here) is a mixture of Near_East and a sister group of MA1.***

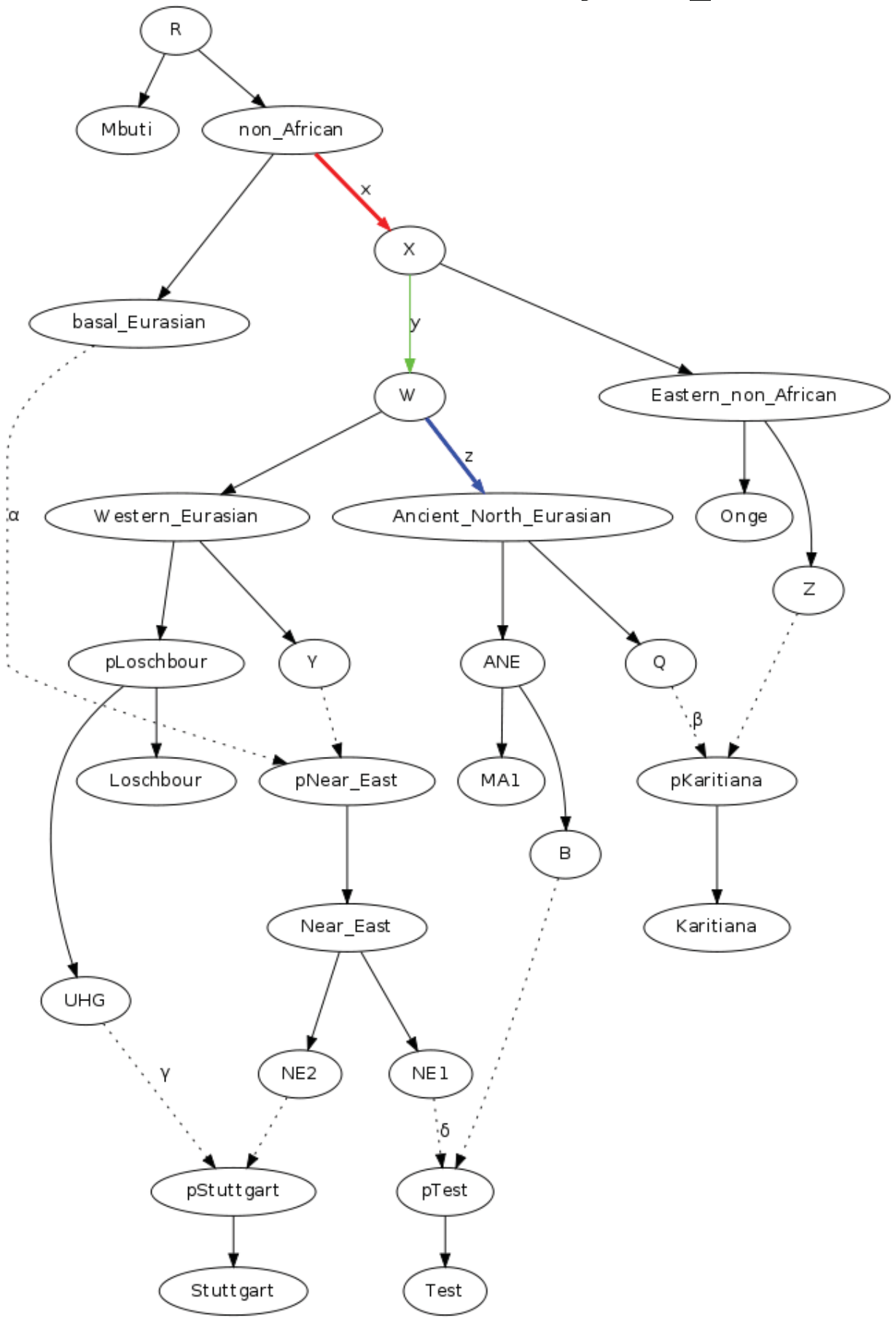

***Table S14.15: Admixture proportions for Near Eastern populations fit as a mixture of the ancient East and Ancient North Eurasians.*** *A lower bound that can be obtained via the ratio $f_4$(Test, Stuttgart; Karitiana, Onge) / $f_4$(MA1, Stuttgart; Karitiana, Onge) is also indicated.*

| | Near East | ANE (fitted) | ANE (lower bound) |
|---|---|---|---|
| Abkhasian | 0.814 | 0.186 | 0.157 ± 0.052 |
| Chechen | 0.730 | 0.270 | 0.244 ± 0.049 |
| Cypriot | 0.867 | 0.133 | 0.097 ± 0.056 |
| Druze | 0.882 | 0.118 | 0.047 ± 0.055 |
| Lezgin | 0.712 | 0.288 | 0.261 ± 0.049 |

It is also possible to derive a lower bound of ANE ancestry from the model of Fig. S14.20 by the $f_4$-ratio $f_4(Test, Stuttgart; Karitiana, Onge) / f_4(MA1, Stuttgart; Karitiana, Onge)$.

For the denominator:

$$f_4(MA1, Stuttgart; Karitiana, Onge) = \beta(z+\alpha(1-\gamma)y) \qquad \text{(S14.4)}$$

This expectation reflects the fact that the paths MA1→Stuttgart and Karitiana→Onge overlap only when Karitiana descends from Ancient_North_Eurasian (fraction $\beta$). The segment with length $z$ is always traversed by both paths in that case, but the segment with length $y$ is only traversed when Stuttgart has Basal_Eurasian ancestry (fraction $\alpha(1\text{-}\gamma)$).

For the numerator:

$$f_4(Test,\ Stuttgart;\ Karitiana,\ Onge) = \beta(z+\alpha(1-\gamma)y)(1-\delta) - \beta\delta\gamma\alpha y \qquad \text{(S14.5)}$$

The first term in (2) is the same as in (1) multiplied by 1-$\delta$, since 1-$\delta$ fraction of the Test population's ancestry descends from ANE. For the portion of Test's ancestry $\delta$ that comes from Near East, the path Test→Stuttgart does not overlap with Karitiana→Onge except in the case that Stuttgart descends from UHG ($\gamma$ fraction) and the Test population descends from Basal Eurasian ($\delta a$); in all other cases, the path Test→Stuttgart only passes through the Western_Eurasian subtree and is uncorrelated to the Karitiana→Onge one. By dividing (S14.4) by (S14.3) we thus obtain $f_4(Test,\ Stuttgart;\ Karitiana,\ Onge)\ /\ f_4(MA1,\ Stuttgart;\ Karitiana,\ Onge)$ = $1\text{-}\delta\text{-}\delta\gamma\alpha y/(z+\alpha(1\text{-}\gamma)y) \leq 1\text{-}\delta$. The lower bound obtained for these five populations is also shown in Table S14.15.

An implication of this analysis is that ANE-related ancestry may be particularly high in the Northeast Caucasus, as both fitted and lower bound values for Lezgins and Chechens exceed inferred ANE values for Europeans (compare Table S14.10 and Table S14.15). The high affinity of the Northeast Caucasus to MA1 is also demonstrated in Extended Data Fig. 6, where the statistic $f_4(Test,\ Chimp;\ MA1,\ Loschbour)$ exhibits highest values in the region. In light of our other results, it is not surprising that these populations would have high ANE-related ancestry. They are at the northern end of the Near Eastern cline (Fig. 1B) and have the highest values of common genetic drift with MA1 among Near Eastern populations (Extended Data Fig. 4), as measured by $f_4(Test,\ Stuttgart;\ MA1,\ Chimp)$. However, the high MA1-related admixture in Northeast Caucasians seemingly contradicts Extended Data Fig. 4 which shows many Europeans to have even higher values of the statistic.

This is not in fact a contradiction, however, because for Europeans the statistic can be written as:

$$\begin{aligned} f_4(European,\ Stuttgart;\ MA1,\ Mbuti) = {} & \alpha_{EEF} f_4(Stuttgart,\ Stuttgart;\ MA1,\ Mbuti) \\ & + \alpha_{WHG} f_4(Loschbour,\ Stuttgart;\ MA1,\ Mbuti) \\ & + \alpha_{ANE} f_4(B,\ Stuttgart;\ MA1,\ Mbuti) \end{aligned} \qquad \text{(S14.6)}$$

The first term vanishes, and both other terms are positive, since B and MA1 are sister clades and Loschbour and MA1 share drift that Stuttgart lacks because of its basal Eurasian admixture, with $f_4(Loschbour,\ Stuttgart;\ MA1,\ Mbuti)$ = 0.004573 (Z= 6.799).

By contrast for North Caucasians:

$$\begin{aligned} f_4(North\ Caucasian,\ Stuttgart;\ MA1,\ Mbuti) = {} & \alpha_{ANE} f_4(B,\ Stuttgart;\ MA1,\ Mbuti) \\ & + \alpha_{Near_East} f_4(NE1,\ Stuttgart;\ MA1,\ Mbuti) \end{aligned} \qquad \text{(S14.7)}$$

The second term is negative, because $f_4(NE1,\ Stuttgart;\ MA1,\ Mbuti) = -\alpha\gamma(x+y)$.

Intuitively, the shared drift shared between a test population and MA1 is diluted by Near Eastern ancestry (because of the Basal Eurasian ancestry in the Near East), and augmented by WHG ancestry (because of the lack of Basal Eurasian ancestry in Loschbour).

We have conveniently labeled MA1-related ancestry "Ancient North Eurasian" because of the provenance of MA1 in Siberia, but at present we cannot be sure whether this type of ancestry originated there or was a recent migrant from some western region.

Conversely, we do not currently know whether the signal of admixture observed in the Near East and Caucasus reflects an arrival of MA1-related ancestry from the east, or alternatively dilution of native MA1-related ancestry by an expansion of a Near Eastern population carrying Basal Eurasian admixture, associated perhaps with the expansion of Levantine/Mesopotamian early agriculturalists who seem to have influenced the Y-chromosome distribution of the region[27]. Future studies of ancient Central Eurasians may help resolve such questions of migration timing and directionality.

**Concluding Remarks**

We chose to model the 3-way admixture as taking place in the order (Early European Farmers, (West European Hunter Gatherers, Ancient North Eurasians)), but we caution that the order is unknown and may become apparent as later samples from Europe and elsewhere provide ancient DNA for study. Different combinations of the three ancestral populations may well have contributed to the formation of modern Europeans. Nonetheless, our co-fitting of population pairs (Fig. S14.15 and Fig. S14.16) reveals that the WHG/(WHG+ANE) ratio is fairly narrowly constrained over many European populations, so the chosen order seems reasonable. In addition, the consistency of the estimates with those from SI17, which do not specify a branching order, provides further confidence regarding our estimates of ancestry proportion.

A geographically parsimonious hypothesis would be that a major component of present-day European ancestry was formed in eastern Europe or western Siberia where western and eastern hunter-gatherer groups could plausibly have intermixed. Motala12 has an estimated WHG/(WHG+ANE) ratio of 81% (Fig. S14.8), higher than that estimated for the population contributing to present-day Europeans (Fig. S14.15). Motala and Mal'ta are separated by 5,000km in space and about 17,000 years in time, leaving ample room for a genetically intermediate population. The lack of WHG ancestry in the Near East (Extended Data Fig. 6, Fig. 1B) together with the presence of ANE ancestry there (Table S14.15) suggests that the population who contributed ANE ancestry there may have lacked substantial amounts of WHG ancestry, and thus have a much lower (or even zero) WHG/(WHG+ANE) ratio.

It is important to remember that the amount of WHG ancestry indicated in Tables S14.7 and S14.8 is not the total amount of European hunter-gatherer present in these populations, since Early European Farmers also possessed some such ancestry (SI13). Conversely, we assumed that "Hunter" was composed only of WHG/ANE ancestry, but it is possible that the actual population that admixed with EEF may have already possessed EEF ancestry itself. Our results point to three major ancestral source populations for most modern Europeans (Fig. S14.16). However, in the absence of ancient DNA from later periods of European history we cannot determine whether this process of admixture was simple and corresponds to an archaeologically visible event, or was more protracted over time. The fact that late Neolithic farmers still resembled Stuttgart (Fig. 1B) and Early Bronze Age Central Europeans resembled modern Europeans, at least mitochondrially[28], suggests the hypothesis that at least part of the admixture occurred over a relatively short period of time.

Some of our modeling is surely too simplistic and will need to be modified as newer ancient DNA samples become available and make it possible to constrain the model even further. Nevertheless, we are encouraged by the fact that admixture estimates presented in SI17 that do not require modeling of deep history tend to agree with the ones derived here under an explicit model.

In the spirit of parsimony we chose to limit the number of admixture edges to 2 for the main model (Fig. S14.6), as a model with only as many edges could fit the ancient samples, and modern European populations could be accommodated easily in this scaffold (Fig. S14.13 and Fig. 2A). More complex models with 3 or more admixture events could be devised, but cannot be constrained fully by our data as the number of ancient genomes is still small and limited in space and time, with crucial periods and places missing. The study of archaic humans has revealed an ever-increasing complexity of admixture and unexpected links across time and space[8,29-32], and as more ancient DNA samples became available, and it is likely that the story of our more immediate prehistoric ancestors will be shown to be even more complex.

# Supplementary Information 15

## *MixMapper* Analysis of Population Relationships

Mark Lipson*, Iosif Lazaridis and David Reich

* To whom correspondence should be addressed (lipsonm@mit.edu)

To explore models of European and western Eurasian population history involving admixture, we used the *MixMapper* tree-fitting software[1]. *MixMapper* is similar to ADMIXTUREGRAPH[2] in that it builds phylogenetic models of population relationships based on *f*-statistics, but unlike ADMIXTUREGRAPH, it does not require the specification of the tree topology by the user; instead, it determines the best-fitting topology automatically, including the sources of gene flow for admixed populations.

*MixMapper* works in two phases. First, the program uses *f*-statistics to assist in the selection of a *scaffold tree* of populations to be modeled as unadmixed. Then, admixed populations are added to the scaffold tree with optimized mixture parameters. In order to minimize over-fitting, the program only considers simple models: either a single two-way admixed population or a three-way admixed population with ancestry related to a specified two-way admixed population. The uncertainty in all parameter estimates is measured by block bootstrap resampling of the SNP set (100 replicates with 50 blocks). All ranges given in this note represent 95% confidence intervals.

More methodological details can be found in the original *MixMapper* publication[1]. Below, we describe the results of fitting ancient and modern European and related populations.

**Scaffold tree selection**

We attempted to build a scaffold tree containing a subset of the following: four ancient Eurasian populations (Loschbour, MA1, Motala, and Stuttgart); present-day Europeans; Native Americans (represented by Karitiana, a population with no evidence of recent European admixture); eastern Eurasians (represented by the unadmixed Onge); and sub-Saharan Africans (represented by Mbuti, a population that is to first approximation unadmixed relative to non-Africans).

At the first step, we found that all present-day European populations have at least one significantly negative $f_3$ statistic ($Z < -2$, indicating admixture), so we removed them from consideration for the scaffold tree. From among the remaining seven populations, we required Mbuti and Onge to be in the scaffold as outgroups. With this constraint, all possible subsets of four or more populations (the minimum necessary for fitting admixtures) yielded significantly non-additive scaffold trees (analogous to a non-zero $f_4$ statistic) except for three consistent with perfect additivity: {Mbuti, Onge, Loschbour, Motala}, {Mbuti, Onge, MA1, Motala}, and {Mbuti, Onge, Loschbour, MA1}.

Of these three possible most additive scaffold trees, we eliminated the first because Loschbour and Motala are closely related, meaning that this scaffold contains too few diverged populations for our purposes. To choose between the remaining two possible scaffolds, we attempted to fit Motala as admixed onto the {Mbuti, Onge, Loschbour, MA1} scaffold tree and Loschbour as admixed onto the {Mbuti, Onge, MA1, Motala} scaffold tree, with the rationale that the more robust fit would indicate which of Motala and Loschbour is better modeled as unadmixed and which as admixed.

(i) With a scaffold tree of {Mbuti, Onge, Loschbour, MA1}, Motala fit as admixed with 53-81% Loschbour-related ancestry and 19-47% MA1-related ancestry (100% bootstrap support).
(ii) With a scaffold tree of {Mbuti, Onge, MA1, Motala}, Loschbour fit as admixed with 58-79% Motala-related ancestry and 21-42% basal-Eurasian-like ancestry (89% bootstrap support).

The first of these fits is consistent with the observation that $f_4$(Motala, Loschbour; MA1, Mbuti) is significantly positive, as reported in the main text. Conversely, statistics of the form $f_4$(EE, Chimp; Loschbour, Motala), where EE is an eastern Eurasian population, would be expected to be negative if Loschbour were a mixture between components related to Motala and basal Eurasians, but in fact they are consistent with zero or slightly positive (Table S15.1).

***Table S15.1: No evidence for basal Eurasian admixture in Loschbour.***

| Eastern Eurasian pop (EE) | $f_4$(EE, Chimp; Loschbour, Motala) | Z-score |
|---|---|---|
| Onge | 0.000141 | 0.255 |
| Ami | 0.000593 | 1.087 |
| Atayal | 0.000227 | 0.404 |
| Han | 0.000314 | 0.596 |
| Naxi | 0.000174 | 0.326 |
| She | 0.000425 | 0.782 |

In light of these observations, we used the four-population scaffold tree {Mbuti, Onge, Loschbour, MA1} for all subsequent admixture-fitting analyses (Figure S15.1). We also confirmed that our results were similar when using the alterative scaffold tree with Motala in place of Loschbour.

***Figure S15.1: Scaffold tree used to fit admixtures.*** *All tree figures are plotted in units of* $f_2$ *distances*.

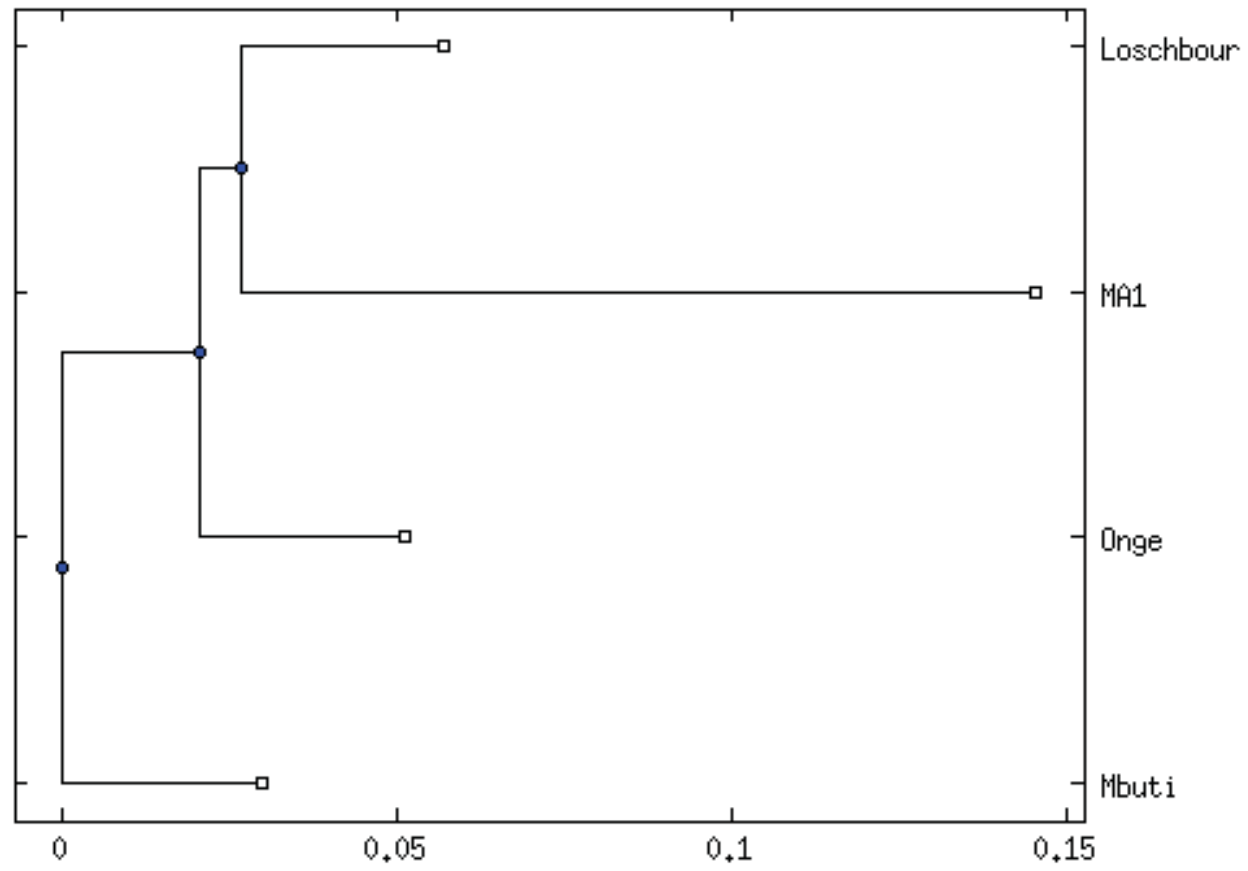

**Motala, Stuttgart, and Karitiana have robust two-way admixture fits**

Of the other populations of interest, we found that Motala, Stuttgart, and Karitiana have confident fits as two-way admixtures (Figure S15.2). As mentioned above, Motala fit as admixed with 53-81% Loschbour-related ancestry and 19-47% MA1-related ancestry (100% bootstrap support for the topology). The final ancient sample, Stuttgart, fit as admixed with 47-79% Loschbour-related ancestry and 21-53% “basal Eurasian” ancestry (99% bootstrap support for the topology). Karitiana, meanwhile, fit as admixed with 34-68% eastern Eurasian (Onge-related) ancestry and 32-66% MA1-related ancestry (100% bootstrap support for the topology).

We note two additional features of these results.

First, the fits are very similar to those obtained with ADMIXTUREGRAPH (see SI14), increasing our confidence in the qualitative inferences.

Second, the inferred branching positions of the mixing populations are informative about historical relationships. Whereas the MA-1 related components in Motala and Karitiana are consistent with being from the same ancestral (Ancient North Eurasian) source, the Loschbour-related components in Stuttgart and Motala appear to be from different ancestral (hunter-gatherer) sources (Figure S15.2).

***Figure S15.2: Three populations fit as two-way admixtures.*** *Dotted lines depict admixture events, shaded triangles are 95% confidence intervals for the split points of mixing populations, and terminal branches for admixed populations have length equal to half of the total estimated off-tree drift.*

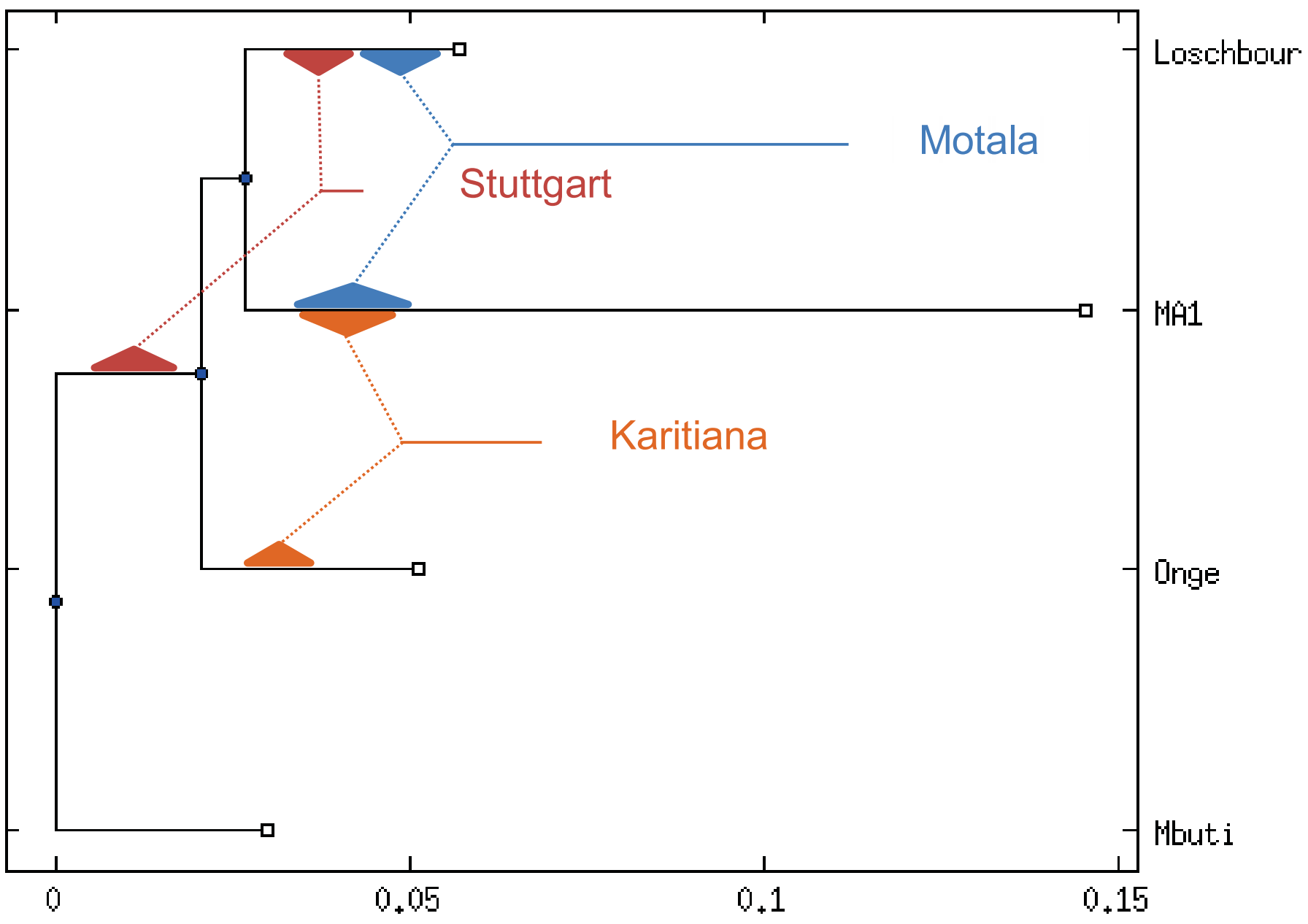

**Three distinct ancestry components for present-day Europeans**
Finally, we searched for the optimal admixture models for representative present-day European populations, for which purpose we selected Sardinian, Tuscan, and French. We began by fitting the Europeans as two-way admixed. For all three populations, the bootstrap replicates were divided between fitting them as having Loschbour-related and basal Eurasian ancestry and having Loschbour-related and Mbuti-related ancestry (Table S15.2). While Sardinians are known to have a very small proportion of recent sub-Saharan African ancestry, the Mbuti-related fits are not plausible. In our experience, this low confidence in the split points, particularly the “spilling over” of ancestry from the basal Eurasian branch to the Mbuti branch, often indicates a more complex history than two-way admixture. Thus, we viewed the two-way fits as being of questionable accuracy.

Next, we explored three-way admixture models. We used a new feature in *MixMapper* v2.0 that allows us to determine whether a test population fits better as two-way or three-way admixed, taking into account the different parameterizations of the models[3]. For the three modern European populations, we compared their fits as two-way admixed versus three-way admixed with one ancestral component related to either Stuttgart or Motala, and for all six tests, the two-way fit was better than the three-way fit. Given that the two-way fits were not convincing, we also concluded that a three-way model with ancestry directly related to Stuttgart or Motala plus one other unadmixed source is not sufficient to explain the data.

***Table S15.2: Questionable two-way admixture fits for present-day Europeans.*** *Split points are given in fractions of branch lengths in the scaffold tree, where 0 is the top of a branch (older) and 1 is the bottom (younger). "Basal Eurasian" is the common ancestral branch of the three non-African groups.*

| European population | Ancestral mixing branch 1 (split point / total branch) | Ancestral mixing branch 2 (split point / total branch) | Bootstrap support | Branch 1 ancestry |
|---|---|---|---|---|
| Sardinian | Loschbour (0.23-0.37) | Basal Eurasian (0.10-0.70) | 93% | 70-87% |
| Sardinian | Loschbour (0.23-0.27) | Mbuti (0.03-1.00) | 7% | 88-95% |
| Tuscan | Loschbour (0.17-0.27) | Basal Eurasian (0.05-0.65) | 67% | 73-90% |
| Tuscan | Loschbour (0.17-0.20) | Mbuti (0-1.00) | 33% | 89-97% |
| French | Loschbour (0.20-0.23) | Basal Eurasian (0.05-0.60) | 14% | 82-92% |
| French | Loschbour (0.17-0.23) | Mbuti (0.03-1.00) | 86% | 93-99% |

While the simple three-way models were not optimal, we examined more carefully the best-fit three-way topologies of the form Stuttgart-related + other, since the two-way fits (Table S15.2) show a close relationship between present-day Europeans and Stuttgart. We found that the three-way admixture models have a very suggestive, bimodal pattern, whereby some bootstrap replicates are optimized with Stuttgart-related and Loschbour-related ancestry, and the rest are optimized with Stuttgart-related and MA1-related ancestry (Table S15.3). *MixMapper* is limited to three-way admixtures based on an intermediate two-way admixed population (here, Stuttgart), but this pattern provides evidence that modern Europeans might be well-modeled with Stuttgart-related ancestry plus both additional Loschbour-related ancestry and MA1-related ancestry (Figure S15.3).

We note that the split points for the extra Loschbour-related ancestry are inferred to overlap with the hunter-gatherer ancestry in Stuttgart (Figure S15.2 and Table S15.3). Thus, the most plausible admixture model for present-day Europeans appears to involve three ancestral components, which can be defined either as Stuttgart-, Loschbour-, and MA1-related (EEF, WHG, and ANE, as in Figure 2), or as basal Eurasian, total hunter-gatherer, and MA1-related (with the first two in different relative proportions than in Stuttgart). Moreover, while the inferred mixture proportions in Table S15.3 come from three-way fits and are thus not exactly correct, the relative values indicate a gradient of ancestry, with Sardinians having the most Stuttgart-related and least MA1-related ancestry, in agreement with our full estimates (e.g., Figure 2B). Our analysis here does not indicate the sequence of the mixture events, so the history could have taken several different forms, perhaps involving admixture between early farmers (with Loschbour-related and basal Eurasian ancestry) and a second admixed population (with its own Loschbour-related ancestry component, plus MA1-related ancestry), as in Figure 2A.

## Conclusions

Using *MixMapper*, we have shown that our diverse set of Eurasian populations is best modeled with Onge, Loschbour, and MA1 as unadmixed; Karitiana, Motala, and Stuttgart as two-way admixed; and present-day Europeans as three-way admixed. This history recapitulates our inferences from ADMIXTUREGRAPH (Figure 2A, Figure S14.11), making us more confident that our admixture graphs and parameter estimates are robust and accurate.

***Table S15.3: Three-way admixture fits for modern Europeans.*** *Split points are as defined in Table S15.2. All models assume one ancestry component related to the admixed Stuttgart branch.*

| European population | Ancestral mixing branch 3 (split point / total branch) | Bootstrap support | Stuttgart-related ancestry |
|---|---|---|---|
| Sardinian | Loschbour (0.17-0.70) | 80% | 74-92% |
| Sardinian | MA1 (0.82-1.00) | 20% | 98-99% |
| Tuscan | Loschbour (0.07-0.23) | 55% | 60-78% |
| Tuscan | MA1 (0.03-0.51) | 45% | 84-96% |
| French | Loschbour (0.17-0.30) | 71% | 54-64% |
| French | MA1 (0.23-1.00) | 29% | 92-98% |

***Figure S15.3: The most plausible history for modern Europeans involves three components.*** *Dashed lines show sources of ancestry; all other notations are as defined in Figure S15.2.*

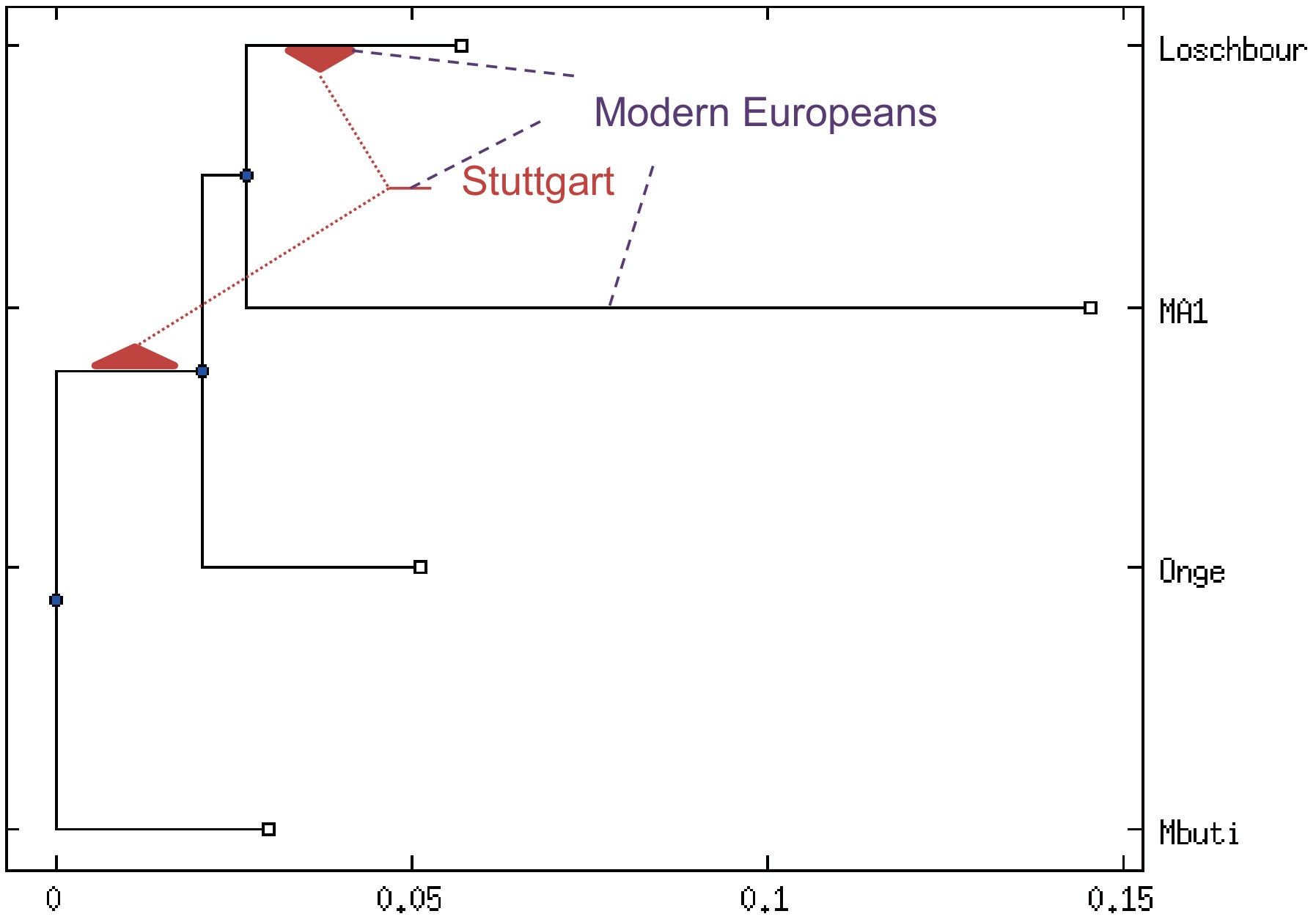

# Supplementary Information 16

***TreeMix* analysis of population relationships**

Gabriel Renaud†*, Iosif Lazaridis†, Janet Kelso and David Reich

* To whom correspondence should be addressed (gabriel_renaud@eva.mpg.de)
† Contributed equally to this section

We used the *TreeMix*[1] software to develop models of population relationships that also allow for multiple admixture events. *TreeMix* takes as input SNP genotype data from any number of populations, and then identifies a phylogenetic tree incorporating a specified number of admixture events that minimizes the difference between the observed and predicted *f*-statistics.

A limitation of *TreeMix*—along with the *MixMapper* method (SI15)—is that it does not allow users to explicitly specify models of population relationships and formally test the goodness of their fit to data.

An advantage of *TreeMix* and *MixMapper* is that they are unsupervised procedures, and hence are less vulnerable to the concern that the prior expectations about human history of the researchers using them will bias the results. Unsupervised methods can also be used to infer models of relationships for more populations than ADMIXTUREGRAPH, as ADMIXTUREGRAPH requires manual exploration of model space (SI14).

Here we apply *TreeMix* both to Human Origins genotype data and to whole genome sequence data[2].

**Human Origins genotype data**

We applied *TreeMix* to Loschbour, Stuttgart, Motala12, MA1[3], LaBrana[4] and the Iceman[5], and also included Karitiana, Onge and Mbuti as representatives of non-West Eurasians. We restricted to 265,521 sites after excluding SNPs where there were no-calls in any of the studied individuals. In order to retain enough SNPs, we did not include the lower-coverage ancient samples in Fig. 1B.

With no admixture edges (Fig. S16.1, top), *TreeMix* groups Onge and Karitiana (eastern non-Africans) and places MA1 in a basal position with respect to West Eurasians. Loschbour groups with LaBrana to the exclusion of Motala12, consistent with these two samples being part of a West European Hunter-Gatherer meta-population. Stuttgart groups with the Iceman, consistent with being part of an Early European Farmer meta-population. However, the residuals indicate a poor fit for the allele frequency differentiation statistics $f_2$*(Stuttgart, Mbuti)* and $f_2$*(Iceman, Mbuti)* for the ancient farmers, and $f_2$*(Karitiana, MA1)* and $f_2$*(MA1, Motala12)* involving MA1 and Karitiana/Motala12.

Allowing a single admixture event (Fig. S16.1, middle) finds evidence of mixture in the Karitiana. Karitiana is grouped with MA1 but with an estimated 48% admixture from a population related to the Onge; this resolves the poor fit of $f_2$*(Karitiana, MA1)*, and is consistent with the ADMIXTUREGRAPH and $f_4$-ratio estimates of ancestry proportion in SI14. Stuttgart, the Iceman and Motala12 continue to fit poorly.

Allowing two admixture events (Fig. S16.1, bottom) preserves the MA1→Karitiana mixture (48%) and also adds a basal Eurasian (42%) mixture into the (Stuttgart, Iceman) ancestral population which resolves the poor-fitting $f_2$ statistics involving these early farmers. The most striking discrepancy in the plot of residuals involves the statistic $f_2$*(MA1, Motala12)*.

Allowing three admixture events (Fig. S16.2, top) preserves the Onge→Karitiana mixture (48%) and also basal Eurasian→(Stuttgart, Iceman) mixture (32%), and also adds an MA1→Motala12 admixture (15%). Thus, *TreeMix* recapitulates the three main features of the data also found in our primary analysis. The relationship of Loschbour and Motala12 is the most discrepant based on the residuals.

Allowing four admixture events (Fig. S16.2, middle) preserves the basal Eurasian admixture (38%) into (Stuttgart, Iceman), the MA1 admixture into Motala12 (18%), the MA1 admixture into Karitiana (50%), and also suggests that LaBrana, while forming a clade with Loschbour, may have 11% basal Eurasian admixture itself. The finding that LaBrana belonged to the Y-chromosome haplogroup C-V20, which is extremely rare in present-day Europeans, could in principle be consistent with its possessing a deeply diverged Eurasian ancestry lacking in Loschbour and Motala12, which both belong to the fairly common Y-chromosome haplogroup I (SI5). However, a direct test fails to confirm a signal of more Basal Eurasian ancestry in LaBrana than in Loschbour. Specifically, we computed D-statistics of the form *D(LaBrana, Loschbour; Dai, Chimp)* and *D(LaBrana, Loschbour; Papuan, Chimp),* which should be non-zero if there is Basal Eurasian ancestry in LaBrana. These are non-significant ($|Z|<1.1$ in all comparisons) (Extended Data Table 2). We speculate that the signal of Basal Eurasian ancestry in LaBrana may instead be driven by the history of admixture between WHG-related hunter-gatherers and Near Eastern farmers that formed the EEF (SI13), a signal that is not detected by *TreeMix*. If the mixing WHG were more closely related to Loschbour than to LaBrana, as indeed is suggested by the fact that the statistic *D(LaBrana, Loschbour; Stuttgart, Chimp)* is $Z = -2.6$ (Extended Data Table 2), this could cause *TreeMix* to compensate by modeling Basal Eurasian gene flow into LaBrana. However, the same statistic is $Z = -1.8$ when restricting to transversions in whole genome sequencing data that are not affected by C→T and G→A errors (Extended Data Table 2), which is not formally significant. The study of additional ancient genomes from pre-Neolithic Europeans should improve power to determine which Mesolithic Europeans were most closely related to the WHG-related ancestors of Stuttgart.

For the sake of completeness, we also include the results of the *TreeMix* analysis for five admixture events (Fig. S16.2, bottom). This infers a small (~3%) admixture from LaBrana into the Iceman lineage, which historically would not be surprising given the opportunity for some WHG-related gene flow into European farmers in the thousands of years after they arrived. This, however, must have been relatively small, both based on the percentage inferred by TreeMix and the fact that the Iceman formally fits as a clade with Stuttgart in the analysis of SI14.

***Figure S16.1: TreeMix analysis with 0, 1 and 2 admixture events for Human Origins data***

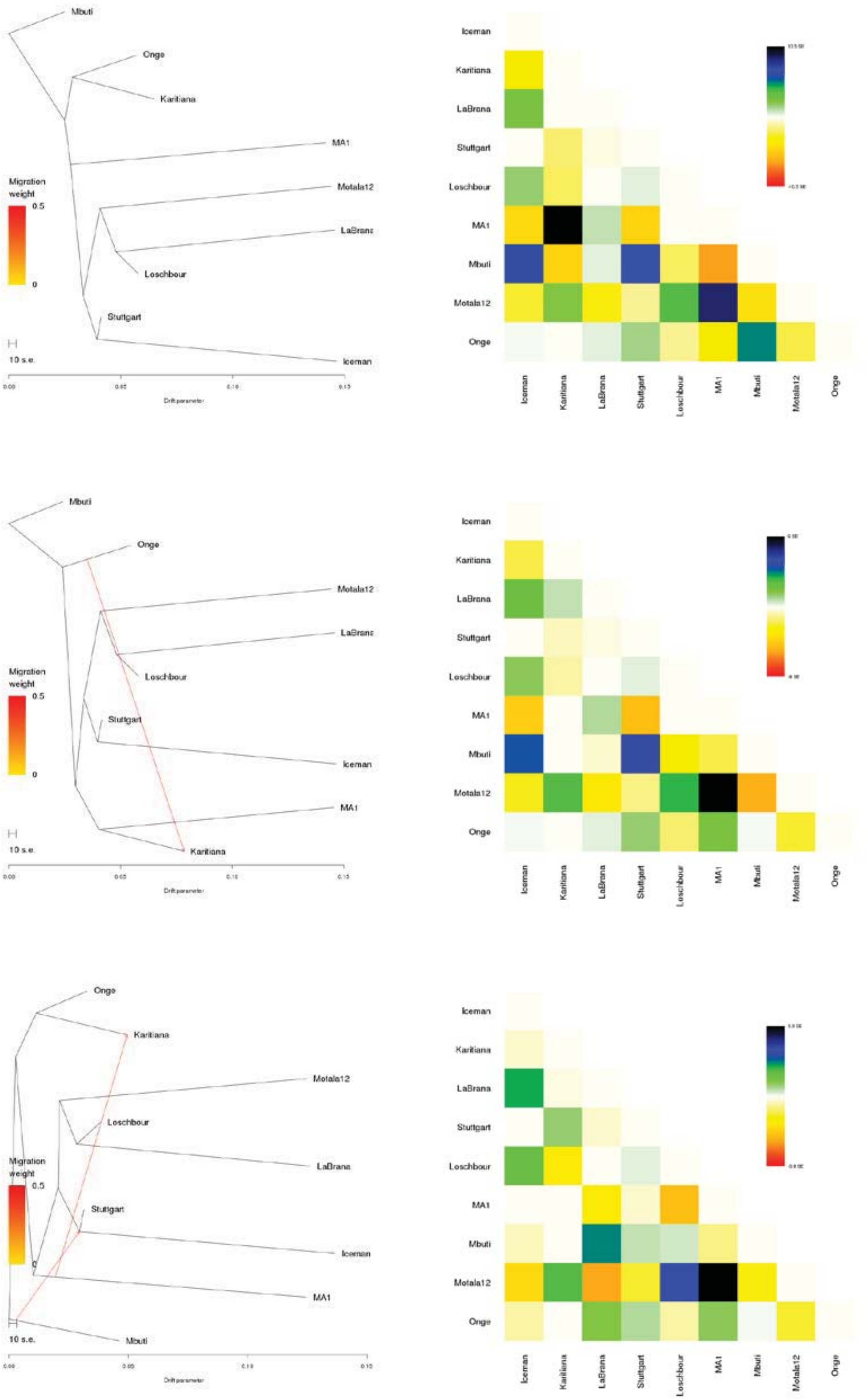

***Figure S16.2: TreeMix analysis with 3, 4 and 5 admixture events for Human Origins data***

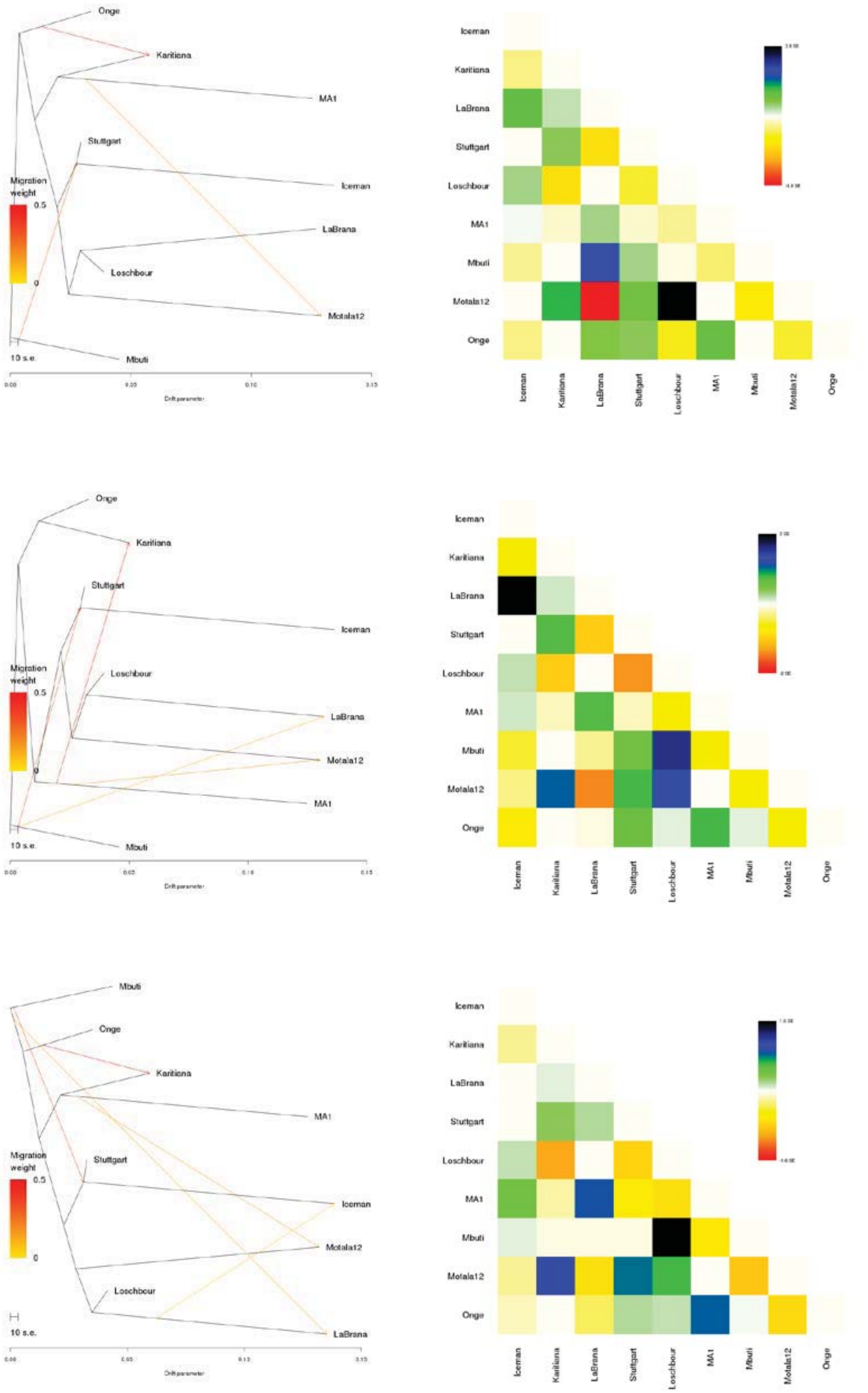

**Whole genome sequence data**
In the previous section, we inferred population relationships using the Human Origins genotyping dataset. However, such genotyping datasets are affected by ascertainment bias due to how the SNPs were chosen for the array[6]. They can sometimes also be limited in statistical power in comparison to whole genome shotgun sequencing datasets. We therefore used the *TreeMix* framework as an opportunity to test the robustness of our inferences about population history using whole genome sequencing data.

We used the genotypes extracted from the Stuttgart, Loschbour and Motala whole genome sequences as described in SI2 to infer allele information for the ancient samples. To represent present-day humans, we used the Mbuti, Dai, Karitiana individuals from the B panel of present-day humans described in Prüfer et al[2], generating diploid genotype calls on these individuals using the same method as for the ancient samples. In addition, we genotyped (again, using the same methodology described in SI2) the ancient MA1 Upper Paleolithic Siberian[3] and Iberian LaBrana[4] samples. We restricted analysis to sites that were biallelic across the analyzed samples. Due to the limited amount of Motala data, we pooled Motala individuals into a single population. We used chimpanzee to root the tree, and obtained the chimpanzee allele call from EPO[7] alignments derived from Ensembl v69. We obtained allele frequencies from the PHRED likelihood (PL) fields of the VCF files. If the most likely genotype is ~2000 times (corresponding to a difference in PHRED likelihood of 33) more likely than all remaining ones (e.g.,homozygous for the reference being more likely than heterozygous or homozygous alternative), we used this genotype for allele frequency inference. The frequency was derived from the genotype by counting 2 alleles for homozygous sites and 1 for heterozygous sites. If a genotype could not be inferred with sufficient confidence, we inferred a single allele if the allele represented by one of the two homozygous genotypes was ~2000 times more likely than the second one. We used default parameters for *TreeMix*, and varied the number of admixture events from 0 to 5. Out of 1.1 billion sites that passed filters, we retained 384,115 segregating sites as input for *TreeMix*.

When we do not allow any admixture events (Figure S16.3), the Loschbour, LaBrana and Motala samples cluster together while Stuttgart is an outgroup. The clearest signals are the underestimation of $f_2$*(MA1, Karitiana)*, $f_2$*(MA1, Motala12)*, and $f_2$*(Stuttgart, Mbuti)* by the fitted model as indicated by the plot of residuals.

When we allow for a single admixture event (Figure S16.3), the program adds gene flow from a population basal to non-Africans to the Stuttgart sample (47%). This new migration edge accounts for the excess of eastern non-African related alleles in all hunter-gatherers compared to Stuttgart.

When we allow for two migrations (Figure S16.3), we infer a migration from MA1 to Karitiana (40%) in addition to the Basal Eurasian admixture into Stuttgart (49%).

When we allow three migrations (Figure S16.4), we observe an extra Basal Eurasian migration edge into LaBrana (17%) as well as MA1 admixture into Karitiana (39%) and Basal Eurasian admixture into Stuttgart (52%) with the remainder (48%) from the (Loschbour, LaBrana) common ancestor.

When we allow four migrations (Figure S16.4), there is 44% Basal Eurasian admixture into Stuttgart with the remainder 56% of Stuttgart's ancestry from the ((Loschbour, LaBrana), Motala) common ancestor. We also observe 41% of MA1 admixture into Karitiana, 15% Basal Eurasian admixture into LaBrana, and a new admixture from MA1 into Motala (11%). As mentioned above, we do not find compelling evidence that LaBrana and Loschbour differ in their relationship to eastern non-Africans using a formal D-statistic test.

When we allow five migration events, we observe a small 1.5% admixture from MA1 into Dai.

We were concerned that ancient DNA deamination, which causes an elevated rate of C→T and G→A errors in ancient DNA data, could lead *TreeMix* to incorrect inferences of population relationships (e.g., it could result in clustering of ancient samples). Restring our dataset to 114,187 transversion

SNPs, we ran *TreeMix* again for 0 to 5 migrations (data not shown). This analysis recapitulates MA1→Karitiana and Basal Eurasian→Stuttgart admixture, but explains greater affinity of MA1 to Motala than to Loschbour and LaBrana by postulating that the (Loschbour, LaBrana) WHG group has Basal Eurasian ancestry itself (which indirectly makes MA1 more similar to Motala than to WHG). We do not find this evidence compelling because if WHG had Basal Eurasian ancestry that Motala (SHG) lacked, then statistics of the form $f_4$*(Loschbour or LaBrana, Motala12; Eastern non-African, Chimp)* should be negative. However, they appear consistent with zero in either the comprehensive search in SI14, or the comparison of D-statistics of the same form in Extended Data Table 2 using whole genome transversions in either genotype or whole genome sequence data.

Taken together, the *TreeMix* analyses on the whole genome sequencing data are qualitatively consistent with those in the genotyping data, as well as the results inferred in the other sections of this paper.

## Comparison of admixture estimates

Table S16.1 compares admixture estimates when allowing for m=4 migration edges in the *TreeMix* analysis. For the sake of comparison, we also include estimates of the same quantities obtained using $f_4$-ratio estimation (SI14, Fig. 2A), model-fitting (SI14) and *MixMapper* analysis (SI15).

***Table S16.1: Comparison of admixture estimates obtained with different methods.***

| | ***TreeMix*** | | **$f_4$-ratio on genotype** | **ADMIXTUREGRAPH** | ***MixMapper*** |
|---|---|---|---|---|---|
| | **Genotype** | **Genome** | **(SI14)** | **(SI14; Fig. S14.7, S14.8)** | **(SI15)** |
| Basal Eurasian→Stuttgart | 0.38 | 0.44 | 0.44 | 0.34 | 0.21-0.53 |
| MA1→Karitiana | 0.50 | 0.39 | 0.41 | 0.47 | 0.32-0.66 |
| MA1→Motala | 0.18 | 0.11 | | 0.19 | 0.19-0.47 |
| Basal Eurasian→LaBrana | 0.11 | 0.17 | | | |

Perfect agreement between the different methods is not expected given (i) the different methodology, (ii) the different data used (e.g., the whole genome-based analysis uses Dai instead of Onge to represent eastern non-Africans due to the lack of whole genome sequence data from the latter), and (iii) the inherent statistical uncertainty in estimating admixture proportions.

## Conclusion

The *TreeMix* analyses on genotype and sequence data agree with each other and with ADMIXTUREGRAPH (SI14) and *MixMapper* (SI15) in inferring the major events discussed in this paper (Basal Eurasian admixture into early farmers, MA1 admixture into Native Americans, and Ancient North Eurasian admixture into Motala).

The *TreeMix* analysis also raises additional possibilities about further gene flows. These should be possible to investigate further as UDG-treated data become available from southern European samples related to LaBrana and the Iceman.

We caution that the methods used in this paper for inferring population relationships are far from independent, as they all rely on the study of *f*-statistics. Nevertheless, we are encouraged that they arrive at qualitatively similar inferences using different model-fitting methodologies.

***Figure S16.3: TreeMix analysis for 0, 1 and 2 admixture events using whole genome data***

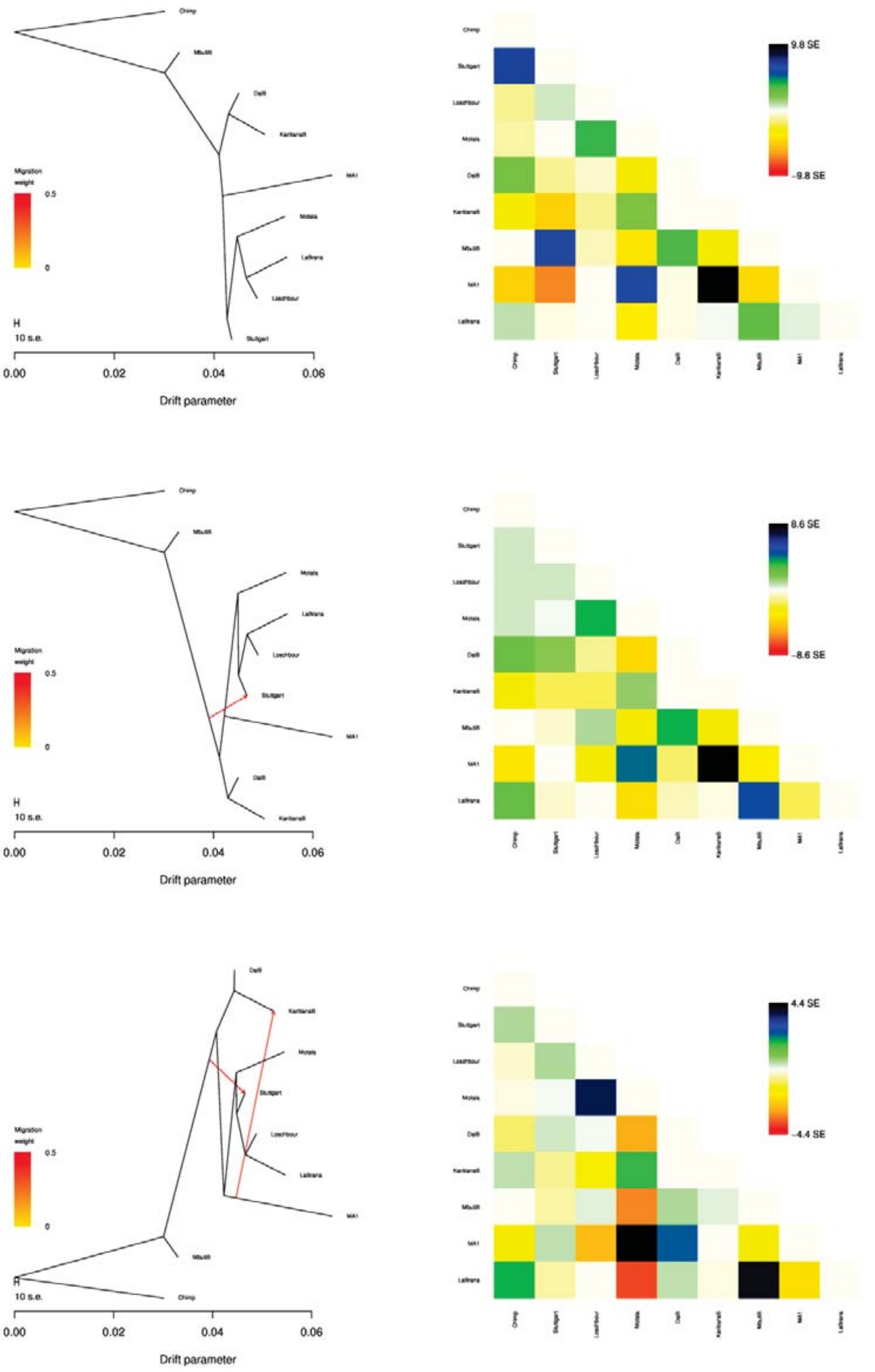

***Figure S16.4: TreeMix analysis for 3, 4 and 5 admixture events using whole genome data***

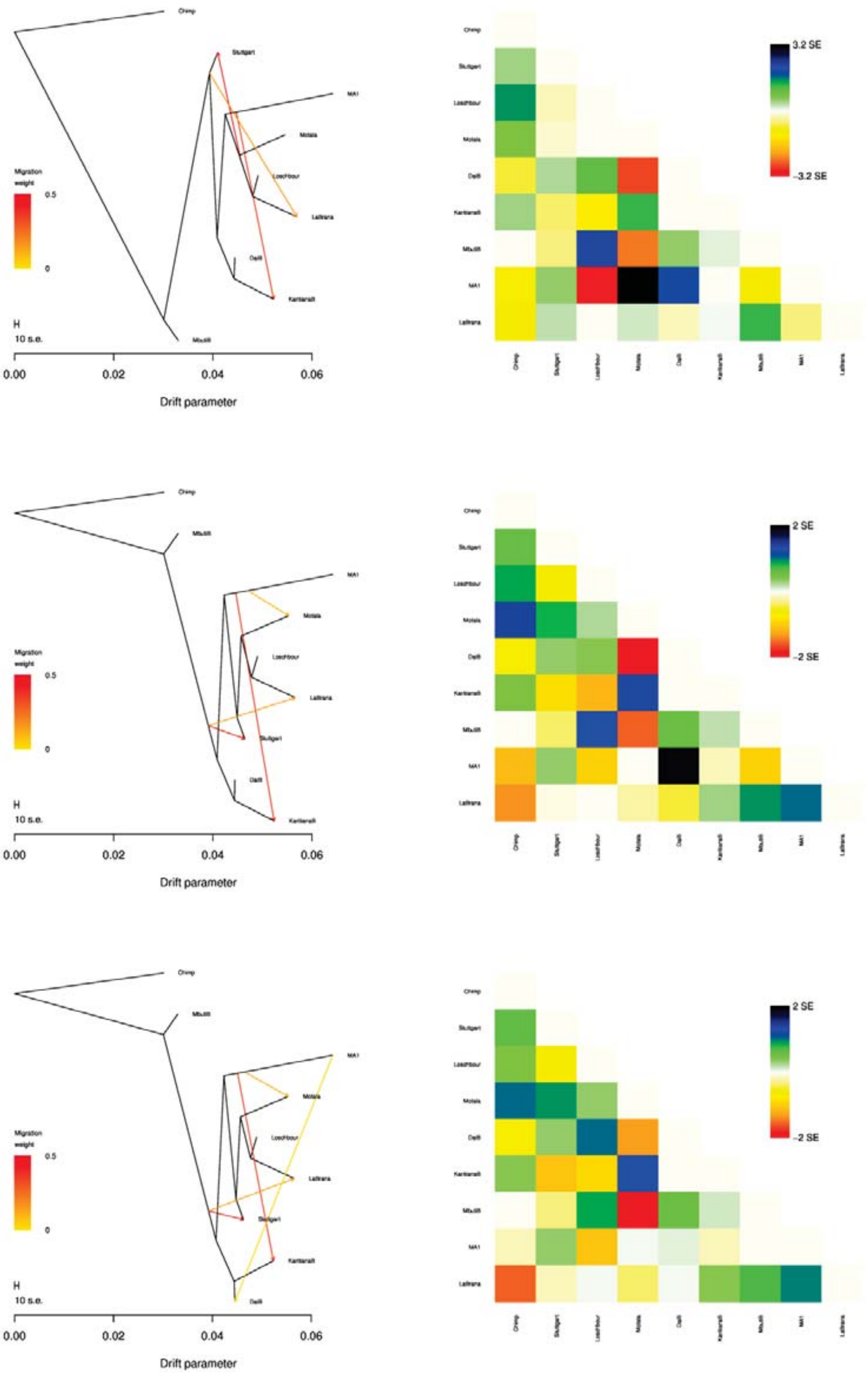

# Supplementary Information 17

## Admixture estimates that do not require phylogenetic modeling

Iosif Lazaridis*, Nick Patterson and David Reich

* To whom correspondence should be addressed (lazaridis@genetics.med.harvard.edu)

In SI14 we identify a plausible model of the relationships of deeply diverged non-African populations that does not contradict the data to within the limits of our resolution, and then use this model to derive admixture proportions. One consequence of our modeling is to show that a range of puzzling observations can be reconciled with the evidence if one postulates that at least one "ghost" population ("Basal Eurasians") contributed to present-day West Eurasian populations. In SI14 we also show that another such "ghost" population ("Ancient North Eurasians") can be reconciled with the recently published Paleolithic MA1 sample from Siberia[1].

In this section we estimate mixture proportions for European populations in a way that does not require making assumptions about the deep phylogenetic relationships among non-African populations. One advantage of this is that it avoids errors that might arise due to forcing a set of populations into an explicit model. A second advantage is that it can be applied over a large number of world populations without precisely modeling events taking place outside West Eurasia.

We first estimate admixture proportions of European populations in terms of the two prehistoric Europeans (Loschbour and Stuttgart). Loschbour-related admixture appears to be general across Europe, on the basis of (i) the intermediate position of Europeans between Loschbour and the Near East (Fig. 1B), (ii) the fact that population pairs of the form (X=Loschbour, Y=Near East) often produce the lowest $f_3$*(European; X, Y)* statistics (Table 1, Extended Data Table 1, SI11), and (iii) the fact that Europeans have a positive $f_4$*(European, Stuttgart; Loschbour, Chimp)* statistic (Extended Data Fig. 4). Stuttgart-related admixture is a reasonable starting hypothesis because of (i) the geographical importance of the Linearbandkeramik as the first food producing culture in large parts of continental Europe[2], (ii) mtDNA evidence suggesting substantial persistence of early farmer lineages in present-day Europeans[3], (iii) the fact that many Europeans have very negative $f_3$*(European; Stuttgart, MA1)* statistics (Table 1, Extended Data Table 1), (iv) the existence of Stuttgart/Sardinian-like individuals from a wide geographical range in Europe and from different times,[4,5] and (v) the existence of the "European cline" in Fig. 1B which strongly suggests that many European populations were formed by admixture of a Stuttgart/Sardinian-like population and an unknown element mostly concentrated today in northern Europe.

Our approach (Fig. S17.1) is to study statistics of the form $f_4$*(European, Stuttgart;* $O_1$*,* $O_2$*)* where $O_1$, $O_2$ are two non-West Eurasian populations from a set of 15 populations without any evidence of recent European admixture (SI9). This assumption is necessary because this statistic can be interpreted[6] as the drift path overlap between *European→Stuttgart* and $O_1 \rightarrow O_2$. If, say, $O_1$ has recent admixture from a French source, then the value of the statistic will be higher when *European=French* than when *European=Russian*, because of the additional common drift shared with the French, and not because the French and the Russians are differentially related to the non-recently mixed portion of $O_1$. A similar problem arises if a test European population has recent admixture from $O_1$, or $O_2$. For example, recent Native American admixture ancestry will result in the statistic's value not only being affected by the relationship of the constituent elements to Native Americans, but also by the substantial common drift that ensued in the Americas down to the present.

In Extended Data Fig. 4, we plot the statistics $f_4$*(West Eurasian, Stuttgart; MA1, Chimp)* vs. $f_4$*(West Eurasian, Stuttgart; Loschbour, Chimp)*. Both Near Eastern and European populations are often positive for the first statistic (suggesting MA1-related gene flow in both Europe and the Near East), but only Europeans are positive for both, consistent with the hypothesis that Europeans have pre-Neolithic hunter-gatherer related ancestry. Europeans form a cline of increasing common drift with both Loschbour and MA1, so we will derive them as a mixture of the following elements:

- 1-$\alpha$ fraction of ancestry of early European farmers (EEF): a sister group of Stuttgart

- $\alpha$ ancestry fraction of "Hunter", a population itself a mixture of:
    - $\beta$ ancestry fraction of Loschbour-related west European hunter-gatherers (WHG)
    - 1-$\beta$ ancestry fraction of MA1-related Ancient North Eurasians (ANE)

Thus, we can write:

$$\text{European} = (1\text{-}\alpha)\ \text{EEF} + \alpha(\beta\text{WHG}+(1\text{-}\beta)\text{ANE}) \qquad \text{(S17.1)}$$

The above equation describes the fraction of ancestry inherited from a population of Stuttgart-like Early European Farmers and Loschbour/MA1-like pre-Neolithic hunter-gatherers, but this does not necessarily correspond to the actual populations historically involved. It is possible that this admixture took place in stages, so that, for example, the actual population responsible for the WHG/ANE ancestry in Europe already had some EEF ancestry. It should be possible to gain insight into the populations that were actually involved in these mixtures through ancient DNA studies of later periods of European history. However, our estimated $\alpha$ and $\beta$ should correctly correspond to the ancestry proportions from the deep ancestors even in this case.

***Figure S13.1: Admixture estimation that makes minimal assumptions about phylogeny.*** *We assume only that the three admixing populations (WHG, ANE, EEF) are sister groups of the ancient individuals (Loschbour, MA1, Stuttgart) and that these are related in complex (but not modeled) ways with a set of outgroups. By exploiting correlations of* $f_4$*-statistics involving the ancient individuals and outgroups, we can estimate admixture proportions.*

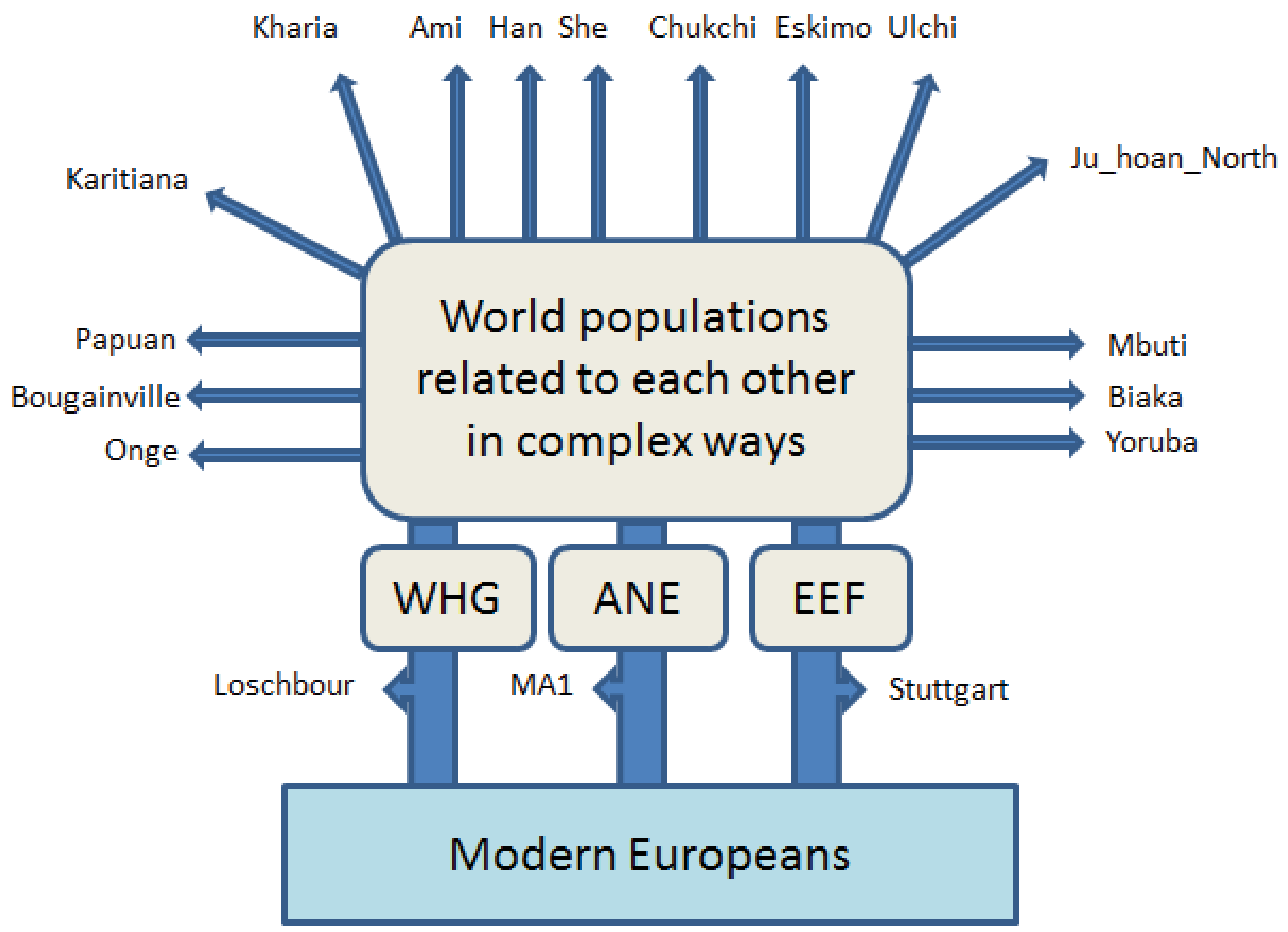

We can write down an $f_4$-statistic involving Europeans and Stuttgart on the left-hand side as follows:

$$f_4(\text{European, Stuttgart}; O_1, O_2) = \qquad \text{(S17.2)}$$
$$= \alpha\beta\ f_4(\text{Loschbour, Stuttgart}; O_1, O_2) + \alpha(1\text{-}\beta)\ f_4(\text{MA1, Stuttgart}; O_1, O_2)$$

The 1-$\alpha$ term has vanished because EEF and Stuttgart form a clade so their allele frequency differences are uncorrelated to any of the outgroups. Using the 15 non-West Eurasians, we obtain 105=15×14/2 $(O_1, O_2)$ pairs and thus 105 equations of the above form. We can then fit using least squares for the coefficients $A=\alpha\beta$ and $B=\alpha(1-\beta)$ and thus estimate $\beta_{est} = 1/(1+B/A)$ and $\alpha_{est} = A+B$. The estimated mixture proportions are EEF=1- $\alpha_{est}$, WHG= $\alpha_{est}\beta_{est}$, ANE= $\alpha_{est}(1-\beta_{est})$. We estimate standard errors using a Block Jackknife[7] dropping one chromosome at a time[8].

The results are shown in Extended Data Table 3 together with other mixture estimates. We observe no systematic bias compared with the model-based estimates of SI14, as revealed by the number of standard errors by which the two estimates differ. None of the estimates differ by more than 2.1 standard errors. The mean and standard deviation of the estimate differences for the different ancestral populations are 0.8±0.73 (EEF), -0.58±0.7 (WHG), and 0±0.74 (ANE) standard errors.

We conclude that the method presented in this note and the fully model-based method presented in SI14 produce similar estimates for these populations, suggesting that the simple model devised in SI14 using Mbuti, Onge, Karitiana as the only non-west Eurasian populations and only two admixture events (basal admixture in Stuttgart and Ancient North Eurasian admixture in Karitiana) may capture some essential features of deep Eurasian prehistory.

Extended Data Table 2 includes, for completeness, aberrant estimates for seven populations. We discuss the evidence for East Eurasian ancestry in Finns, Mordovians, and Russians in SI14; such ancestry is not accounted for in Equation S17.1, which assumes that all the ancestry of populations is EEF/WHG/ANE-related. The effect on the parameter fit is to produce negative EEF admixture; this is not surprising in view of Extended Data Fig. 6 which shows that Finns, Mordovians, and Russians differ from Stuttgart and most Europeans in sharing additional drift with Han. The $f_4$-statistics used by our method are influenced both by the distant relationship of EEF/WHG/ANE to East Asians, and the more recent common drift shared by Finns, Mordovians, and Russians with some of them. Estonians exhibit the greatest discrepancy between the ancestry estimates from the full phylogenetic modeling in SI14 and the minimal phylogenetic modeling reported in this note (2.1 standard errors less EEF ancestry). Their geographic proximity to the other far-northeastern European populations, combined with the weakly significant signal, suggests that they may harbor some of this ancestry as well.

Three other populations produce anomalous estimates in Extended Data Table 2: Ashkenazi Jews, Sicilians, and Maltese. We observed in SI14 that these populations cannot be co-fit in the same admixture graph with most other Europeans, and this suggests that they do not fully trace their ancestry to the same EEF/WHG/ANE elements as most of Europe. Further evidence for this is presented in Extended Data Fig. 4 where all three populations have a negative value of $f_4(Test, Stuttgart; Loschbour, Chimp)$, and thus are inconsistent with them being populations of Stuttgart-related ancestry with additional Loschbour-related input, since such populations would have a zero or positive value of the statistic, as most Europeans do. All three populations strongly deviate towards the Near East in Extended Data Fig. 4 and Fig. 1B, and it is likely that they possess Near Eastern ancestry that is not mediated via Stuttgart. Finally, the Spanish produce a barely negative -0.015 ± 0.165 estimate of WHG ancestry which is suggestive that their ancestry is also not fully accounted by the EEF/WHG/ANE admixture; in SI14 we show that the Spanish may possess ancestry (likely from Africa) that may contribute to this discrepancy.

In conclusion, the admixture estimates reported in this note show reasonable concordance with the fully model-based ones of SI14 for populations that have no evidence of additional ancestry beyond that which is represented by Stuttgart, Loschbour, and MA1. Additionally, populations that produce anomalous results in the present estimation coincide with those that fail to fit the model-based one, giving us more confidence in the results of both methods.

# Supplementary Information 18

## Segments identical due to shared descent between present-day and archaic samples

Joshua G. Schraiber*, Montgomery Slatkin

*To whom correspondence should be addressed (jgschraiber@berkeley.edu)

We analyzed the sharing of tracts of identity by descent (IBD) between present-day and ancient samples by using the POPRES SNP genotyping dataset[1], along with sequence data generated for the analysis of the Denisova individual[2].

For every SNP in the POPRES dataset, we used the genotype calls for Loschbour and Stuttgart generated by the Genome Analysis Toolkit (GATK)[3] (SI 2). We detected likely segments of IBD using RefinedIBD as implemented in BEAGLE 4[4] with the settings "ibdtrim=20" and "ibdwindow=25". We kept all IBD tracts spanning at least 0.5 centimorgans (cM) and with a LOD score > 3. We note that in fact we are detecting segments that are identical by state (IBS), but previous studies have shown that they correlate strongly to IBD segments[5].

We quantified IBD sharing in two ways. First, we measured the average number of IBD blocks shared between two populations, $P_i$ and $P_j$,

$$S_{ij} = \frac{\sum_{k \in P_i} \sum_{l \in P_j} N_{kl}}{n_i n_j} \qquad \text{(S18.1)}$$

where $n_i$ is the number of individuals in population *i, k* and *l* index individuals, and $N_{kl}$ is the number of IBD blocks shared between individuals *k* and *l*.

As a second quantification of IBD sharing, we measured the average length of IBD blocks shared between populations $P_i$ and $P_j$,

$$S_{ij} = \frac{\sum_{k \in P_i} \sum_{l \in P_j} L_{kl}}{n_i n_j} \qquad \text{(S18.1)}$$

where again $n_i$ is the number of individuals in population *i, k* and *l* index individuals, and $L_{kl}$ is the number of IBD blocks shared between individuals *k* and *l*.

We detected substantial IBD sharing between present-day populations, replicating the results of ref.[5]. In addition, our method inferred IBD sharing between the ancient and present-day samples, as measured by both quantifications of IBD (Figures S18.1, S18.2).

We examined in detail the distribution of IBD sharing between present-day and ancient populations, and in Tables S18.1 and S18.2 report the top 10 present-day populations that share IBD blocks with Loschbour and Stuttgart according to number of IBD blocks or length of IBD blocks, respectively. According to ref.[5], most IBD sharing between present-day populations is due to ancestors living in the last 2-3 thousand years. On the surface, our results suggest that IBD sharing can potentially last for substantially longer.

We hypothesize that our detection of segments of IBD beyond the threshold of the population separation time highlighted in ref.[5] is likely due to these being segments of the genome that have very low recombination rates, allowing signals of IBD to persist over longer times (as a larger physical distance span is available for detection).

Alternatively, it is possible that some of the evidence for IBD is artifactual due to shared selective sweeps in a common ancestral population (IBS), which result in false-positive signals of IBD sharing as it is in fact difficult to detect real differences between any haplotypes in the region.

***Figure S18.1. Histogram of IBD sharing between ancient and present-day samples.*** *In each panel, a histogram of the average number of IBD blocks shared between either Loschbour (panel A, mean = 3.18) or Stuttgart (Panel B, mean = 3.01) and present-day populations is shown.*

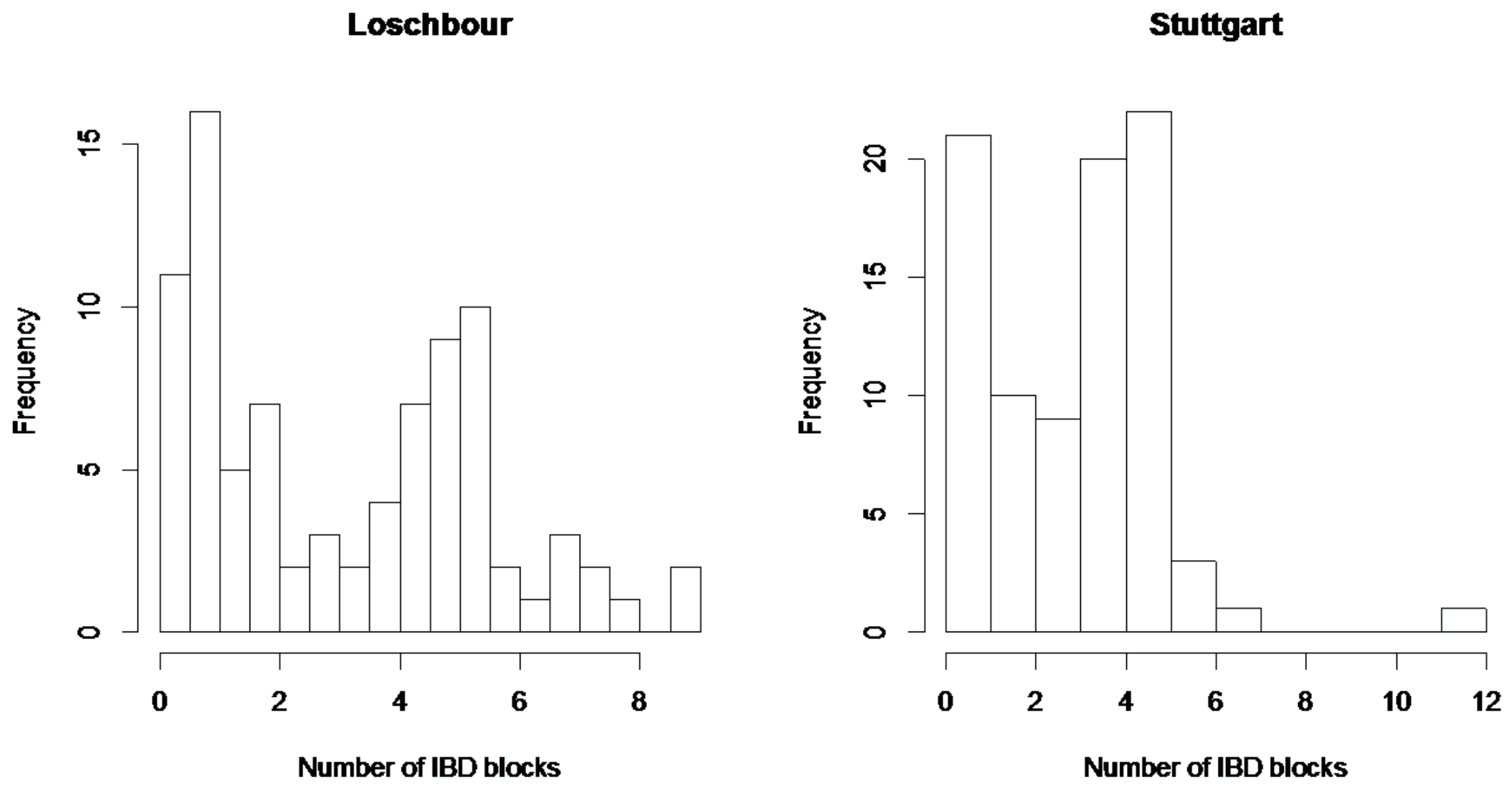

***Figure S18.2. Histogram of IBD sharing between ancient and present-day samples.*** *In each panel, a histogram of the average length of IBD (in base pairs) shared between either Loschbour (panel A, mean = 4.45x10$^{6}$) or Stuttgart (Panel B, mean = 4.12x10$^{6}$) and present-day populations is shown.*

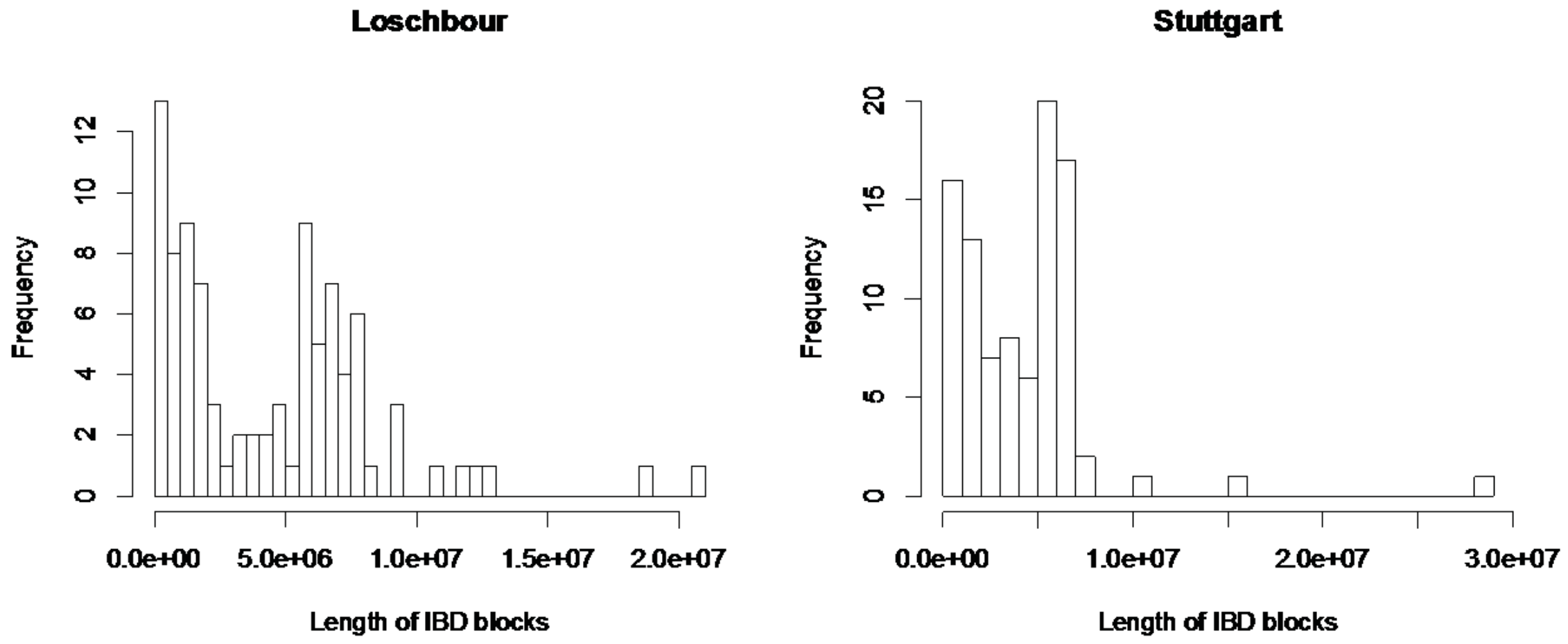

***Table S18.1. The 10 populations that share the most IBD blocks with Loschbour and Stuttgart.***

| Loschbour | | Stuttgart | |
|---|---|---|---|
| **Present-day Population** | **Mean number of shared IBD blocks** | **Present-day Population** | **Mean number of shared IBD blocks** |
| Denmark | 9 | Sardinian | 12 |
| European immigrants to North America | 9 | Slovakia | 7 |
| Finland | 8 | European immigrants to Zimbabwe | 6 |
| Ukraine | 7.5 | Macedonia | 5.5 |
| European immigrants to South Africa | 7.5 | Slovenia | 5.5 |
| French | 7 | Bulgaria | 5 |
| Sweden | 6.6 | Ukraine | 5 |
| Scotland | 6.6 | Latvia | 5 |
| Russia | 6.2 | Cyprus | 5 |
| Latvia | 6 | Swiss-Italian | 4.8 |

*Note: For each modern population listed, we report the average number of IBD blocks per individual*

***Table S18.2. The 10 populations that share the most IBD length with Loschbour and Stuttgart.***

| Loschbour | | Stuttgart | |
|---|---|---|---|
| **Present-day Population** | **Mean IBD sharing (in base pairs)** | **Present-day Population** | **Mean IBD sharing (in base pairs)** |
| European immigrants to Zimbabwe | $2.06 \times 10^7$ | Sardinian | $1.57 \times 10^7$ |
| French | $1.9 \times 10^7$ | Slovakia | $1.04 \times 10^7$ |
| European immigrants to North America | $1.25 \times 10^7$ | Macedonia | $7.67 \times 10^6$ |
| Denmark | $1.16 \times 10^7$ | Kosovo | $7.51 \times 10^6$ |
| Scotland | $1.07 \times 10^7$ | Austria | $6.87 \times 10^6$ |
| Lebanon | $9.18 \times 10^6$ | Serbia | $6.76 \times 10^6$ |
| European immigrants to South Africa | $9.03 \times 10^6$ | Bosnia-Herzegovina | $6.61 \times 10^6$ |
| Ukraine | $9.01 \times 10^6$ | Portugal | $6.56 \times 10^6$ |
| Netherlands | $8.29 \times 10^6$ | Finland | $6.52 \times 10^6$ |
| Sweden | $7.97 \times 10^6$ | Swiss-Italian | $6.40 \times 10^6$ |

*Note: For each modern population listed, we report the average length of IBD blocks shared per individua*

Whatever the explanation for the detected segments of shared IBD, we explored whether the ordering of populations based on the inferred IBD segments mirrored the genetic relationships we inferred from other aspects of the data. We observe areas of notable concordance.

- Evidence for deep relatedness of Loschbour and Stuttgart. The patterns of IBD sharing of Loschbour and Stuttgart to other world populations are positively correlated (Figures S18.3, S18.4). This is consistent with these two populations being deeply related so that they have correlated levels of shared IBD to non-West Eurasian populations (e.g. Africans or eastern non-Africans). Loschbour shares a slightly higher number of IBD tracts with the present-day populations that happen to be in the POPRES dataset than does Stuttgart (3.18 vs. 3.01, respectively). In addition, slightly more of Loschbour's genome can be found in IBD tracts ($4.45x10^6$ bp vs. $4.13x10^6$ bp, respectively).

- Evidence that Loschbour is genetically closer to northern Europeans and that Stuttgart is genetically closer to southern Europeans. The top 10 populations in terms of IBD sharing with Loschbour tend to be in northern Europe or migrants from northern Europe. The top 10 populations in terms of IBD sharing with Stuttgart tend to be in southern Europe or migrants from southern Europe. These patterns are consistent with relatively higher proportions of WHG ancestry in both Loschbour and northern European populations, and higher proportions of EEF ancestry in both Stuttgart and southern European populations. This is further visible in the fact that Sardinians stand out as a strong outlier, sharing significantly more IBD tracts with Stuttgart than Loschbour, consistent with the evidence that Sardinians are among the modern European populations most closely related to the EEF.

***Figure S18.3. IBD blocks shared by Loschbour and Stuttgart are correlated.*** *Each point corresponds to a present-day population, plotted according to one of two IBD estimates. (a) We plot each samples' average sharing with Loschbour (x-axis) and Stuttgart (y-axis); Spearman rank correlation = 0.59, slope of best fit line = 0.39. (b) We plot each sample's length of sharing with Loschbour (x-axis) and Stuttgart (y-axis). Spearman rank correlation = 0.55, slope of best fit = 0.27.*

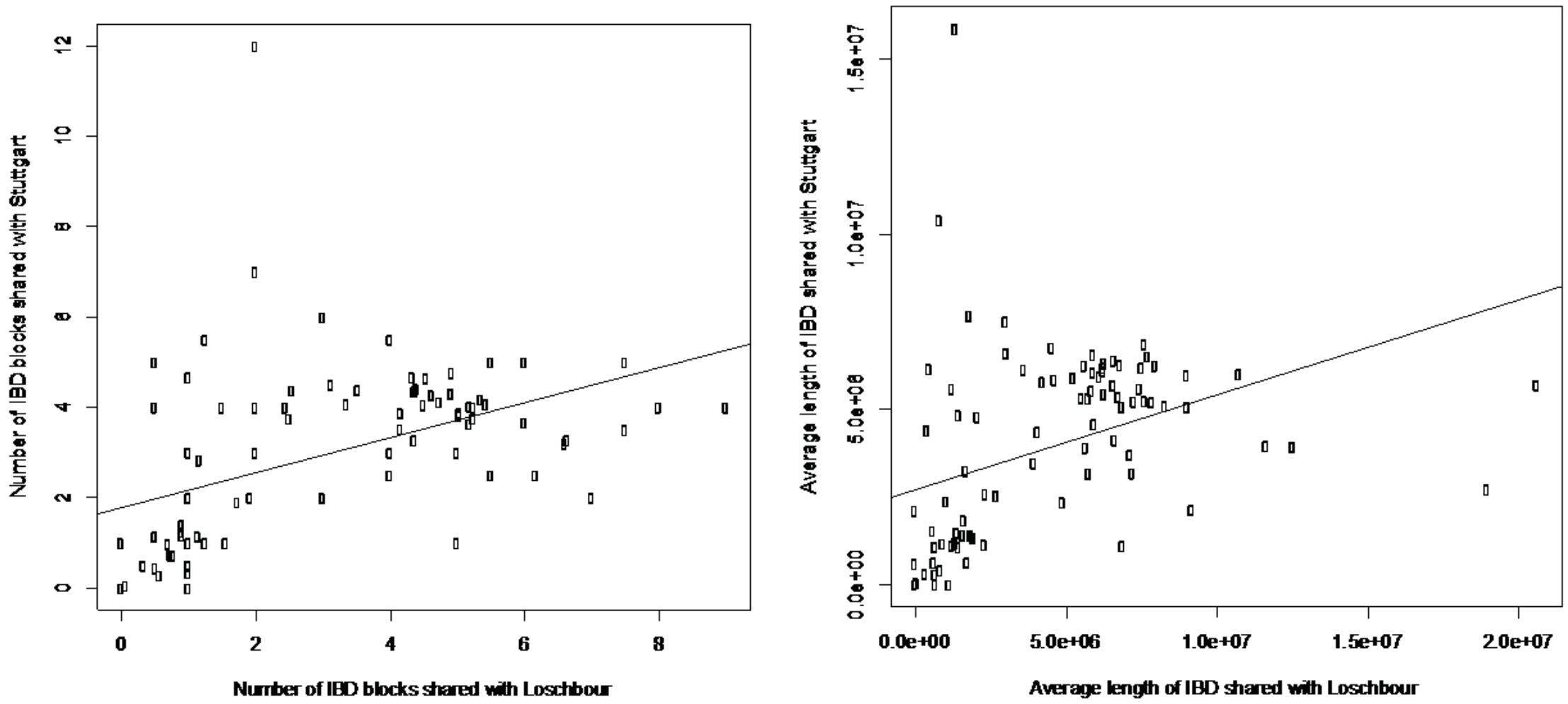

# Supplementary Information 19

**ChromoPainter/fineSTRUCTURE analysis**

Iosif Lazaridis* and David Reich

* To whom correspondence should be addressed (lazaridis@genetics.med.harvard.edu)

We used ChromoPainter[1] to study the population structure of West Eurasian populations in the Human Origins dataset, also including the high quality ancient diploid individuals (Loschbour and Stuttgart) sequenced in this study.

ChromoPainter requires an initial phasing/imputation step. To avoid imputing alleles into sites for which data was missing in the ancient genomes, we restricted analysis to a set of 495,357 sites which were complete in all 779 included samples (777 modern West Eurasians of Fig. 1B, Loschbour, and Stuttgart). We phased the data using BEAGLE 4[2] with parameters *phase-its*=50 and *impute-its*=10. We phased each of chromosomes 1-22 separately, and then combined results using ChromoCombine[1].

ChromoPainter estimates the number of "chunks" of ancestry inherited by a population from a "donor" population, reporting all pairwise choices of donor and recipient. In Fig. S19.1 we plot the number of chunks of ancestry inherited by present-day West Eurasian populations from Loschbour and Stuttgart as donor populations. The Figure is broadly reminiscent of Fig. 1B in that it shows two parallel European and Near Eastern clines bridged partially by a number of Mediterranean and Jewish populations. European populations have an excess of chunks donated by Loschbour compared to Near Eastern populations, and they vary in the number of chunks donated by Stuttgart, with Sardinians being a clear outlier showing an excess of chunks donated by Stuttgart. This is consistent with Fig. 1B which identifies them as the population most closely clustering with Early European Farmers.

***Figure S19.1: Chunks donated by Stuttgart and Loschbour to present-day West Eurasians***

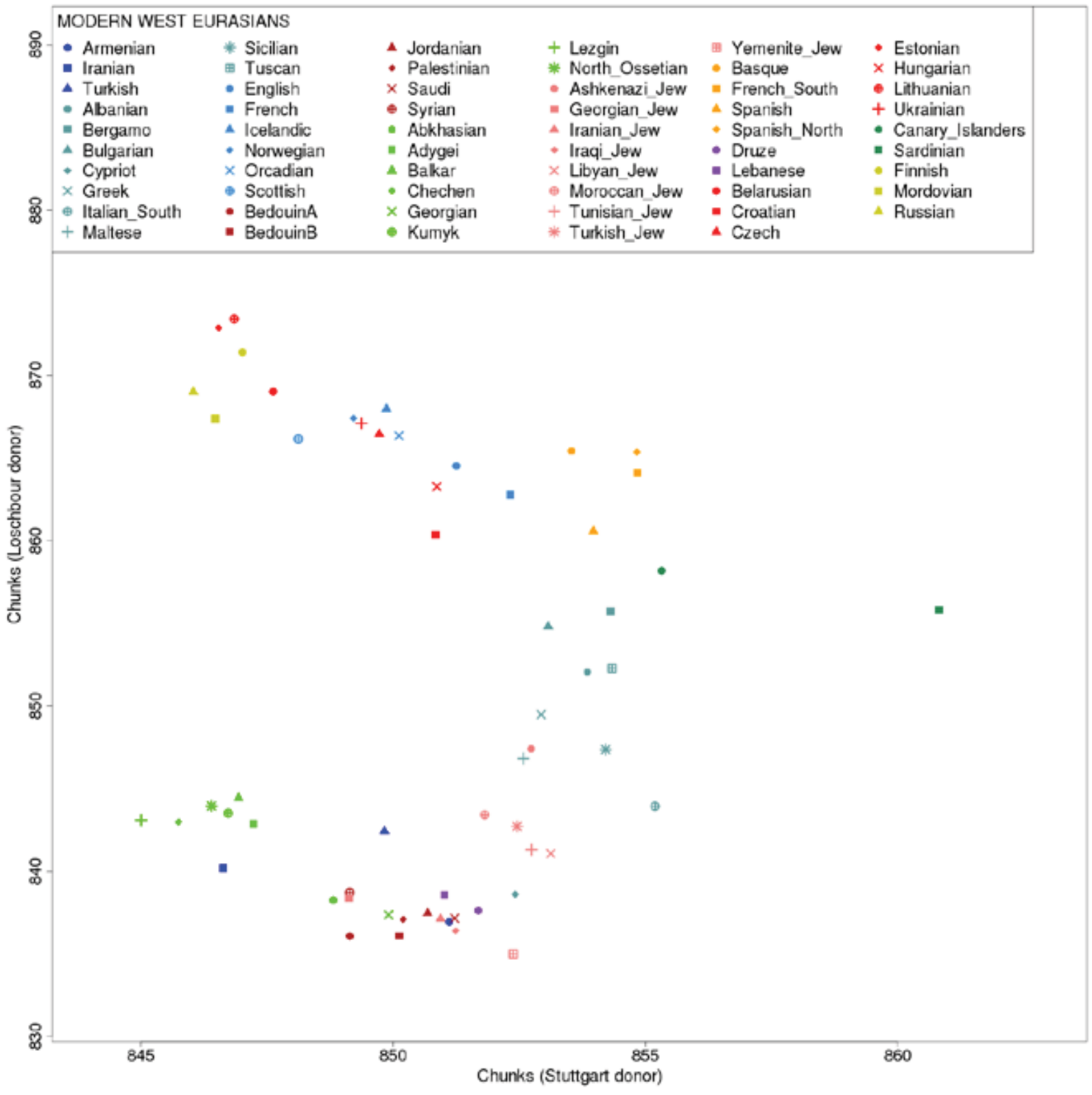

This analysis is qualitatively consistent with the IBD analysis of SI 18 which is performed on a different dataset. Both analyses would benefit from a high quality genome of an individual harboring substantial Ancient North Eurasian ancestry from a more recent period; this would allow the study of haplotype sharing between such an individual and present-day West Eurasian populations.

We also processed the ChromoPainter/ChromoCombine output with fineSTRUCTURE[1] using 250,000 burnin and 2,500,000 runtime MCMC iterations. Fig. S19.2 shows a Principal Components Analysis by fineSTRUCTURE which strongly resembles that of Fig. 1B.

***Figure S19.2: Principal Components Analysis***

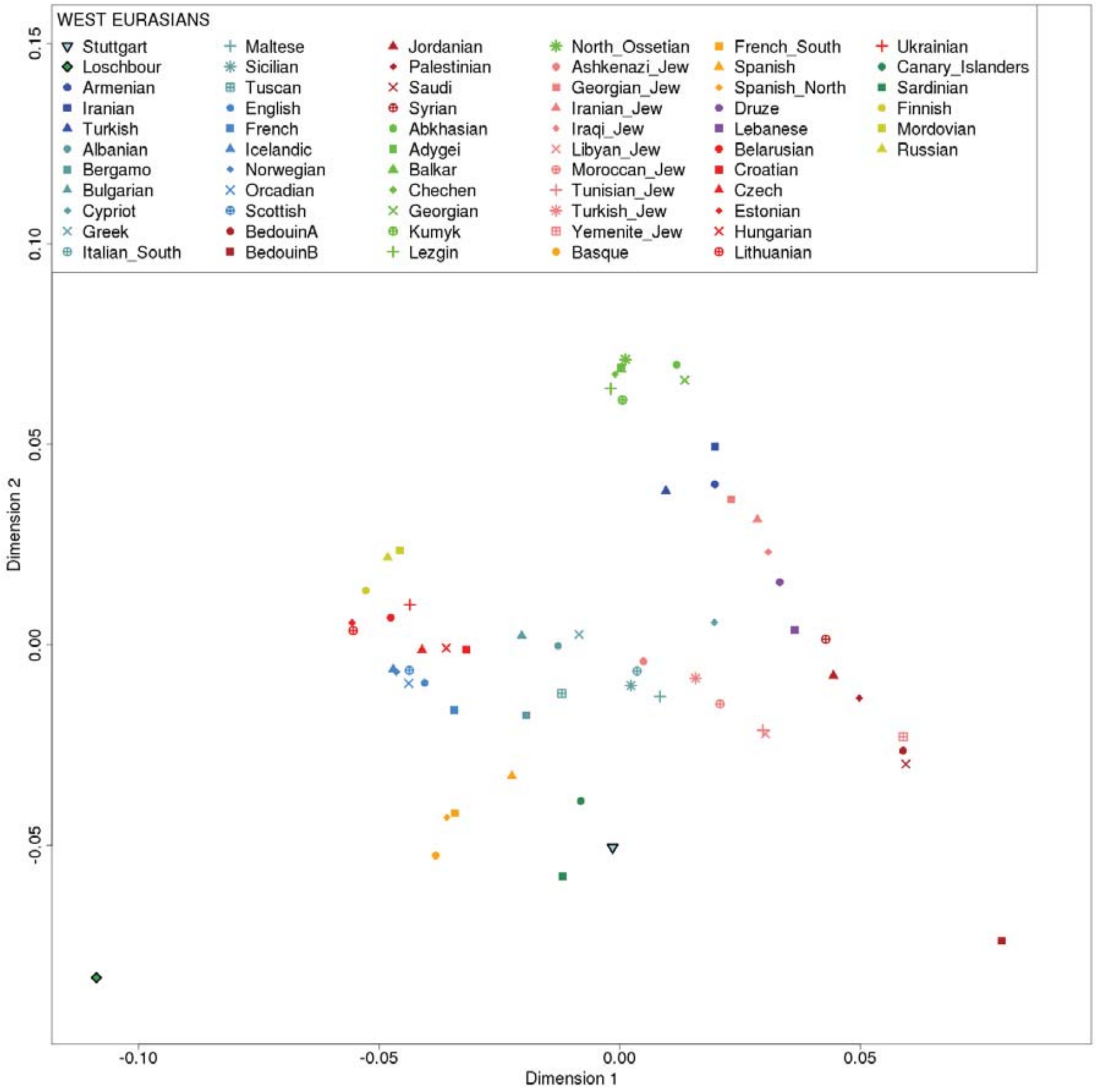

The co-ancestry matrix (Fig. S19.3) confirms the ability of this method to meaningfully cluster individuals. We highlight two clusters: Stuttgart joins all Sardinian individuals in cluster A and Loschbour joins a cluster B that encompasses all Belarusian, Ukrainian, Mordovian, Russian, Estonian, Finnish, and Lithuanian individuals. These results confirm Sardinians as a refuge area where ancestry related to Early European Farmers has been best preserved, and also the greater persistence of WHG-related ancestry in present-day Eastern European populations. The latter finding suggests that West European Hunter-Gatherers (so-named because of the prevalence of Loschbour and La Braña) or populations related to them have contributed to the ancestry of present-day Eastern European groups. Additional research is needed to determine the distribution of WHG-related populations in ancient Europe.

***Figure S19.3: Coancestry matrix heat map.*** *Details of the closest relatives of Stuttgart (A) and Loschbour (B) are magnified on the left.*

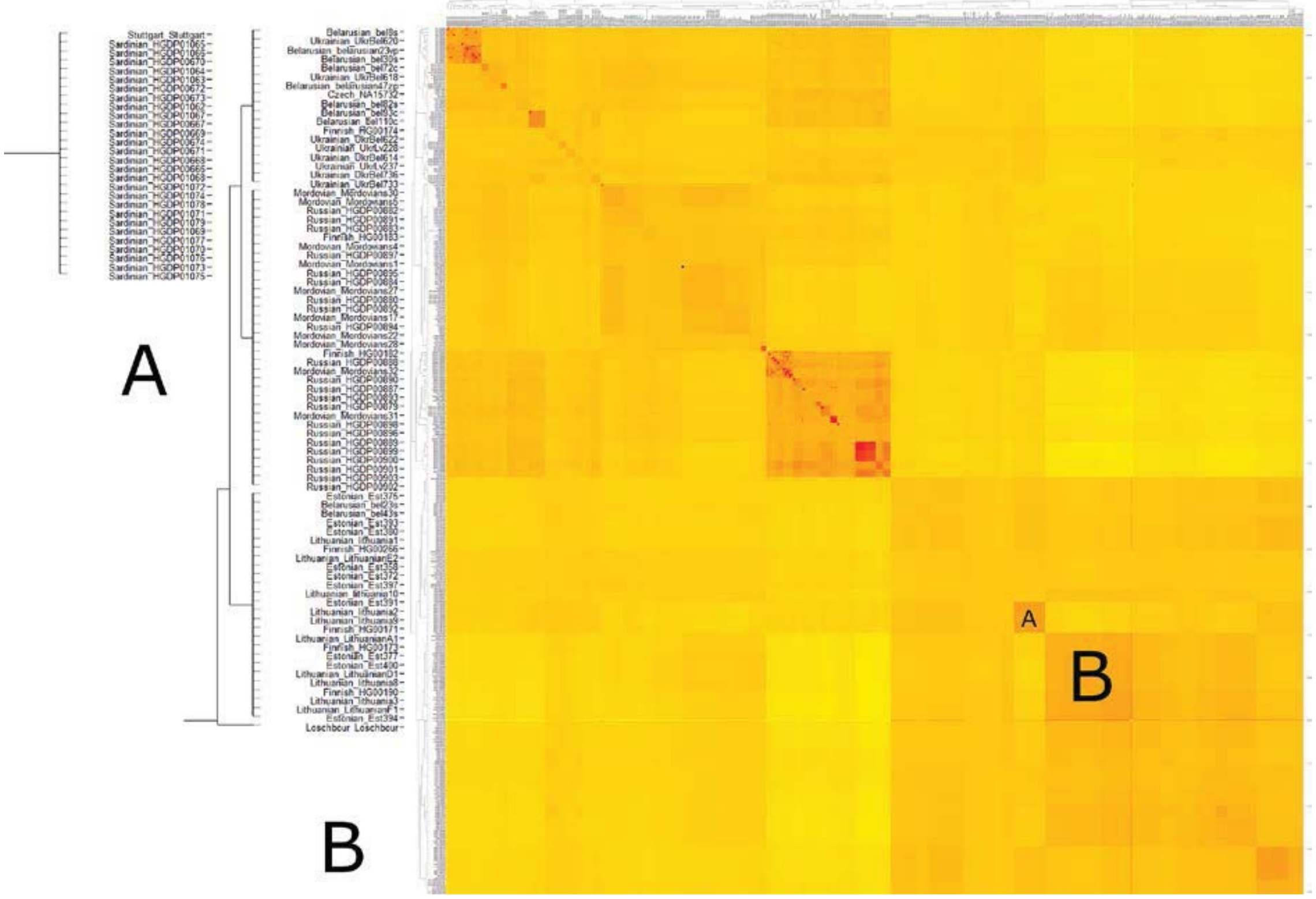